%% file: thesis_husne.tex
\documentclass[a4paper]{book}
\usepackage[titletoc]{appendix}
\usepackage{filecontents}
\usepackage{longtable}
\usepackage{textcomp}
\usepackage{gensymb}
\usepackage{rotating}
\usepackage{float}
\usepackage[linktocpage]{hyperref}
\usepackage[utf8]{inputenc}
\usepackage{color}
\usepackage{graphicx}
\usepackage{amsmath}
\usepackage{amsfonts}
\usepackage{amssymb}
\usepackage{pdfpages}
\usepackage{afterpage}
\DeclareTextSymbol{\degre}{OT1}{23}
\usepackage{geometry}
\providecommand{\U}[1]{\protect\rule{.1in}{.1in}}
\usepackage{tikz}
\usepackage{array}
\usetikzlibrary{decorations.pathmorphing,calc}
\usetikzlibrary{shapes.arrows}
\usetikzlibrary{arrows,shapes,backgrounds}
\usetikzlibrary{backgrounds,fit}
\usetikzlibrary{automata,positioning}
\usepackage{setspace}
\newcommand{\angstrom}{\textup{\AA}}
\newcolumntype{C}[1]{>{\centering\let\newline\\\arraybackslash\hspace{0pt}}m{#1}}
\newcolumntype{L}[1]{>{\raggedright\let\newline\\\arraybackslash\hspace{0pt}}m{#1}}

\begin{document}

\frontmatter

\begin{titlepage}
\begin{center}
\tikz[overlay, remember picture]\draw (0.5cm,2.5cm) node[below]{\parbox[t]{1.1\textwidth}{\center \textbf{\Large{UNIVERSITE NICE SOPHIA ANTIPOLIS  \\ European Community Erasmus Mundus Joint Doctorate \\ International Relativistic Astrophysics Ph.D. Program}}}};

\vspace{-0.5cm}
\begin{tikzpicture}[overlay]
  \node (myfirstpic) at (0,-3) {\includegraphics[width=8cm]{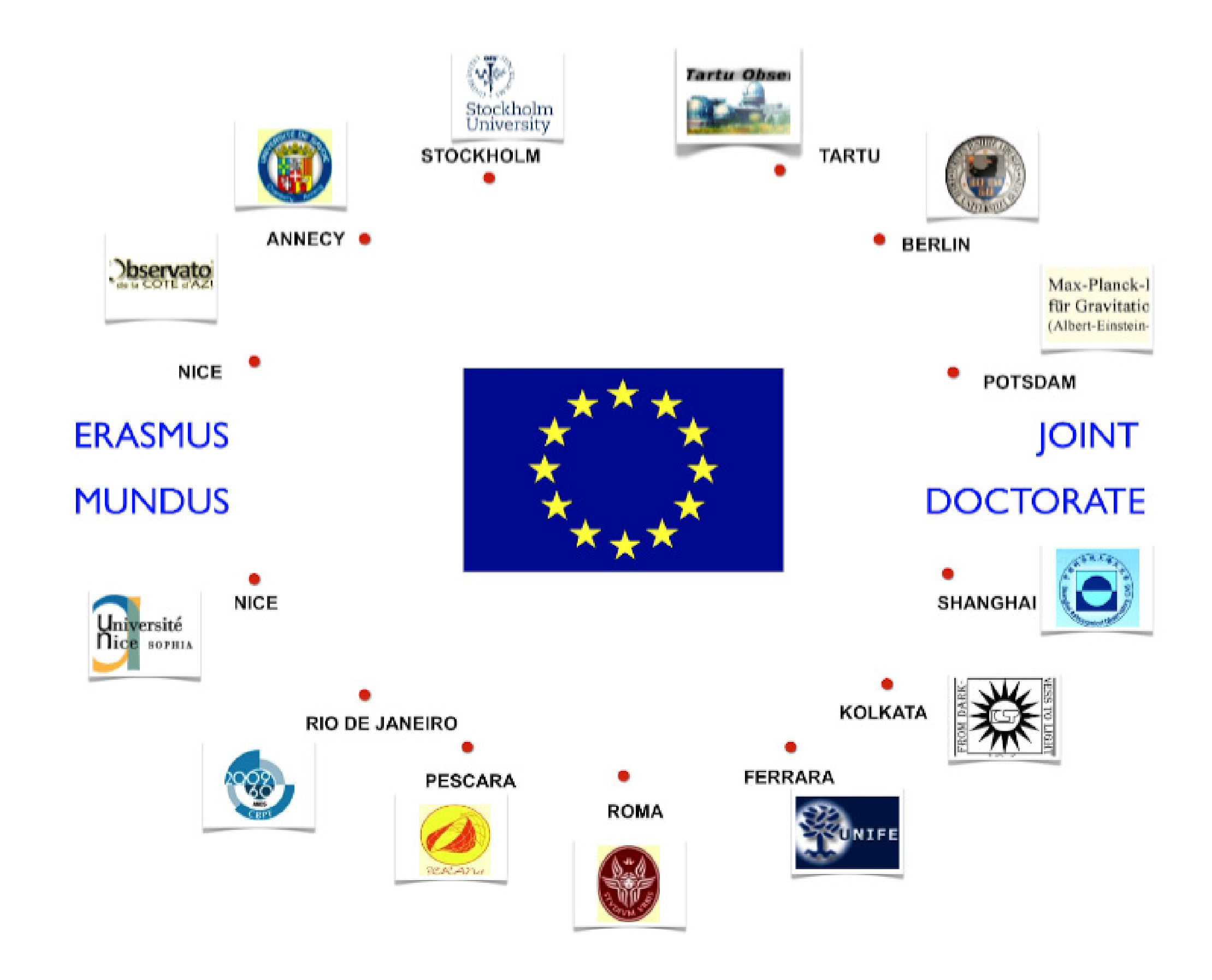}};
\end{tikzpicture}

\vspace{7cm}
\textbf{\Huge{ {Study of a Population of Gamma-ray Bursts with Low-Luminosity Afterglows} }}

\vspace{1cm}
~
~

\Large{Presented and defended by\\}
\vskip 0.3cm
\textbf{\huge{H\"usne DERELI}}
\\
\vskip 0.1cm
\Large{Erasmus Mundus Fellow}

\vspace{0.5cm}

\Large{Defended on \\ the 16$^{\text{th}}$ of December 2014}

\vspace{-0.3cm}
\end{center}
\vskip 0.5cm
\vspace{0.8cm}

\vspace{0.2cm}
\tikz[overlay, remember picture]\draw (8cm,-0.5cm) node[below]{\parbox[t]{1.1\textwidth}{\Large{
\textbf{\Large{Commission:}}
~
\center
\begin{tabular}{lll}
Aysun Aky\"uz & Professor & Reporter\\
Lorenzo Amati & Researcher, INAF & Examiner\\
Jean-Luc Atteia & Astronomer & Reporter\\
Michel Bo\"er & Director of Research, CNRS ~~~~ & Supervisor \\
Pascal Chardonnet & Professor & Examiner \\
Massimo Della Valle ~~~& Professor & Co-supervisor\\
Bruce Gendre & Visiting Professor & Examiner \\
\end{tabular}
}}};

\end{titlepage}

\newpage

\chapter{Summary}

Gamma-ray bursts (GRB) are extreme events taking place at cosmological distances. Their origin and mechanism have been puzzling for decades. They are crudely classified into two groups based on their duration, namely the short bursts and the long bursts.

Such a classification has proven to be extremely useful to determine their possible progenitors: the merger of two compact objects for short bursts and the explosion of a (very) massive star for long bursts. Further classifying the long GRBs might give tighter constraints on their progenitor (initial mass, angular momentum, evolution stage at collapse) and on the emission mechanism(s).

The understanding of several aspects of GRBs has greatly advanced after the launch of the \text{Swift} satellite, as it allows for multi-wavelength observation of both the prompt phase and the afterglow of GRBs. On the other hand, a world collaboration to point ground optical and radio telescopes has allowed many breakthroughs in the physics of GRBs, for instance with several detections of supernova in the late afterglow phase.

In my thesis, I present evidence for the existence of a sub-class of long GRBs, based on their faint afterglow emission. These bursts were named low-luminosity afterglow (LLA) GRBs. I discuss the data analysis and the selection method of these bursts. Then, their main properties are described (prompt and afterglow). Their link to supernova is strong as 64\% of all the bursts firmly associated to SNe are LLA GRBs. This motivated the study of supernovae in my thesis.

Finally, I present additional properties of LLA GRBs: the study of their rate density, which seems to indicate a new distinct third class of events, the properties of their host galaxies, which show that they take place in young star-forming galaxies, not different from those of normal long GRBs.

Additionally, I show that it is difficult to reconcile all differences between normal long GRBs and LLA GRBs only by considering instrumental or environmental effects, a different ejecta content or a different geometry for the burst. Thus, I conclude that LLA GRBs and normal long GRBs should have different properties.

In a very rudimentary discussion of the possible progenitor, I indicate that a binary system is favored in the case of LLA GRB. The argument is based on the initial mass function of massive stars, on the larger rate density of LLA GRBs compared to the rate of normal long GRBs and on the type of accompanying SNe.

Such a classification of GRBs is important to constrain their emission mechanisms and possible progenitors, which are still highly debated. However, more multi-wavelength observations of weak bursts at small redshift are required to give tighter constraints on the properties of both the burst and its accompanying supernova if present.

\chapter{Acknowledgements}

\input{acknowledgement.tex}
\tableofcontents
\listoffigures
\listoftables

\mainmatter
\input{Chapter1}
\input{Chapter2}

\input{Chapter3}
\input{Chapter4} 
\input{Chapter5}

\input{Chapter6}

\bibliographystyle{unsrt}
\bibliography{thesis_husne.bib}

\begin{appendices}
\input{appendix1.tex}
\input{appendix2.tex}
\input{appendix3.tex}
\end{appendices}

\end{document}

%% file: acknowledgement.tex
I would like to thank many people for their help and support while I was making this work.

 My first thanks will go to the European Commission for the support through the Erasmus Mundus Joint Doctorate Program, to the program coordinator Prof. Pascal Chardonnett and to Dr. Catherine Nary Man director of ARTEMIS laboratory and Prof. Farrokh Vakili director of C\^ote d'Azur Observatory, as well as to the secretaries of ARTEMIS, Seynabou Ndiaye and David Andrieux for providing all comfortable conditions for my thesis study.
 
~ 

Secondly, I would like to thank the director of the Astronomical Observatory of Capodimonte Prof. Massimo Della Valle, also my co-supervisor who provided me good conditions during my mobility period. I must also thank several persons who contributed directly or indirectly for my studies during that period. My first gratitude will go to Dr. Maria Teresa Botticella, for her understanding, patience, help, effort as well as encouragements. My second gratitude will be for Dr. Massimo Dall'ora, Dr. Stefano Valenti, Dr. Andrea Pastorello, Prof. Stefano Benetti, Dr. Luc Desart, Dr. Eda Sonba\c{s}, Dr. Korhan Yelkenci, Dr. Sinan Alis for their help and helpful comments for my studies on the Supernovae topic. Lastly, I would like to thank Cristina Barbarino for her friendship.

~

I also would like to thank to Prof. Remo Ruffini, director of ICRANet, for his encouragements. I must also thank Dr. Luca Izzo for his collaboration during that time. And I should not forget to thank all my colleagues in the Erasmus Mundus Joint Doctorate and IRAP Programs, especially Liang Li and Yu Wang for their collaboration. A special thank you will go for two very special persons, Jonas Pedro Pereira and Onelda Bardho for their warm and helpful friendships in any condition. 

~

I would like to express my gratitude to my supervisor, Dr. Michel Bo\"er, whose expertise and understanding added considerably to my thesis study. I am grateful for his guidance throughout my study. My special thanks are also to Dr. Bruce Gendre for his collaboration, whose expertise, advices, valuable comments, also added to my Ph.D. experience. I also would like to thank my colleagues of the GRB group, Dr. Tania Regimbau, Onelda Bardo, Karelle Siellez and Duncan Meacher for useful discussions. Additionally, special thanks will go to Dr. Alian Klotz and Dr. Lorenzo Amati for their collaboration. Finally, I wish to thank my friends students at the C\^ote d'Azur Observatory for their warm friendships.    
 
~

Finally, my special gratitude will be for Damien B\'egu\'e, my husband, whose expertise, helpful comments and discussions improved my knowledge during my study and whose love, understanding, patience supported me to finish my Ph.D. I also would like to thank his parents for their love and support. My last special thanks will go to each member of my family for their love and specially to my sister Sunduz \c{C}i\c{c}ek and my best friend Sevin\c{c} Mantar for their endless support.   

~

I must thank Jean-Louis Sougn\'e for proofreading my thesis. Finally, I acknowledge my thesis reporters Prof. Dr. Aysun Aky\"uz and Dr. Jean-Luc Atteia for their helpful comments. 

~

H\"usne Dereli is supported by the Erasmus Mundus Joint Doctorate Program by Grant Number 2011-1640 from the EACEA of the European Commission. She completed her Ph.D. thesis in the laboratory of ARTEMIS directed by Nary Man Catherine which is located in the OCA, C\^ote d'Azur Observatory directed by Farrokh Vakili.

%% file: Chapter1.tex
\chapter[Introduction]{\parbox[t]{\textwidth}{Introduction}}
\chaptermark{Introduction}
\section{Diversity of GRBs Progenitors}

\begin{figure}[!ht]
\begin{center}
\includegraphics[width=7cm]{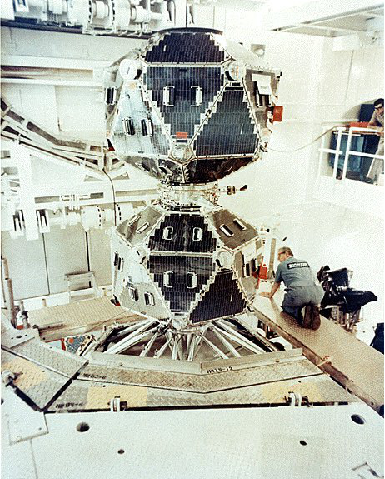}
\caption{The Vela 5b satellite before its launch. Image Credit: NASA
\label{vela5b_2}}
\end{center}
\end{figure}

Gamma-ray bursts (GRBs) are extraordinary events appearing randomly in the sky. They were discovered in 1967 \cite{klebesadel1973}, by the US Vela military satellites which were designed to monitor nuclear tests in the atmosphere and in the outer space after the Limited Nuclear Test Ban Treaty signed between the USSR and the USA. Within three years, 16 events were recorded by the satellites Vela 5 and 6 (see Figure \ref{vela5b_2}). However, it was clear very quickly that these events originated from the sky and not from the Earth: since there were more than one satellite, it was possible to measure the time delay between two satellites to get the direction of the events. 

A GRB appears as a sudden burst of high energy photons (keV - GeV), lasting from a few tens of milliseconds up to several minutes. Thanks to the results of BATSE \cite{paciesas1999} it is known that they are distributed isotropically in the sky (see Figure \ref{BATSE_all_sky}) which is an indication of their cosmological origin, later confirmed by redshift measurements. Since they are located at cosmological distances \cite{costa1997, vanParadijs1997}, the recorded flux at the Earth implies that they are powerful emitters and release an enormous isotropic equivalent energy ranging from $10^{48}$ to $10^{54}$ ergs. The variability observed in the radio emission (hours to weeks after the prompt emission) shows that the emission region has a stellar size in the order of 10$^{17}$~cm  \cite{frail1997}, see Figure~9 of \cite{piran2004}. When these two pieces of information are combined, they point towards a phenomenon running extreme physics near a compact object (stellar mass black hole or young magnetar).

For a very long time, the amount of energy needed for this phenomenon puzzled the theoreticians, and several models were built without considering it. The models were operating a central (compact) engine and the questions were about how to provide the energy. Nowadays, several objects are supposed to be able to trigger a gamma-ray burst.

\begin{figure}[!ht]
\begin{center}
\includegraphics[width=9cm]{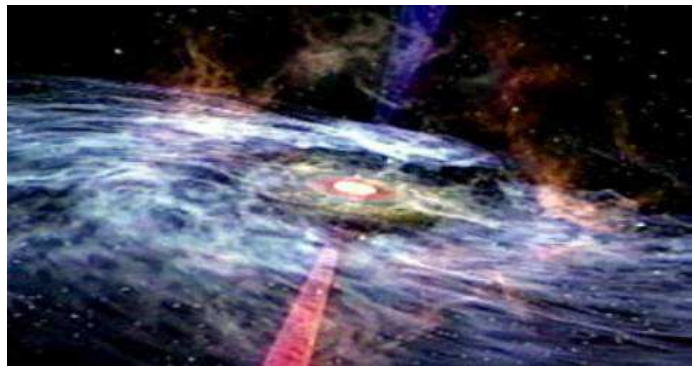}
\caption{Artist close view of a GRB, with the remaining of the progenitor being accreted by the central engine. Image Credit: NASA.
\label{artist_GRB}}
\end{center}
\end{figure}

Massive stars (for example Wolf-Rayet stars) are the first possibility. Indeed, they can collapse to form a black hole. The remaining of the star is accreted by a newly born black hole which powers the GRB \cite{woosley1993} (see Figure \ref{artist_GRB}).

Another possible progenitor is a binary system formed of two compact objects (Neutron Star (NS)-NS, Black Hole (BH)-NS, BH-BH), merging together as a result of orbital angular momentum lost when gravitational wave radiation is emitted \cite{eichler1989}.

The last possibility is a Magnetar which is a neutron star with a large magnetic field and a high rotation speed. The rotational energy is extracted by the rotating magnetic field, slowing down the neutron star. As a result it can eventually collapse into a black hole \cite{usov1992}.

All these objects are formed by the death of stars: GRBs seem to be linked to the final act of the stellar evolution. This idea is even strengthened by the observation of a GRB-Supernova (SN) association. Other important observations link GRBs and massive stars in the place of birth of long GRBs. 

\section{The Last Stages of Stellar Evolution}

\subsection{How do stars die?}
The first stars were created 13.8 billion years ago, shortly after the Big Bang \cite{salvaterra2009, tanvir2009}. The lifetime of the most massive star is only a few million years (for example, it is 3 millions years for a 60~$M_{\odot}$), while it can be up to trillions of years ($10^{12}$) for the least massive stars \cite{bertulani2013}. The star evolution towards its end depends on the physical parameters of star: initial mass, metallicity, mass loss rate, rotation speed, etc. Below, the fate of stars is presented as a function of increasing initial mass.
 
 Brown and red dwarfs represent extremely low massive ($0.072 M_{\odot} \leq M \leq 0.8 M_{\odot}$) stars, which are mainly observed in binary systems \cite{lane2001, liu2010}. They are thought to be formed by a collapsing cloud of gas and dust like \textit{normal} stars, and they begin to burn their hydrogen. However, in the case of brown dwarfs, their surface temperature and luminosity never stabilize since their mass is not big enough to maintain thermonuclear fusion. Thus, they cool down as they age. \cite{burrows1993}. On the other hand, for red dwarfs, helium (He) is produced and constantly remixed with hydrogen (H) in the whole volume of the star, thus avoiding the creation of a He core. They develop slowly for a long time, expected to be much larger than the age of universe. Thus, advanced red dwarfs are not observed.

\begin{figure}[!ht]
\begin{center}
\includegraphics[width=7cm]{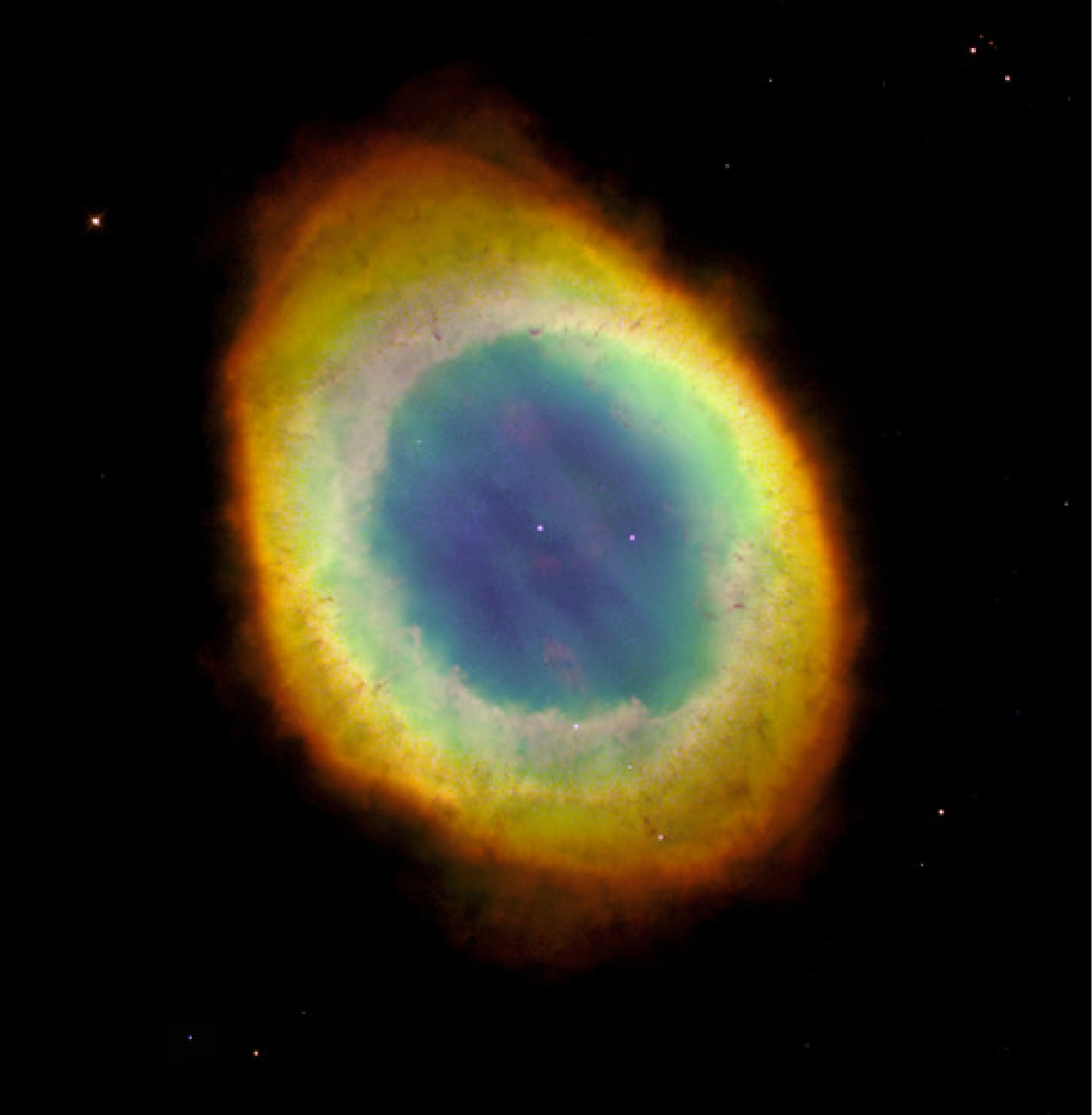}
\caption{ The Ring Nebula in Lyra, of size about 1.3 light year (to be compared to $7.3 \times 10^{-8}$ light year for the size of the Sun), is a giant shell of gas surrounding a central star. Image Credit: NASA/Hubble Heritage \label{Lyra_nabula}}
\end{center}
\end{figure}

\begin{figure}
\begin{center}
\begin{tikzpicture}[scale=0.8]
\draw (7cm,0cm) node[above,opacity=0.95]{\includegraphics[width=6.4cm]{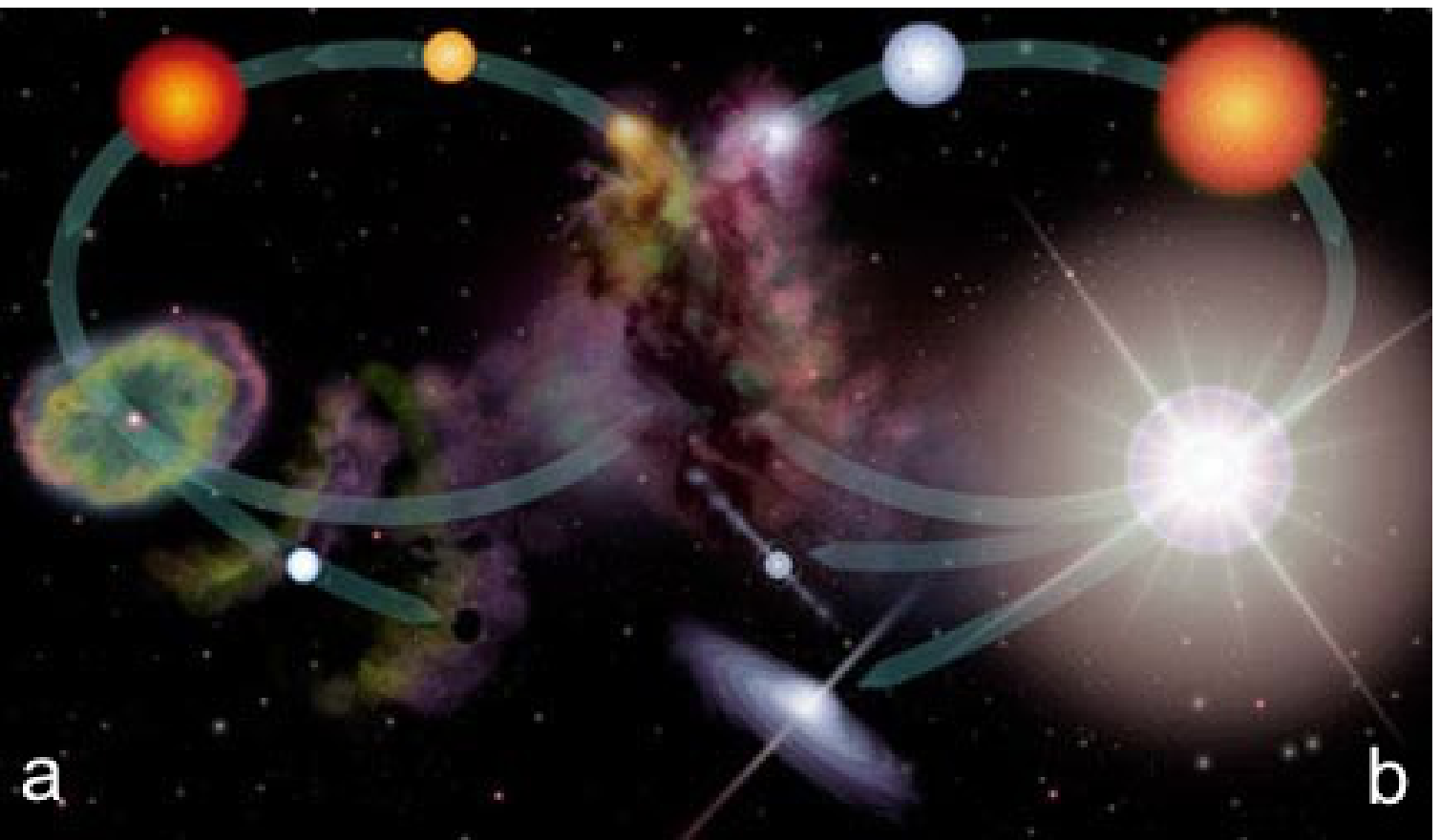}};
\path (1cm,4.5cm) coordinate (leftline1);
\path (7,3.5) coordinate (nebula);
\path (7,5) coordinate (nebulap);
\path (3,5) coordinate (nebulap1);
\path (4,4.3) coordinate (redSPGlow);
\path (1cm,3.5cm) coordinate (leftline2);
\path (3.7,2.5) coordinate (planetaryneb);
\path (1cm,2.5cm) coordinate (leftline3);
\path (4.6,1.65) coordinate (whiteD);
\path (1cm,1.5cm) coordinate (leftline4);
\path (5.55,1.32) coordinate (blackD);
\path (1cm,0.5cm) coordinate (leftline5);
\path (10.1,4.2) coordinate (redSPGh);
\path (12cm,4.5cm) coordinate (rightline1);
\path (9.8,2.2) coordinate (SNe);
\path (12cm,3.5cm) coordinate (rightline2);
\path (11.5cm,1.7cm) coordinate (interm1);
\path (7.4,1.7) coordinate (NS);
\path (12cm,2.5cm) coordinate (rightline3);
\path (11.5cm,1.3cm) coordinate (interm2);
\path (7.7,1) coordinate (BH);
\path (12cm,1.5cm) coordinate (rightline4);

\draw [thick, green] (nebula)--(nebulap)--(nebulap1)--(leftline1);
\draw [thick, green] (redSPGlow)--(leftline2);
\draw [thick, green] (planetaryneb)--(leftline3);
\draw [thick, green] (whiteD)--(leftline4);
\draw [thick, green] (blackD)--(leftline5);
\draw [thick, green] (redSPGh)--(rightline1);
\draw [thick, green] (SNe)--(rightline2);
\draw [thick, green] (NS)--(interm1)--(rightline3);
\draw [thick, green] (BH)--(interm2)--(rightline4);

\filldraw[green] (leftline1) circle (0.001cm) node[anchor=east] {\color{blue}{Nebula phase}};
\filldraw[green] (leftline2) circle (0.001cm) node[anchor=east] {\color{blue}{Red supergiant}};
\filldraw[green] (leftline3) circle (0.001cm) node[anchor=east] {\color{blue}{Planetary nebula}};
\filldraw[green] (leftline4) circle (0.001cm) node[anchor=east] {\color{blue}{White dwarf}};
\filldraw[green] (leftline5) circle (0.001cm) node[anchor=east] {\color{blue}{Black dwarf}};
\filldraw[green] (rightline1) circle (0.001cm) node[anchor=west] {\color{blue}{Red supergiant}};
\filldraw[green] (rightline2) circle (0.001cm) node[anchor=west] {\color{blue}{Supernova}};
\filldraw[green] (rightline3) circle (0.001cm) node[anchor=west] {\color{blue}{Neutron star}};
\filldraw[green] (rightline4) circle (0.001cm) node[anchor=west] {\color{blue}{Black hole}};
\end{tikzpicture}
\end{center}
\caption{The life cycle of a low mass star (a) and a high mass star (b). Both scenarios start from the nebula phase at the center of the figure. Depending on the initial mass, two evolution paths can be followed, which end by a black dwarf for small initial mass (towards the left) and a neutron star or a black hole for large initial mass (towards the right). Image Credit: NASA}
\label{life_cycles}
\end{figure}

The low mass stars between 0.8 and 8~$M_{\odot}$ take a few billion years to burn out their hydrogen \cite{kahn1974}. At the end of hydrogen burning, the size of the star decreases (because the radiation pressure decreases) and  the central density and temperature increase. Under this condition, helium fusion begins (see \cite{larson2003, mckee2007} for reviews) converting it to oxygen and carbon and increasing the radius of the star: this is the red supergiant phase. In the next contraction stage the density becomes so high that the size of the star stabilizes due to the electron degeneracy pressure: electrons are Fermi particles, so two of them cannot be at the same energy level, leading to a pressure-like strength at very high density. During this stage the outer layers of the star are expelled by strong stellar winds and form a planetary nebula, composed of hot gas, ionized by ultraviolet radiation from the core of the star, see one example in Figure \ref{Lyra_nabula}. At the end, a white dwarf is formed, which cools down to become a black dwarf. The life cycle of a low massive star is summarized in Figure \ref{life_cycles}(a). 
However, this kind of compact object cannot create a GRB, because the accretion onto the white dwarf, when it is forming, is very small. Thus the accretion disk cannot power a GRB and a black hole cannot be created since white dwarf cannot reach the Chandrasekhar limit (1.4~$M_{\odot}$).

\begin{figure}[!ht]
\begin{center}
\includegraphics[width=9cm]{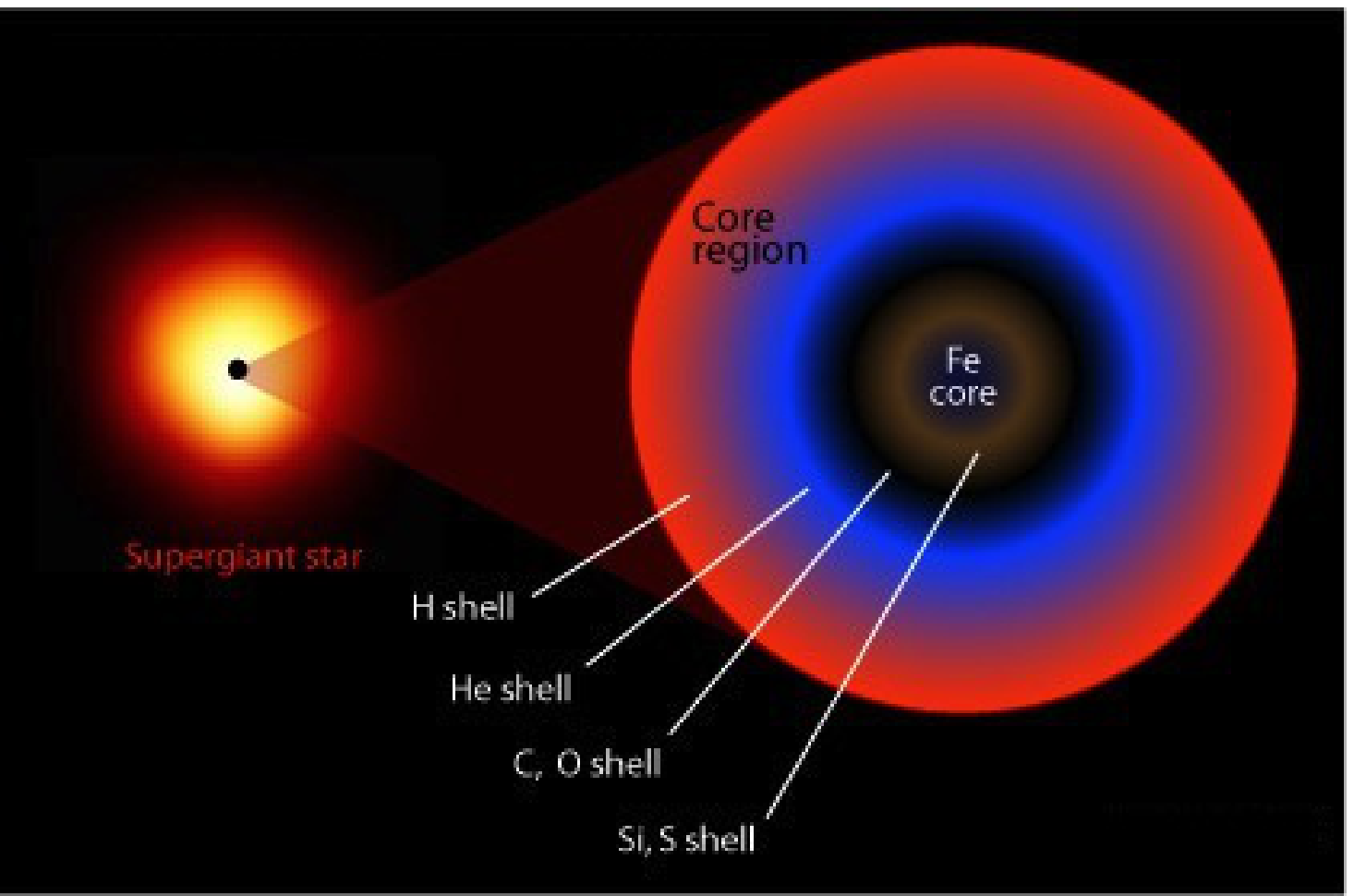}
\caption{Illustratration of the internal shell structure of a supergiant on its last days as an example of a massive star which has synthesized elements up to iron. Image Credit: CSIRO \label{element_synt}}
\end{center}
\end{figure}

The massive stars with initial mass larger than 8~$M_{\odot}$ cannot reach such a stable level: they start the CNO cycle, and they can burn all heavier elements starting with carbon and oxygen, up until the iron which is the most stable element (see Figure \ref{element_synt} for various burning stages). They acquire an \textit{onion} structure with the lighter elements composing the outer layers.

\begin{figure}[!ht]
\begin{center}
\includegraphics[width=0.8\textwidth]{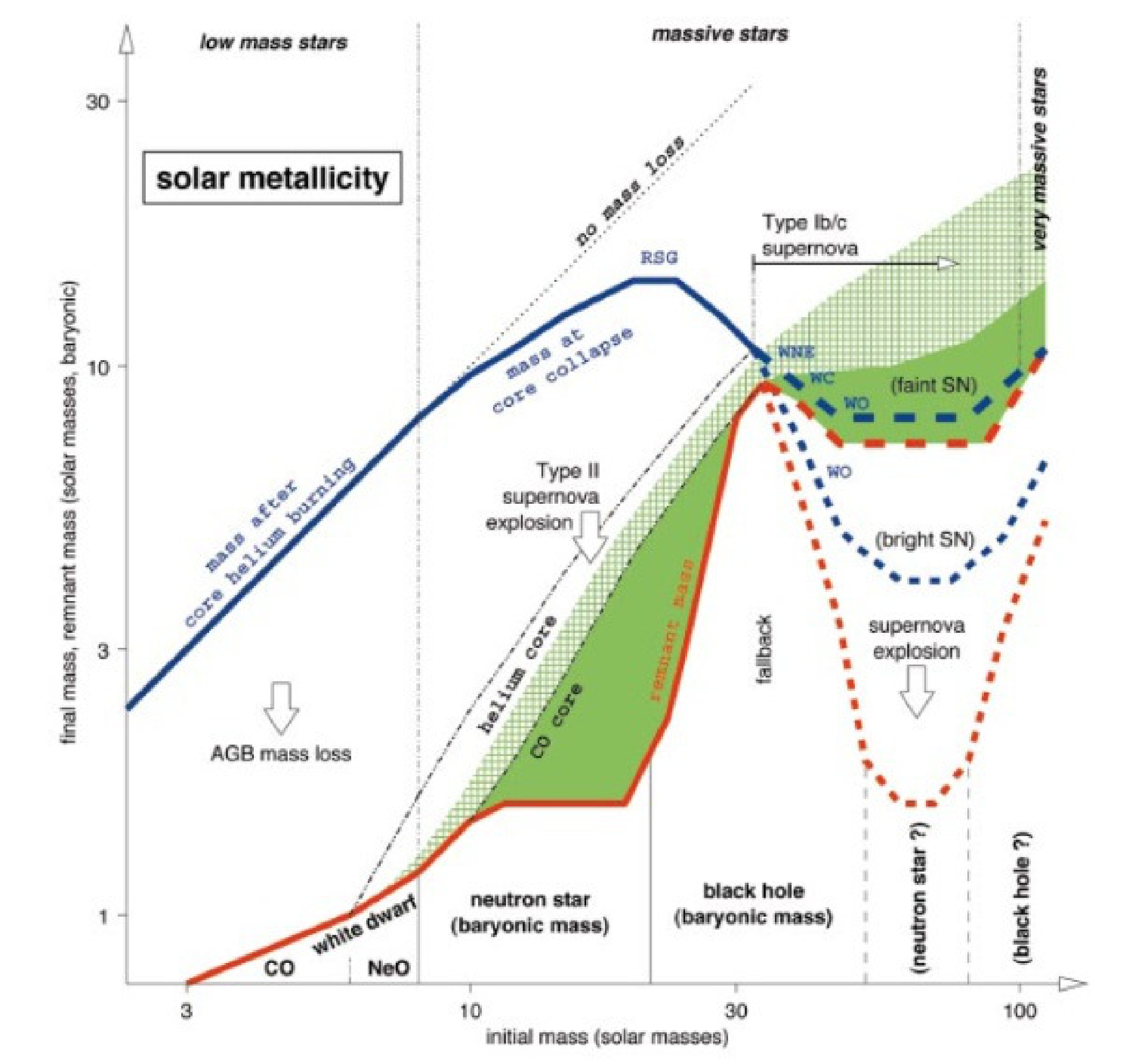}
\caption{Initial and final mass function of non-rotating stars of solar composition, from \cite{woosley2002}. The blue curve shows the mass of the star without its envelope (if lost) while the red curve shows the final mass of the remnant. Heavy elements are produced in the dark and light green region. Below 8~$M_{\odot}$ (initial mass), in the asymptotic giant branch (AGB), a star loses its envelope and leaves a CO (carbon-oxygen) or NeO (neon-oxygen) white dwarf. Above 8~$M_{\odot}$ and below 25~$M_{\odot}$, the H envelope and part of the He core are ejected. Then, He is converted to C and O. Finally, heavier elements are synthesized, leading to the creation of a neutron star, as the star explodes as a type II SN. Above 25~$M_{\odot}$ and below 33 ~$M_{\odot}$, a black hole forms either by fallback of the ejecta from the star or directly by the iron core collapse, leading to a type Ib/c SN explosion. Above 33~$M_{\odot}$, this explosion (type Ib/c) is the result of a Wolf-Rayet (W-R) star evolution. In this region, there are two possibilities. First, the low W-R mass loss rate, represented by the blue and red tick dashed lines. It can synthesize heavy elements, leading to different type of W-R stars: WNE (early nitrogen in its spectra), WC (carbon in its spectra), WO (oxygen in its spectra). The collapse of these W-R stars gives a faint SN. Second, there are high W-R mass loss rate, represented by the thin dashed lines (blue and red). In this region, W-R stars collapse and form a bright SN. The final remnant is a black hole. 
\label{nonrotating_stars}}
\end{center}
\end{figure}

When the iron core reaches its Chandrasekhar mass, electron degeneracy pressure cannot balance the gravitational pull. The core collapses, and in the meantime, the iron is photo-disintegrated into free nucleons and alpha particles. Finally, the collapse stops as the central density reaches 10$^{14}$ g.cm$^{-3}$, which allows for neutron degeneracy pressure, which is similar to electron degeneracy pressure. At the end, a proto-neutron star is formed, of dimensions in the order of some tens of kilometers. The sudden deceleration launches a shock wave 20 km from the center which travels upstream through the in-falling matter left of the iron core, which is accreted onto the proto-neutron star after being shocked. The shock loses energy as it fights its way out through the inflow, but neutrinos emitted by the cooling of the proto-neutron star pushes it on. For a few seconds, the proto-neutron star emits an enormous amount of energy mostly in neutrino and gravitational waves. A tiny fraction of this energy (approximately 10$^{51}$ ergs) is sufficient to make the star explode as a supernova (SN) by the shock wave which propogates through the outer layer of the star at thousands of kilometers per second (\textit{e.g.} \cite{woosley2002} for a review).
Below initial masses of $\sim$ 25~$M_{\odot}$, neutron stars are formed. Above that, black holes form, either in a delayed manner by fallback of the ejected matter or directly during the iron-core collapse (above $\sim$ 40~$M_{\odot}$) as it can be seen in Figure \ref{nonrotating_stars} \cite{woosley2002}. The life cycle of a very massive star is summarized in Figure \ref{life_cycles}(b).

However, the end of extremely massive stars is different. For example, above 100~$M_{\odot}$, stars suffer electron-positron pair instability after carbon burning. This begins as a pulsational instability of the helium cores of $\sim$ 40~$M_{\odot}$. As its mass increases, the pulsations become more violent, ejecting any remaining hydrogen envelope and an increasing fraction of the helium core itself. An iron core can still form in hydrostatic equilibrium in such stars, but it collapses to a black hole. The pair-instability supernovae can produce all elements from oxygen to iron through nickel.
Depending on the initial mass function, the creation of the Ni$^{56}$ isotope might be possible. Indeed, it requires the synthesis of at least five solar mass of Nickel (Ni$^{58}$), which is produced by the silicon burning process \cite{woosley2002}. All of these massive stars (8~$M_{\odot}$ $\le$ $M$ $\le$ 260~$M_{\odot}$) are thought to produce a cataclysmic event at the end of their life, which can be classified according to their observed properties. In brief, deaths of non-rotating massive stars is summarized in Table \ref{massive_stars} \cite{woosley2002}. These discussions are summarized on Figure \ref{nonrotating_stars} \cite{woosley2002} which also displays the type of the corresponding expected SNe.

\begin{table}
\centering
  \caption{Deaths of non rotating massive helium stars \label{massive_stars} \cite{woosley2002}}
  \begin{tabular}{ccC{0.3\textwidth}}
  \hline
   Main Sequence Mass & He Core & Supernovae Mechanism \\
    \hline
8~$M_{\odot}$ $\le$ $M$ $\le$ 95~$M_{\odot}$ & 2.2~$M_{\odot}$ $\le$ $M$ $\le$ 40~$M_{\odot}$ & Fe core collapse to neutron star or a black hole \\
95~$M_{\odot}$ $\le$ $M$ $\le$ 130~$M_{\odot}$ & 40~$M_{\odot}$ $\le$ $M$ $\le$ 60~$M_{\odot}$ & Pulsational pair instablity followed by Fe core collapse \\
130~$M_{\odot}$ $\le$ $M$ $\le$ 260~$M_{\odot}$ & 60~$M_{\odot}$ $\le$ $M$ $\le$ 137~$M_{\odot}$ & Pair instability supernovae \\
\hline
$M$ $\ge$ 260~$M_{\odot}$ & $M$ $\ge$ 137~$M_{\odot}$ & direct collapse \\
 & & to a black hole \\
\hline
\end{tabular}
\end{table}

\begin{figure}[!ht]
\begin{center}
\includegraphics[width=9cm]{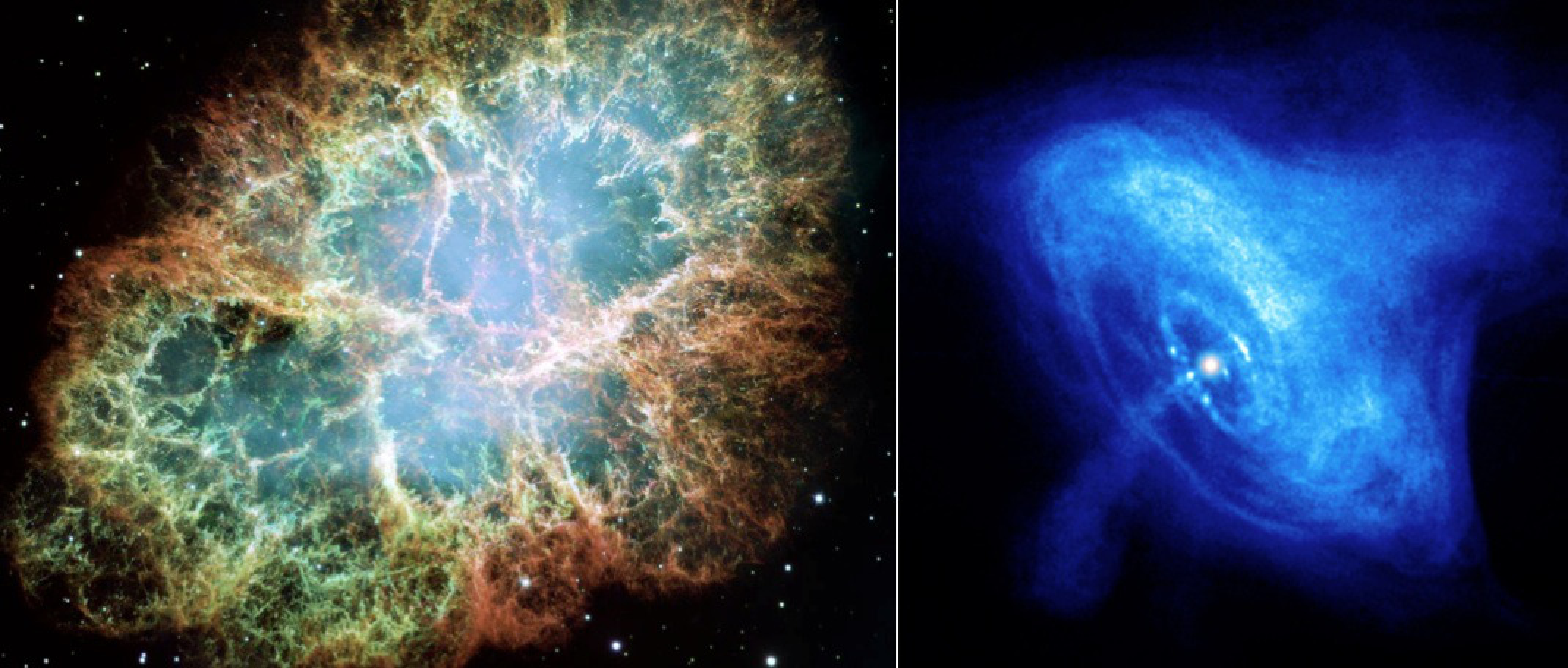}
\caption{The remnant SN~1054 (the Crab nebula in visual band and X-ray with clearly seen crab pulsar). Image Credit: NASA \label{crab_nebula}}
\end{center}
\end{figure}

All these SNe will leave a remnant: a nebula composed of the outer layers of the massive star expelled during the collapse, with a compact object (black hole or neutron star) at its center. For instance, one of the first observed and recorded supernova was SN~1054 from the year of its explosion. In its remnant, called the Crab nebula, there is a young neutron star: the Crab Pulsar, see Figure \ref{crab_nebula}.

\subsection{Supernovae (SNe) classification}

\begin{figure}[!ht]
\begin{center}
\includegraphics[width=0.38\textwidth]{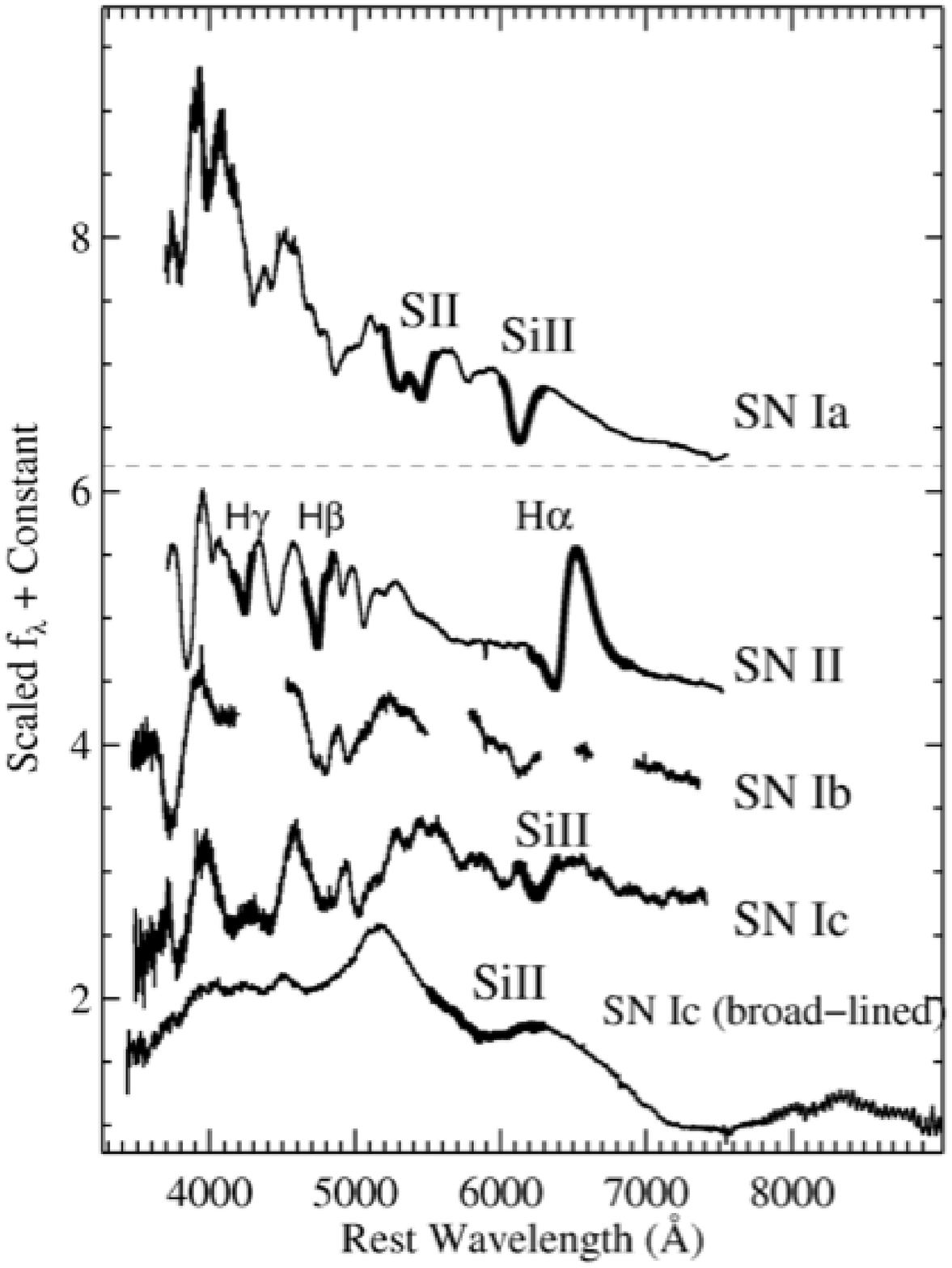}
\includegraphics[width=0.54\textwidth]{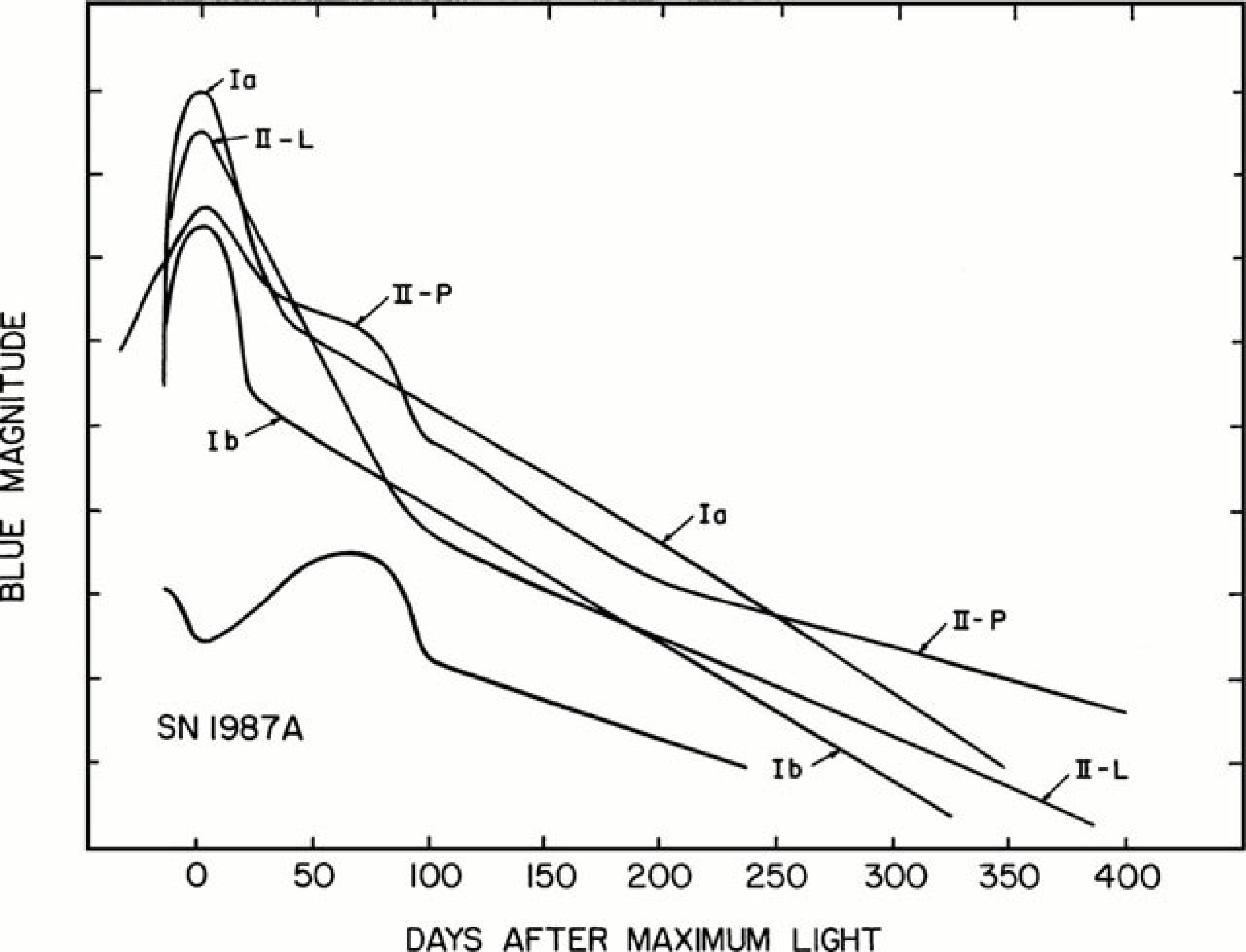}
\caption{Left: Comparison of the spectra of different SNe types: SN~1994D for type Ia, SN~1999em for type II, SN~2004gq for type Ib, SN~2004gk for type Ic. The broad line Ic SN ($\textit{e.g.}$ SN~1998bw) does not show H or He lines in its optical spectra and its lines widths are 2-3 times larger than the normal type Ic SN which is indicative of high velocity ejecta \cite{modjaz2008}. Right: The light-curve of different type of SNe \cite{filippenko1997}.
\label{spectra_ofSN_type}}
\end{center}
\end{figure}

Supernovae (SNe) are violent explosions associated with the death of stars. They are characterized by a large increase in brightness up to -20 magnitudes. Observationally, by identifying different elements in their spectra, they can be divided into different subclasses \cite{filippenko1997}. This classification depends mainly on the composition of the star, so on its initial mass \cite{woosley2002}. An early classification of SNe was proposed, in which type I and type II are differentiated by the absence (type I) or presence (type II) of hydrogen (H) lines \cite{minkowski1941}. Figure \ref{spectra_ofSN_type} displays the spectra (left) and light-curves (right) of different types of SNe.

Type II SNe are usually mentioned in the literature with four subclasses: type IIP (plateau in their light-curves), type IIL (linear decline in their light-curves) which constitutes the majority of type II SNe (\textit{e.g.} \cite{barbon1979}), type IIn with narrow line in their spectra, and type IIb which is an intermediate type of SNe, with early features of type II SNe which are replaced by type Ib features at late times.
 
\begin{figure}[!ht]
\begin{center}
\includegraphics[width=12cm]{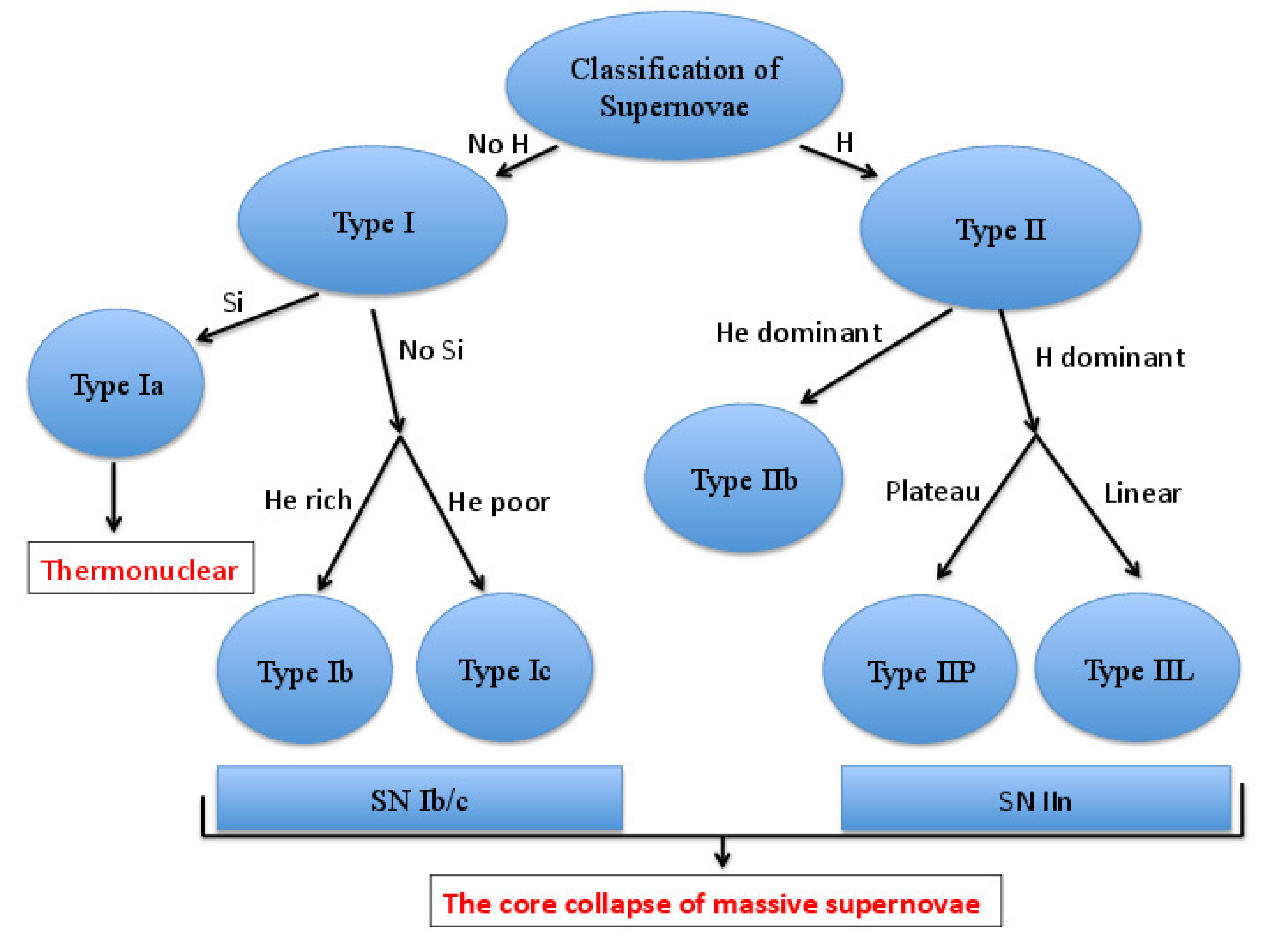}
\caption{Schematization of supernovae classification \label{supernova_class}}
\end{center}
\end{figure}

 Moreover, type I SNe are divided into three subclasses. Type Ia SNe show silicon lines in their spectra and are thought to originate from the thermonuclear explosion of an accreting white dwarf (\textit{e.g.} \cite{branch1995}). Type Ib SNe show helium lines but do not show silicon lines in their spectra. Finally, type Ic SNe are very similar to type Ib SNe: they only lack He lines. The basic properties of SNe and their classification are shown in Figure \ref{supernova_class}. Type Ib and Ic SNe have many similarities, therefore they are sometimes jointly called Type Ib/c supernovae. Together with type II SNe, they are referred to as core-collapse SNe, because they are believed to be produced by the core collapse of massive stars ($\textit{e.g.}$ \cite{li2008}). However, type Ib/c and type II SNe have different progenitors (differently evolved stars).
 
Type Ib/c SNe have lost their hydrogen (Ib) and helium (Ic) envelopes before the explosion; thus they are called stripped-envelope SNe. The mass loss is thought to arise from the increased mass of the progenitor relative to the progenitors of type II SNe. This kind of progenitors show stronger stellar winds which can blow away the outer layers of the progenitor star. Moreover, the higher mass loss can also be explained by line-driven winds due to the increased metal content of the progenitor \cite{puls1996, kudritzki2000, mokiem2007}. Thus, the progenitors of type Ib/c SNe could also have higher metallicity than type II SNe and their most likely progenitors are believed to be massive W-R stars surrounded by a dense wind envelope \cite{gaskell1986}. Such winds are at the origin of broad emission lines in the spectrum of the stars and change the surface composition, reflecting the presence of heavy elements created by nuclear burning (see Figure \ref{nonrotating_stars}). These kinds of stars are thought to be at the origin of type I SNe because they have little H (as in WNL: nitrogen in late type spectra) which can even be completely absent, like in WC and WO. These N, C, and O subtypes of W-R stars indicate the presence of strong lines of nitrogen, carbon, or oxygen in their spectra \cite{maheswaran1994}.

In addition, most massive stars are in binary systems. Because of the gravitational interactions within the binary, H can be removed from one of the star, which will end as a type Ib/c SN. 

\subsection{Extreme case : Gamma-Ray Burst (GRBs)}

Interestingly enough, all GRBs which are associated to SNe were identified to the stripped-envelope SNe Ib/c. If the released isotropic energy of GRBs is between $10^{48}$-$10^{54}$ ergs, the emission is often assumed to be collimated in jets in order to decrease the required energy. In the collapsar model, the GRB is assumed to be emitted by a massive star experiencing a gravitational collapse \cite{woosley1993}. Therefore, very special conditions are required for a star to evolve all the way to a gamma-ray burst: it should be very massive (probably at least 40~$M_{\odot}$ on the main sequence) and rapidly rotating to form a black hole and an accretion disk of mass around few 0.1~$M_{\odot}$. GRBs can be powered from the black hole by at least two different ways:
\begin{itemize}
\item neutrino annihilation: $\nu_e + \nu_e^-  = e^+ + e^-$ \cite{mochkovitch1993, macFadyen1999}.
In this process, neutrinos are efficiently produced in the accretion disk because of its high temperature and high density. These neutrinos escape the accretion disk and annihilate in the polar region of the black hole. As large amounts of neutrinos are produced, the created electron positron plasma is optically thick and it is at the origin of the fireball. 

\item magnetically dominated instability (Blandford-Znajek mechanism) \cite{blandford1977}.
In this process, the magnetic field of the black hole and of the accretion disk extracts the rotational energy of the black hole. 
\end{itemize}

In other models, the merger of two compact objects (two neutron stars or a neutron star with a black hole) is involved \cite{eichler1989}. In these models, the energy can be extracted thanks to neutrino-antineutrino annihilation produced in the hot and dense torus and this energy available for the relativistic jet is sufficient to power a short GRB if a modest beaming is assumed. 

\section{Observational Properties of GRBs}

\subsection{Spatial properties}
\begin{figure}[!ht]
\begin{center}
\includegraphics[width=12cm]{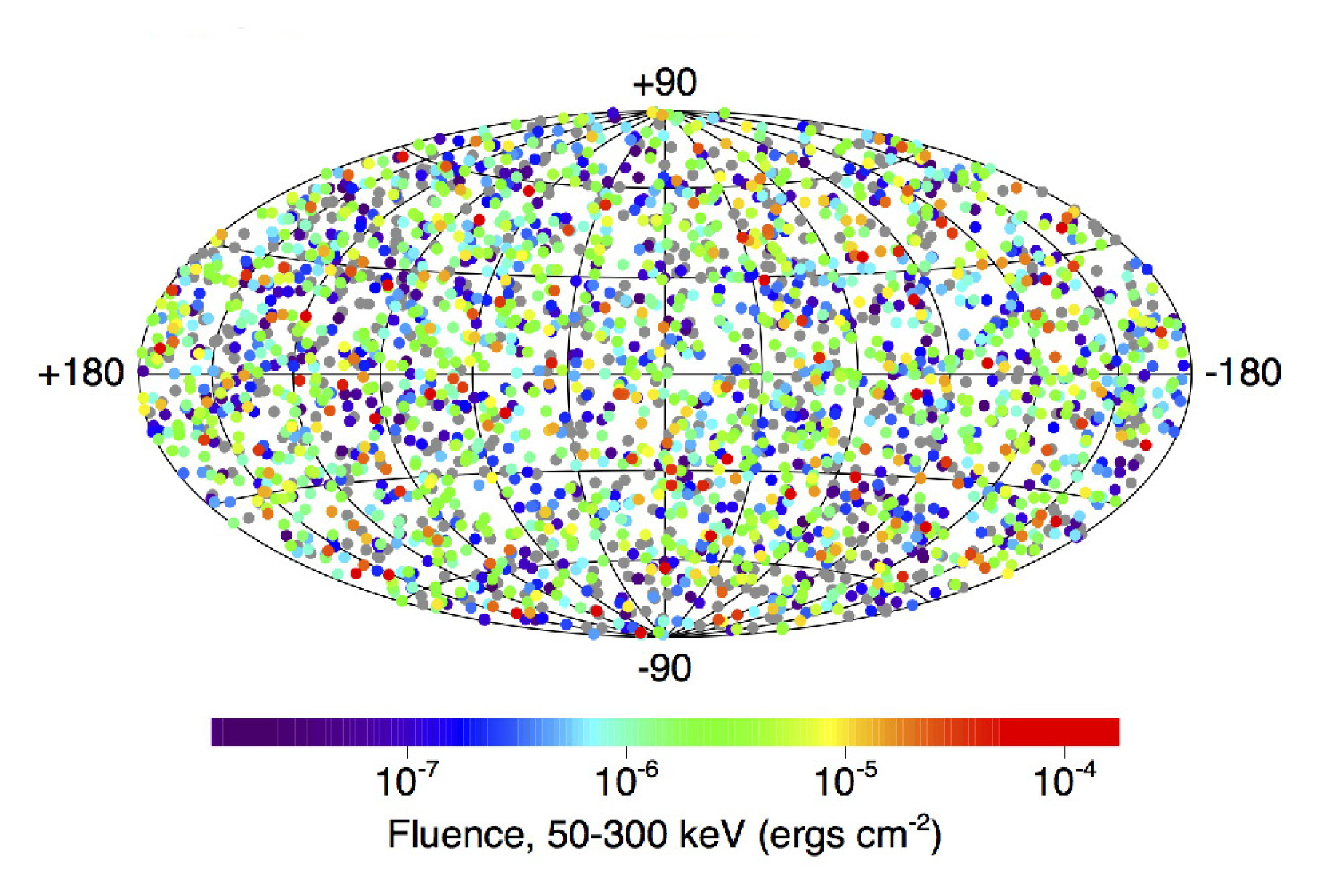}
\caption{The distribution of all 2704 GRBs detected by BATSE satellite: they are clearly isotropically distributed \cite{paciesas1999}. Image Credit: NASA/BATSE \label{BATSE_all_sky}}
\end{center}
\end{figure}

After the discovery of GRBs and before the discovery of their afterglows, they were explained by more than 100 different theoretical models \cite{ruderman1975}, the reason being that their distance was not known. The first clue about the distance came with the result of BATSE: it was observed that GRBs are distributed isotropically on the sky (see Figure \ref{BATSE_all_sky} for all observed burst by the BATSE satellite). It implies that GRBs originate at cosmological distances \cite{fenimore1993} or from a large halo around our galaxy.

This controversy was solved by the redshift measurements of GRB~970228 at redshift z = 0.835 \cite{metzger1997} which was made possible by the BeppoSAX satellite. It provided the localization of the burst with less than an arcminute uncertainty in radius \cite{costa1997}, which allowed a follow-up at optical \cite{vanParadijs1997} and other wavelengths.

\subsection{Temporal properties}
\label{sec_temporal}

Based on the observations, two main times of emission can be defined, namely the prompt phase and the afterglow. The prompt phase can be associated to the central engine activity; the afterglow is a long-lasting emission, gradually decreasing, coming after the prompt phase. It is thought to be the result of the interaction of a relativistically expanding plasma with the environment of the progenitor, with the possible contamination of emission from late central engine activity.

\begin{figure}[!ht]
\begin{center}
\includegraphics[width=8cm]{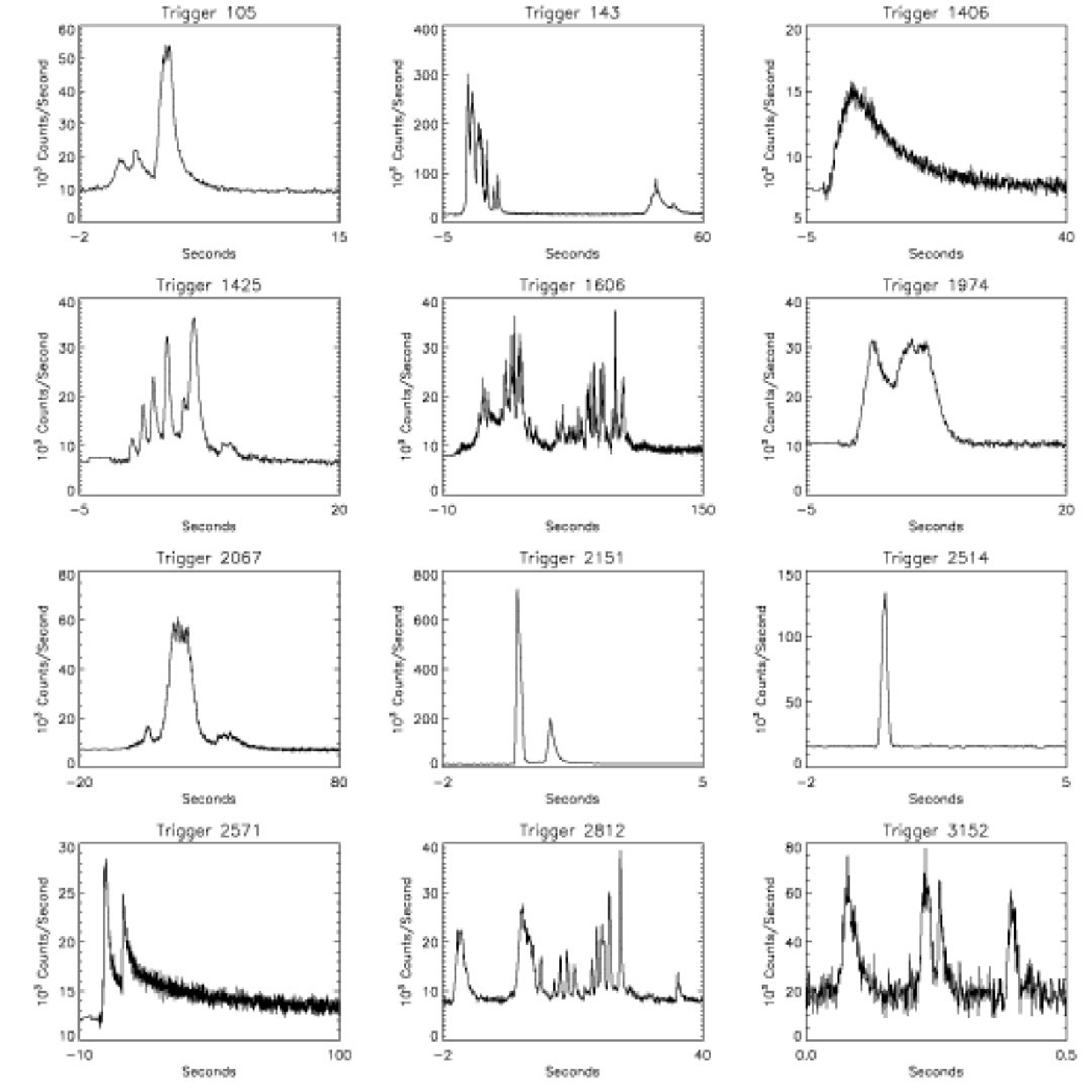}
\caption{Different light-curves from the GRBs prompt phase, as observed by BATSE \cite{fishman1995, meszaros2006}. \label{BATSE_light_curve}}
\end{center}
\end{figure}

\begin{figure}[!ht]
\begin{center}
\includegraphics[width=10cm]{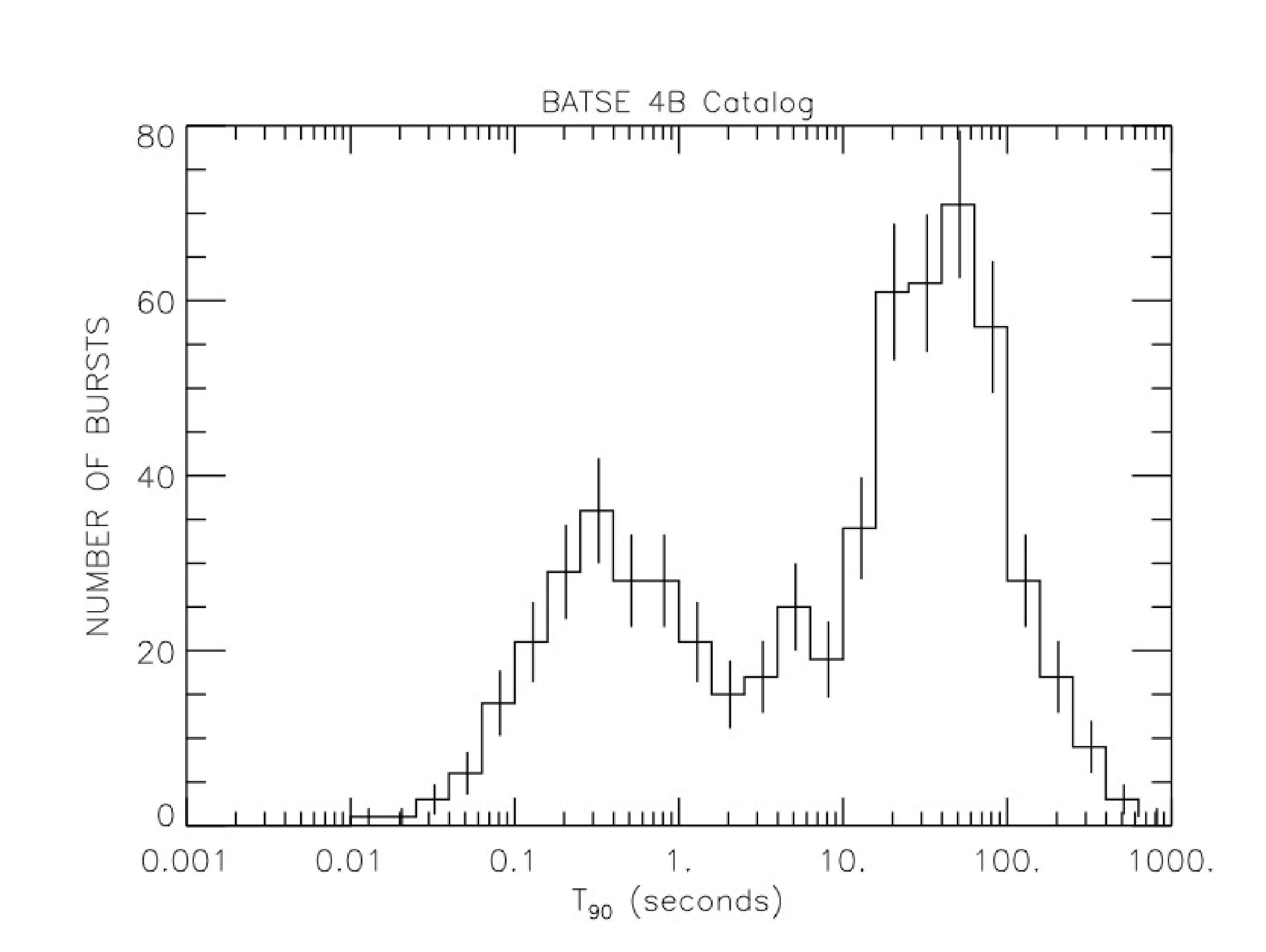}
\caption{Duration distribution of short-long GRBs observed by BATSE \cite{meegan1997}. Image Credit: NASA/BATSE \label{BATSE_short_long}}
\end{center}
\end{figure}

The prompt phase is characterized by a high flux of gamma-ray photons (keV - MeV). In this phase, each source shows different behavior and trend in its light-curve: single peak, double peaks, multiple peaks, smooth or spiky light-curves, see Figure \ref{BATSE_light_curve}. The observed variability is high, down to 10 miliseconds \cite{golkhou2014}. The duration of the prompt phase called T$_{90}$ corresponds to the time during which 90\% of the energy is emitted in the keV - MeV range. In 1993, with the data collected by BATSE, It was found that the T$_{90}$ distribution of all sources shows two separated groups, see Figure \ref{BATSE_short_long}. Specifically, the short bursts have a duration T$_{90}$$<$ 2 s with the average observed value around 0.73 seconds while the long bursts have a duration T$_{90}$$>$ 2 s with the average observed value around 17 seconds \cite{kouveliotou1993, dezalay1996, ghirlanda2009}. 

\subsection{Spectral properties of the prompt emission}
The spectrum of the prompt phase is well-fitted by the Band model \cite{band1993}:
\begin{equation}
\begin{split}
&  N_{E} ( E ) = 
A \begin{cases} 
\biggl(\frac{E}{100 \ \rm keV }\biggr)^{\alpha} \exp \biggl[- \frac{E}{ E_{0} } \biggr], \ \ \ \ \ (\alpha - \beta)
E_{0}  \geq {E} \\
\biggl( \frac{(\alpha - \beta)E_{0}}{ 100 \ \rm keV } \biggr)^{ \alpha - \beta } \exp(\beta -\alpha) \times \biggl[ \frac{E}{100 \ \rm keV} \biggr]^{\beta },  \ \ \ \ \  (\alpha - \beta)
E_{0} \leq {E}, \\
\end{cases}
\end{split}
\label{Eq.band}
\end{equation}
where the four parameters are the amplitude \emph{A}, the low and high energy spectral indexes, $\alpha$ and $\beta$ respectively, and the spectral break energy $E_{0}$. This function is made of two broken power law smoothly joined. 

The mean values of the spectral indexes are $\alpha$ = - 0.92 $\pm$ 0.42 and $\beta$ = - 2.27 $\pm$ 0.01 for the long GRBs and $\alpha$ = - 0.4 $\pm$ 0.5  and $\beta$ = - 2.25 \cite{nava2011} for the short GRBs \cite{ghirlanda2002, ghirlanda2009}. For some bursts, the low energy spectral index is larger than - 2/3, which is not compatible with the synchrotron emission theory (the so-called synchrotron line of death) \cite{preece1998}. 

There is also another difference between short and long GRB spectra. The peak energy of short GRBs is on average larger (harder spectrum) than the peak energy of the long ones (mean values 398 keV and 214 keV respectively \cite{ghirlanda2009}).

The spectra of the prompt phase have been fitted with other models \cite{ryde2005, guiriec2013}, for instance band plus black body, to understand and separate the different emission mechanisms (thermal or non-thermal).

Very high energy photons (GeV) have been detected in several GRBs. The spectrum is well-fitted by a power law and a break in energy can sometimes be seen. The main properties of that feature are that it can be delayed for some seconds compared to the prompt phase and it lasts much longer. The emission mechanisms producing these photons are still puzzling and strongly debated.

\subsection{Afterglow}
 
\begin{figure}[!ht]
\begin{center}
\includegraphics[width=10cm]{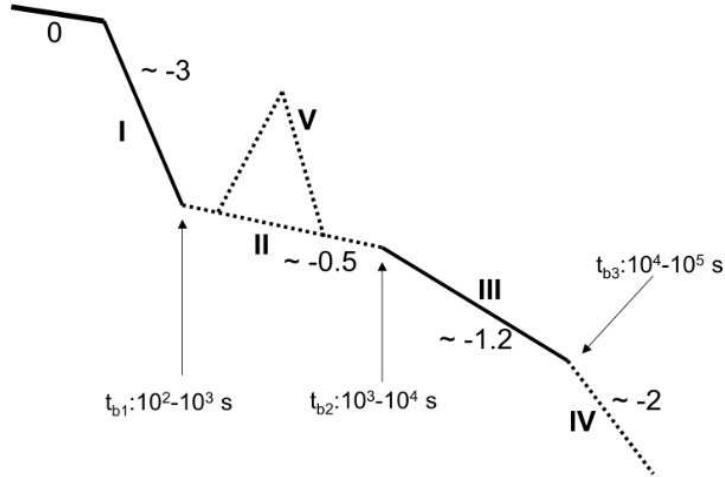}
\caption{The schema of the various afterglow phases in long GRBs shown in the log flux-time plane \cite{zhang2006a}. The prompt phase (0) is generally followed by a steep decline afterglow (I) which can then break to a shallower decline called plateau (II), a standard afterglow phase (III), comes after and possibly, a jet break (IV). Sometimes X-ray flares can be seen (V) \cite{zhang2006a}. \label{Swift_canonical_light_curve}}
\end{center}
\end{figure}

The afterglow is observed at all wavelengths: X-ray \cite{costa1997}, optical \cite{vanParadijs1997}, IR, and radio \cite{frail1997}. Thanks to its low variability and observed time range (from minutes to weeks after the GRB event), a canonical X-ray light-curve for the afterglow was defined from the result of \textit{Swift}/BAT-XRT instruments. It is displayed on Figure \ref{Swift_canonical_light_curve}: the 0 symbol indicates the prompt phase, and the four remaining segments, with their corresponding temporal indexes, are associated two by two and identified as early and late afterglow \cite{rees1992, meszaros1997, panaitescu1998}: I and II (respectively the steep and shallow decay), and III and IV (respectively a standard afterglow and a jet break) \cite{nousek2006}. Part I and III, marked by solid lines, are most common and the other three components, marked by dashed lines, are only observed in a fraction of all bursts. Part I is thought to be associated  with the prompt phase \cite{zhang2006a, willingale2007} when the central engine is still active; the rest of the afterglow is due to the dynamics of the interaction between the jet and the surrounding medium. 

The spectra in the keV range of all phases but the prompt are well-fitted with a simple power-law model: 
\begin{equation}
N ( E ) = A \left( \frac{ E }{ 100  \ \rm keV} \right )^{-\beta }
\label{simple_PL}
\end{equation}
where \emph{A} is the amplitude and $\beta$ is the spectral index.

The spectral index is constant throughout parts II, III, IV and is in the order of $\beta\sim1$. However, it is softer for parts I and V (flares which are thought to be the extended emission of the prompt phase \cite{zhang2014}).

Even if the detection rate is high \cite{chandra2012}, the number of well-sampled light-curves is smaller in other wavelengths: $\sim$300 in X-ray \cite{evans2009, dereli2014} $\sim$68 in optical \cite{zaninoni2013} and $\sim$6 in radio. 
These light-curves can be compared with the X-ray light-curves to see if they share common properties. The most important result from the optical photometric observations is the optical bump in the light-curve between 10-25 days after the GRB explosion, which is interpreted as an SN explosion and indicate the connection between GRBs and SNe. On the other hand, the optical spectral observations provide the redshift by the measurement of absorption and emission lines.

\subsection{GRB-SN association}

The first direct evidence for a GRB-SN association was made when GRB~980425 was spectroscopically and photometrically linked with type Ic-broad line SN~1998bw \cite{galama1998},  as seen in Figure \ref{spectra_ofSN_type}. This connection was also predicted theoretically by the collapsar model \cite{woosley1993}.
 
This connection was further confirmed in 2003 with the spectroscopic and photometric association between GRB~030329 and SN~2003dh \cite{stanek2003, hjorth2003}, and GRB~031203 and SN~2003lw \cite{malesani2004}. These three bursts provided clear evidence that the progenitors of some, if not all long GRBs, are associated to the explosion of a massive star. Since then, other SNe were observed, either spectroscopically or identified as an optical bump in the late optical light-curve around 10 days after the burst.

\subsection{Different kinds of GRBs}
\begin{figure}[!ht]
\begin{center}
\includegraphics[width=8cm]{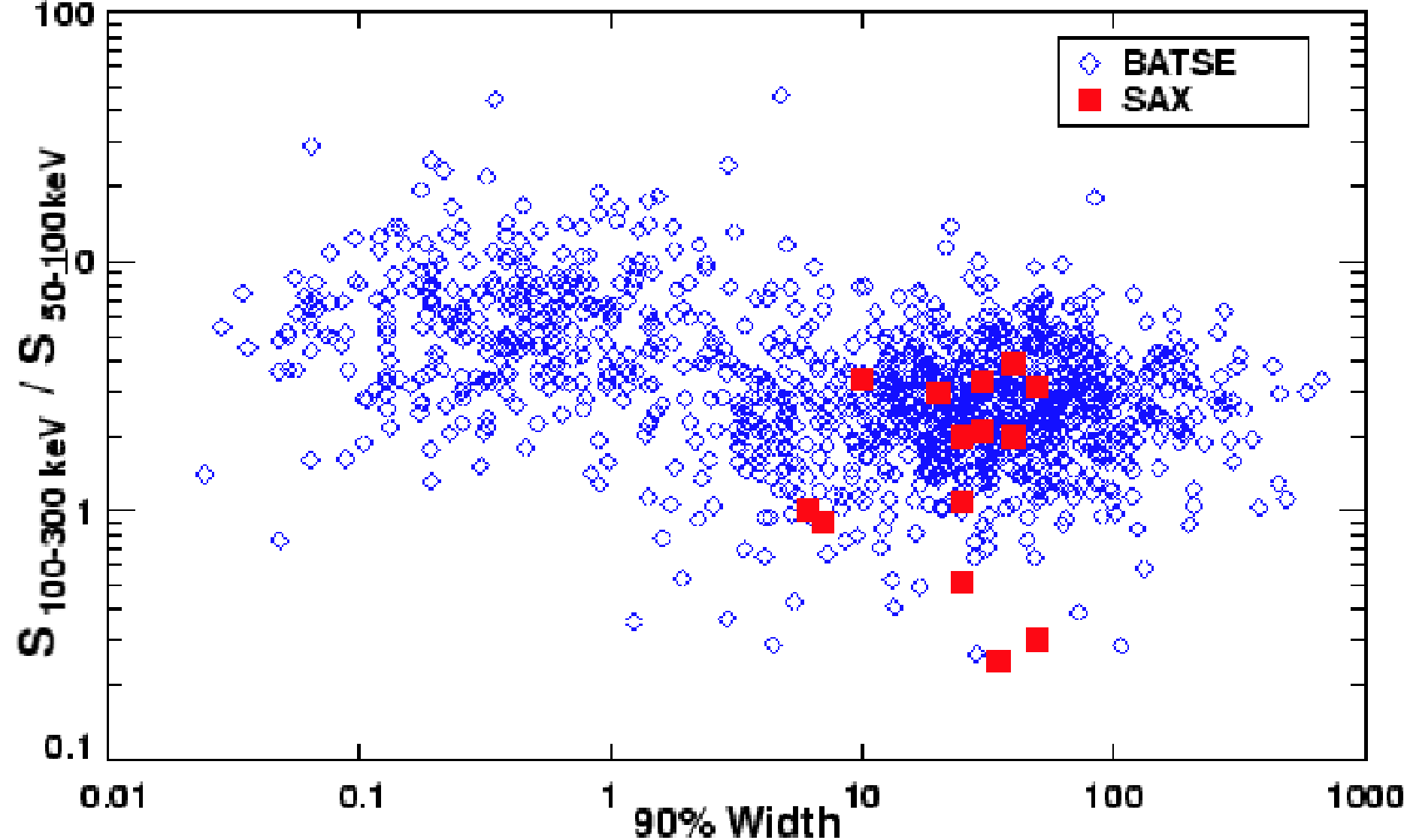}
\caption{Hardness ratio of short - long GRBs obtained by BATSE and BeppoSAX \cite{meegan1997}. Image Credit: NASA/BATSE} 
\label{hardness_ratio}.
\end{center}
\end{figure}

\paragraph{Short bursts vs. long bursts:}
The study of the duration of the prompt phase has shown the existence of two classes of GRBs, namely the short and the long bursts, see subsection \ref{sec_temporal}. They are also differentiated by the hardness ratio \cite{kouveliotou1993, dezalay1996} defined as the ratio of the bolometric fluence measured in two different bands: $S_{100-300 \text{keV}}/S_{50-100 \text{keV}}$, see Figure \ref{hardness_ratio}. The hardness ratio of short GRBs is slightly larger than that of long GRBs.

In addition, \textit{Swift} has shown that the redshift distributions of short and long bursts are not consistent \cite{gehrels2009}, which strengthen the theory that at least two different progenitors are responsible for GRBs. Another strong hint comes from the associated host galaxies. Long bursts are associated to star forming regions with high star forming rate in spiral or irregular galaxies with low mass and low metallicity. On the other hand, short bursts seem to be associated to regions with low star formation rate, either inside a galaxy or a low star-forming elliptical galaxy characterized by high mass and high metallicity or even have evaluated end left the central part of their galaxy. These findings strongly suggests old stars or stellar remnants as the progenitors of short bursts. 

\paragraph{Ultra long burst:}
Within the diversity of GRBs, one of the extreme case is the so-called ultra long GRBs of typical duration in the order of thousands of seconds (\textit{e.g.} 25000s for GRB~111209A \cite{gendre2013}). They are thought to originate from the collapse of a low-metallicity blue supergiant in rapid rotation  \cite{gendre2013, levan2014, piro2014}.

\paragraph{X-ray flashes (XRF):}
A class of sources, called X-ray flashes or XRFs, was characterized by the HETE-2 (High Energy Transient Sources Experiment) satellite in 2001 \cite{heise2001}. They have similar soft spectrum and duration to that of long GRBs \cite{vanderspek1999}. Their total energy is small in the prompt phase and they are brighter in the X-ray emission. They appear to be correlated with supernova explosions. As a result, they might have the same origin as GRBs. Possibly, they are either a lower energetic subset of the same phenomena or the differences are due to their orientation relative to our line of sight.

\paragraph{Dark bursts:}
Another kind of GRBs is composed of the so called dark GRBs. They are defined by a large absorption (gas) and extinction (dust) in their host galaxies: they are characterized by a hydrogen column density larger than $10^{23} \text{cm}^{-2}$ and by an extinction larger than 2.6 mag.

\paragraph{Long bursts without SNe:}
Moreover, several long/soft GRBs (\textit{e.g.} GRB~060614) without an associated supernova have been discovered by \textit{Swift} \cite{dellaValle2006}. These events open the door for a so far unknown third class of GRBs, which challenges the idea of collapsing massive stars and binary mergers being the only progenitors of GRBs and suggesting that the tidal disruption of a star by a black hole would be an ideal way to power a long duration GRB \cite{lu2008}.

~

To conclude, the diversity of GRBs can reflect multiple progenitors and different types of interactions with their environments.

\subsection{Progenitors of GRBs} 

Since the discovery of GRBs, many possible progenitors have been proposed. However, only a few remain, mainly because of the enormous energy budget required. 
Among the massive star progenitor models, two are popular: the first one is the collapsar model \cite{woosley2002, zhang2004, woosley1993, macFadyen1999}, and the second one is  the millisecond magnetar model \cite{usov1992, meszaros1997, thompson1994, wheeler2000}.

\subsubsection{Collapsar model}

A collapsar is a fast-rotating massive star about the collapse. It lost its H layer during its evolution by stellar winds. As it collapses onto a black hole, it creates an accretion disk. The nickel at the origin of the SN is created by the accretion disk, while the jet is created by accretion of the matter of the disk onto the black hole, and is collimated by the material of the star. 

Not all SNe and not even all most energetic ones (hypernovae) produce GRBs. This could be explained by the requirements on metallicity and rotational speed or by the fact that the jet is not pointing towards the Earth. However, it is yet unclear if all long GRBs are accompanied by an SN. For example, a coincident supernova for GRB~060614 has to be 100 times fainter than a \textit{standard} SN, as imposed by the flux limits. This may indicate that long GRBs are composed of different populations.

\subsubsection{Millisecond Magnetar model}

A magnetar is a fast-rotating neutron star  \cite{usov1992}. Its high magnetic field and high angular momentum provide the energy for the GRB and the SN. As the produced outflow is highly magnetized, a high radiative efficiency is expected \cite{vanPutten2011}. 

Most GRB light-curves show a plateau and some times flares. They are interpreted by energy injection in the outflow long time after the prompt phase. The millisecond magnetar model is able to explain this energy injection by accretion of matter onto the BH.

\subsubsection{Binary of Compact Object}
Two compact objects lose their energy and orbital angular momentum by gravitational wave radiation and merge as a result \cite{eichler1989}. These binaries are composed by Neutron Star (NS) - NS, Black Hole (BH)-NS, BH-BH \cite{zhang2004, eichler1989, fryer1999, zhang2001}.

Enough energy to power a GRB can be created in a binary merger. The duration of the GRB is comparable to the lifetime of the accretion disk. So in the binary mergers, the duration of the resulting GRB is expected to be some miliseconds while it is longer in the collapsar model.

\section{Emission Theory of Gamma-Ray Bursts: The Fireball Model}

\subsection{General description}
Nowadays, the standard model for the emission mechanisms is the so-called fireball model \cite{meszaros1993a}. It considered a large amount of energy (10$^{51} - 10^{54}$ ergs) released in a very small region (r$_0$ $\sim$ 10$^6$ or 10$^7$ cm) as implied by the observed variability in the prompt phase. 

From the volume and the energy, it is possible to show that the region of emission is optically thick for pair creation: this is the so-called compactness problem \cite{piran1999}. It can be solved by considering that the emitting object is moving relativistically towards the observer. This is possible when the energy is much larger than the rest mass energy. The released energy can be dominated either by thermal energy or by magnetic energy, leading to two different acceleration mechanisms \cite{piran1993, drenkhahn2002}. The Lorentz factor of the outflow is increasing up to the saturation radius. In the case of radiation-dominated outflows, it is defined as $r_s = r_0 \times \eta$, where $\eta = E_0 /(M_0 c^2)$ is the ratio between the total energy $E_0$ and the rest mass energy $M_0 c^2$ of the ejecta. The coasting Lorentz factor is estimated to be as large as $\Gamma$ = 100 - 1000. Once all energy has been converted to kinetic energy, it is necessary to find a way to convert it back to radiation with a high efficiency. This is achieved by shocks.  

This model is summarized on Figure \ref{fireball_model} (for the typical radius and associated emission mechanisms) and Figure \ref{lorentz_factor} (for the corresponding evolution of the Lorentz factor). The different possible processes are discussed in details below. 

\subsubsection{Photospheric emission}
\begin{figure}[!ht]
\begin{center}
\includegraphics[width=12cm]{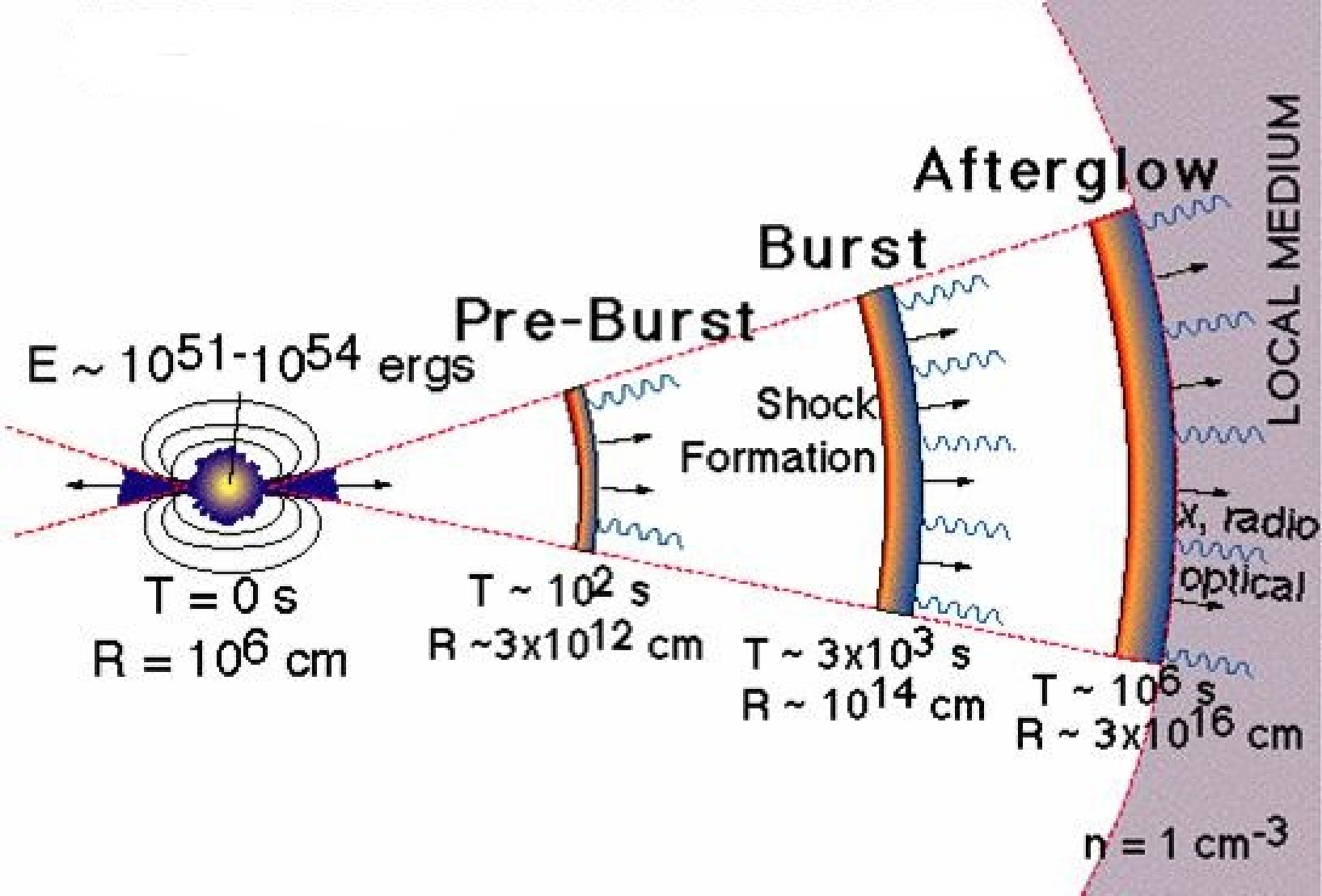}
\caption{Standard fireball model. Image Credit: NASA/\textit{Swift} \label{fireball_model}}
\end{center}
\end{figure}

Under the conjugate action of the density decrease and the increase of the Lorentz factor, the opacity for Compton scattering is decreasing. As it drops below the unity, all photons initially trapped within the outflow are set free to reach a distant observer. This is called the photospheric emission. The outflow becomes transparent typically a radius of some 10$^{12}$ cm (see Figure \ref{fireball_model}).

The efficiency is between 5 and 30\% of the total energy of the burst, and can even be larger in some specific cases (for example GRB~090902B \cite{bromberg2011}). The spectrum of this emission is thermal (it has a nearly black-body shape) and is characterized by its normalization and temperature $T_{obs}$. However, a black-body spectrum is too narrow to account for most observations. Sub-photospheric (below the radius at which the plasma becomes transparent) dissipation were also studied as a way to broaden the expected thermal spectrum, and further increase the efficiency of the photosphere.

\subsubsection{Internal shock model}
As the emission of the outflow at the center is not steady, different parts of the plasma are moving with different velocities. When two parts of the plasma with different speeds collide, internal shocks form. They accelerate part (if not all) of the electrons in a power law and locally increase the strength of the magnetic field, at the expense of the kinetic energy. The electrons, which are accelerated by the 2nd order Fermi mechanism, radiate synchrotron emission in the induced magnetic field. Additionally, the synchrotron photons can be inverse Comptonized to even higher energy (synchrotron self-Compton).

The typical  radius for internal shock, $r_{is} \sim \eta^2 \ r_{0}$, is in the order of 10$^{14}$ cm (see Figure \ref{fireball_model}). The main problem of the internal shock model is that the efficiency is low (5\% to 20\%), while the observations indicate an efficiency larger than 50\% for most bursts \cite{cenko2006}.

\begin{figure}[!ht]
\begin{center}
\includegraphics[width=8cm]{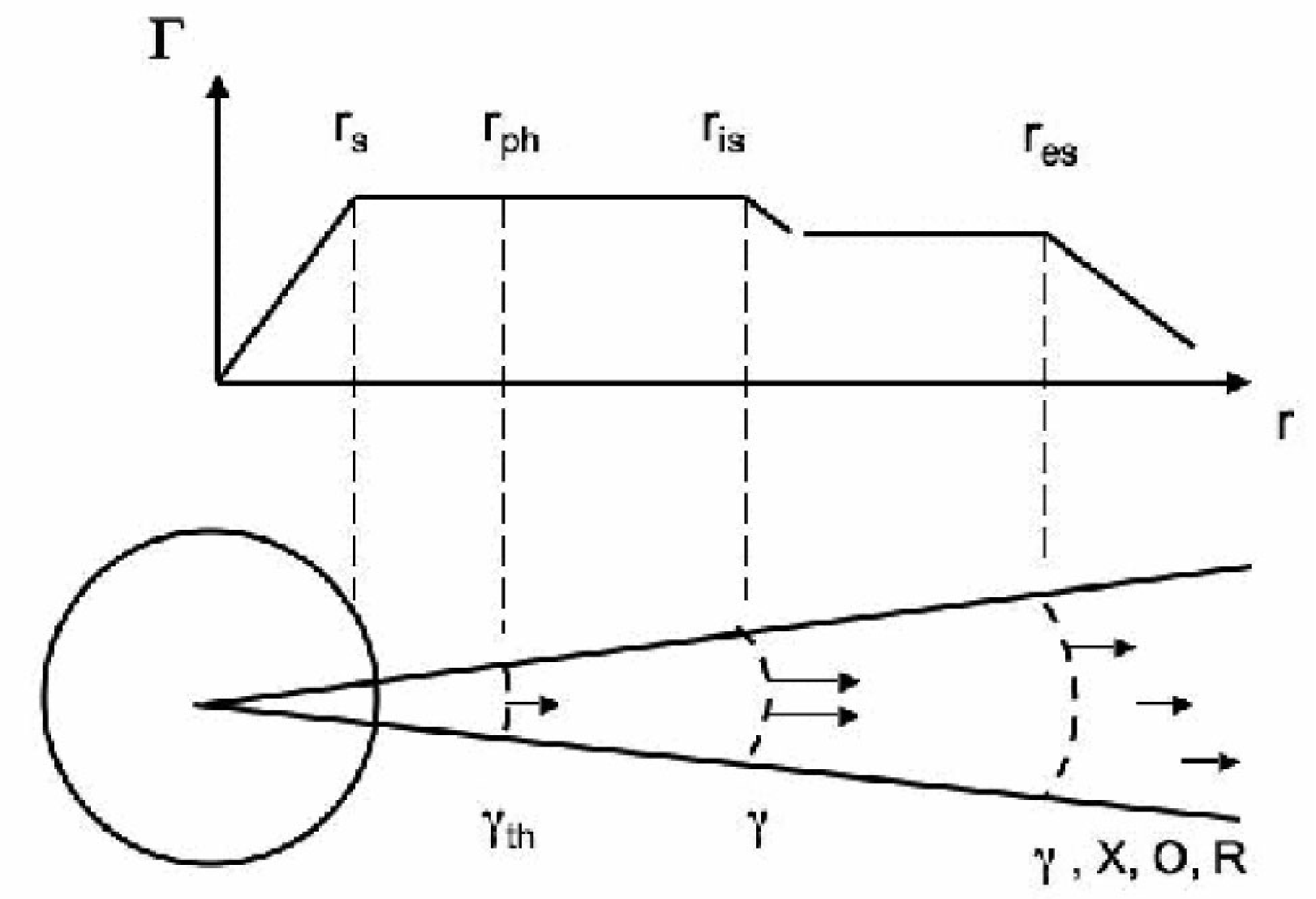}
\caption{Schematic evolution of the jet Lorentz factor and examples of symbolic locations of radius: the saturation radius $r_s$, photospheric radius $r_{ph}$, internal shock radius $r_{is}$ and external shock $r_{es}$ \cite{meszaros2006}. \label{lorentz_factor}}
\end{center}
\end{figure}

\subsubsection{External shock model}

The remaining kinetic energy (which was not dissipated by internal shock) is released when the outflow is decelerated by the interstellar medium (ISM), which can be either of constant density $n \sim$ 1 cm$^{-3}$ or the wind of the progenitor, $n  \propto r^{-2}$, where $r$ is the radius. This deceleration creates the reverse shock (which propagates back into the ejecta \cite{meszaros1993}) and the forward shock (which propagates into the external medium): these compose the so-called external shock model.

These two shocks are created as soon as the outflow starts to expend at the central radius $r_0$, but they become efficient only when the kinetic energy can be efficiently reduced, that is to say when the swept ISM mass is large enough. Such a condition is fulfilled at typical radii 10$^{16}$ - 10$^{17}$ cm \cite{meszaros1993} (see Figure \ref{fireball_model}). As for the internal shocks, the magnetic field is increased by plasma instabilities and accelerated particles radiate synchrotron in the induced magnetic field.

\subsubsection{GeV emission in External shock}
The GeV emission which is seen in the spectrum of some GRBs can be explained by different models. The first possibility is inverse Compton emission of synchrotron photon produced in the external shock and scattered by accelerated electrons (synchrotron self-Compton) \cite{kumar2009}. The second possible model is the pair loading of the surrounding medium of the burst by prompt photons. These photons are scattered by cold electrons of the ISM, then they interact with other photons of the prompt emission to create electron-positron pairs. The local ISM density is increased by a large factor \cite{beloborodov2014}. These pairs are accelerated to relativistic speeds by the huge radiative pressure and inverse Comptonize the prompt MeV photons. Another explanation might be hadronic processes \textit{e.g.} proton-proton interaction, neutron decay, proton-photon interaction \cite{asano2007}.

\subsubsection{Jet break}
\label{jet_break}
In addition, GRBs are assumed to be collimated into jets. The relativistic beaming produces a visibility cone with an opening angle of $1/\Gamma$, which increases as the outflow decelerates (see Figure \ref{lorentz_factor}). When the opening angle is equal to the jet half-opening angle (1/$\Gamma = \theta_j$) the afterglow light-curve should break and decline more rapidly: this is called the jet break. 

The jet half-opening angle can be computed from the time of the jet break by assuming the standard model for the afterglow. There are two possibilities: 
\begin{itemize}
\item the afterglow interacts with the ISM of constant density,
\item the afterglow interacts with the wind of the progenitor whose density decreases proportionally to $r^{-2}$, where $r$ is the radius.  
\end{itemize}

For the ISM of constant density, the half-opening angle is given by \cite{levinson2005, sari1999b}:
\begin{equation}
\label{eq_opening}
\theta (t_{b}, E_{iso} ) = 0.161\left(\frac{t_{b,d}}{1+z}\right)^{3/8}n^{1/8}\left(\frac{\eta_\gamma}{E_{iso,52}} \right)^{1/8}\text{,}
\end{equation}  
where the standard values for the number density of the medium $n=1~\text{cm}^{-3}$ is assumed. It also takes into account the radiative efficiency of the prompt phase, $\eta_\gamma$. $E_{iso}$ is the isotropic equivalent energy. In this study, each quantity expressed as $X_n$ is such that $X = X_n \ 10^n$. z is the redshift of source. Finally, the break time $t_{b}$ is in days, $t_{b,d}$. 

\subsection{Detailed afterglow theory}
\label{detailed_afterglow_theory}

\subsubsection{Blast-wave evolution}
The following derivations can be found with more detailed in \cite{dermer2009}.
GRB afterglows can be succesfully explained by the interaction between the outflow and the ISM, resulting in the decrease of the outflow speed \cite{rees1992}. The kinetic energy lost by the outflow is converted to kinetic energy of the shocked ISM and to internal energy. A fraction $\epsilon_B$ of this internal energy is assumed to be in the form of magnetic field while a fraction $\epsilon_e$ is assumed to be in the form of kinetic energy of electrons. Blandford and McKee solved the relativistic hydrodynamic equations of motion considering adiabatic and radiative relativistic blast waves \cite{blandford1976}. When the internal energy created in the collision is completely emitted, the evolution is said to be fully radiative, otherwise, it is adiabatic. 
The evolution of the Lorentz factor in a constant ISM for the adiabatic case is given by:
\begin{equation}
\label{eq_gamma_ad}
\Gamma(t) \cong 5.4 \left(\frac{1+z }{ t_{day}}\right)^{3/8} \left(\frac{E_{52} }{n_{0}}\right)^{1/8}\text{,}
\end{equation}  
where $t_{day}$ is the time measured by the observer in days after the GRB, $n_{0}$ is the ISM proton number density and $E_{52}$ is the total explosion energy.
The evolution of the Lorentz factor in a constant ISM for in the radiative case is given by \cite{dermer2009}:
\begin{equation}
\label{eq_gamma_rad}
\Gamma(t) \cong 2.3 \left(\frac{1+z}{t_{day}}\right)^{3/7} \left(\frac{E_{52} }{n_{0} \Gamma_2}\right)^{1/7}\text{,}
\end{equation}  
where $\Gamma$ is the initial Lorentz factor of the outflow. 

\subsubsection{Electron distribution}

The shocks, created by the interaction of the relativistic outflow with the ISM, accelerate the electrons in a power law between the Lorentz factors $\gamma_{min}$ and $\gamma_{max}$. The comoving distribution function of the shocked electrons is assumed to be:
\begin{align}
\label{eq_electron_spect}
\frac{dN_e(\gamma) }{ d\gamma} & \propto \gamma^{-p} & \gamma_{min} \leq \gamma \leq \gamma_{max}\text{,}
\end{align}  
where $p$ is the electron injection index.

Since the fraction of internal energy given to the electrons $\epsilon_e$ is prescribed, and since the electron distribution function gives the kinetic energy of the electrons by integration, the minimum electron Lorentz factor can be computed and is given by:  
\begin{equation}
\label{eq_gammamin}
\gamma_{min} = \epsilon_e \ \frac{m_p }{ m_e} \ \frac{p-2 }{p-1} \ \Gamma\text{,}
\end{equation}  
when assuming that all electrons are accelerated, $\gamma \gg 1$, $p > 2$ and $\gamma_{max} \gg \gamma_{min}$, and where $m_p$ and $m_e$ are the mass of a proton and of a electron respectively, and $\Gamma$ is the Lorentz factor of the outflow. 

The electrons gain energy by the second-order Fermi acceleration process while they lose their energy by radiating synchrotron photons. To compute $\gamma_{max}$, the acceleration rate is set equal to the radiation-loss rate, which yields:
\begin{equation}
\label{eq_gammamax}
\gamma_{max} = 2 \times 10^8 \frac{ \epsilon_{max}^{1/2}}{\epsilon_B^{1/4} \ n_0 \ \Gamma^{1/2}} \text{,}
\end{equation} 
where $\epsilon_{max}$ is a constant in the order of the unity, (see Equation 11.44 in \cite{dermer2009}).

Finally, the cooling Lorentz factor $\gamma_c$ corresponds to the Lorentz factor for which the rate of energy lost by synchrotron emission is equal to the rate of energy lost by adiabatic cooling:
\begin{equation}
\label{eq_gammac}
\gamma_{c} = \frac{9 \ m_e \ (1+z)}{128 \ m_p \ \sigma_T \  \epsilon_B \ n_0 \ c \ \Gamma^3 \ t_{day}} \text{,}
\end{equation} 
where $\sigma_T$ is the Thompson scattering cross section, $c$ is speed of light and $t$ is the observer frame dynamical time. 

Taking into account all different cooling rates, the electron distribution function is approximated by: 
\begin{equation}
\label{eq_electron_dist}
\frac{dN}{d\gamma} \propto  \gamma_0^{s-1} \left \{
\begin{aligned}
&\gamma^{-s} & \text{for ~~} \gamma_0 \leq \gamma \leq \gamma_1 \text{,} \\ 
&\gamma_1^{-s} \left (\frac{\gamma}{\gamma_1} \right)^{-(p+1)} & \text{~~~~~~~~~~for ~~} \gamma_1 \leq \gamma \leq \gamma_{max} \text{.}
\end{aligned}
\right.
\end{equation} 

\begin{figure}[!ht]
\begin{center}
\includegraphics[width=0.45\textwidth]{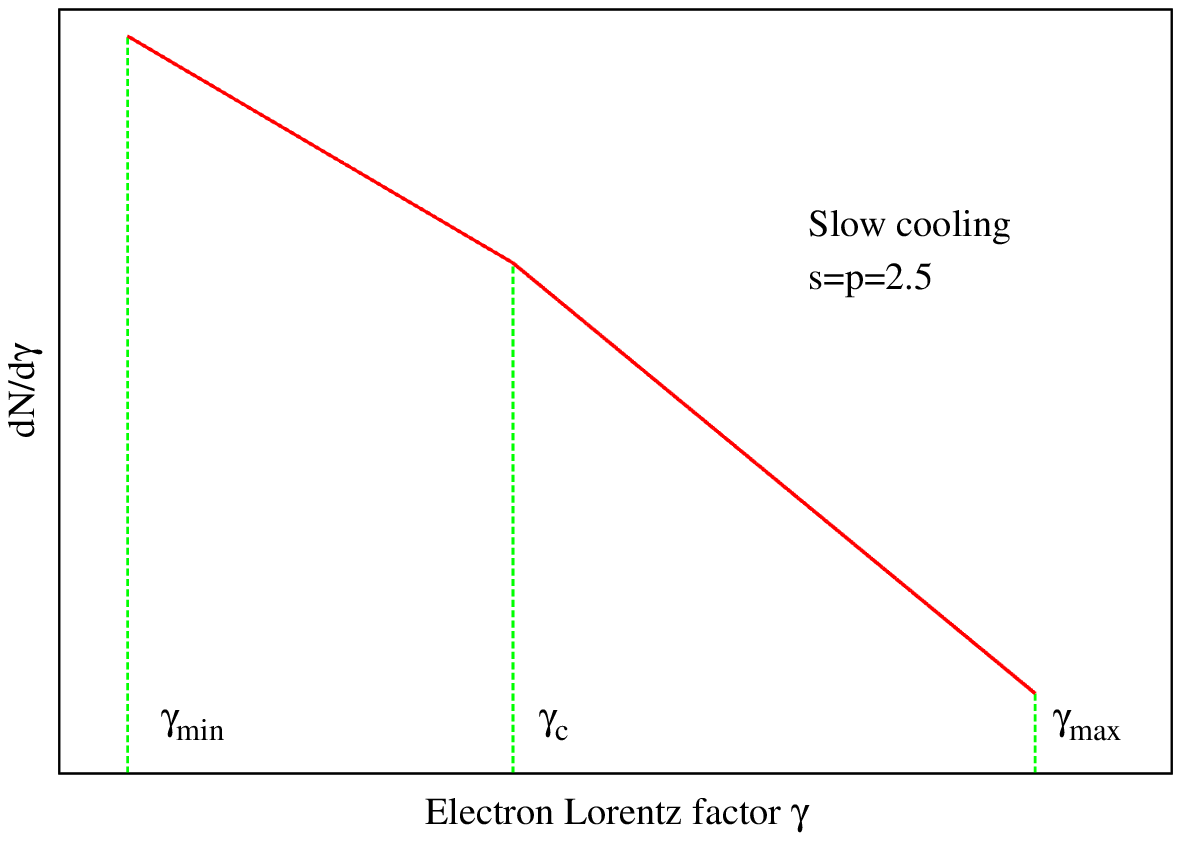}
\includegraphics[width=0.45\textwidth]{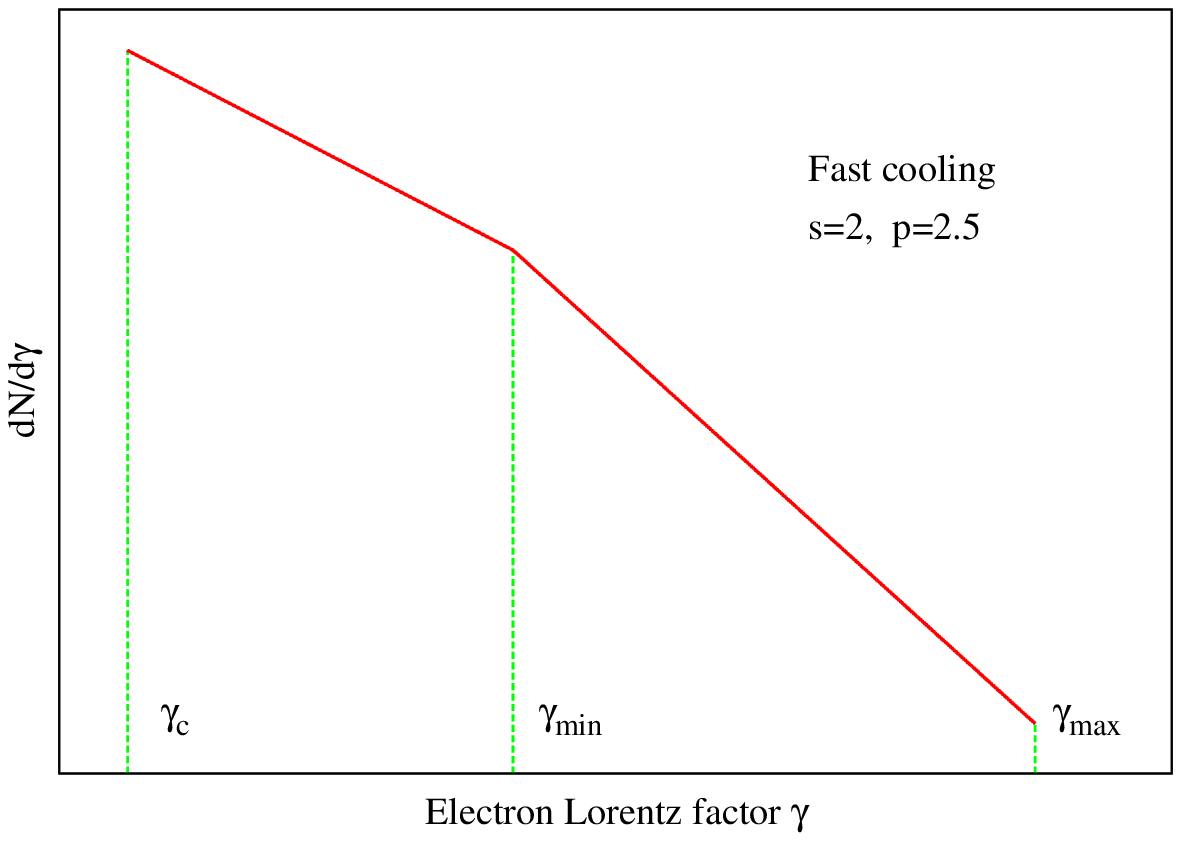}
\caption{Electron distribution function for the slow (left) and fast-cooling (right) assuming $p$ = 2.5. \label{slow_fast_gamma}}
\end{center}
\end{figure}

In the so-called $\textit{slow-cooling regime}$, the magnetic field is weak so that the electrons do not cool below $\gamma_{min}$ by emitting synchrotron radiation: the cooling affects only the electrons in the high-energy tail of the distribution. In this case, $\gamma_0 = \gamma_{min}$, $\gamma_1 = \gamma_c$ and $s = p$. Conversely, if the magnetic field is strong enough, the electrons are cooled by synchrotron radiation to Lorentz factor smaller than $\gamma_{min}$ on the dynamical time scale of the system. This regime, called $\textit{fast-cooling regime}$, is characterized by $\gamma_0 = \gamma_{c}$, $\gamma_1 = \gamma_{min}$ and $s = 2$. Both regimes are summarized on Figure \ref{slow_fast_gamma}. 

\subsubsection{Synchrotron emission}

The observed characteristic synchrotron frequency is given by \cite{rybicki1979}:
\begin{equation}
\label{eq_nu0}
\nu = \frac{e \ B \ \gamma^2 \ \Gamma}{2 \ \pi \ m_e \ c \ (1+z)}\text{,}
\end{equation} 
where $e$ is the elementary charge, $\gamma$ is the Lorentz factor of the electron, $B$ is the comoving magnetic field and $\Gamma$ is the Lorentz factor of the outflow. 

Using Equation \ref{eq_nu0}, the cooling frequency, $\nu_c$, is obtained from the corresponding cooling Lorentz factor $\gamma_{c}$ for an adiabatic blast wave: 
\begin{equation}
\label{eq_nuc}
\nu_{c} = (9 \times 10^{12}\text{Hz}) (1+z)^{-1/2} \epsilon_{B}^{-3/2} n^{-1} E_{52}^{-1/2} t_{d}^{-1/2}\text{,}
\end{equation} 

The typical synchrotron frequency (or injection frequency), $\nu_m$, is calculated by following the same way and considering $\gamma_{min}$: 
\begin{align}
\label{eq_num}
\nu_{m} = (6 \times 10^{15} \text{Hz}) (1+z)^{1/2}  \left (\frac{p-2}{p-1}\right)^2 \epsilon_{e}^{2} \ \epsilon_{B}^{1/2} \ E_{52}^{1/2} \ t_{d}^{-3/2}\text{,}
\end{align} 

Another characteristic frequency is introduced: $\nu_a$ is the transition frequency below which the photons are absorbed by synchrotron self-absorption. It is called the synchrotron self-absorption frequency. Its evolution is given by $\nu_a \propto t^{-1/2}$ for an adiabatic blast wave and $\nu_a \propto t^{-4/5}$ for a radiative blast wave in the fast-cooling regime. However, the slow-cooling regime is characterized $\nu_a \propto const$ in the adiabatic case and $\nu_a \propto t^{-3/35}$ in the radiative case. As an example, for adiabatic expansion in the slow-cooling regime, $\nu_a$ is given by:
\begin{equation}
\label{eq_nua}
\nu_{a} = (2 \times 10^{9}\text{Hz}) (1+z)^{-1} \epsilon_{e}^{-1}  \epsilon_{B}^{1/5} n^{3/5} E_{52}^{1/5} \text{,}
\end{equation} 

The synchrotron spectrum is composed of four segments separated by the typical frequencies, $\nu_c$, $\nu_m$ and $\nu_a$. In the slow-cooling regime corresponding to $\nu_m < \nu_c$, the spectrum is given by \cite{meszaros2006}: 
\begin{equation}
\label{eq_fnu_slow}
F_{\nu} = F_{\nu,max} \left \{
\begin{aligned}
&(\nu_a / \nu_m)^{1/3} (\nu / \nu_a)^2 &  ~~~~~~~~~~ \nu < \nu_a \text{,} \\ 
&(\nu / \nu_m)^{1/3} & ~~~~~~~~~~ \nu_a \leq \nu < \nu_m \text{,} \\
&(\nu / \nu_m)^{-(p-1)/2} & ~~~~~~~~~~ \nu_m \leq \nu < \nu_m \text{,} \\
&(\nu_c / \nu_m)^{-(p-1)/2} (\nu / \nu_c)^{-p/2}& ~~~~~~~~~~ \nu_c \leq \nu < \nu_{max} \text{,}\\
\end{aligned}
\right.
\end{equation} 
where $F_{\nu,max} = (20 mJy) (1+z) \epsilon_B^{1/2} E_{52} \ d^{-2}_{L,28}$ and $\nu_{max}$ is the maximum synchrotron frequency computed from $\gamma_{max}$. It is also usually assumed that $\nu_a < \text{min}(\nu_m, \nu_c)$.

In the fast-cooling regime corresponding to $\nu_m > \nu_c$ , the spectrum is given by \cite{meszaros2006} 
\begin{equation}
\label{eq_fnu_fast}
F_{\nu} = F_{\nu,max} \left \{
\begin{aligned}
&(\nu_a / \nu_c)^{1/3} (\nu / \nu_a)^2 &  ~~~~~~~~~~ \nu < \nu_a \text{,} \\ 
&(\nu / \nu_c)^{1/3} & ~~~~~~~~~~ \nu_a \leq \nu < \nu_c \text{,} \\
&(\nu / \nu_c)^{-1/2} & ~~~~~~~~~~ \nu_c \leq \nu < \nu_m \text{,}\\
&(\nu_m / \nu_c)^{-1/2} (\nu / \nu_m)^{-p/2}& ~~~~~~~~~~ \nu_m \leq \nu < \nu_{max} \text{.} \\
\end{aligned}
\right.
\end{equation} 

\begin{figure}[!ht]
\begin{center}
\includegraphics[width=0.45\textwidth]{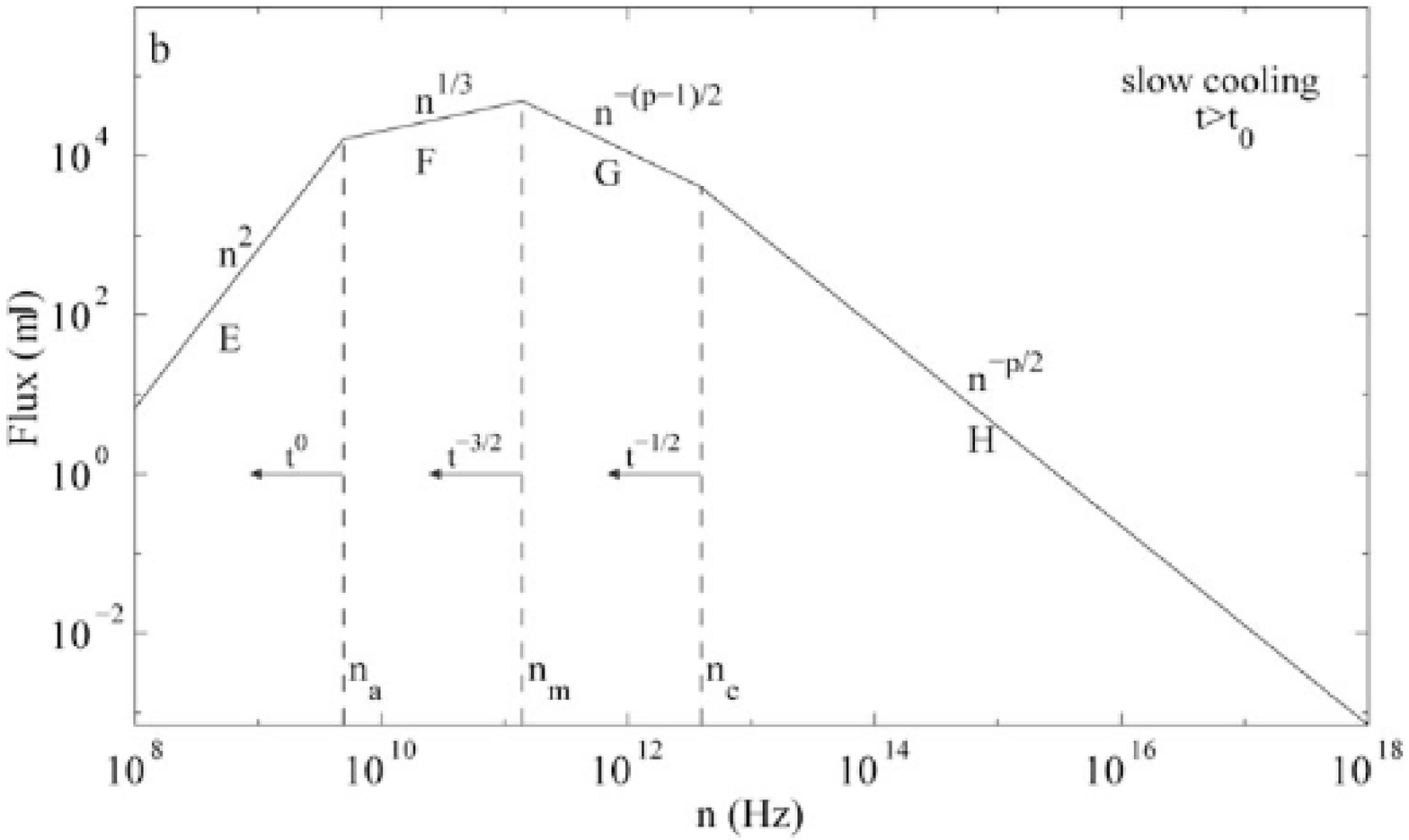}
\includegraphics[width=0.45\textwidth]{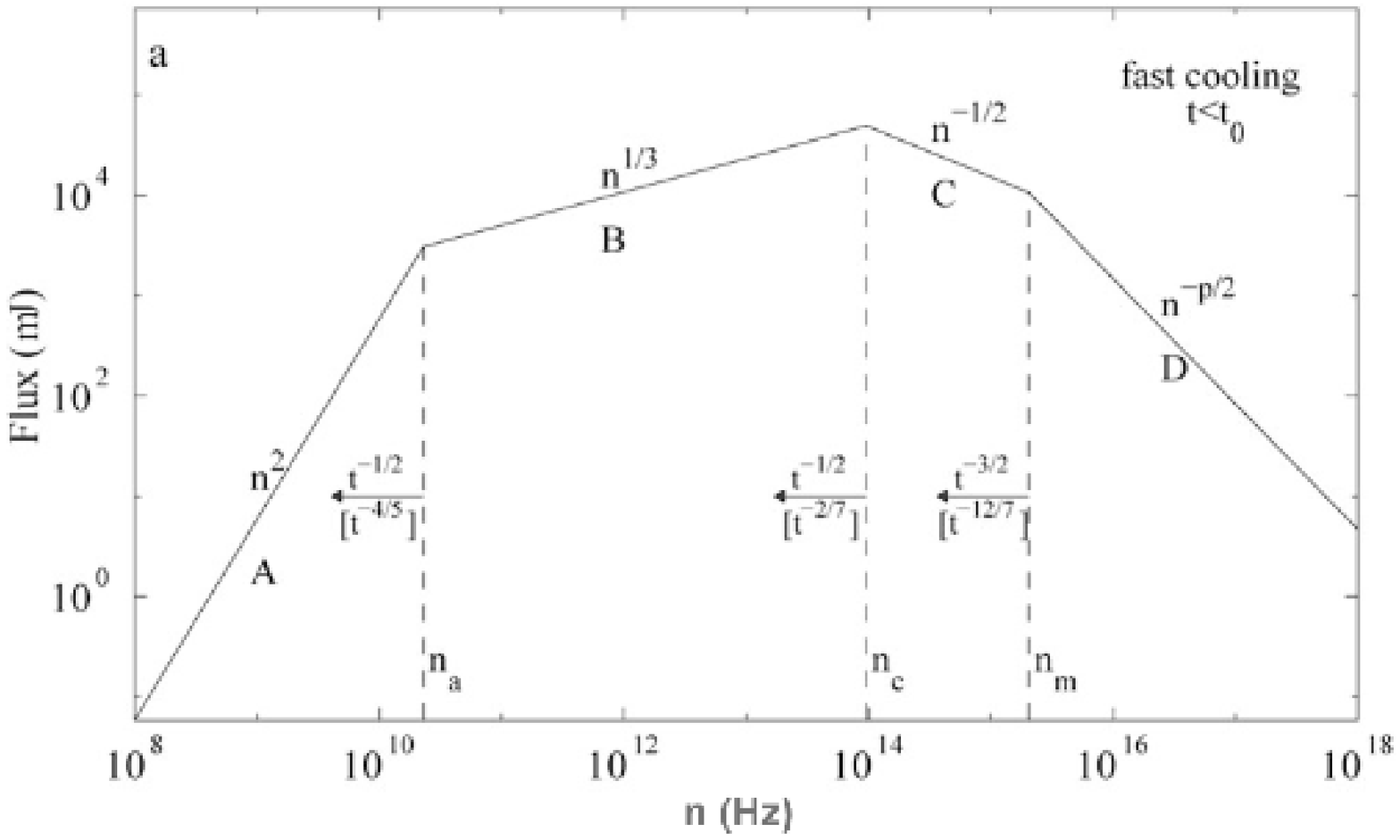}
\caption{Synchrotron spectrum in the two regimes: slow-cooling (left) and fast-cooling (right) \cite{sari1998}. \label{spectrum_fast_slow}}
\end{center}
\end{figure}

The spectra for the two regimes are illustrated in Figure \ref{spectrum_fast_slow}. 

In the dynamical evolution, the outflow goes from initially fast to slow-cooling regime. The light-curves at a given frequency $\nu_*$ can be computed by considering the time $t_0$ which is the transition time between the fast and slow-cooling regimes. It is computed by setting $\nu_c = \nu_m$ and is given by \cite{meszaros2006}:
\begin{equation}
\label{eq_t0}
t_0 = \left \{
\begin{aligned}
&210 \ \epsilon_B^{2} \ \epsilon_e^{2} \ E_{52} \ n_1 \ \text{days,}&  ~~~~~~~~~~ ~~~~~~~~~~ \text{adiabatic,}  \\ 
&4.6 \ \epsilon_B^{7/5} \ \epsilon_e^{7/5} E_{52}^{4/5} \gamma_2^{-4/5} n_1^{3/5} \ \text{days,}& ~~~~~~~~~~ ~~~~~~~~~~ \text{radiative.} 
\end{aligned}
\right.
\end{equation} 

The corresponding frequency $\nu_0$ is such that $\nu_0 = \nu_c (t_0) = \nu_m (t_0)$ and is given by \cite{meszaros2006}: 
\begin{equation}
\label{eq_nu_0}
\nu_0 = \left \{
\begin{aligned}
&1.8 \times 10^{11} \epsilon_B^{-5/2} \epsilon_e^{-1} E_{52}^{-1} n_1^{-3/2} \ \text{Hz,}&  ~~~~~~~~~~ \text{adiabatic,}  \\ 
&8.5 \times 10^{12} \epsilon_B^{-19/20} \epsilon_e^{-2/5} E_{52}^{-4/5} \gamma_2^{4/5} n_1^{-11/10} \ \text{Hz,}& ~~~~~~~~~~ \text{radiative.} 
\end{aligned}
\right.
\end{equation} 

\begin{figure}[!ht]
\begin{center}
\includegraphics[width=0.45\textwidth]{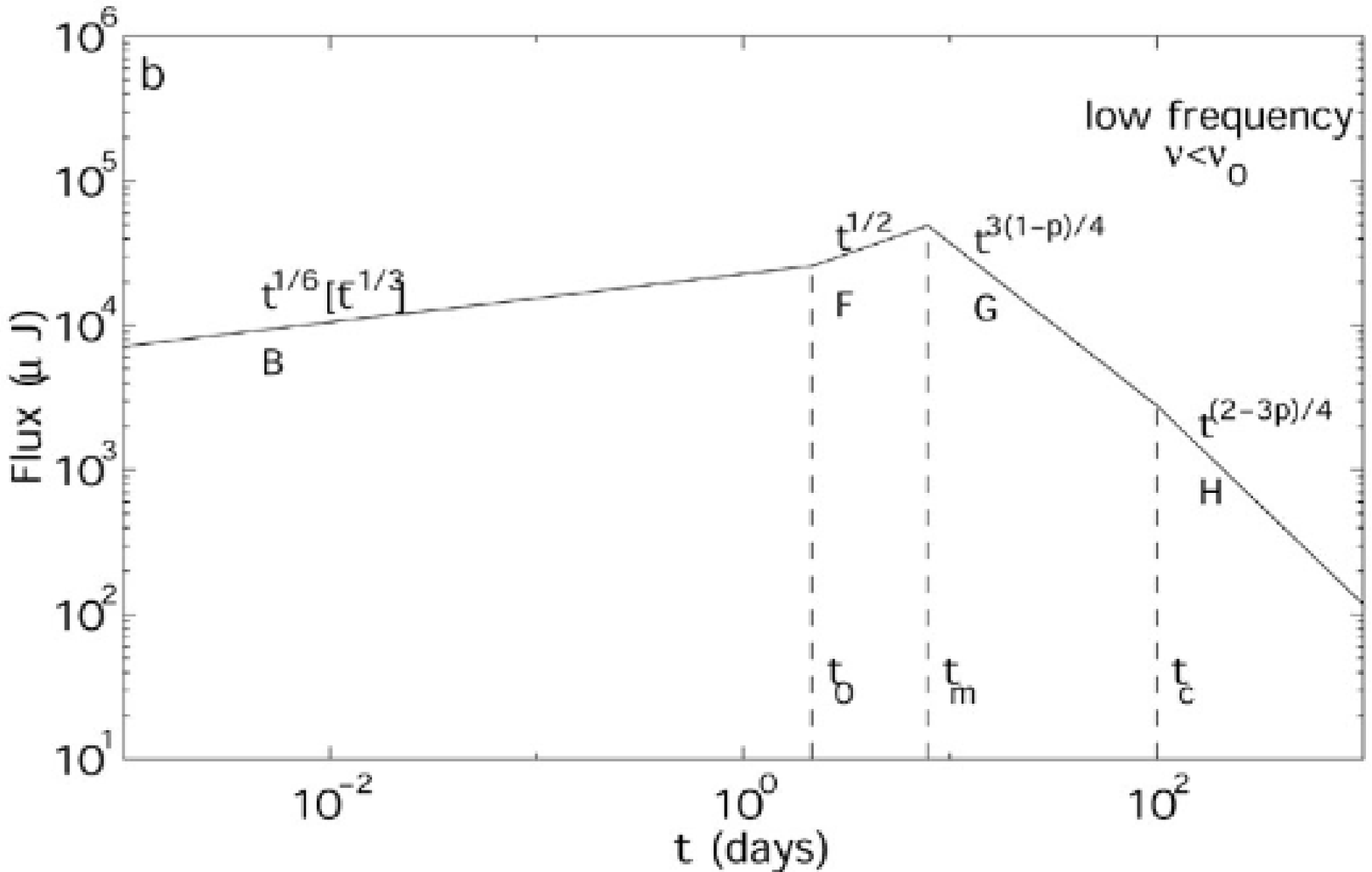}
\includegraphics[width=0.45\textwidth]{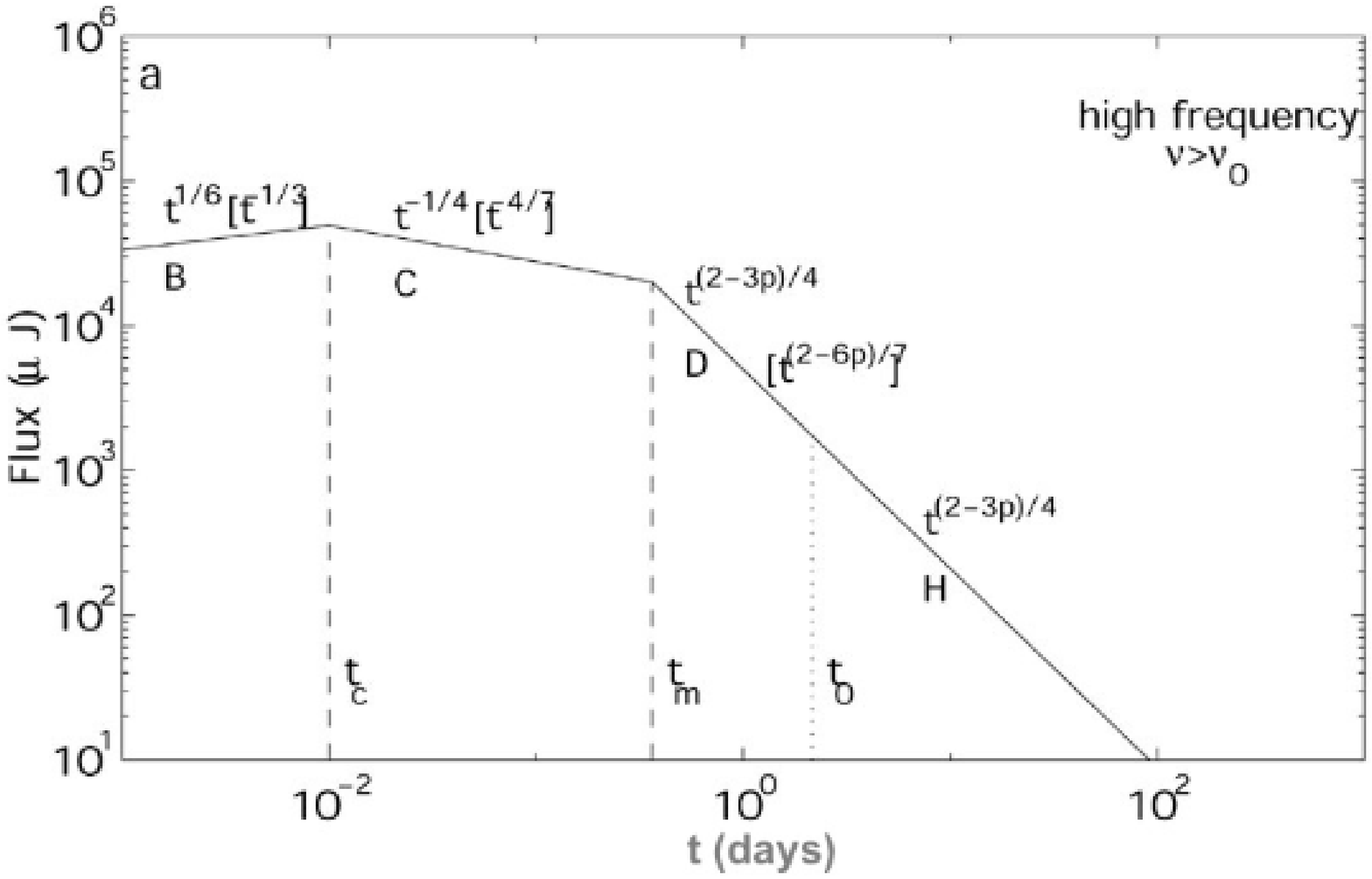}
\caption{Synchrotron light-curves in the two regimes: low frequency (left) and high frequency (right) \cite{sari1998} \label{LC_low_high_nu}.}
\end{center}
\end{figure}

Neglecting synchrotron self-absorption, there are two possibilities: $\nu_* > \nu_0$ or $\nu_* < \nu_0$. The first possibility (referred to as high-frequency light-curve) implies $t_0 > t_m > t_c$, where $t_m$ and $t_c$ correspond to the times when $\nu_* = \nu_m$ and $\nu_* = \nu_c$ respectively. The light-curve at frequency $\nu_*$ is composed of four segments, respectively:
\begin{itemize}
\item $\nu_* < \nu_c$ (labeled B, fast-cooling), 
\item $\nu_c <\nu_* < \nu_m$ (labeled C, fast-cooling),
\item $\nu_* > \nu_m > \nu_c$ (labeled D, fast-cooling),
\item $\nu_* > \nu_c > \nu_m$ (labeled H, slow-cooling: because $\nu_c > \nu_m$).
\end{itemize}
The second possibility, $\nu_* < \nu_0$, referred to as low-frequency light-curve implies $t_0 < t_m < t_c$. In a similar way, the light-curve at frequency $\nu_*$ is composed of four segments, respectively:
\begin{itemize}
\item $\nu_* < \nu_c < \nu_m$ (labeled B, fast-cooling),
\item $\nu_* < \nu_m < \nu_c$ (labeled F, slow-cooling),
\item $\nu_m < \nu_* < \nu_c$ (labeled G, slow-cooling),
\item $\nu_m < \nu_c < \nu_*$ (labeled H, slow-cooling).
\end{itemize}
The light-curve is represented in Figure $\ref{LC_low_high_nu}$, taken from \cite{sari1998}.

In all calculations above, it is assumed that the surrounding medium is a uniform ISM. 
In the case of a wind, which corresponds to $n \propto r^{-2}$, all equations can be established in a similar way. The shape of the spectrum is unchanged, only the time evolution of the cooling, injection and absorption frequencies are changed \cite{panaitescu2000}:
 \begin{equation}
\label{Eq:coolingfreqwind}
\nu_c = 1.1 \times 10^{14} \ E_{52}^{1/2} A_*^{-2} \epsilon_{B,-2}^{-3/2} \ T_d^{1/2} \ {\rm Hz} \text{,} 
\end{equation}
where $E_{52}$ is the isotropic energy in units of $10^{52}$ ergs, $A_*$ is the initial number density in the wind, $T_d$ is the time expressed in days after the burst.
 
\begin{align}
\nu_{m} & = 0.6 \times 10^{13}E_{52}^{1/2} \ \epsilon_{e,-1}^{2} \ \epsilon_{B,-1}^{-3/2} \ T_d^{1/2} \ {\rm Hz} \text{,} \label{Eq:SSabsorptionfreqwind} \\
\nu_a & = \left \{
\begin{aligned}
&3.5 \times 10^{12} E_{52}^{-2/5} A_*^{11/5} \ \epsilon_{B,-2}^{6/5} \ T_{d,-2}^{-8/5} \ {\rm Hz} &  ~~~~~~~~~~ \text{radiative,}  \\ 
&9.3 \times 10^{9} E_{52}^{-2/5} A_*^{6/5} \epsilon_{e,-1}^{-1} \ \epsilon_{B,-2}^{1/5} \ T_{d}^{-3/5} \ {\rm Hz} & ~~~~~~~~~~ \text{adiabatic.} 
\end{aligned}
\right.
\end{align} 

Being able to distinguish between two environments is important to determine the nature of the progenitor of GRBs.

\paragraph{Closure relations:}
The broadband spectrum and corresponding light-curve of synchrotron radiation can be derived from a power-law distribution of electrons accelerated by relativistic shocks as explained above. The power-law parameters of the spectrum and the light-curve can be obtained by modeling the flux $F_\nu$ as:

 \begin{equation}
 \label{Eq_Fnu_alpha_beta}
F_\nu \propto t^{-\alpha} \nu^{-\beta} \text{,}
\end{equation}
where $t$ is the observer time, $\nu$ is the observer frequency, $\alpha$ and $\beta$ are respectively the temporal and the spectral indexes, see Equation \ref{eq_fnu_slow} and Equation \ref{eq_fnu_fast}.

These spectral and temporal decay indexes can be combined into several closure relations for synchrotron emission \cite{meszaros1998, sari1998, sari1999a, chevalier2000, zhang2004}. These relations can be used to compare theoretical models with the afterglow observations in order to investigate the geometry of the burst, its surrounding medium, the microphysics of the fireball, and its cooling state. In Table \ref {table_closure}, the closure relations are recalled for an homogeneous interstellar medium, for a wind and for a jet in the slow and fast-cooling regimes, taking into account $p > $2 (standard decay index for the accelerated electrons).

\begin{table}[!ht]
\centering
  \caption{Closure relations, temporal index $\alpha$ and spectral index $\beta$ in various afterglow models where the convention $F_{\nu}\propto t^{\alpha} {\nu}^{\beta}$ is adopted \cite{zhang2004}. And validation of closure relations in X-ray and optical band. \label{table_closure}}
  \begin{tabular}{ccccc}
  \hline
                                             & $p > $2 & X-ray & Optical \\
   \hline
  ISM, slow-cooling              & $\alpha(\beta)$ & &\\
  ${\nu}_a<{\nu}<{\nu}_m$ & $\alpha$ = 1/2, $\beta$ = 1/3  & no & yes \\
  ${\nu}_m<{\nu}<{\nu}_c$   &$\alpha=3\beta/2$ & yes& yes \\
  ${\nu}>{\nu}_c$                 &$\alpha=(3\beta+1)/2$ & yes& no\\
\hline
 ISM, fast-cooling               & & &\\
  ${\nu}_a<{\nu}<{\nu}_m$ & $\alpha$ = 1/6, $\beta$ = 1/3  & no & yes \\
  ${\nu}>{\nu}_m$               &$\alpha=(3\beta+1)/2$& yes& no\\
  \hline
   Wind, slow-cooling        & & & \\
   ${\nu}_a<{\nu}<{\nu}_m$ &$\alpha$ = 0, $\beta$ = 1/3  & no & yes \\
  ${\nu}_m<{\nu}<{\nu}_c$ &$\alpha=(3\beta-1)/2$ & yes& yes \\
  ${\nu}>{\nu}_c$               &$\alpha=(3\beta+1)/2$ & yes& no \\
  \hline
  Wind, fast-cooling          & & & \\
  ${\nu}_a<{\nu}<{\nu}_m$ & $\alpha$ = -2/3, $\beta$ = 1/3  & no & yes \\
  ${\nu}>{\nu}_m$             &$\alpha=(3\beta+1)/2$ & yes& no \\ 
  \hline
  Jet, slow-cooling            & & & \\
  ${\nu}_a<{\nu}<{\nu}_m$ &$\alpha$ = -1/3, $\beta$ = 1/3  & no & yes \\
  ${\nu}_m<{\nu}<{\nu}_c$ &$\alpha=2\beta-1$ & yes& yes\\
  ${\nu}>{\nu}_c$               &$\alpha=2\beta$ & yes& no \\
  \hline
\end{tabular}
\end{table}


~
\newpage

\section{Purpose of the Thesis}

The work presented in this thesis attempts to strengthen our understanding of the diversity of GRBs which can result from their progenitors or their environments. 

It is known and widely accepted that GRBs are associated to SNe. Unlike SNe, which are precisely classified into different types, mainly depending on the composition of the star, GRBs are only grossly classified into two sub-types depending on the duration. However, a more detailed classification would allow to better understand the diversity of GRBs and better constrain their environment(s) and progenitor(s).

 It is thought that the long GRBs originate from the core collapse of a massive star. Core collapse SNe are very diverse, depending on the mass of the progenitor, its size and the remaining elements in its outer layers. Thus, different kinds of GRBs might be determined based on the same criteria: the mass of the progenitor, its size and the remaining elements in its outer layers. Indeed, many differences can be found, which have led to the creation of groups: Ultra-long GRBs (based on the duration), X-ray flashes (based on the spectral hardness of the prompt emission) and dark bursts (based on the absorption and extinction of the afterglow light). These differences (and these groups) are also seen in the subclass of GRBs firmly associated to SNe.
 
~

Even with deep observations of bright GRBs, their origin is still unknown. In this thesis, I present the results of my work, dedicated to a better understanding of the GRB diversity. In particular, I studied the faint GRBs.
I defined a new sample of GRBs, based on the X-ray afterglow luminosity. The bursts in that sample are called low-luminosity afterglow (LLA) GRBs. The highlight was set on contrasting the properties of LLA GRBs to those of normal long GRBs.

Indeed, using data from BeppoSAX, it was discovered that the X-ray light-curves of long GRBs define several well-separated groups once the distance effects are corrected \cite{boer2000}. This was later on confirmed by extending to the samples from the XMM-Newton and Chandra data \cite{gendre2005}. Three groups were determined based on the empirical properties of the X-ray afterglows, namely group I (mean flux: $7.0 \times 10^{-12} \text{erg.s}^{-1} \text{cm}^{-2}$ at one day), group II (mean flux: $3.1 \times 10^{-13} \text{erg.s}^{-1} \text{cm}^{-2}$ at one day) and group III (all other bursts less luminous than group II events) \cite{gendre2008}. The focus was put only on groups I and II, in order to explain the observed clustering. Indeed, group III events were too few in order to perform a meaningful statistical study. However, thanks to the \textit{Swift} satellite \cite{gehrels2004}, the sample of available bursts has grown and it was possible to resume the study of  group III namely low-luminosity afterglow (LLA) GRBs \cite{dereli2014}.

~

~

Hereafter, I give the outline of my thesis. 

In Chapter 2, the data reduction processes of Type IIb SN~2004ex on the photometric and spectroscopic data (in the optical band) are presented.  The results are compared with the prototypical spectra and light-curves of type II SNe (SN~1993J, SN~2008ax) and with the light-curves of type Ib (SN~2007Y) and type Ic (SN~1994I) SNe. The spectral lines of SN~2004ex are identified by using the spectral similarities with SN~1993J. The velocity, mass and kinetic energy of the ejected H are calculated. Finally, the physical properties of SN~2004ex are computed by using the similarities with the light-curve of SN~2008ax.

 In Chapter 3, the data processing of GRBs in the X-ray band is explained. The selection method of the LLA GRBs sub-sample is given. Their spectral and decay indexes are computed. 
 
 In Chapter 4, the statistical study of LLA GRBs is presented. First, the redshift distribution is analyzed and it is found that LLA GRBs are in average closer than normal lGRBs. Second, different selection effects are discussed and it is proved that LLA GRBs do not suffer from them. Third, their afterglow properties are compared by using the closure relations in both X-ray and optical bands. Finally, the prompt properties are discussed in light of the Amati correlation.
 
 In Chapter 5, I discuss many of the other properties of LLA GRBs: rate density, host galaxy and possible progenitor. The conclusions of the thesis and the perspectives follow in Chapter 6. 


%% file: Chapter2.tex

\chapter[Observation and Data Reduction Applied to SN~2004ex]{\parbox[t]{\textwidth}{Observation and Data Reduction Applied to SN~2004ex}}
\chaptermark{Optical data reduction}

\label{Chapter_SN2004ex}

In this chapter, I will present  the photometric and spectroscopic analysis of the type IIb supernova 2004ex during the transition $(30~\text{days} \le t \le60~\text{days})$ and nebular ($t \ge 60~\text{days}$ past explosion) phases. 
The purpose of this work was to complete the analysis of the data from SN~2004ex taken by the 1.82~m Copernico Telescope of Mt. Ekar, 3.5~m Telescopio Nazionale Galileo (TNG) and 2.2~m MPG/ESO telescope.

The results are compared with the prototypes of type~IIb:  SN~1993J and SN~2008ax respectively.  I find that the light-curve of SN~2004ex is very similar to the two prototypes light-curves, however it is closer to the one of SN~2008ax. On the other hand, its helium lines (He~I $\lambda$ = 4394, 6561, 6914, 7168~$\angstrom$) are still well-detected 4 weeks after the explosion. However, the hydrogen (H$_{\alpha}$) at 6286~$\angstrom$ is weaker than the one found for SN~1993J at a similar period.

\section{The intermediate characteristics of type II SNe}
Different kinds of SNe can be identified by performing spectroscopic analysis and considering the elements synthesized during the SN evolution and explosion. An other criterion is about the brightness of the SN which can be obtained by performing a photometric analysis. 

SNe were classified into type I and type II according to the absence or the presence of H lines respectively \cite{minkowski1941}. Type II SNe only occur in spiral galaxies. They are thought to be the result of the core collapse of massive young stars \cite{barbon1999}. The progenitors of core-collapse SNe are massive stars, either single or in binary systems, which have completed the nuclear burning stage. Different types of core-collapse SNe (IIb, IIP, IIL, Ib, Ic) have been identified based on their spectroscopic and photometric properties, forming a sequence explained by the progenitor mass loss history.

Recently, several SNe with intermediate characteristics have been discovered, suggesting a smooth transition and requiring the introduction of the hybrid classes  type IIb, Ic and Ib/c SNe, the latter being associated with GRBs \cite{tiengo2003}. 
The spectra of IIb SNe are dominated by H lines at early times while He lines become prominent at late times.
On the one hand, spectra of type Ic SNe show neither He~I lines nor H$_\alpha$,  while the spectra of Ib SNe show HeI lines and no H$_\alpha$.

Furthermore, the similarity between the light-curves of type~IIb with those of type Ib events \cite{drout2011}, in addition to the known spectral similarities at late times and the similar peak radio luminosities also suggests that these two types of events might come from similar progenitor systems. Their association with GRBs also supports the idea that SNe~Ic and Ib/c are related. Thus, the intermediate characteristics of type-IIb~SNe supports the idea of a continuous sequence of SNe having different envelope mass (\textit{i.e.} types II, IIb, Ib, Ib/c, Ic \cite{dellaValle2005}). 

However,  it was investigated with the type~IIb and type~Ib that they could be associated with interacting binaries \cite{arcavi2012}. This idea was also studied by several other authors:  Blinnikov et al. \cite{blinnikov1998} showed the relevance of this possibility by the modelling of SN~1993J and Bersten et al. \cite{bersten2012} tested this idea on SN~2011dh by using single and binary progenitors modellings of type IIb~SN. This issue is also widely discussed by Dessart et al. \cite{dessart2011, dessart2012} who simulated binary-star models for the production of SNe~IIb/Ib/Ic and proposed that the progenitors of SNe~IIb and Ib should have main-sequence masses smaller than 25 $M_{\odot}$ and be in a binary system whose stars are close to each other. However, SNe~Ic should be the result of a more massive single star because of the lack of He~I lines in their spectra.

\section{Introduction of SN~1993J and SN~2008ax: comparative sample}

\subsection{Properties of SN~2008ax}
SN~2008ax has been discovered just 18 days before its maximum peak in the NGC~4490 at redshift 0.001855. It shows a rapidly declining light-curve up to 40~days.
The amount of $^{56}$Ni synthesized in the explosion is between 0.07 and 0.15~$M_{\odot}$. The kinetic energy of SN~2008ax estimated to be $1- 6 \times 10^{51}$~ergs, while its total ejecta mass is $2 - 5~M_{\odot}$. The ejecta velocity is in the range of 23~000 - 26~000 km.s$^{-1}$. The ejected element H$_{\alpha}$ has very high early-time velocity of 13~500~km.s$^{-1}$, which is rapidly decreasing until the day 14 in velocity evolution (for more information, see \cite{pastorello2008, chornock2011}).

From the spectroscopic and photometric evolution of SN~2008ax, the progenitor constrained to be a W-R star of type WNL, which is either a single massive W-R star or a lower-mass W-R star (main-sequence mass 10 -14~$M_{\odot}$) in an interacting binary system \cite{pastorello2008, chornock2011}.

\subsection{Properties of SN~1993J}

The  first ever SN~IIb event identified was SN~1987K \cite{filippenko1988};  another more recent example is SN~1993J which is a nearby event observed in the M81/NGC 3031 galaxy at redshift 0.000113 \cite{barbon1995}. The light-curve of SN~1993J was unusual with a narrow peak (shock break out) followed by a secondary maximum (optically thick region), which is similar to the one of SN~1987A. After a rapid luminosity decline around 50 days after the explosion, the light-curve of SN~1993J showed almost well-fitted exponential tail (optically thin region) with a decline rate faster than normal SNe~ II and similar to that of SNe~Ia. It is remarkable that the late time spectra, except for the presence of He~I lines, appear similar in all core-collapse SNe types. 

The progenitor of SN~1993J was a supergiant star of spectral type K \cite{aldering1994}. It was the massive member of a binary system and its mass was in the range 12 to 17~$M_{\odot}$. In this binary system the two stars had comparable main sequence masses \cite{maund2004}. 

\section{Supernovae: SN~2004ex}	
\begin{figure}[!ht]
\centering
\includegraphics[width=0.45\textwidth]{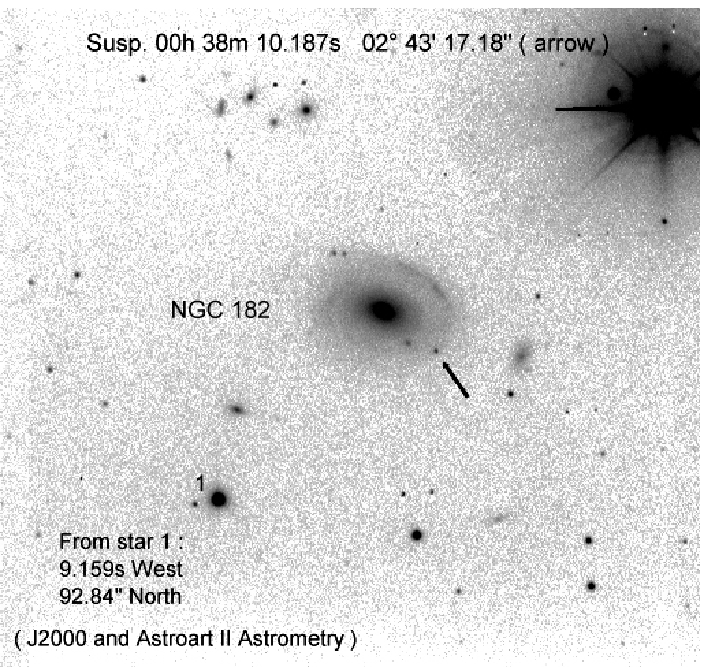}
\includegraphics[width=0.425\textwidth]{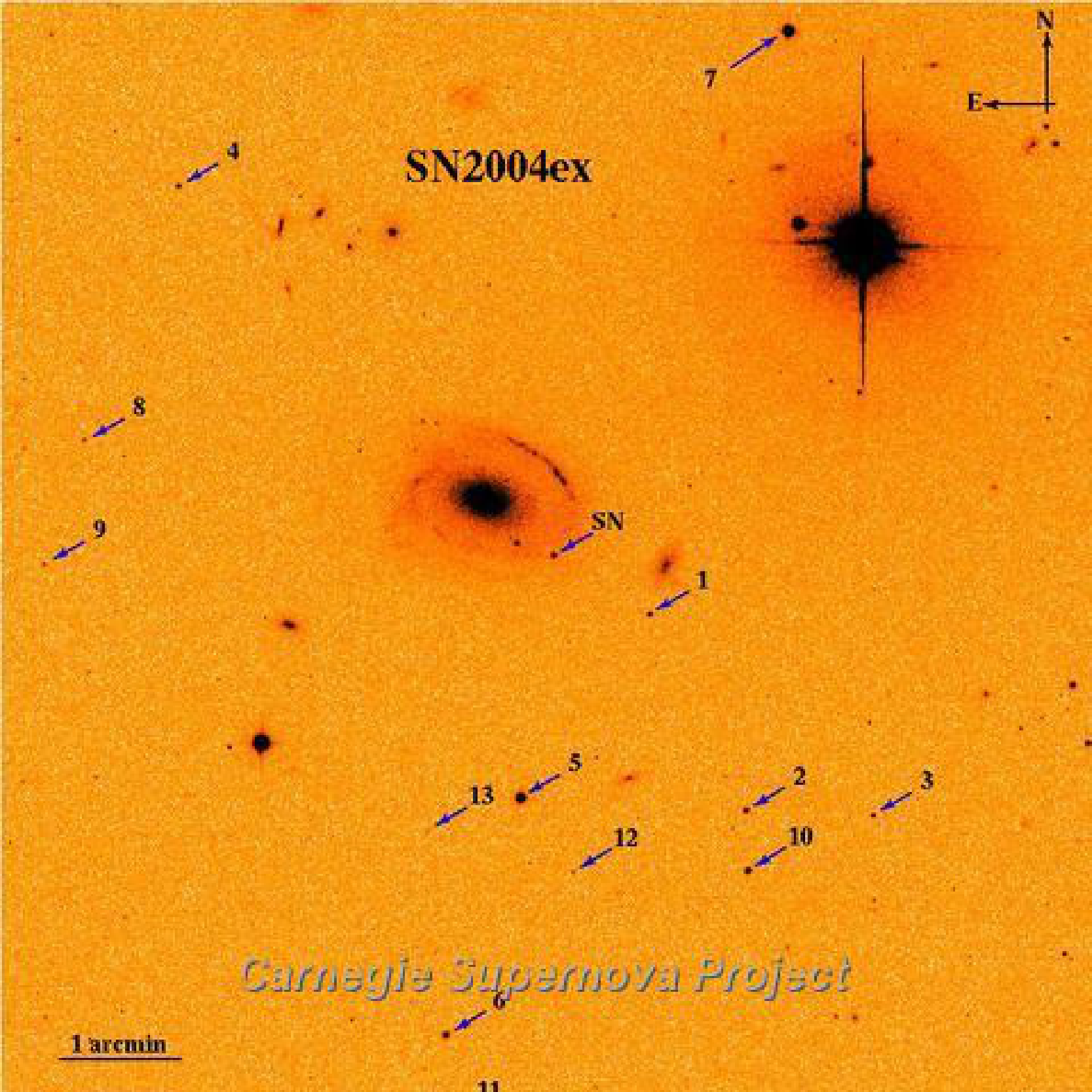} 
\caption{Optical image of SN~2004ex with its host galaxy, NGC 182, (left) and the map of possible standard stars (right) used for the magnitude calibration.
\label{NGC182_SN2004ex_standard_stars}}
\end{figure}

	SN~2004ex discovered in the galaxy NGC~182 on October 10$\textsuperscript{th}$, 11.34 UT, 2004. The SN was firstly discovered by Tenagra II, a 0.81~m telescope, with an unfiltered magnitude of about -17.7. It  was  later observed by the LOSS telescope on October 13.33, UT 2004. Its celestial coordinates are $\alpha$= 00h38m10.19s, $\delta$= +02h43m17.2s (equinox 2000.0). It is located 33'' West and 25.3'' South of the nucleus of its host galaxy  \cite{jacques2004}, which is a SBa barred spiral galaxy, at redshift $z = 0.01755$ obtained by the host galaxy spectral observation, as it can be seen in Figure \ref{NGC182_SN2004ex_standard_stars} (left).

	The optical photometric datasets (in BVRI bands) of SN~2004ex were acquired in 6 post-explosion epochs, while spectroscopic observations were performed for 4 post-explosion epochs. They were obtained using the following instruments:

(i) Device Optimized for the LOw RESolution (DOLORES), with a scale of 0.252~arcsec.px$^{-1}$ and a field of view of $8.6 \times 8.6 \text{arcmin}^{2}$, at the 3.58 m Telescopio Nazionale Galileo (TNG) (La Palma, Spain),

(ii) The Wide Field Imager (WFI), with a scale of 0.238~arcsec.px$^{-1}$ and a field of view of $34 \times 33 \text{arcmin}^{2}$, at the 2.2 m European Southern Observatory Telescope (La Silla, Chile),

(iii) The Asiago Faint Object Spectrograph and Camera (AFOSC), with a scale of 0.46~arcsec.px$^{-1}$, a field of view of $7.8 \times 7.8 \text{arcmin}^{2}$ and range 355-780~nm, resolution 2.4~nm, mounted on the 1.82~m Copernico Telescope of Mt. Ekar (Asiago, Italy).

The observations and the data reductions are presented in the subsection \ref{sec_photo} for photometry and \ref{sec_spect} for spectroscopy. The results are compared with two other type~IIb~SNe: SN~1993J and SN~2008ax. The conclusion follows. 

\section{Photometry}
\label{sec_photo}
\subsection{Data analysis}

	The photometric datasets were collected in the wavelength range from 3650 to 8060 $\angstrom$ which corresponds to the Johnson-Bessel filters UBVRI at all 6 epochs. Table 1 shows the results of the observations as well as the telescopes used. 
	 
	 The datasets were pre-reduced according to the classical prescriptions (\textit{i.e.} overscan, trim, bias, flat-field correction etc.) in order to remove the instrumental effects. The photometric analysis was performed by using the Queen's University supernova Belfast Archive (QUBA) semi-automatic pipeline Point Spread Function (PSF) module based on the Image Reduction and Analysis Facility (IRAF) \cite{valenti2011}. 
The dataset acquired for the photometric night of December 17$\textsuperscript{th}$, 2004 obtained by the WFI instrument, was calibrated by using  the Stetson extension of the Ru149 Landolt catalog  \cite{landolt1992, landolt2009}. A list of five local standard stars was built. They are labeled by 1, 2, 4, 10, 12 in Figure \ref{NGC182_SN2004ex_standard_stars} (right). This list was used for the calibration of the other non-photometric nights and to calculate the magnitude of the supernova by considering the contribution of the galaxy to the background. It should also be taken into account that the contribution of the galaxy is spacially rapidly changing when the SN is on one of its arm (see Figure \ref{NGC182_SN2004ex_standard_stars}). Indeed, the contribution of the galaxy is larger if the SN is on one of its arm than when it is not.
 
\subsection{The light-curve of SN~2004ex}	
	
\begin{figure}[!ht]
\centering
\includegraphics[width=0.7\textwidth]{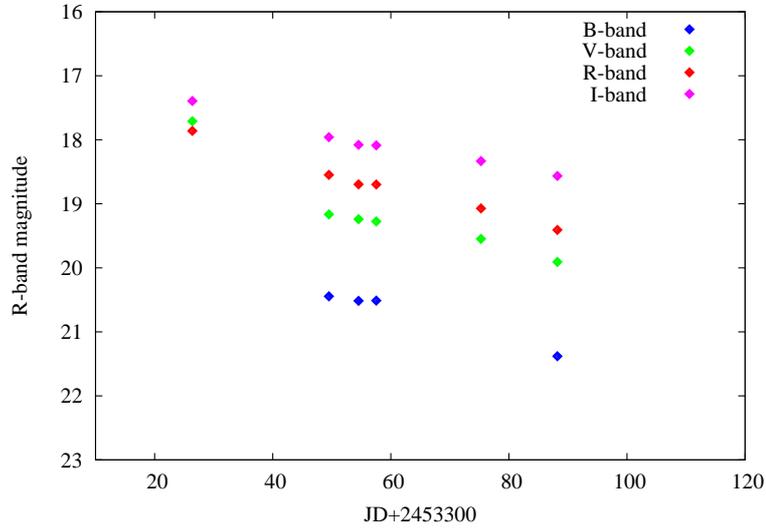}
\caption{The 4-band optical light-curve of SN~2004ex. The x-axis is in JD days and the y-axis is the apparent magnitude.
\label{mag_4band}}
\end{figure}

	The output of the PSF-fitting with the QUBA pipeline on the Johnson-Bessell (UBVRI) photometry of SN~2004ex is shown on Figure \ref{mag_4band} and is presented in Table \ref{phot_results},  together with the error on the magnitude, estimated by the PSF-fitting technique. 

\begin{table*}[!ht]
\centering
  \begin{tabular}{ccccccc}
  \hline
 Date&JD + 2453300&B&V&R&I&Inst.\\%
 \hline
       Now. 16$\textsuperscript{th}$ &26.39& &17.71$\pm$0.002&17.86$\pm$0.04&17.39$\pm$0.05&Ekar \\%
       Dec. 9$\textsuperscript{th}$ &49.52&20.45$\pm$0.04&19.17$\pm$0.03&18.55$\pm$0.02&17.96$\pm$0.02&WFI \\%
       Dec. 12$\textsuperscript{th}$ &54.55&20.52$\pm$0.03&19.24$\pm$0.03&18.7$\pm$0.03&18.08$\pm$0.02&WFI \\%
       Dec. 17$\textsuperscript{th}$ &57.54&20.51$\pm$0.03&19.28$\pm$0.03& 18.7$\pm$0.03&18.09$\pm$0.02&WFI \\%
       Jan. 1$\textsuperscript{th}$ &75.28&&19.07$\pm$0.08&19.55$\pm$0.07&18.33$\pm$0.07&Ekar\\%
       Jan. 17$\textsuperscript{th}$ &88.21&21.38$\pm$0.15&19.91$\pm$0.08&19.41$\pm$0.07&18.57$\pm$0.06&Ekar\\%
       \hline
\end{tabular}
\caption{The photometric observations. The Johnson-Bessell UBVRI photometric results were obtained with the PSF-fitting technique. \label{phot_results}}
\end{table*}

\begin{figure}[!ht]
\centering
\includegraphics[width=0.7\textwidth]{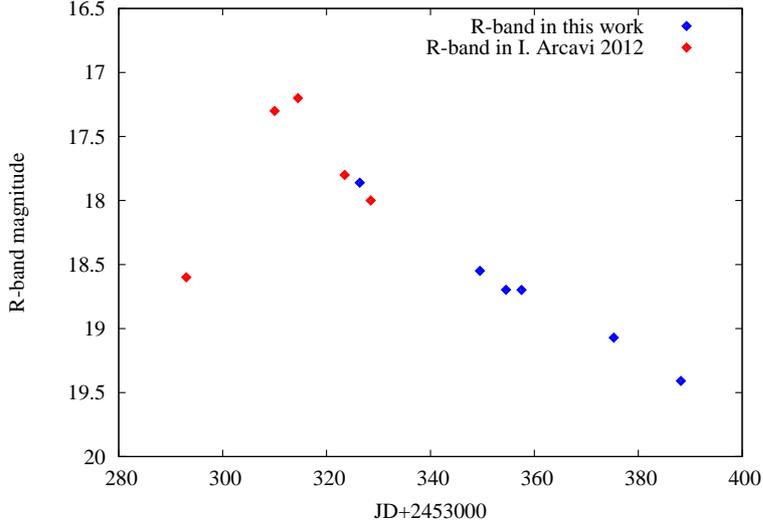}
\caption[The optical light-curve of SN~2004ex in the R-band]{The optical light-curve of SN~2004ex in the R-band. The photometric results from Arcavi et al. \cite{arcavi2012} are in red, while the results of this analysis are in blue.\label{mag_Rband}}
\end{figure}

Since data used for this analysis was taken during the transition $(30~\text{days} \le t \le60~\text{days})$ and the nebular ($t \ge 60~\text{days}$ past explosion) phases, it is not possible to obtain important properties: luminosity peak, width of the peak, the nickel mass, energy of the ejecta. However, the light-curve of SN~2004ex is complete in the R-band when adding the results from Arcavi et al. \cite{arcavi2012} obtained during the photospheric phase which is in the optically thick region ($t \le 30~\text{days}$ past explosion \cite{arnett1982}) as shown in Figure \ref{mag_Rband}. As a result, the peak magnitude is $-17.2$ and the brightness of the source decreases rapidly after 21.5~days. The explosion date is 53289~MJD (October 11$\textsuperscript{th}$, 2004).

The light-curves are usually fitted by a polynomial function of order~4. Here it is performed using the Legendre polynomials base defined as: 

\begin{equation}
f_n(x) = a_1 \ F_1(x) + a_2 \ F_2(x) + \cdots + a(n) \ F_n(x)
\end{equation}
where the $a_n$ are coefficients to be fitted for, the $F_n(x)$ are the Legendre polynomials containing terms of order x$^{(n-1)}$.
\begin{equation}
\begin{aligned}
   & F_1(x) = 1  \text{,} & \\
   & F_2(x) = x  \text{,} & \\
   & F_n(x) = \left[(2n-1) \ F_{n-1}(x) \ x - (n-1) \ F_{n-2} (x)\right]/n  \text{.} &
   \end{aligned}
\end{equation}
 
\begin{figure}[!hb]
\centering
\includegraphics[width=0.45\textwidth]{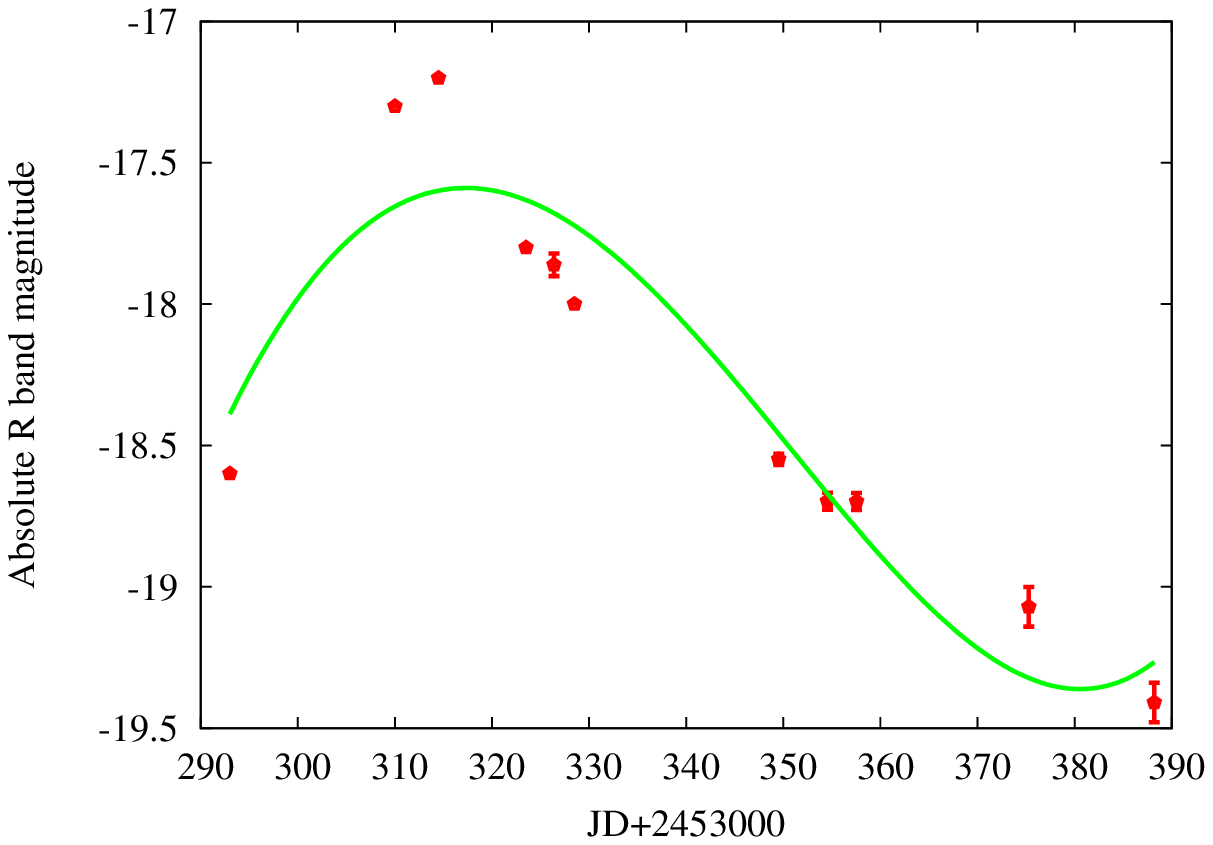}
\includegraphics[width=0.45\textwidth]{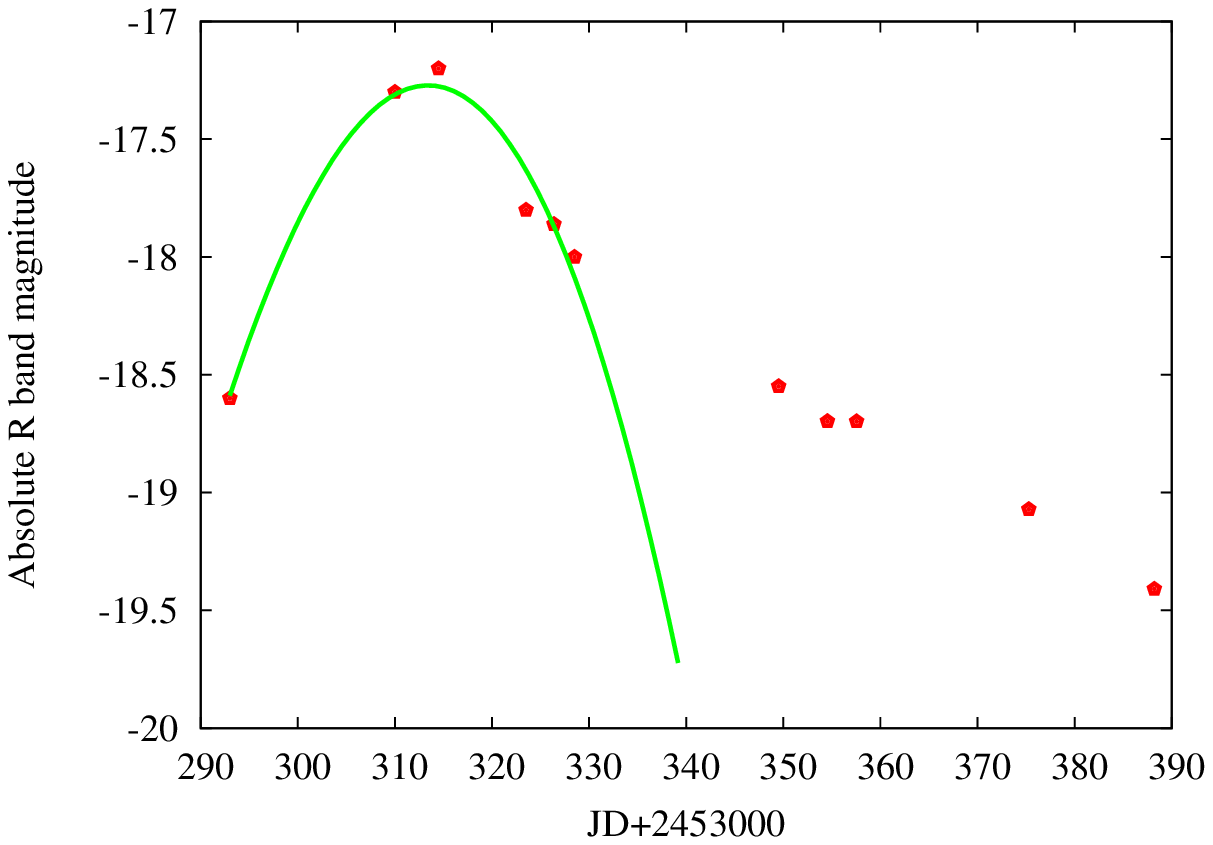}
\caption{The fitted optical light-curve of SN~2004ex. Optical R-band light-curve of SN~2004ex (red dots) together with the best fit from a Legendre function (green continuous line). To the left (right), the fourth-order (third-order) Legendre polynomial is considered.
 \label{mag_Rband_fit}}
\end{figure}

The fourth and third-order Legendre polynomials fit parameters are presented in Table \ref{parameter_Legendre_pol}. The best fit to the light-curves are shown in Figure \ref{mag_Rband_fit}. The third-order polynomial was used because the fit obtained by the fourth order was not good and especially the peak could not be properly reproduced. 
\begin{table}[!ht]
\centering
  \caption{The fourth and third-order Legendre polynomial fit parameters and maximum magnitude. \label{parameter_Legendre_pol}}
  \begin{tabular}{lcccclll}
  \hline
Legendre & $a_1$ & $a_2$ & $a_3$ & $a_4$ & Magnitude$_{max}$ \\
polynomial&&&&& mag  \\
    \hline
fourth order & -1.85$\pm$0.5 & 0.01$\pm$0.002  & (-4.3$\pm$1.2)$\times$10$^{-5}$ & (9.4$\pm$2.7)$\times$10$^{-9}$ & -17.6  \\
third order & -1.23$\pm$0.12 & 0.05$\pm$0.0005 &(-1.1$\pm$0.12)$\times$10$^{-5}$& & -17.27 \\
\hline
\end{tabular}
\end{table}
The maximum magnitude of SN~2004ex ($-17.27$~mag) obtained by the third-order Legendre polynomial is compatible with the result ($-17.2$~mag) obtained by \cite{arcavi2012}.

\subsection{The comparison with other type IIb SNe}

\begin{figure}[!ht]
\centering
\includegraphics[width=0.7\textwidth]{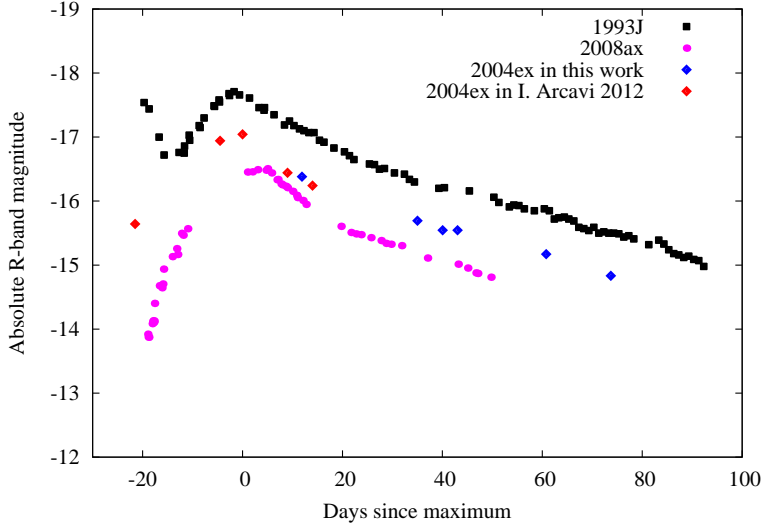}
\caption{The optical light-curves of SN~2004ex compared with the light-curves of type IIb SN~1993J \cite{richmond1994} and SN~2008ax \cite{pastorello2008} in the R-band. \label{mag_Rband_comp_IIb}}
\end{figure}

In order to better constrain the source, the light-curve is compared with the ones of two other sources: Figure \ref{mag_Rband_comp_IIb} presents the comparison of SN~2004ex  with the prototypical type~IIb SN~1993J and the well-known type~IIb SN~2008ax. These light-curves were corrected by the distance modulus which is given by
\begin{equation}
\mu = M - m = 5 log_{10}(d) - 5 + A_V
\end{equation}
where $M$ and $m$ are the absolute and the apparent magnitudes respectively and $d$ is the distance. $A_V$ is the absorption and it is given by: $A_V = 3.2 \times E(B-V)$ where $E(B-V)$ interstellar reddening and it is 0.022 for SN~2004ex. A distance estimation gives $d=70.55$ Mpc for NGC 182 (SN~2004ex) from the redshift calculation with the recession velocity of NGC~4490 (SN~2008ax) corrected for Local Group infall into the Virgo Cluster of 5240 $\text{km.s}^{-1}$ \cite{crook2007} by adopting a cosmological model with $H_o = 70$ $\text{km.s}^{-1}.\text{Mpc}^{-1}$.  The distance modulus for NGC~4490 is ${\mu}$ = 29.92$\pm$0.29 mag ($d$ = 9.6~Mpc) \cite{pastorello2008} and ${\mu}$ = 28.06 mag ($d$ = 4.11~Mpc) for M81/NGC~3031 (SN~1993J). The time was also rescaled by defining the peak time as $t_{p}$ = 0 for each SN. The comparison is made in the R-band. 

As it is shown that SN~2004ex is similar to SN~1993J regarding its spectral properties and SN~2008ax and when it comes to its photometric properties, the initial mass of its progenitor can be between 10~-~17~$M_{\odot}$. SN~2004ex is a stripped-envelope core-collapse SN. 
Because of the many similarities with SN~2008ax, it is natural to consider that the same amount of $^{56}$Ni was synthesized in the explosion (between 0.07 and 0.15~$M_{\odot}$). An upper limit for the kinetic energy of SN~2004ex can be then be estimated as being the kinetic energy of SN~2008ax: $1 - 6 \times 10^{51}$~ergs.

\begin{table}[!ht]
\centering
  \caption{The physical parameters of three SNe (SN~1993J, SN~2008ax and SN~2004ex). \label{compare_SNtype}}
  \begin{tabular}{llllllll}
  \hline
SN & Type & M$_{B,max}$ & E(B-V)$_{tot}$ & $^{56}$Ni$_{mass}$ & Ejecta mass & E$_{kinetic}$ & Ref. \\
&&mag&&~$M_{\odot}$& ~$M_{\odot}$ &10$^{51}$ erg & \\
    \hline
SN~2008ax & IIb & -17.32$\pm$0.50 & 0.4$\pm$0.1 & 0.07 - 0.15 & 2 - 5 &1 - 6 & \cite{taubenberger2011} \\
SN~2004ex & IIb & -17.71$\pm$0.002 & 0.022& $\sim 0.07 - 0.15$& $\sim 2 - 5$& $\sim 1-6$ & this work \\
SN~1993J & IIb & -17.23$\pm$0.50 &0.2 & 0.10 & 1.3 & 0.7 &\cite{richardson2006} \\
\hline
\end{tabular}
\end{table}

\subsection{The comparison with other types of SNe}

\begin{figure}[!ht]
\centering
\includegraphics[width=0.45\textwidth]{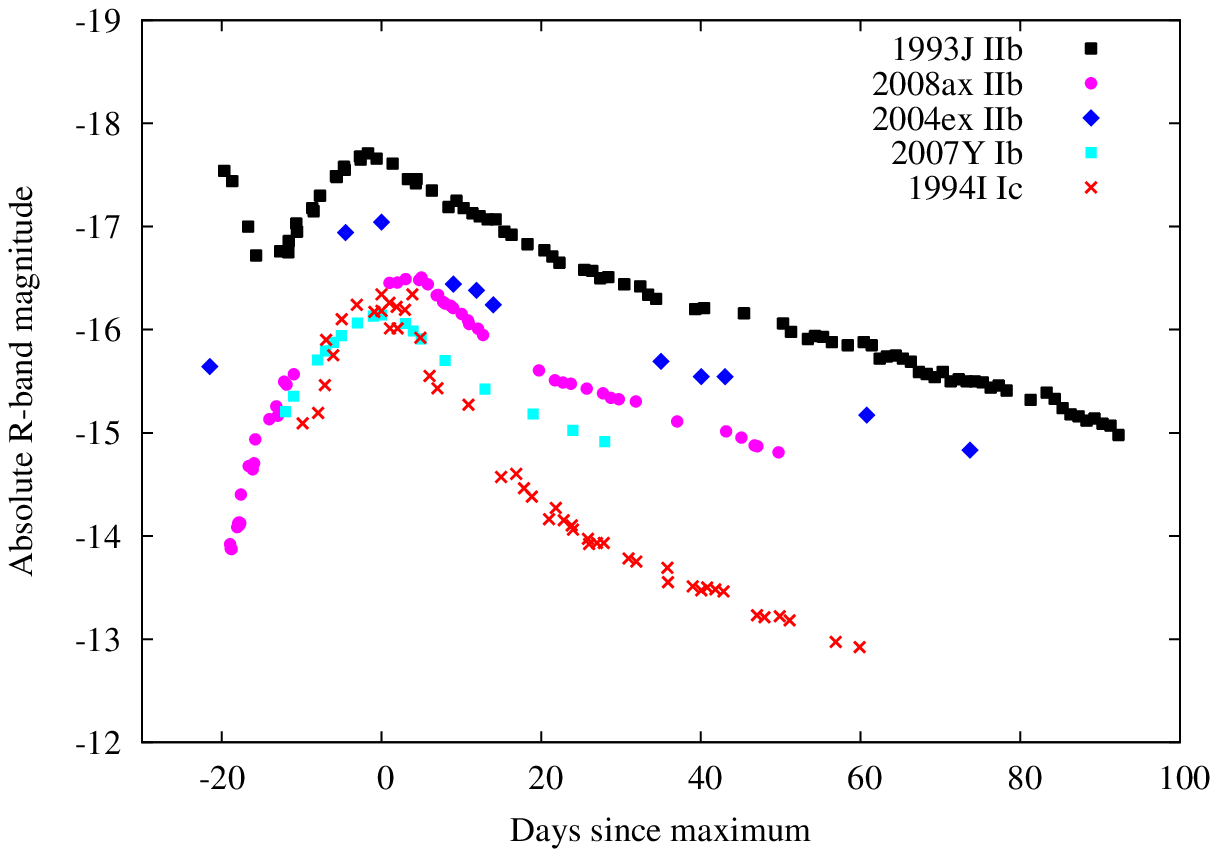}
\includegraphics[width=0.45\textwidth]{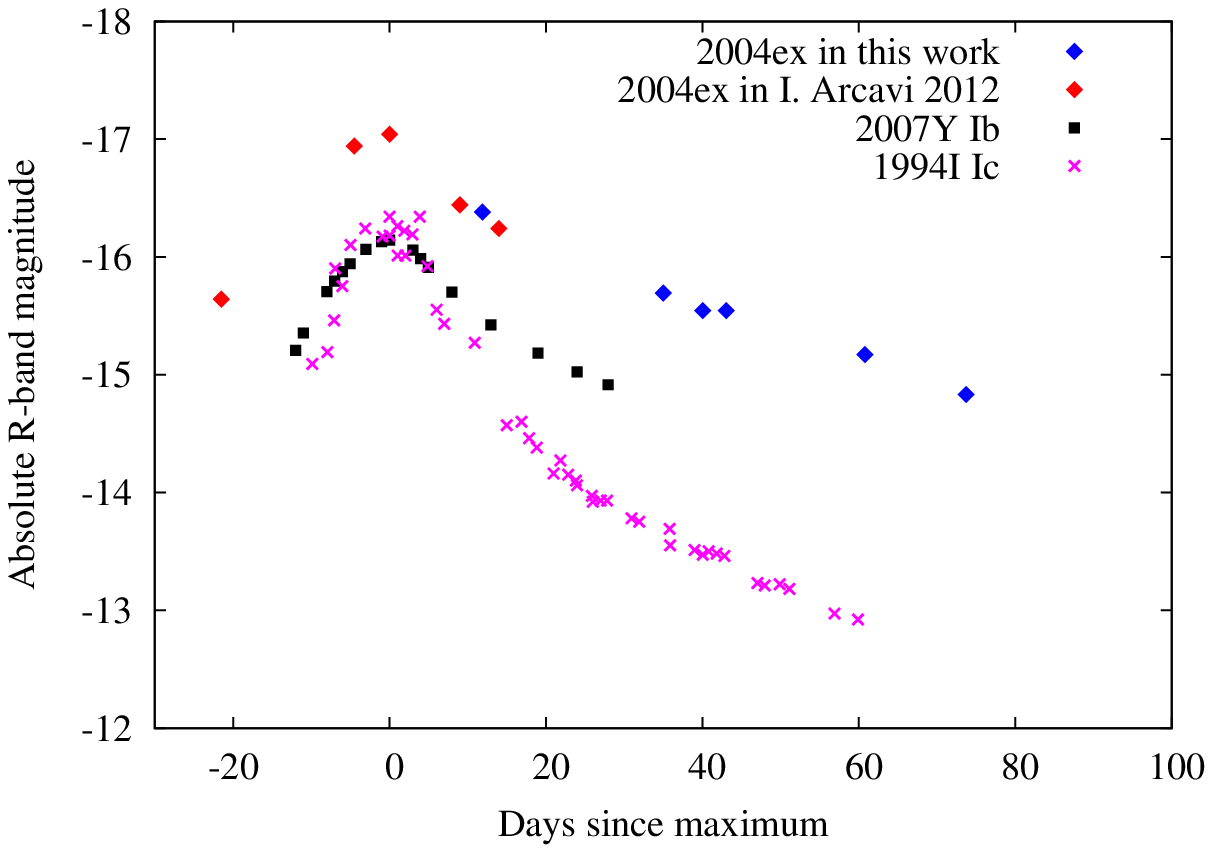}
\caption{Left: The optical light-curves of SN~2004ex (blue diamonds)(blue diamonds) compared with four light-curves of type~IIb (SN~1993J, (black squares) and SN~2008ax (magenta circles), type~Ic SN~1994I (red crosses) \cite{richmond1996} and type Ib/c SN~2007Y (cyan squares) \cite{stritzinger2009} in the R-band. Right: The optical light-curves of SN~2004ex (red and blue diamonds) compared with two light-curves of type~Ic SN~1994I (black square) \cite{richmond1996} and Ib/c SN~2007Y (magenta crosses) \cite{stritzinger2009} in the R-band. \label{mag_Rband_comp_Ibc}}
\end{figure}

In Figure \ref{mag_Rband_comp_Ibc} the light-curve of SN~2004ex is compared to the light-curves of type~Ic SN~1994I \cite{richmond1996} and Ib/c SN~2007Y \cite{stritzinger2009} \footnote{The type of SN~2007Y is still debated: type~Ib \cite{stritzinger2009} or type~Ib/c \cite{drout2011}}. The distances 6.2 \cite{lee1995} and 19.31~Mpc \cite{stritzinger2009} were adopted for each of them respectively. The light-curve of SN~2004ex is similar to that SN~2007Y, as expected since type~IIb SNe and Ib SNe have similar light-curves. 

\section{Spectroscopy}
\label{sec_spect}
\subsection{Data analysis}

Spectroscopic datasets have been acquired at 4 epochs, respectively 36, 52, 60 and 86 days after the explosion, by using the 1.82~m Ekar and the 3.5~m TNG telescopes (see Table \ref{spect_observations}). The  datasets consist of 2 spectra collected in the transition phase and 2 during the nebular phase. 
	
\begin{table}[!ht]
   \centering
      \begin{tabular}{c|c|c|c|c|cl}  
           \hline
      Date&JD&Day&Inst.&Grism& Phase\\
	\hline
      Nov. 16$\textsuperscript{th}$ &2453326.39&36&Ekar&Gr4& transition \\
      Dec. 1$\textsuperscript{th}$ &2453341.39&52&TNG &Gr4& transition\\
      Dec. 9$\textsuperscript{th}$ &2453349.52&60& Ekar&Gr4& nebular\\
      Jan. 4$\textsuperscript{th}$ &2453375.32&86&Ekar&Gr4& nebular\\
          	\hline
   \end{tabular}
   \caption{Spectroscopic observations of SN~2004ex. \label{spect_observations}}
   \label{tab:pare}
\end{table}

	Spectra were pre-reduced and calibrated according to standard methods using IRAF routines ($\textit{onesdspec, ccdred and specred packages}$): the raw images were de-biased, overscan-corrected, trimmed, and flat-field-corrected after normalization of the flat-field image along the dispersion axis before the extraction of the spectra. 
During de-bias, firstly the master bias image was created by combining all available bias images in order to improve the statistic and reduce the random noise; then it was subtracted to the source images. The overscan correction was made firstly by determining the mean bias level of each image in the overscan region and then by removing it from the all images. After the overscan correction, the trim  was applied to cut off the affected region from the edge. At the end, the flat-field correction was made by using a master flat which is a combination of all flat images, in order to increase the signal-to-noise. Lastly, sky-flats were used to remove artifacts such as cosmic rays and stars. 
	
	The one-dimensional spectra were obtained by mean of optimized extraction across the dispersion and by subtracting the galaxy contribution with the \textit{apall} package. After the definition of both the aperture and the background region at one wavelength, the task traces the position of the aperture on the bidimensional image. With this procedure, the night sky lines are also removed. The extraction of lamp spectra (usually Ne-HgCd, He-Ar lamps), obtained in the same instrumental configuration, allows the wavelength calibration with the \textit{apsum} package. In most cases, the error on the wavelength calibration is less than 2 $\angstrom$. 
	
	The next step is the flux calibration. The response curve for the given instrumental configuration has been obtained by spectroscopically observing standard stars with the same telescopes. The flux calibration is typically accurate within 20\%. 
	 
\begin{figure}[!ht]
\centering
\includegraphics[width=0.5\textwidth]{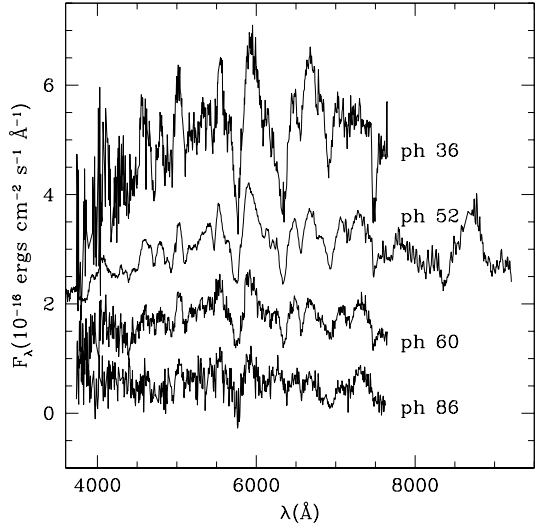}
\caption{The spectroscopic datasets of SN~2004ex at four phases.\label{spectra_all_ph}}
\end{figure}

	 In my analysis, the spectra were corrected for the redshift (z~=~0.01755) of the host galaxy, the reddening was eliminated using the $\textit{deredden}$ task in IRAF program and the external extinction value (host galaxy reddening, E(B-V)~=~0.022) has been calculated according to the celestial coordinates and observation time \cite{schlegel1998}.  The spectral evolution is shown on Figure \ref{spectra_all_ph} at each epoch (phases 36, 52, 60, 86). In these figures, the evolution of the elements can be followed during each phase. The dominant elements, especially He~I, are present until the phase~86. 
	 
The identifications of spectral lines of SN~2004ex were originally performed the absorption lines and some emission lines by using \textit{\text{the} sarith \text{,} identify and splot} tasks which are in \textit{\text{the} specred} package in IRAF.  

\subsection{The spectral result of SN~2004ex}
Then the position of the lines are compared with the typical identifications given \cite{barbon1995} for SN~1993J and for SN~2008ax  \cite{pastorello2008, taubenberger2011}. These lines are presented in Table \ref{tab:pare} and shown in Figures \ref{spect_36ph_comp_1993J}, \ref{spect_52ph_comp_1993J}, \ref{spect_60ph_comp_1993J}, \ref{spect_86ph_comp_1993J} for each phase respectively, together with a comparison with the spectrum of SN~1993J. Finally, Figure \ref{comp_08ax_all_ph} shows a comparison of the spectra of SN~2004ex and SN~2008ax for all phases.

\subsection{The velocity and mass of hydrogen}
The spectra of SN~2004ex show both prominent He I lines and a relatively faint H$_\alpha$ line, all with P-Cyg profiles \cite{harutyunyan2004}. The wavelengths are shown in Table \ref {spectral_line_results}. Phase~52 of the SN~2004ex spectrum closely resembles that of SN~1993J at 41 days after the explosion \cite{barbon1995}. 
This therefore defined the spectroscopic classifications of SN~2004ex as type IIb supernova and confirmed the result of \cite{harutyunyan2004}. 

The velocity of the element can be calculated by using P-Cyg profiles or by considering balmer $\alpha$ transition (used in my computation):

\begin{equation}
v = z_{dopp} \times c = \frac{\lambda - \lambda_a} {\lambda} \times c 
\end{equation}
where $\lambda$ = 6563 $\text{\angstrom}$ is the wavelength of the balmer $\alpha$ transition, $\lambda_a$ is the measured absorption wavelength, $c$ is the speed of light and $z_{dopp}$ is the doppler redshift.  The wavelength of the H$_{\alpha}$ absorption line at phase 52 is 6286~$\angstrom$. The velocity of the hydrogen envelope is estimated to be as a result 12662~km/s in this phase. However, in the early observation times, the H$_{\alpha}$ absorption minimum velocity found to be around 14000~km/s \cite{filippenko2004}, which is in the range 10000-15000 km/s expected for type~IIb SNe. 

 The mass of the ejected element can be calculated by $E_{kin} = \frac{1}{2} \ m \ v^2$.
When assuming that the kinetic energy of SN~2008ax is a lower limit for the total energy of SN~2004ex, the mass of the hydrogen envelope is estimated to be around 0.6~$M_{\odot}$ (where $E_{kin}$ is 10$^{51}$ erg and  $v_{H_{\alpha}}$ is 12662~km/s).


\begin{table*}[!ht]
   \centering\footnotesize
      \begin{tabular}{|c|c|c|c|c|c|c|c|c|c|c|}
           \hline
      Epochs&$\lambda$&36.&37.&41.&52. &52. &60. &62.&86.&89.\\
      & &2004ex&1993J&1993J&2004ex&1993J&2004ex&1993J&2004ex&1993J\\\hline%
	\hline
      O I& 7254&7144&&&7439&&7288&&7215&\\ \hline
      Ca II& 7202&&&&7238&&7114&&&\\ \hline
      He I&7283&6717&&&7116&7132&6967&7161&6788&\\ \hline
      He I&7065&6573&6882&6844&6878&6892&6765&6904&6660&\\ \hline
     He I&6678&6204&6523&6540&6510&6551&6412&6549&6335&\\ \hline
      H$_\alpha$ & 6563&6004&6343&6346&6286&6353&6187&6358&6154&\\ \hline
      He I+Na I&5876\&5892&5519&5736&5732&5688&5737&5591&5743&5547&\\ \hline
      O I&&&&&5482&&5372&&&\\ \hline
      Fe II&5269&&&5077&5084&5083&5013&5086&&5090\\ \hline
       He I$+$Fe II&5015\&5018&4985&4921&&4925&4923&4834&4915&4769&4925\\ \hline
       He I$+$Fe II&4921\&4925&&4828&4922&&4831&&4837&&\\ \hline
       H$_\beta$&6861&&4720&4833&4715&4724&4631&4337&&4724\\ \hline
       Mg&4571&&&&4612&&&&&\\ \hline
       He l&4471&4619&4429&&4394&4395&4309&4397&4337&4410\\ \hline
       He l&4437&&4116&4723&&&&&&\\ \hline
       H$_\gamma$&4340&&4828&&4225&4237&&4218&&4222\\ \hline
       Ca ll&&&4720&4219&3821&&&&&\\ \hline
   \end{tabular}
    \centering\caption{The spectral lines of each element obtained for all different phases of SN~2004ex and SN~1993J. The intrinsic wavelengths of the line were also given for some elements. \label{spectral_line_results}}
   \label{tab:pare}
\end{table*}

\subsection{The spectral comparison of type IIb SNe}
\begin{figure}[!ht]
\centering
\includegraphics[width=0.5\textwidth]{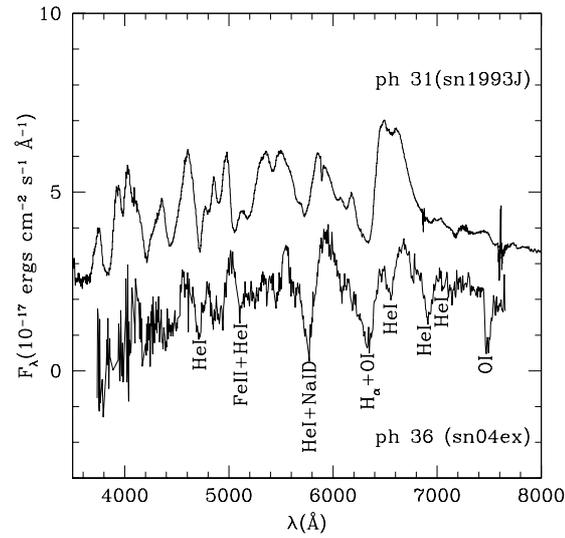}
\caption{The spectrum at +36~days is shown with the identifications of the lines to compare the chemical composition of SN~2004ex and SN~1993J.\label{spect_36ph_comp_1993J}}
\end{figure}

\begin{figure}[!ht]
\centering
\includegraphics[width=0.5\textwidth]{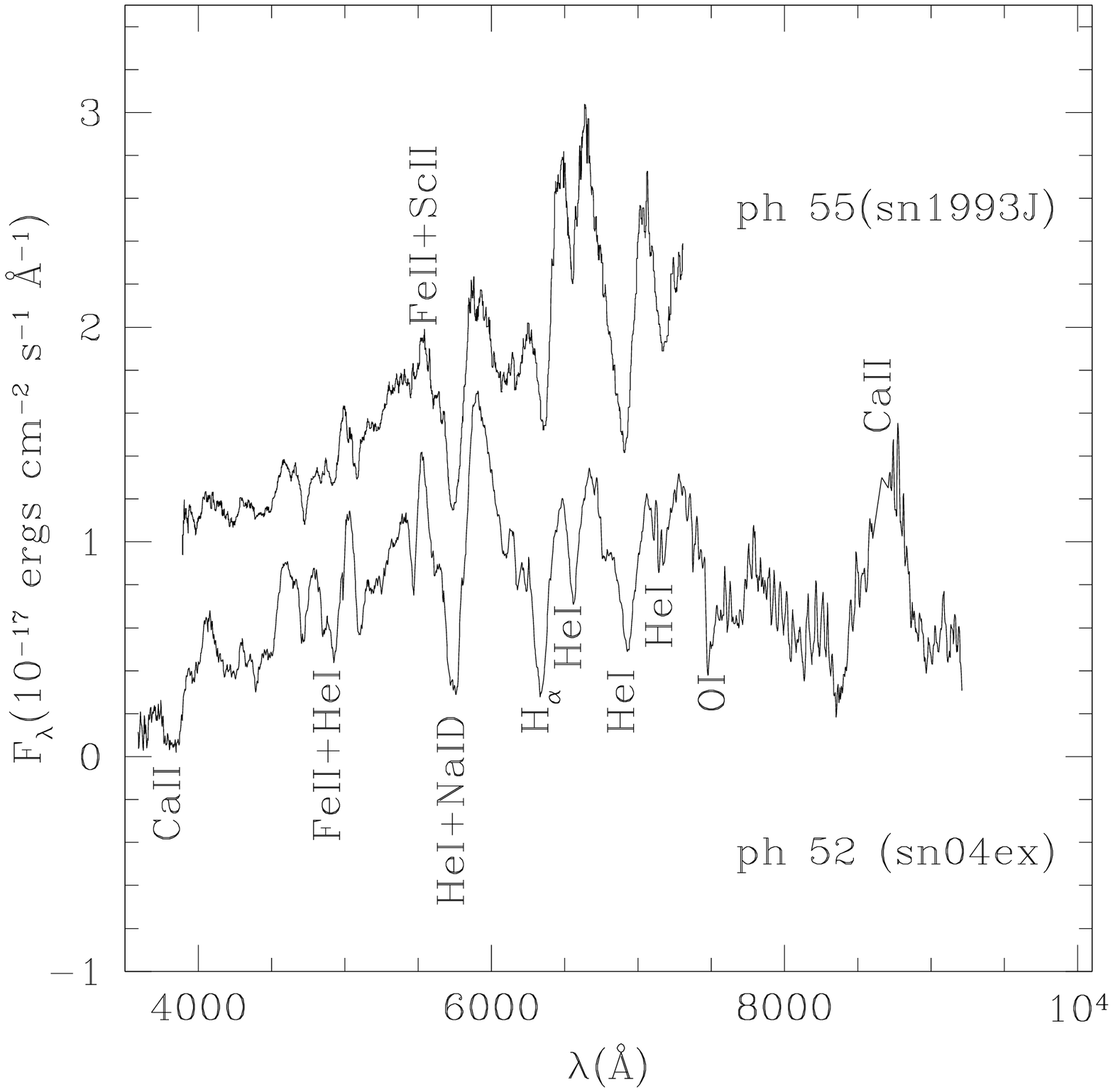}
\caption{The spectrum at +52~days is shown with the identifications of the lines to compare the chemical composition of SN~2004ex and SN~1993J.\label{spect_52ph_comp_1993J}}
\end{figure}

\begin{figure}[!ht]
\centering
\includegraphics[width=0.5\textwidth]{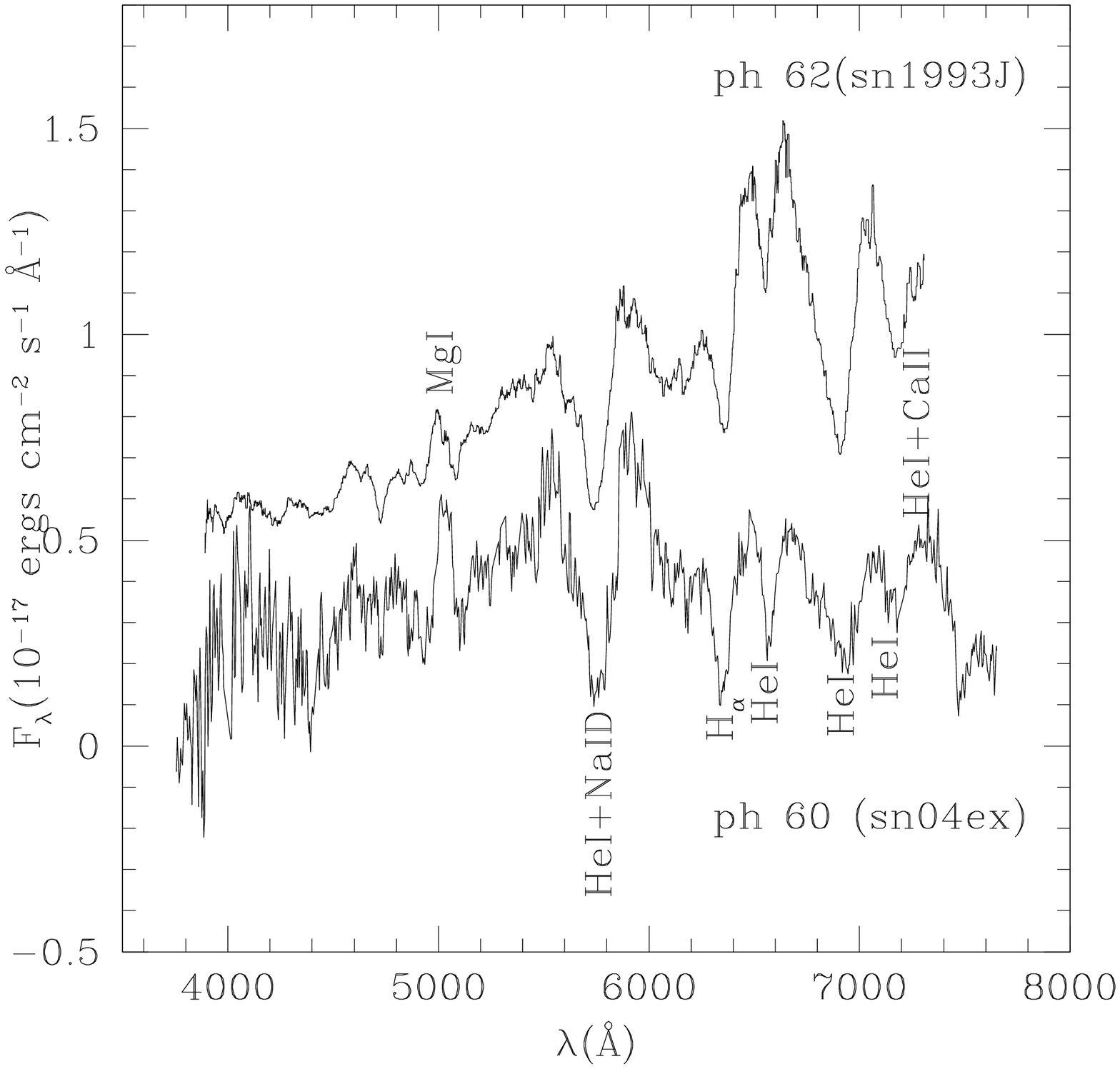}
\caption{The spectrum at +60~days is shown with the identifications of the lines to compare the chemical composition of SN~2004ex and SN~1993J.\label{spect_60ph_comp_1993J}}
\end{figure}

\begin{figure}[!ht]
\centering
\includegraphics[width=0.5\textwidth]{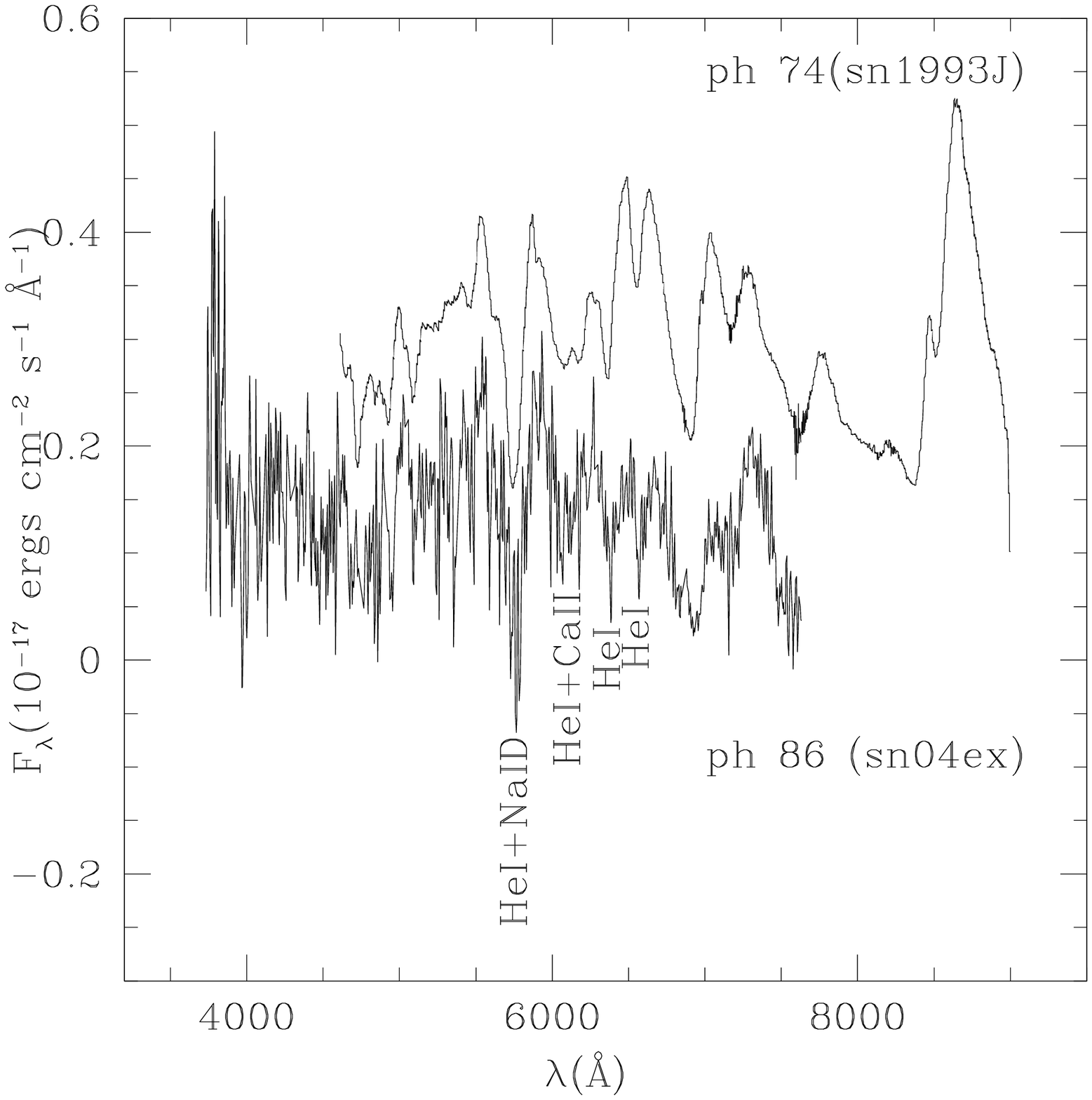}
\caption{The spectrum at +86~days is shown with the identifications of the lines to compare the chemical composition of SN~2004ex and SN~1993J.\label{spect_86ph_comp_1993J}}
\end{figure}

\begin{figure}[!ht]
\centering
\includegraphics[width=0.5\textwidth]{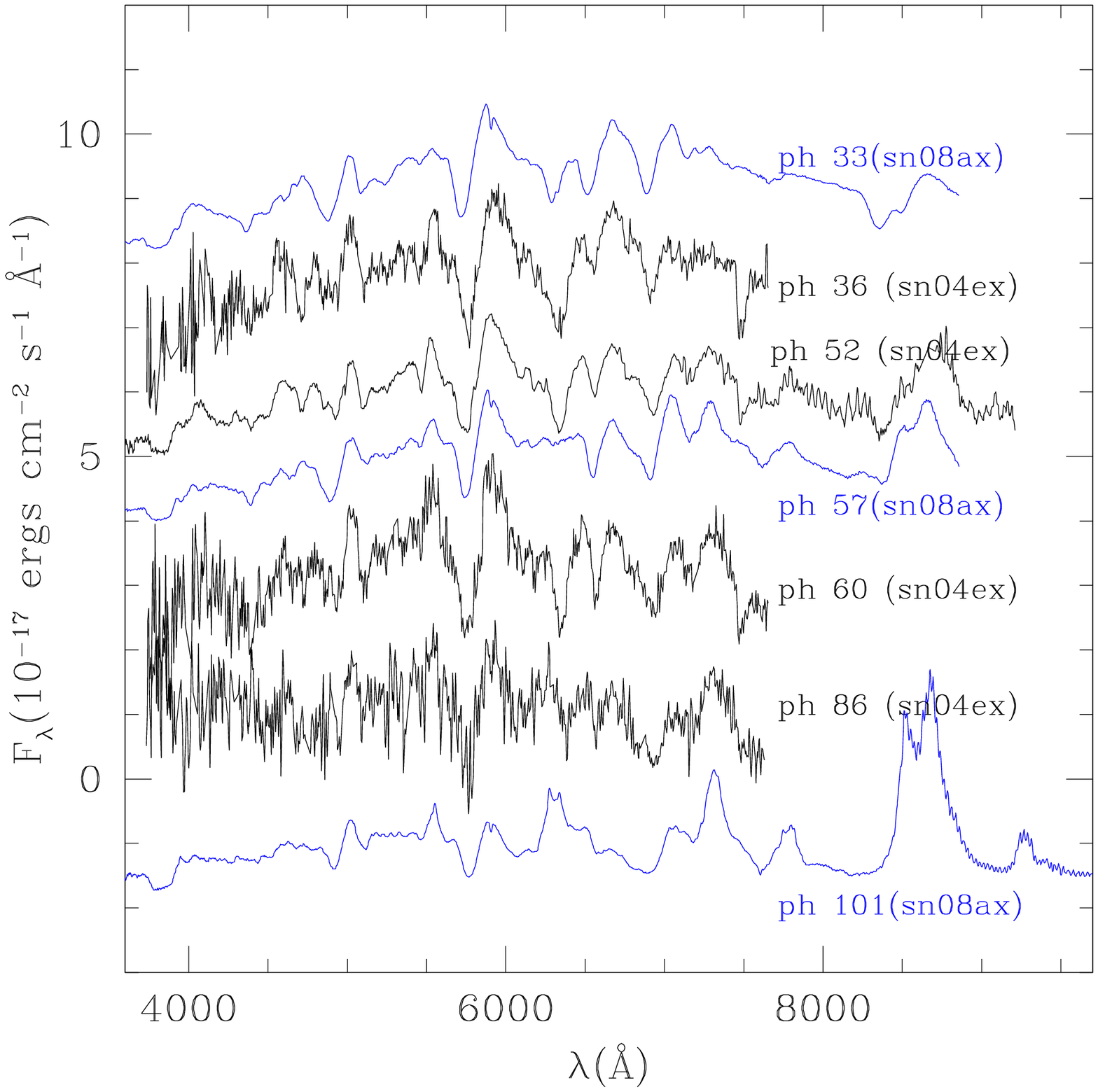}
\caption{The spectra of SN~2004ex at all phases are compared with those of SN~2008ax. \label{comp_08ax_all_ph}}
\end{figure}

The overall characteristics of SN~2004ex reminds those of SN~1993J and SN~2008ax, except for a likely smaller mass for the hydrogen envelope (0.6~$M_{\odot}$, 1.3~$M_{\odot}$, 2-5~$M_{\odot}$ respectively).
Figures  \ref{spect_36ph_comp_1993J}, \ref{spect_52ph_comp_1993J}, \ref{spect_60ph_comp_1993J} and \ref{spect_86ph_comp_1993J} show a comparison of the spectrum of SN~2004ex and SN~1993J. Figure \ref{comp_08ax_all_ph} shows the comparison of spectra between SN~2004ex and SN~2008ax for all phases. In all phases He~I lines are dominant and the H$_{\alpha}$ line is weak. These features are compatible with a type~IIb supernova. 

\section{Discussion}
	The optical light-curves of SN~2004ex displays a temporal decline. From my results, it is not possible to extract information about the Ni mass, or the total energy of the ejecta. Indeed,  the life-time of $^{56}$Ni is around 8.8~days after the explosion; as the first spectrum is at 36 days after the explosion I cannot constrain  $^{56}$Ni. 

According to the classification of SN light-curves \cite{arcavi2012}, SN~2004ex has a rapidly declining light-curve which is a property of type IIb SNe. And it is similar to the light-curves shapes of type~Ib SNe, suggesting that they have similar progenitors. 

	The comparison shows that  the light-curve of SN~2004ex  is very similar in shape to that of SN~2008ax, rather different that of the prototypical type~IIb SN~1993J. Indeed, SN~2004ex shows a faster decline rate at early phase after the peak and lacks the prominent narrow early-time peak which is seen in the light-curve of SN~1993J. Moreover, it shows a faster decline rate at late phase, similar to the light-curve of SN~2008ax. It is known that H-poor core-collapse supernovae display a wide range of behaviors in their light-curves evolution, as seen in the light-curves of all type~IIb SNe. Even if the light-curve of SN~2004ex is quite similar to the light-curve of SN~2008ax, it is significantly brighter than SN~2008ax and fainter than SN~1993J during the whole evolution.	
	
Type~II SNe are generally characterized by their spectral properties rather than their photometric properties. As it can be seen in all phases of SN~2004ex, the evolution of He features are prominent at late times. The nebular phase develops some months after the SN explodes, when the ejecta become transparent and the decrease of density allows for the formation of forbidden emission lines. 

A spectrogram of SN~2004ex obtained at early times shows a blue continuum with relatively broad, P-Cygni H lines similar to the spectra of young type~II supernovae. To conclude, the comparison with SN~1993J and SN~2008ax shows that SN~2004ex strongly is a type~IIb.

\subsection{Possible progenitor}

The nebular phase of an SN is important to obtain information about the inner region of the explosion and its optically thin region. The modeling of this phase can provide information on the ejected mass, the kinetic energy, abundances, geometry (which can hardly be obtained by the other methods). Datasets for this phase were obtained for some type~IIb SNe which are listed in \cite{maurer2010} (\textit{e.g.} SN~1994 and SN~2008ax). SN~2004ex can be added to the list with the results of this work. 

As it is discussed by Maurer et al. \cite{maurer2010}, H$_{\alpha}$ lines at late observation times come from the interaction between the ejecta and the wind, it means that type~IIb SNe should be surrounded by massive stellar winds which can be from the companion star if the binary system is considered. As a result SN~2004ex could be a member of a binary system. 

~

The aim of the study in this Chapter \ref{Chapter_SN2004ex} is to figure out the emission progress of SNe in the optical band and to compare with the different sub type of SNe. This could give the idea that if different kinds of SNe are connected to each other why mainly one type of SN (type Ic SN) observed to be associated to GRBs. 

This work is also encouraged to look for information about associated SNe to the GRBs which were found out during the sampling of low luminosity afterglow GRBs as seen in the following chapters.

%% file: Chapter3.tex

\chapter[Data Reduction Applied to X-ray observations of GRBs]{\parbox[t]{\textwidth}{Data Reduction Applied to X-ray observations of GRBs}}
\chaptermark{X-ray data reduction}

In this chapter, after introducing the \textit{Swift} satellite and describing the data analysis techniques, the selection method of LLA GRBs is explained and discussed. A brief description of the sample follows. All errors are quoted at the 90\% confidence level, and I used a standard flat $\Lambda$CDM model with $\Omega_m$~=~0.3, $\Omega_\Lambda$ = 0.7 and $H_0= 70 \ \text{km} \ \text{s}^{-1} \ \text{Mpc}^{-1}$.

\section{The \textit{Swift} Satellite}
\label{swift_satellite} 
\subsection{Detection techniques for X-rays}

\begin{figure}[!ht]
\begin{center}
 \begin{tabular}{c|c|c}
\includegraphics[width=0.25\textwidth]{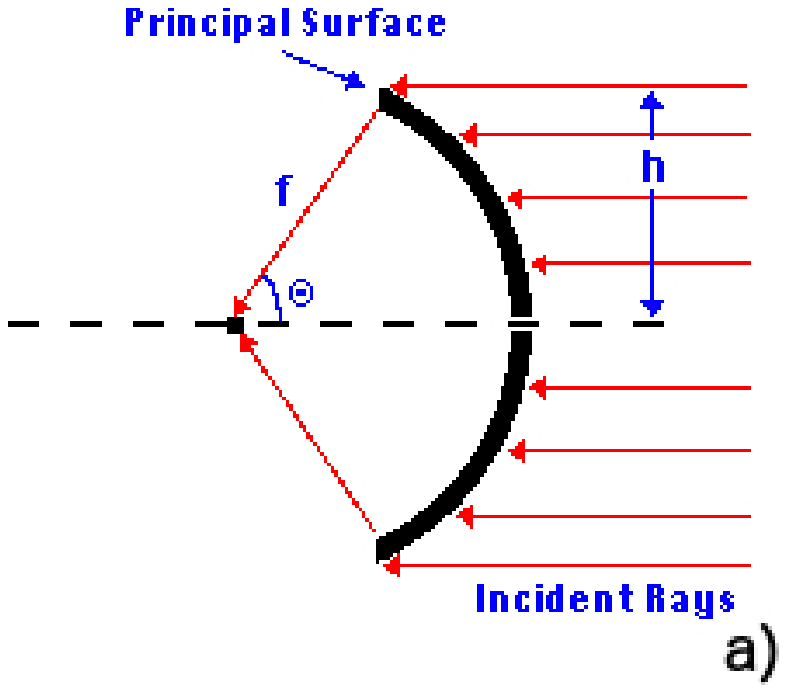} &
\includegraphics[width=0.4\textwidth]{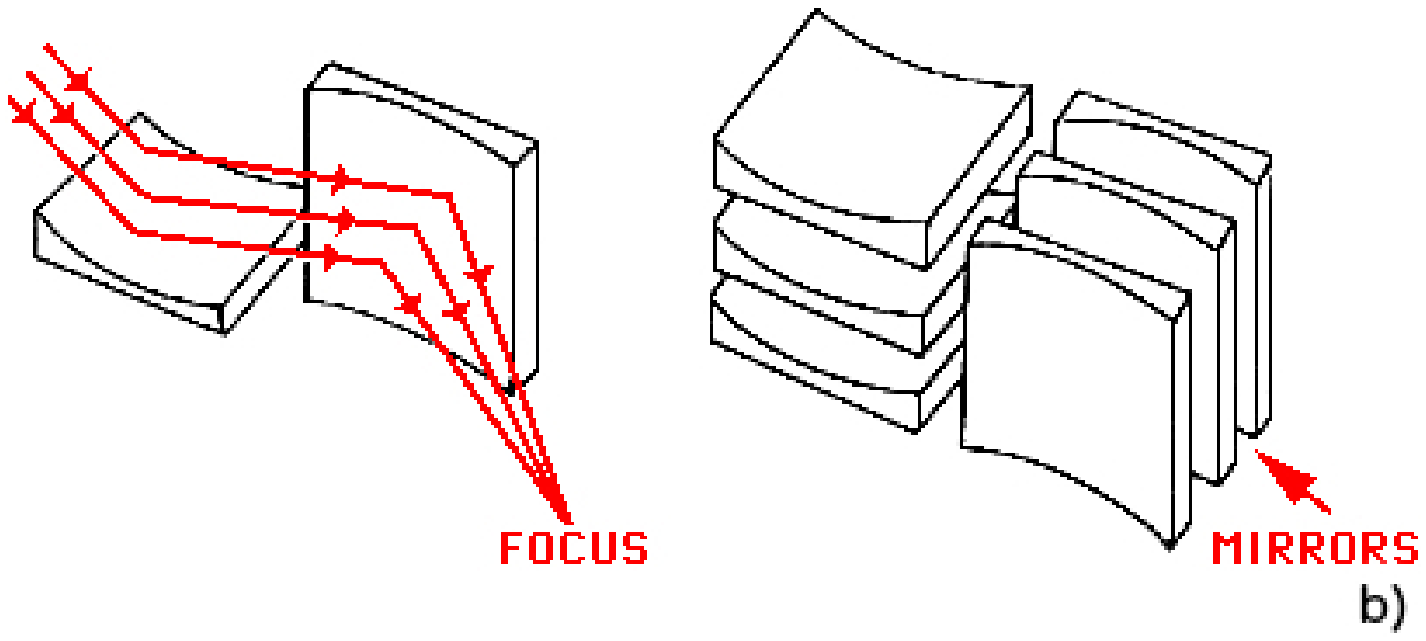} &
\includegraphics[width=0.25\textwidth]{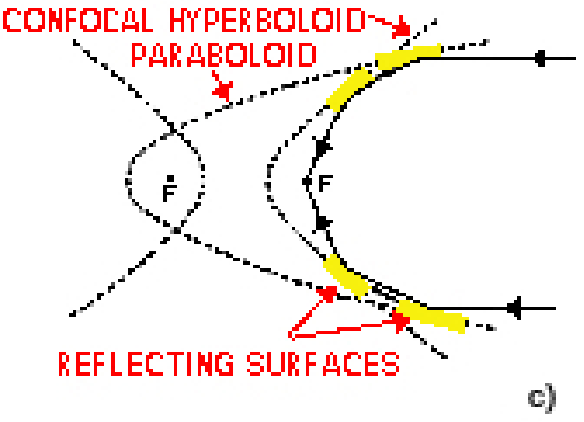}
\end{tabular}
\caption{a) Schematic view of the \textit{Abbe sine} condition. An optical system forms an image of an infinitely distant object as each ray in the parallel beaming emanates from its surface. The quantity $f = h/ sin(\theta)$ should be constant to form a clear image of the object. Here $h$ is the radial distance of the ray from the optical axis and $\theta$ is the angle between the final ray direction and initial ray direction. b) Kirkpatrick-Baez X-ray telescope simplified design. c) Wolter type I X-ray telescope design. Image Credit: NASA \label{X_ray_inst}}
\end{center}
\end{figure}

In 1948, a system was proposed by Kirkpatrick and Baez to focus X-rays. It forms real images and consists of a set of two orthogonal parabolas, on which incident X-rays reflect successively, see Figure \ref{X_ray_inst}b (left). This system satisfies the \textit{Abbe sine} condition. It should be satisfied by every optical system to form clear images of an object at infinity, see Figure \ref{X_ray_inst}a.  

The right part of Figure \ref{X_ray_inst}~b shows the parallel mirrors (usually cylindrically symmetric) which increase the surface area. The most commonly used system is the Wolter Type I system, which is represented in Figure \ref{X_ray_inst}c. It is a simple mechanical configuration and it provides the possibility of nesting several telescopes inside one another in order to increase the effective area.

The X-ray mirrors of Wolter type I systems provide two things: 

\begin{itemize}
\item the ability to determine the location of the arrival of an X-ray photon in two dimensions,
\item simultaneously possessing a reasonable effective area. 
\end{itemize}

These instruments can be made of different materials, gold (the most common) or iridium which can reflect X-ray photons. For instance, the mirrors of the \textit{Swift}/XRT are made of gold which is cooked on Nickel.
 
\subsection{Instrumental properties of \textit{Swift}}
The \textit{Swift} satellite made a ``big bang" effect in the GRBs area, as it updated (and is still updating) all observational knowledge about GRBs. Since its launch on November 20$\textsuperscript{th}$, 2004, it has detected more than about 1000~GRBs and their associated afterglows. Additionally, it has allowed for multi-wavelength observations of both the prompt and the afterglow emission of the bursts in great detail. The description which follows is taken from \cite{gehrels2004}.

\begin{figure}[!ht]
\begin{center}
\includegraphics[width=14cm]{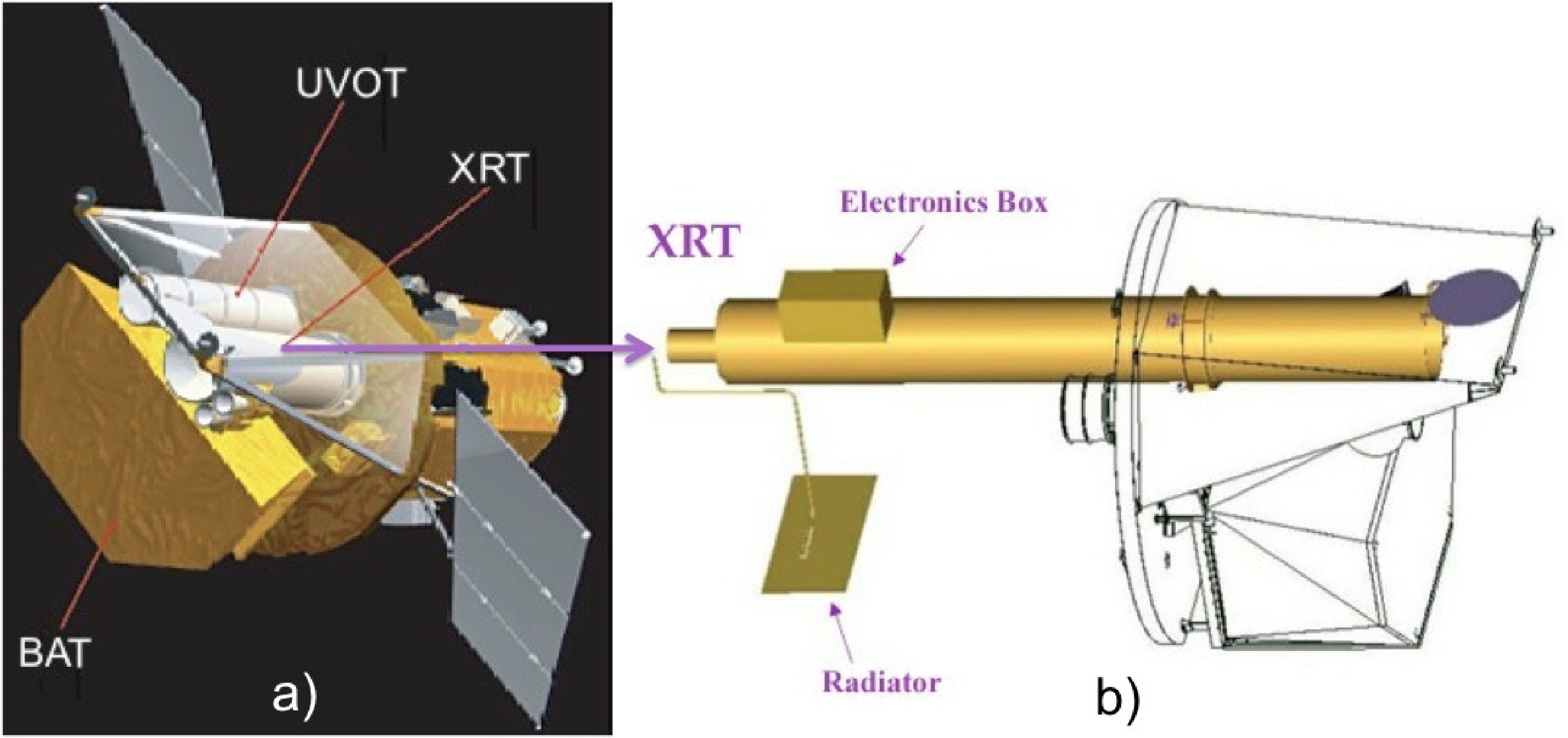}
\caption{a: Schematic view of the \textit{Swift} satellite with its three instruments. b: Overview of the XRT instrument. Image Credit: NASA/GSFC \label{Swift_inst}}
\end{center}
\end{figure}

The \textit{Swift} satellite has three instruments: namely the \textit{Swift}'s Burst Alert Telescope (BAT) observing in the gamma-ray band, the X-Ray Telescope (XRT), and the Ultraviolet/Optical Telescope (UVOT). They are shown on the spacecraft in Figure \ref{Swift_inst} a \cite{gehrels2004}. BAT is dedicated to observing the prompt emission of GRBs and has an energy range of $15 - 150$~keV. It has the capacity to detect weak bursts with its two-dimensional coded aperture mask and large area solid state detector array, as well as to detect bright bursts with its large field of view (1.4~steradians). BAT is able to locate a burst within an arcminute positional accuracy, allowing the satellite to point XRT and UVOT in the direction of the burst in $\sim$100~s, in order to observe the afterglow. It provides spectra and light-curves at X-ray, ultraviolet and optical wavelengths allowing for a concurrent multi-wavelength examination of each burst. 

The \textit{Swift}/XRT is a focusing X-ray telescope with a 110~cm$^2$ effective area, 23.6~$\times$~23.6~arcmin field of view (FOV), 18~arcsec resolution (half-power diameter), and 0.2~-~10 keV energy range. The XRT can locate GRBs to 4~arcsec accuracy within 10 seconds of target acquisition for a typical GRB. It uses grazing incidence Wolter I mirror to focus X-rays onto a CCD. The XRT instrument is specifically designed to study X-ray counterparts of GRBs providing spectra and light-curves over a wide time range beginning 20~-~70 seconds after the burst trigger and continuing for days to weeks thence, covering more than seven orders of magnitude. The layout of the XRT is shown in Figure \ref{Swift_inst} b.

The \textit{Swift}/UVOT instrument has a unique capacity for afterglow studies. It is designed to capture the early UV afterglow, which is not possible from the ground because of the absorption by the atmosphere; as well as optical photons from the afterglow in the $170 - 600$~nm band. Long term observations are usual. This instrument has a modified Ritchey-Chretien optical configuration with a 30~cm primary mirror which provides a good localization of the events with the error radius around 2~arcsec. It has UV/optical grisms and broadband color filters, a 4x magnifier; a clear white-light filter, and a blocking filter.

An important improvement is provided by the quick ground-based follow up. When there is a possible GRB, \textit{Swift} slews to the burst direction and quickly sendsthe position of the event to the ground to allow observations by other observatories. Position is relayed by the GRB Coordinate Network \cite{barthelmy2005}.

\section{X-ray data analysis}
\label{X-ray_data_analysis}
\subsection{Extraction of light-curves and spectra}
\label{ext_LC_spec}
\begin{figure*}[!ht]
\begin{center}
\includegraphics[width=0.6\textwidth]{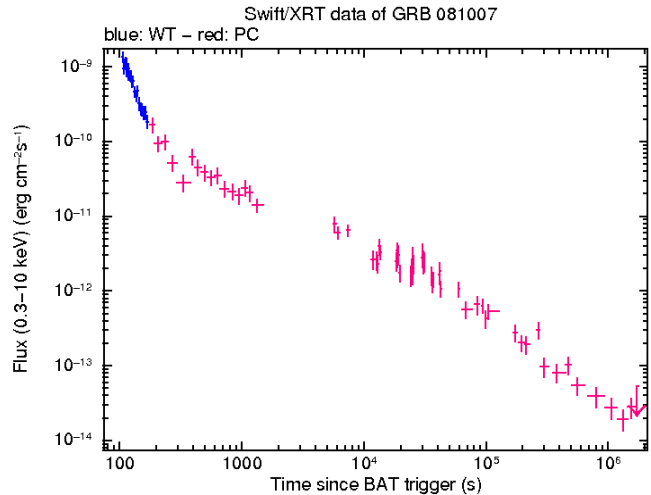} 
\caption{Example of a light-curve (here of GRB~081007) from the \textit{Swift} repository. \label{LC_from_swift_page}}
\end{center}
\end{figure*}

For each source, I analyzed the XRT light-curve in flux unit, which was retrieved from the online {\em Swift} light-curve repository\footnote{http://www.swift.ac.uk/xrt$\_$curves} \cite{evans2007, evans2009}, see one example in Figure \ref{LC_from_swift_page}. However, I did not directly use these light-curves, as the flux calibration is too sensitive for an automated analysis. Because I cannot use directly the data downloaded from the online repository, and needed to estimate independently the spectral index and the count-to-flux conversion factor.
 I first applied, when needed, the latest available calibration to the data. First of all, the early XRT observations\footnote{http://www.swift.ac.uk/swift\_portal/}, considering only photon-counting (PC) mode, were reprocessed at the given position\footnote{http://www.swift.ac.uk/grb\_region/} of each source using the {\it xrtpipeline} tool, which is part of the XRT software, distributed with the HEASOFT package\footnote{http://heasarc.gsfc.nasa.gov/heasoft/}. 

Secondly, I analyzed the data using {\it FTOOLs}\footnote{http://heasarc.gsfc.nasa.gov/docs/software/ftools/ftools\_menu.html} taking the following steps. After the image extraction, I have chosen the source region (30 pixels radius) and the background region (60 pixels radius), see Appendix C (see Figure \ref{source_region} (left)). One difference with the detailed method followed in the Appendix \ref{C} is the consideration of pile-up level. Pile-up occurs when multiple photons enter a pixel of the CCD within the same temporal frame (2.5 sec in the case of PC mode) and are read by the detector as a single photon of energy equal to the sum of all incident photon energies. It generally happens for early observations of bright sources. In this case, the source region has to be reduced by introducing an annulus to remove the region in which the count rate is larger than 0.6 counts.s$^{-1}$, see Figure \ref{source_region} (right).

\begin{figure*}[!ht]
\begin{center}
\begin{tikzpicture}[scale=0.8]
\draw (0cm,0cm) node[above,opacity=0.95]{\includegraphics[width=0.6\textwidth]{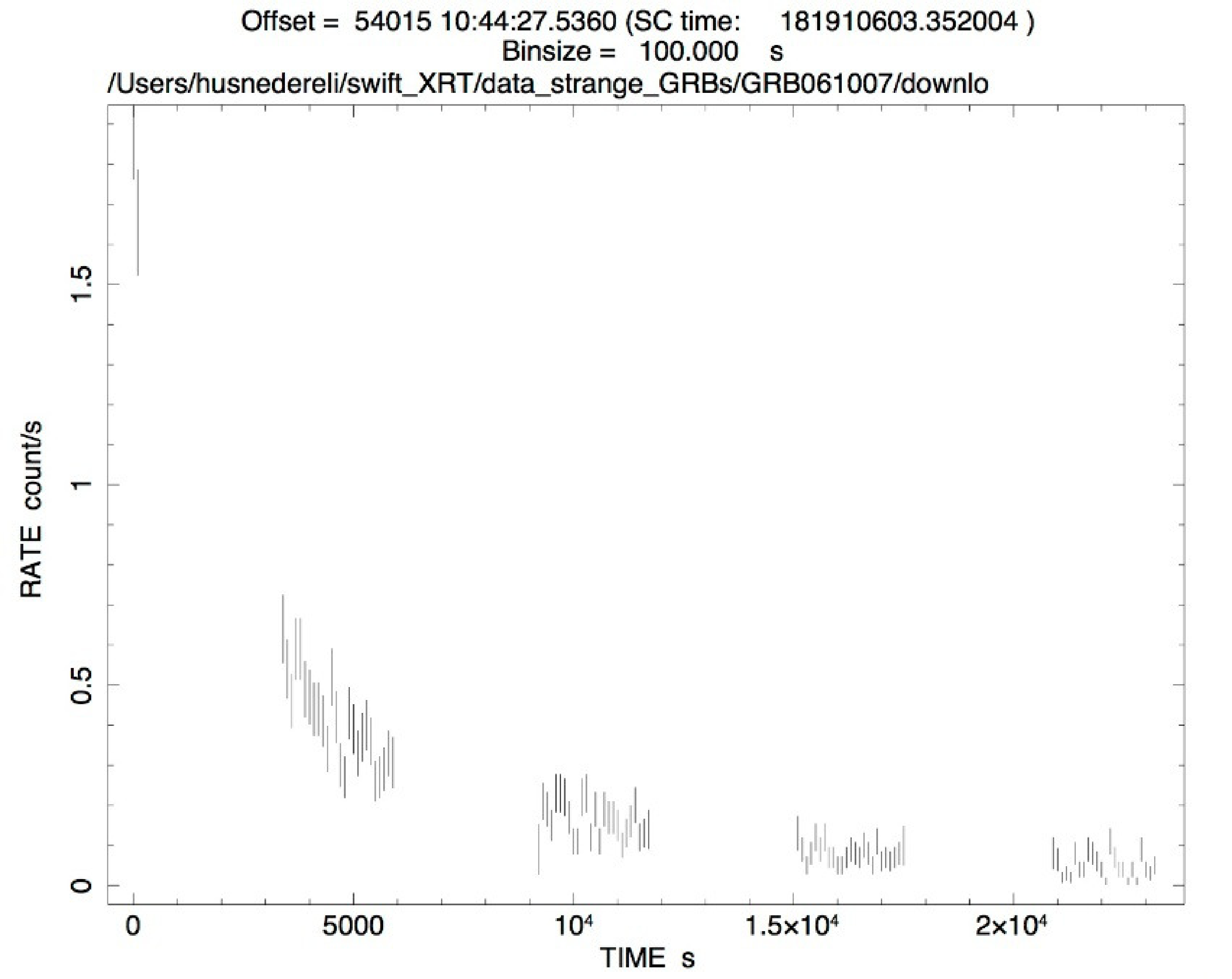}};
\draw [thin, black, dashed] (-2.8cm,0.83cm)--(-2.8cm,7.98cm);
\draw [thin, black, dashed] (-4.55cm,3.2cm)--(5.1cm,3.2cm);
\end{tikzpicture}
\caption{Light-curve of a bright burst (here, GRB~061007) during the data extraction. The horizontal line shows the threshold limit 0.6~counts.s$^{-1}$, such that there is no pile-up. \label{LC_with_100_bin}}
\end{center}
\end{figure*}

However, since most sources in my sample are too weak to be effected by the pile-up, I simply removed data in that time interval when pile-up occurs instead of reducing the region of interest. For that, after filtering the source region, I binned the light-curve by 100 or 500 seconds per bin, depending on the brightness of the source. Then, I extracted the light-curve on the full time interval and I restricted it to the time range for which there is no pile-up, \textit{i.e.} when the count rate within a 30-pixel radius circle region centered on the source was below 0.6~counts.s$^{-1}$ in PC mode \cite{vaughan2006}. This limit is represented by the vertical line in Figure \ref{LC_with_100_bin}, while the horizontal line shows the time limit (which varies for each source). After that, I extracted the spectrum in the time interval free of pile-up. The background light-curve and spectrum are extracted from a region free of pile-up and prompt contamination \cite{vaughan2006}.

I created an Ancillary Response File (ARF) using the full frame event lists in {\it xrtmkarf} and I used the Response Matrix File (RMF) which can be found in the Calibration Database\footnote{CALDB; http://swift.gsfc.nasa.gov/docs/heasarc/caldb/swift/} to fit the data (in Appendix \ref{C}, I show the full analysis of one GRB step by step with the corresponding explanation).

\begin{figure*}[!ht]
\begin{center}
\begin{tikzpicture}[scale=1]
\draw (-2.6cm,0cm) node[above,opacity=0.95]{ \includegraphics[width=5cm]{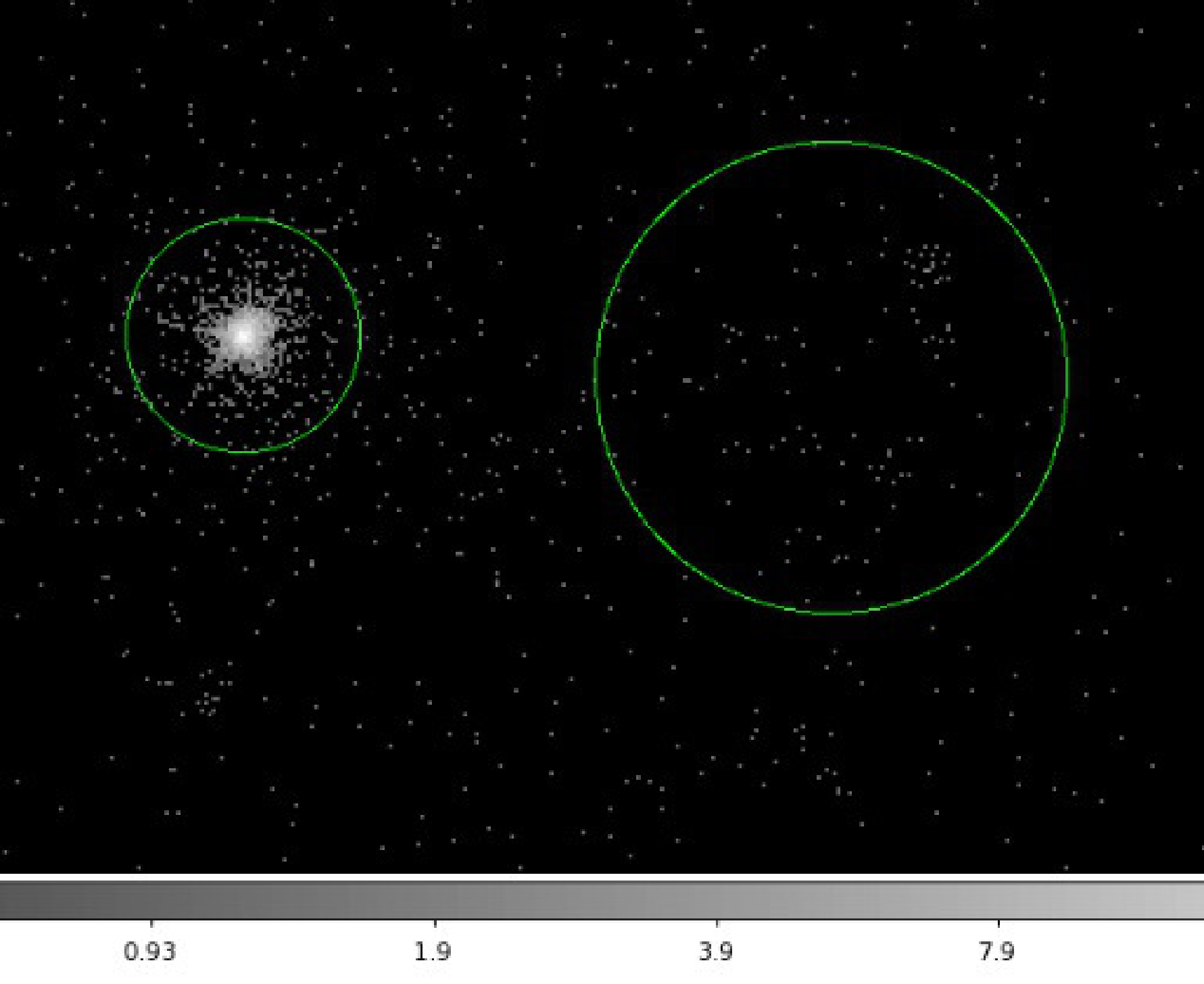}};
\draw (2.6cm,0cm) node[above,opacity=0.95]{\includegraphics[width=5cm]{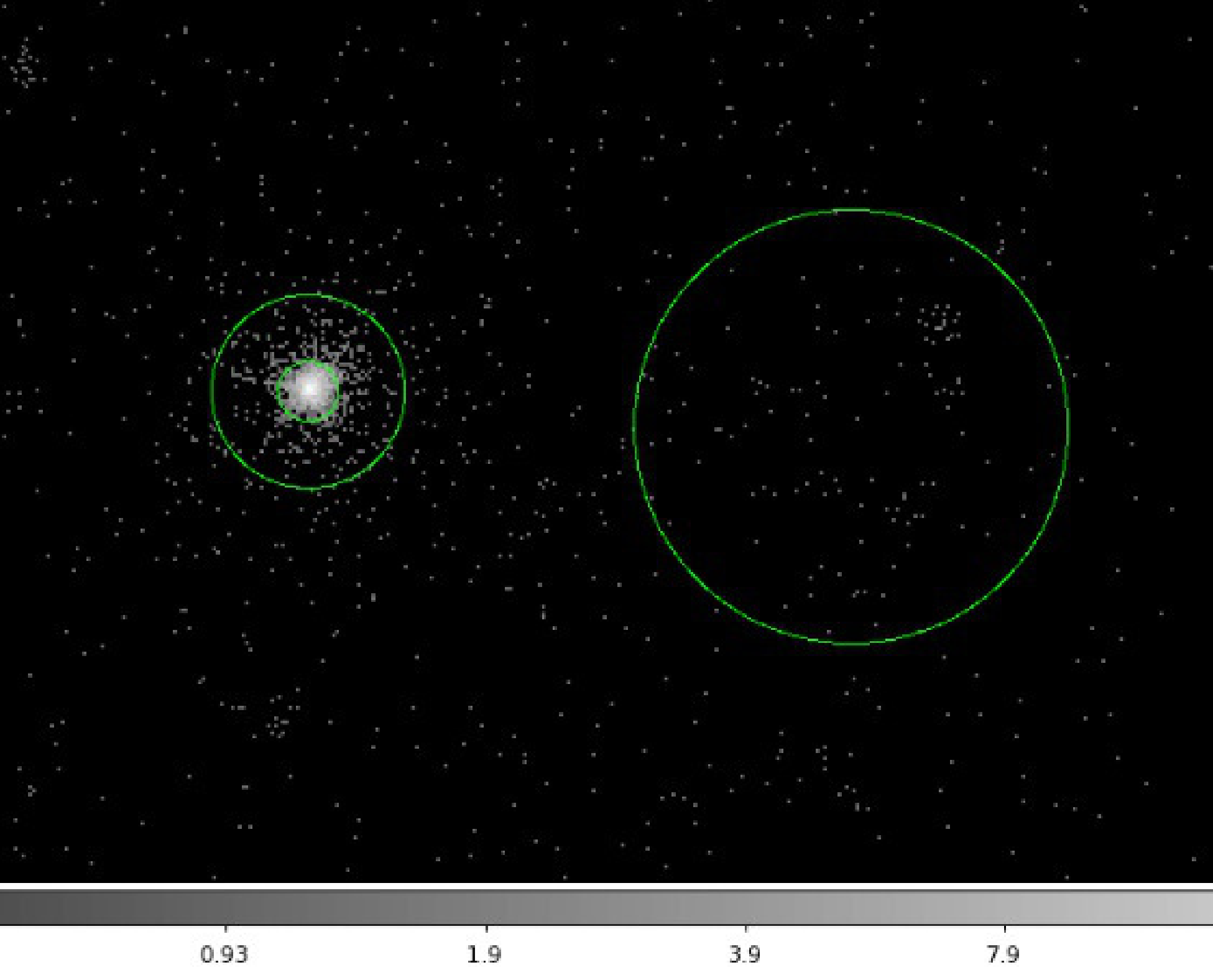}};
\draw [thick, green] (1.36cm,2.55cm) circle (0.12);
\end{tikzpicture}
\caption{Left: The source and background regions are shown by the green circles. Right: The annulus in light green delimits the piled-up region. \label{source_region}}
\end{center}
\end{figure*}

\subsection{Fitting the spectra}
\label{fit_spec}
\begin{figure*}[!ht]
\begin{center}
\includegraphics[width=0.6\textwidth]{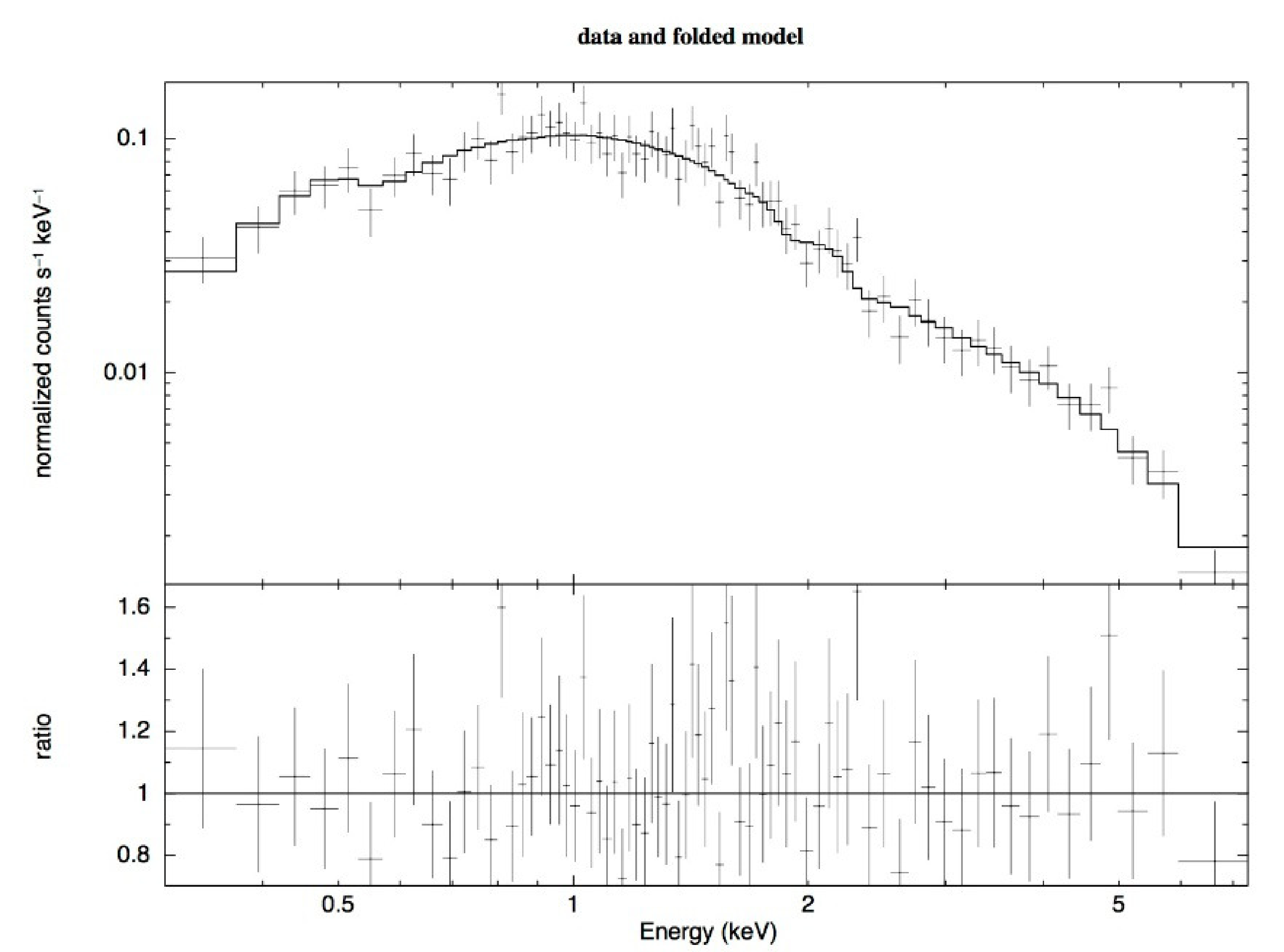}
\caption{Example of an obtained spectrum (here, GRB~061007). The data and the best fit are represented together.\label{spectrum_example}}
\end{center}
\end{figure*}

I fitted the extracted spectrum with \textit{Xspec} using a power-law model, absorbed twice in order to take into account the host galaxy and the Galaxy contributions at low energy $E < 2$ keV. The redshift of the burst was set to the value obtained by optical spectroscopy and the galactic column densities ($N_{H,gal}$) was set to the value from the Leiden/Argentine/Bonn (LAB) Survey of Galactic HI \cite{kalberla2005} (which was calculated by the NH FTOOL\footnote{http://heasarc.gsfc.nasa.gov/lheasoft/ftools}). However, the host column density ($N_{H,X}$) was fitted for. I generally used the $\chi^2$ statistic when there were at least $20 - 25$~counts per bin. A good fit for this statistic is obtained when 0.8~$\leq$ $\chi^2$ $\leq$ 1.1. Otherwise, I used the cash statistic for which at least one count per bin required. This statistic can provide parameters without giving information about the goodness of the fit. As an example, the fit result of the bright GRB~061007 can be seen in Figure \ref{spectrum_example}. Finally, using the fit, I calculated the energy correction factor of each GRB to reduce the energy range of observation from $0.3 - 10$~keV to  $2 - 10$~keV. When this spectral model was found different from the one used in the automated pipeline, I corrected the light-curve according to the parameters (spectral index and energy correction factor) from my analysis (for all these processes, one example can be seen in Appendix~\ref{C}).

\section[Selection of the sample]{Selection of the global sample and corrections of the light-curves}
\chaptermark{Selection of the sample}
\subsection{Selection of long GRBs}
In order to define my sample, I first started with the list of bursts maintained by J. Greiner\footnote{http://www.mpe.mpg.de/$\sim$jcg/grbgen.html}. I took into account all bursts with a measured redshift observed before February 15$\textsuperscript{th}$, 2013, without consideration for the detector triggered by the event and/or observing it. Being interested in the X-ray afterglow, I discarded all bursts not observed by an X-ray instrument (BeppoSAX, Chandra, XMM-Newton, INTEGRAL, \textit{Swift}). They are GRB~050408, GRB~051022, GRB~060712, GRB~061217, GRB~120716A. This leads to a first sample of 283~sources which contains long and short bursts.

I then restricted it to long GRBs only. Searching all sources in the literature, I discarded 12~short GRBs which have $T_{90} < 2$~s. I also took into account cosmological time dilation and removed every intrinsically short burst with ($T_{90, rest} = T_{90} / (1+z) < 2~\text{s}$) \cite{siellez2014}. All discarded short GRBs are presented in Appendix \ref{A}. It leaves only 254~bursts in the sample; among these, 61 were already investigated by \cite{gendre2008}. GRBs of the global sample are presented in Appendix \ref{B}, with some of their parameters: redshift, energy correction factor~(ECF), $\text{log}(T_a)$, (where $T_a$ is the time at which the plateau ends) and spectral index.
This work does not contain any update made in the Greiner webpage or in the Lancaster UK webpage more recent than the February 15$\textsuperscript{th}$, 2013. Moreover, GRBs with uncertain redshift measurements are not considered in this statistical study. Indeed, all redshifts were determined by spectroscopic observations in the optical band. 

\subsection{Selection of the late-time afterglow}
Once the flux calibrations have been checked (and when appropriately corrected, see section \ref{fit_spec}), I selected the late-time afterglow in the light-curves. Indeed, it was clearly stated that any prompt emission cause a broadening of the clustering, preventing any analysis  \cite{gendre2008}. I followed the same method, removing from the light-curves all emission present before the end of the plateau phase using the time $T_a$ at which the plateau ends. $T_a$ is obtained for each GRB by several methods described below. 

\begin{figure*}[!ht]
\begin{center}
\includegraphics[width=0.45\textwidth]{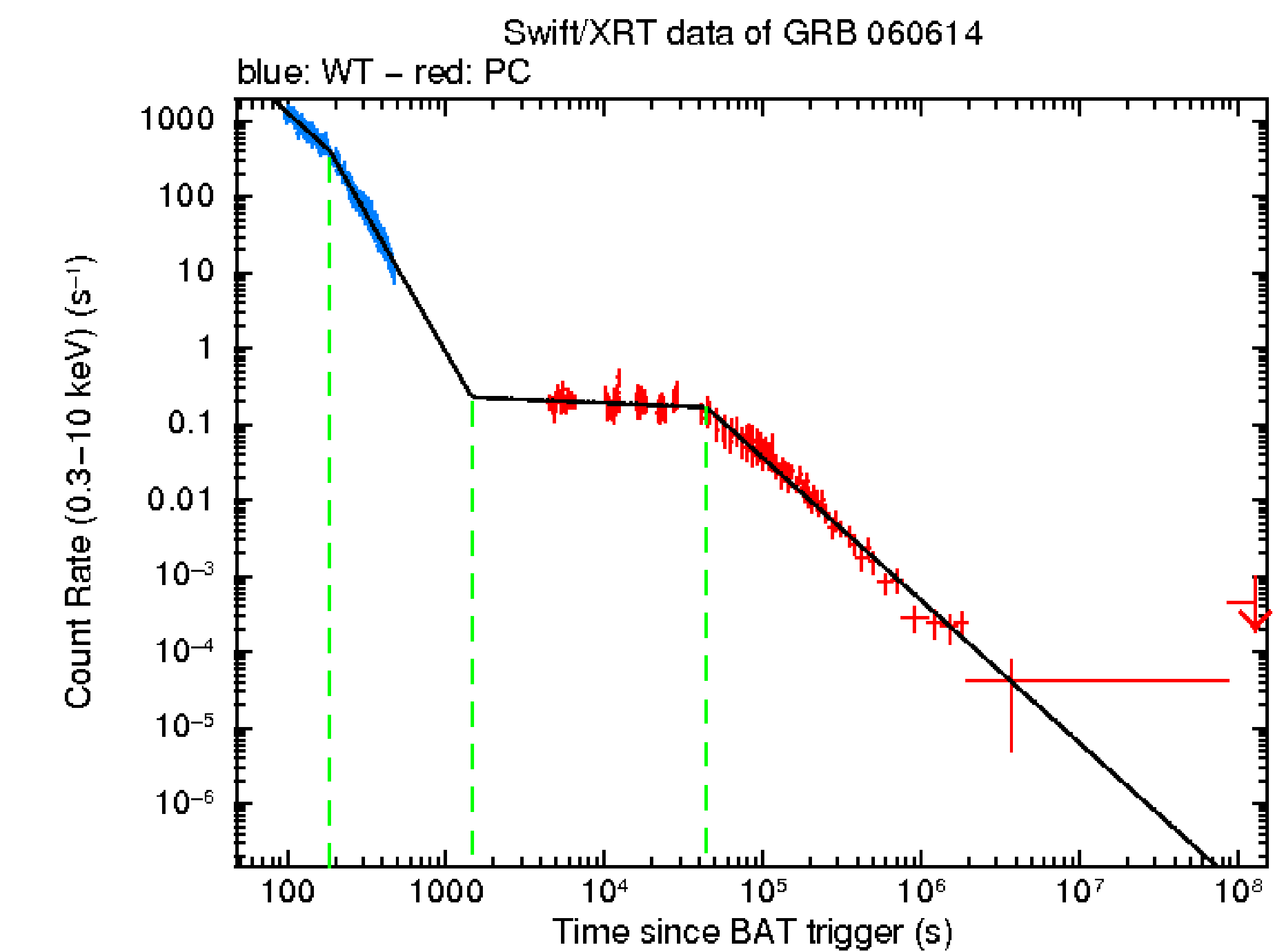}
\includegraphics[width=0.45\textwidth]{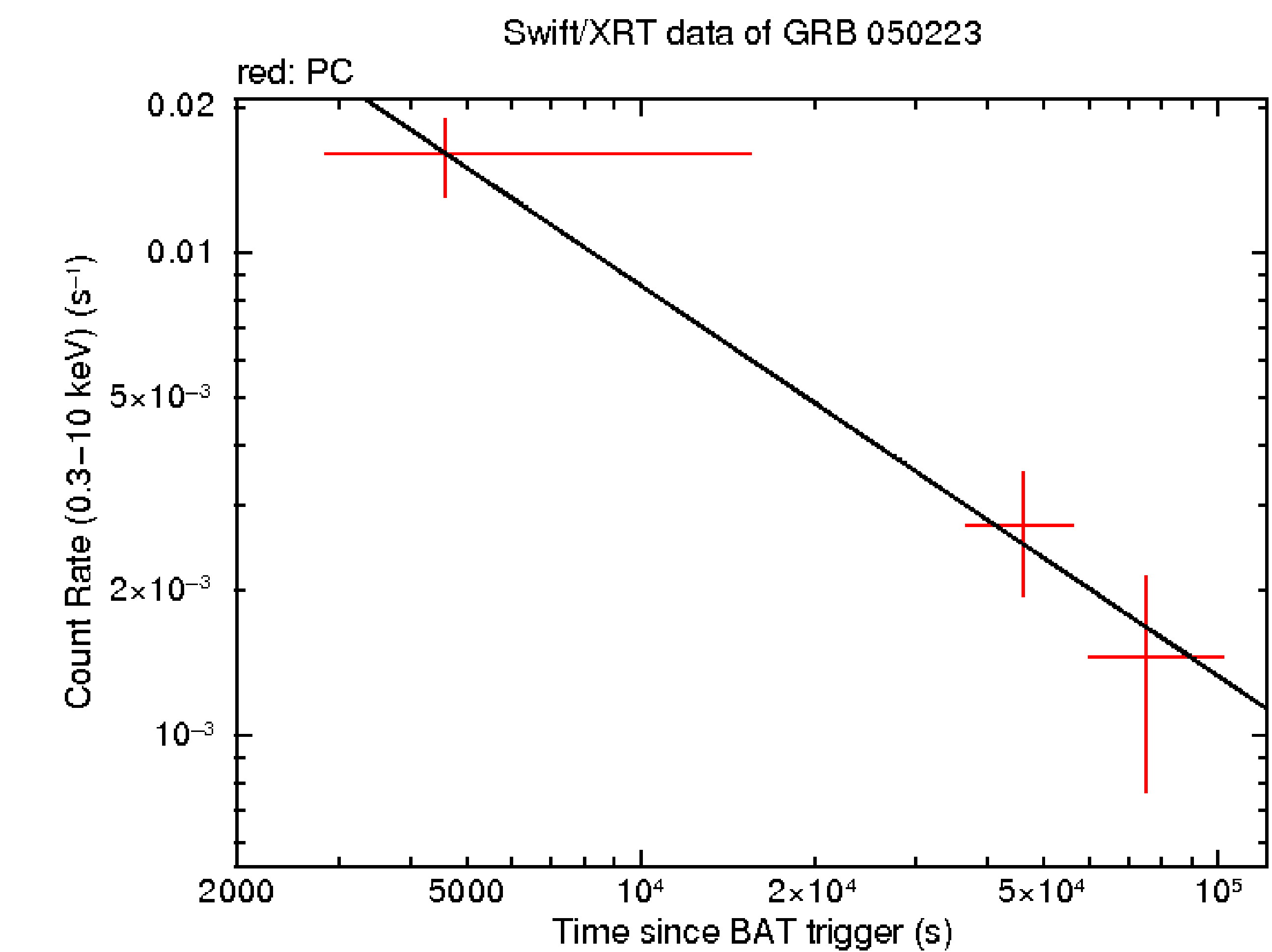}
\caption{Two examples for the determination of $T_a$. Left: There is a clear plateau breaking at 44400~s. Right: when the break is not visible in the light-curve or there is not enough data to provide a meaningful fit, $T_a$ is imposed to be 10$^4$ seconds. \label{Ta_fits_from_Swift_team}}
\end{center}
\end{figure*}

\begin{figure*}[!ht]
\begin{center}
\includegraphics[width=0.45\textwidth]{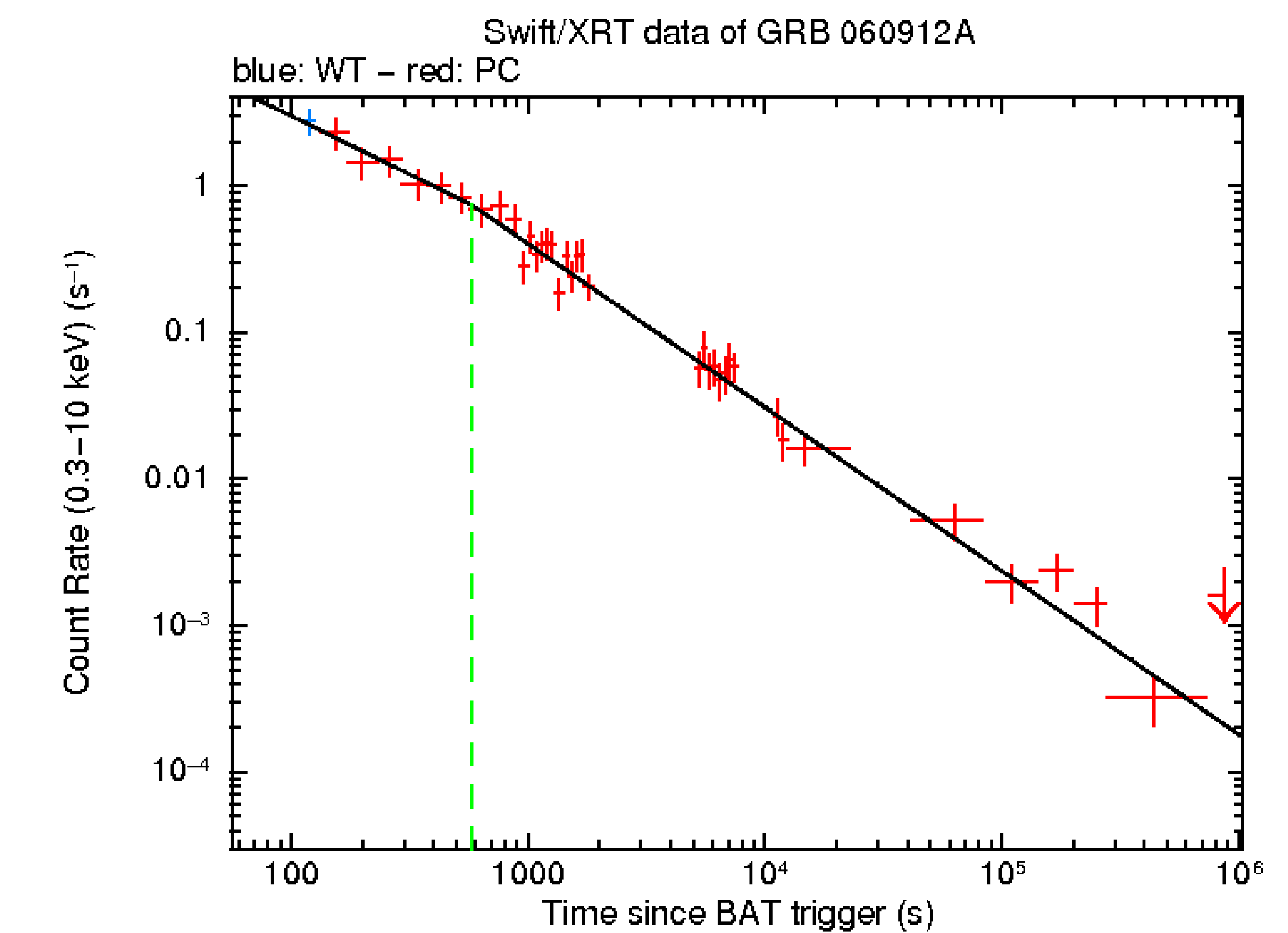}
\includegraphics[width=0.45\textwidth]{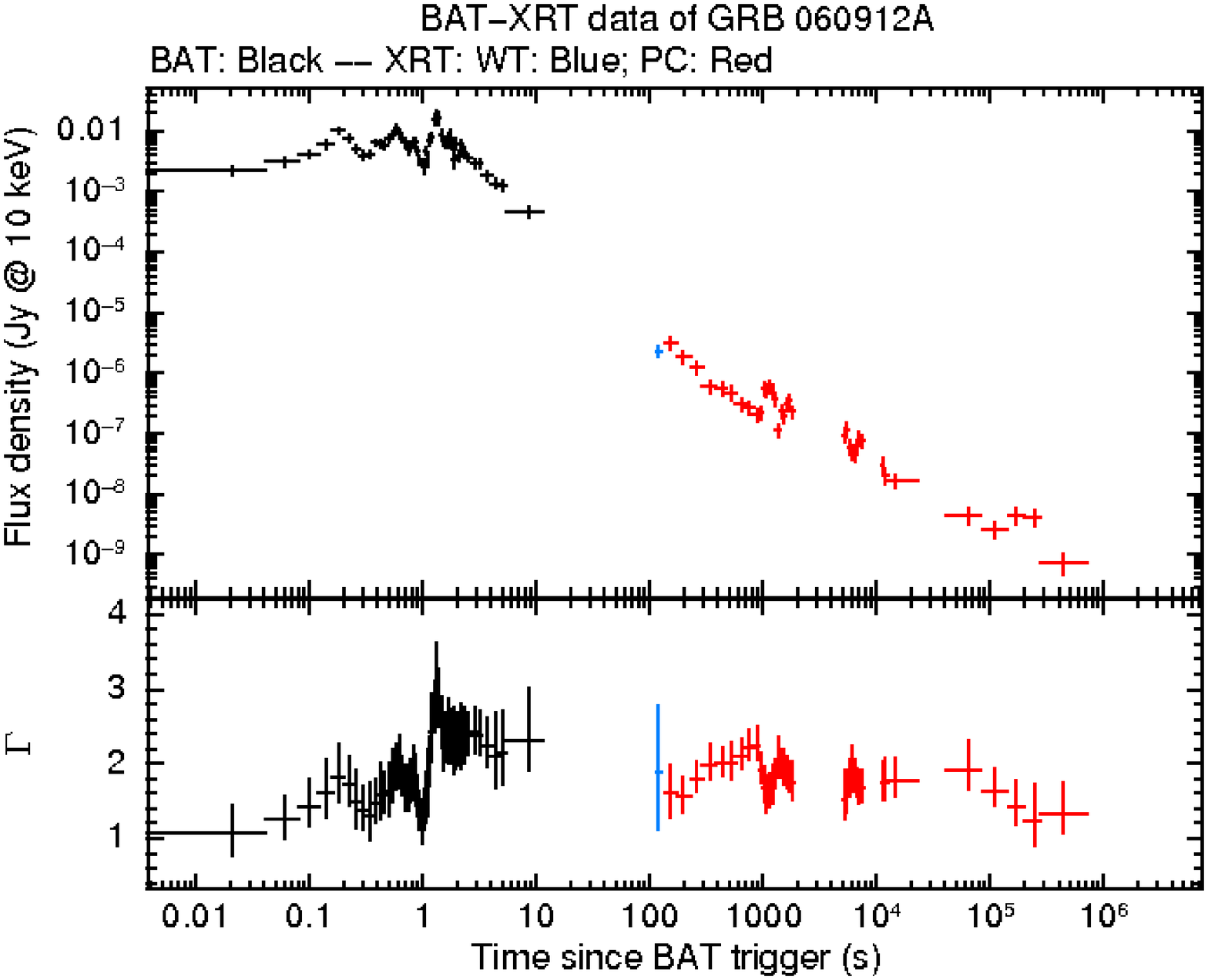}
\caption{Left: Fit of the light-curve obtained from the \textit{Swift} webpage. The break is at around 600~s. Right: BAT and XRT light-curves together. The first observations by XRT seem to be associated to the decay of the prompt emission. After the break at 2000~s, the slope is not close to zero, thus $T_a$ is smaller than 2000~s. \label{Ta_fits_from_Swift_team_compared_BAT}}
\end{center}
\end{figure*}

I mainly used the fit from the \textit{Swift} webpage\footnote{http://www.swift.ac.uk/xrt\_live\_cat/} when there is clear plateau and a break. One example is given in Figure \ref{Ta_fits_from_Swift_team} (left) for this condition. 
When no conclusion could be reached with the light-curve obtained from XRT only, a comparison between the BAT and XRT light-curves\footnote{http://www.swift.ac.uk/burst\_analyser/} was performed. When the decay measured by XRT corresponds to the decay of the prompt phase measured by BAT, this part of the light-curve up to the next change in the slope is considered to be associated to the prompt emission. Since the slope of the plateau is expected to be close to zero, the slope of the next segment in the light-curve determines if it is the plateau or not. If so, $T_a$ is taken as the time at which the next change in the slope happens. Conversely, an upper limit on $T_a$ is obtained. An example is given in Figure \ref{Ta_fits_from_Swift_team_compared_BAT}.

\begin{figure*}[!ht]
\begin{center}
\includegraphics[width=0.7\textwidth]{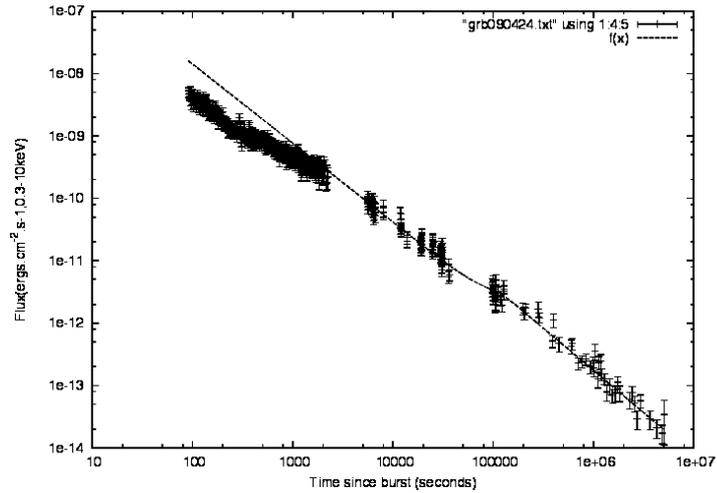}
\caption{One example of an X-ray afterglow, fitted by smoothly joined power laws. As the decay index at late times is not compatible with a jetted afterglow, only the last segment of the light-curve is considered in the analysis. \label{Ta_fits_090424}}
\end{center}
\end{figure*}

When no plateau can be identified, since I am interested only in the late afterglow emission, I have to identify segment 3 and eventually 4, see Figure \ref{Swift_canonical_light_curve}. For that, two smoothly joined power laws (given in Equation \ref{decay_SBPL}) are fitted to the two last segments of the light-curves. If the slope of the very last segment is close to $-2$ and the slope of the previous segment is close to $-1.2$, both parts of the light-curves are kept for the analysis and identified as standard afterglow and jetted afterglow respectively. Conversely, only the last part is considered. An example is given in Figure \ref{Ta_fits_090424} and in Table \ref{table_fit_result_GRB090424} for GRB 090424. As the last decay index is found to be $-1.35$, not compatible with the jetted afterglow, only the last part of the afterglow is considered in the analysis.

\begin{table*}[!ht]
\centering
  \caption{The fit result of GRB~090424.\label{table_fit_result_GRB090424}}
  \begin{tabular}{ccccccc}
  \hline
GRB & $z$ & First &Second &$\chi^2$ & d.o.f & $T_{break}$ \\ 
        &     &index & index &(reduced) &  & (s) \\
    \hline
090424	 & 0.544	&1.16$\pm$0.04  &1.35$\pm$0.03 &1.10 &311 & 10$^5$ \\ 
\hline
\end{tabular}
\end{table*}

Lastly, when the fit is not constrained (mainly for weak in X-ray emission), I imposed $T_a\sim 10000$~s, see one example in Figure \ref{Ta_fits_from_Swift_team}.

I finally removed all flaring emissions from the light-curves, which are thought to be part of the prompt emission \cite{zhang2014}. In addition, GRB~071122 was also removed from the sample because it was not observed enough.

\subsection{Cosmological Scaling} 

In all previous works on the clustering of X-ray afterglow, rather than using the luminosity, the authors used the flux at a common distance (set at $z = 1$) in order to reduce uncertainties on the Hubble constant and spectral parameters. Indeed, each measurement has its own associated error, and the k correction is very sensitive to the spectral index of the afterglow. Using a flux unit has the advantage of reducing the amount of correction applied, and therefore reduces the statistical scattering inserted in the data by the distance correction. To choose the faint sources from the global sample, I followed the same approach for the same reasons.

Firstly, I applied the standard time dilation scaling:

\begin{align}
\label{time_delation_correction}
t_{rest} & = \frac{t_{obs}}{1+z_0} \text{,}  ~~~~~~~~~~\\ 
t_{obs,z=1} & = t_{rest} (1+1) \text{,}  ~~~~~~~~~~ \\
t_{obs,z=1} & = \frac{t_{obs} (1+1)}{1+z_0} \text{,}  ~~~~~~~~~~ 
\end{align}
where $t_{rest}$ is the time measured in the rest frame of the source, $t_{obs}$ is the observed time and $z_0$ is the redshift of the burst.

Using a flux unit has the advantage of reducing the amount of correction applied, and consequently this reduces the statistical scattering inserted in the data by the distance correction.
The total luminosity:

\begin{align}
L & = 4 \pi \ d_L^2 \ F  \\
  & = 4 \pi \ d_{L,z=1}^2 \ F_{z=1}  \text{,}  
\end{align}
where the last equality is obtained by imposing a constant observed luminosity (here $L$ and $F$ are the bolometric luminosity and flux with unit $ergs.s^{-1}.cm^{-2}$). Then, the flux which would have been observed if the burst redshift was $z = 1$ is:  

\begin{equation} 
F_{z=1} = \frac{d_L^2}{d_{L,z=1}^2} \times F_z \text{.} 
\end{equation}

Finally, $d_L$ is the luminosity distance and it is computed by \cite{pen1999}:
\begin{align}
\label{fit_luminosity_distance}
\begin{split}
d_L &=  \frac{c}{H_0}(1+z)\left[\eta(1, \Omega_0) - \eta \left(\frac{1}{1+z}, \Omega_0 \right) \right] \text{,} \\
\eta(a,  \Omega_0) &= 2 \sqrt{s^3 + 1}\left[\frac{1}{a^4} - 0.1540 \frac{s}{a^3} + 0.4304 \frac{s^2}{a^2} + 0.19097 \frac{s^3}{a} + 0.066941s^4 \right]^{-1/8} \text{,} \\
s^3 &= \frac{1-\Omega_0}{\Omega_0} \text{,}
\end{split}
\end{align}
which is a fit of the exact integral expression. It gives a result precise at 4\%. Finally, the parameters are $H_0$ = 70~km.s$^{-1}$.Mpc$^{-1}$, $\Omega_M = 0.3$. 

\subsection{Energy correction} 

Then I restricted the energy range from $0.3 -10$~keV to $2 - 10$~keV band. This allows to neglect absorption effects, so any correction  is needed to take into account the absorption by the ISM. 
To do so, I used the late afterglow spectrum parameters (if not available I used the time averaged spectrum parameters): $N_{H,gal}$, $N_{H,int}$, $z$, photon index obtained from the automatic data analysis\footnote{http://www.swift.ac.uk/xrt\_spectra/}. I calculated the energy correction factor (ECF) by using a Heasarc tool: \textit{WebPIMMS}\footnote{http://heasarc.gsfc.nasa.gov/Tools/w3pimms.html}. The conversion rate was taken to be one since a flux light-curve was used. Finally, the ECF is applied to the observed flux:

\begin{equation}
F_{corr} = F_{obs} \times ECF
\end{equation}

The k-correction takes into account cosmological correction and the down-shift of the gamma-ray energy from the burst to the observer's reference frame since a higher energy component of the source spectrum is redshifted into the sensitivity band of the detector for high redshift sources \cite{coward2012}. It is:

\begin{equation}
\label{k_corr}
k(z) = \frac{\int_{e_1}^{e_{2}} E \ N(E) \ dE}{\int_{(1+z)e_1}^{{(1+z)e_{2}}} E \ N(E) \ dE} \text{,}
\end{equation}
where [$e_1$, $e_2$] is the sensitivity band of the instrument, $N(E)$ is the source photon spectrum fitted with a simple power law for the afterglow of GRBs, see Equation \ref{simple_PL}. Using Equations \ref{simple_PL} and \ref{k_corr}, the k correction can be computed and is applied to all afterglow light-curves: 

\begin{equation}
k =  \frac {(1+z)^\beta}{(1+z)^2} \text{,}  
\end{equation}

where $\beta$ is the spectral index. Rescaling all bursts at redshift one implies that the total correction is:

\begin{equation}
K_{z=1} =  \frac {k(z_0)}{k(1)} = \frac {(1+z_0)^\beta (1+1)^2}{(1+z_0)^2 (1+1)^\beta} \text{,}  
\end{equation}

The k correction is very sensitive to the measured spectral index of the afterglow.
As an example, with a redshift of 4 and a precision of $1.0 \pm 0.3$ for $\beta$, the uncertainty on $k$ is $5 \pm 3$, \textit{i.e.} 60\%. Rescaling to $z = 1$ leads to $1.5 \pm 0.7$, \textit{i.e.} an uncertainty of 16.8\% : this method reduces the scattering induced by the uncertainties on the measurements, allowing for a more precise selection of the sample.

After taking into account the k correction and distance correction, the flux at redshift $z = 1$ is given by:
\begin{equation}
F_{z=1} = \frac{d_L^2}{d_{L,z=1}^2} K_{z=1} \times F_{obs} \times ECF \text{.} 
\end{equation}

\section{Selection of the LLA sample from the global one}

\begin{figure*}[!ht]
\begin{center}
\includegraphics[width=0.9\textwidth]{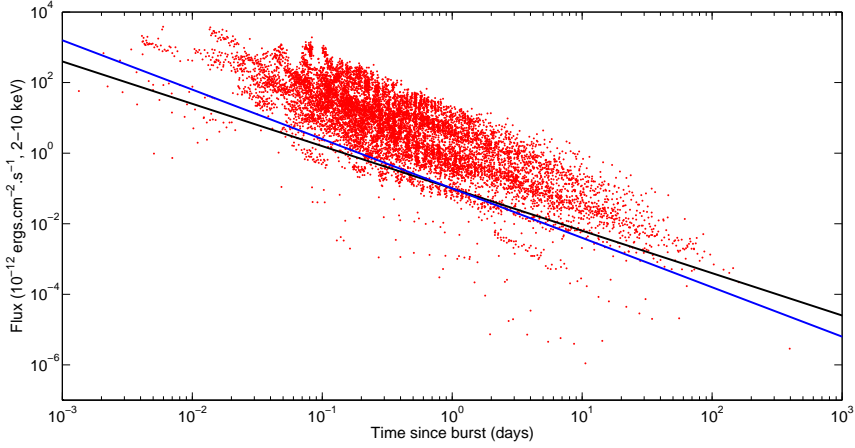}
\caption{The light-curves of all 254 sources, corrected for distance effects (see text) and rescaled at a common redshift $z = 1$. The decay lines 1.2 and 1.4 are presented in black and blue respectively.\label{fig_lines}}
\end{center}
\end{figure*}

\begin{figure*}[!ht]
\begin{center}
\includegraphics[width=0.9\textwidth]{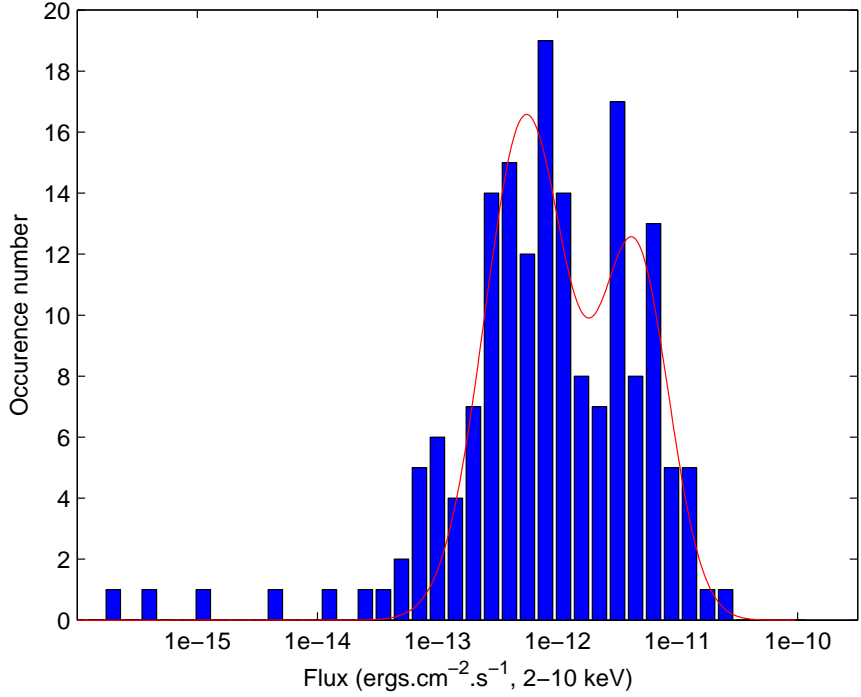}
\caption{The flux distribution at one day fitted by two Gaussians.\label{fig_histo}}
\end{center}
\end{figure*}

All corrected 254~light-curves are displayed in Figure \ref{fig_lines}. In 2008, \cite{gendre2008} found a clear dichotomy in the flux of late afterglow light-curves considering 61~GRBs. The two groups were named group I (high luminous) and group II (middle luminous) and their properties determined. However, now the statistical significance of the complete sample is so high that several bursts can lie within the separation zone of each group. In order to select only group III events (that are neither in group I nor in group II), I had to define the group I and II mean values and standard deviations. For this purpose, I found the flux values for the sources which have measurements before and after at one day after burst then I extracted the flux distribution at z=1 at that time, and fitted it with two Gaussian distributions, assuming that each group is represented by a normal law. The Gaussian distributions are shown in Figure \ref{fig_histo} and their parameters summarized in Table \ref{table_fit}. Subtracting group I and group II leaves only the faint bursts. This way, I determined a flux threshold of $10^{-13}$ erg.s$^{-1}$.cm$^{-2}$ (or 1.4$\sigma$ from the mean value of group II, the 90\% confidence level) at one day. 

However, not all bursts have a flux measurement at the time I performed the fit. I thus finally used the standard decay index $\alpha = 1.2$, expected from the late time of the fireball evolution in a homogeneous medium with the power-law index of the accelerated electrons fixed to $p = 2.3$, to construct a time-varying upper limit for group III. In Figure \ref{fig_lines}, the black line is drawn by using the flux threshold of $10^{-13}$ erg.s$^{-1}$.cm$^{-2}$ at one day  and a decay index of 1.2. When this limit was not enough to choose the sample, I used the mean decay index value of group II events (\textit{i.e.}~1.4) which is presented in blue in the Figure \ref{fig_lines}. I finally discarded all bursts which significantly cross these limits, leaving only 31~LLA~GRBs in sample. They are represented by blue diamonds in Figure \ref{fig_sample} and listed in Table \ref{table_sample}. The bursts which are not in this sample are used as a control sample and they are represented with red points in Figure \ref{fig_sample}.

\begin{table*}[!ht]
\centering
  \caption{Gaussian fit parameters of the flux distribution at one day for group I and II.\label{table_fit}}
  \begin{tabular}{ccc}
  \hline
  Groups & Mean flux & Standard deviation\\
         & (10$^{-12}$ c.g.i.) & (log (10$^{-12}$ c.g.i.))\\
    \hline
I   &   6.3 & 0.40 \\ 
II  &   0.6 & 0.48 \\ 
\hline
\end{tabular}
\end{table*}

As a result, to characterize a sample of GRBs with low luminosity X-ray afterglows, I selected a sample consisting of the 12\% faintest X-ray afterglows from the total population of long GRBs with known redshift.

\section{The spectral index and temporal decay index calculations}
The spectral and temporal decay indexes are obtained from the spectrum and the light-curve after $T_a$, fitted by considering Equation \ref{Eq_Fnu_alpha_beta}. To compute the spectral index, I followed the same method described in section \ref{ext_LC_spec} 
(extraction of the light-curve and the spectrum). Observations were added to the first one up to the point that a good statistic ($\chi^2$ or cash) was obtained. The results of the analysis are presented in Table \ref{spectral_index_fit_results} with their $\chi^2$ or cash statistics. Moreover, the values of the column density $N_{H,X}$ and the spectral indexes (which is equal to the photon index minus one) are summarized in columns 4 and 8 of Table \ref{table_sample}.

\begin{table}
\centering
  \caption{Spectral fit results of LLA GRBs. When the intrinsic column density $N_{H,X}$ was found very small its value was set to zero and the fit performed again. The real values of $N_{H,X}$ are given in parentheses on column~4. \label{spectral_index_fit_results}}
  \begin{tabular}{llccccc}
  \hline
GRB & z & photon & $N_{H,X}$ & $\chi^2$ & d.o.f \\ 
        &     & index & 10$^{22}$ cm$^{-2}$ & (reduced) &  \\
    \hline 
GRB~980425   & 0.0085   &&&& \\
GRB~011121 & 0.36          &&&& \\
GRB~031203   & 0.105    &&&& \\
GRB~050126 & 1.29 &    &&&& \\
GRB~050223   & 0.5915  &&&& \\ 
GRB~050525 & 0.606      & 2.1$\pm$0.4 & 0.38$^{+9.1}_{-0.38}$ & 1.25 (cash) & 9 \\
GRB~050801 &1.38         &   3.1$^{+0.39}_{-0.36}$ & (0.0)  & 0.81 (cash) &108 \\
GRB~050826   & 0.297	&&&& \\
GRB~051006 & 1.059        & 2.5$^{+0.44}_{-0.46}$ &(0.0)  & 0.93 (cash) & 89 \\
GRB~051109B  & 0.08	&&&& \\
GRB~051117B & 0.481  & 2.23 & (0.0)  &1.38 & 2 \\
GRB~060218   & 0.0331	&&&& \\
GRB~060505 & 0.089         & (1.8) & (0.0) &  &   \\
GRB~060614   & 0.125	&&&& \\
GRB~060912A & 0.937	&1.63$\pm$0.18 & 0.0 (0.06) &1.07 & 8 \\
GRB~061021 & 0.3463	&2.02$\pm$0.06 & 0.06$\pm$0.02& 0.93 & 158 \\
GRB~061110A & 0.758	&1.42$^{+0.6}_{-0.6}$ & 0.0 (9.51) &0.88 &6  \\
GRB~061210 & 0.4095 & (1.8) & (0.0) &  &  \\
GRB~070419A & 0.97	        &&&& \\
GRB~071112C & 0.823	&1.83$^{+0.4}_{-0.3}$	& 0.096$^{+0.3}_{-0.1}$ & 0.66	& 6 \\  
GRB~081007 & 0.5295    & 1.99$^{+0.9}_{-0.4}$ & 0.97$^{+6.9}_{-0.97}$   & 0.213 & 1 \\
GRB~090417B & 0.345	& 2.34$^{+0.16}_{-0.15}$ & 2.23$^{+0.31}_{-0.27}$ &0.78	&56 \\
GRB~090814A & 0.696	&1.89$^{+0.62}_{-0.72}$	& 0.0 (6.2e-10)	&0.9 (cash) &35  \\ 
GRB~100316D & 0.059	&1.54$^{+0.45}_{-0.47}$& 0.0 (0.84) &1.25 (cash) & 44 \\
GRB~100418A & 0.6235	&1.87$\pm$0.26 &0.0 (2.6e-02) & 0.96 (cash) &101 \\
GRB~101225A & 0.847 & (1.8)  & (0.0) &  &  \\
 GRB~110106B & 0.618 & 2.32$^{+0.67}_{-0.32}$ & (0.0)  & 0.04  & 2 \\
GRB~120422A & 0.283	 &1.42$\pm$0.3& 0.0 (8.8e-07) &1.08 (cash) &85 \\ 
GRB~120714B & 0.3984	&2.51$^{+0.61}_{-0.66}$& 0.0 (0.0) &1.24 (cash) &14 \\
GRB~120722A & 0.9586	&1.17$^{+4.34}_{-1.98}$ & 28.04$^{+158.4}_{-25.9}$ & 0.29	&1 \\
GRB~120729A & 0.8	   &1.79 $\pm$0.23 & 0.0 (4.16e-7) & 0.57	&7   \\ 
\hline
\end{tabular}
\end{table}

\begin{figure*}[!ht]
\begin{center}
\includegraphics[scale=0.2]{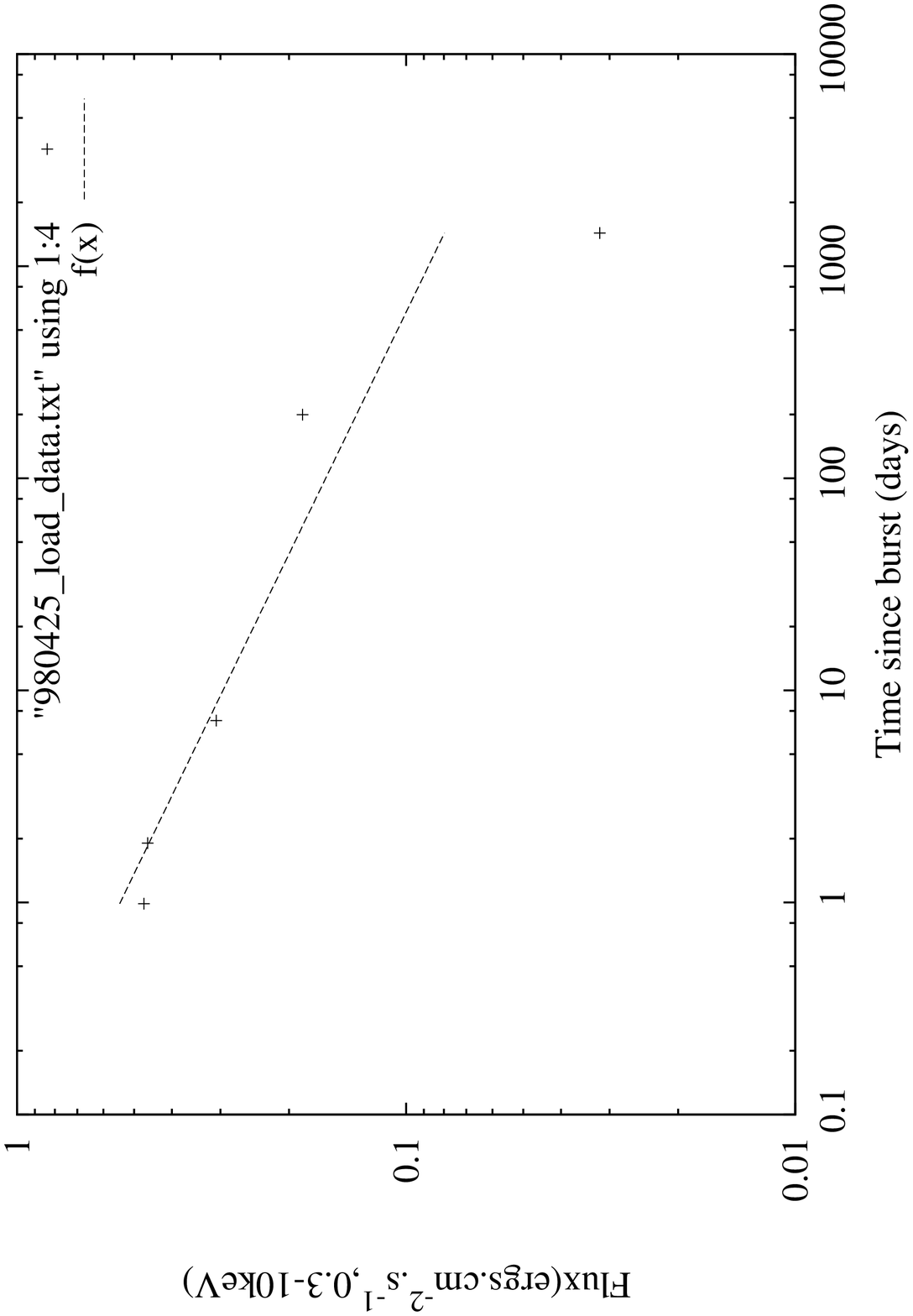}
\includegraphics[scale=0.2]{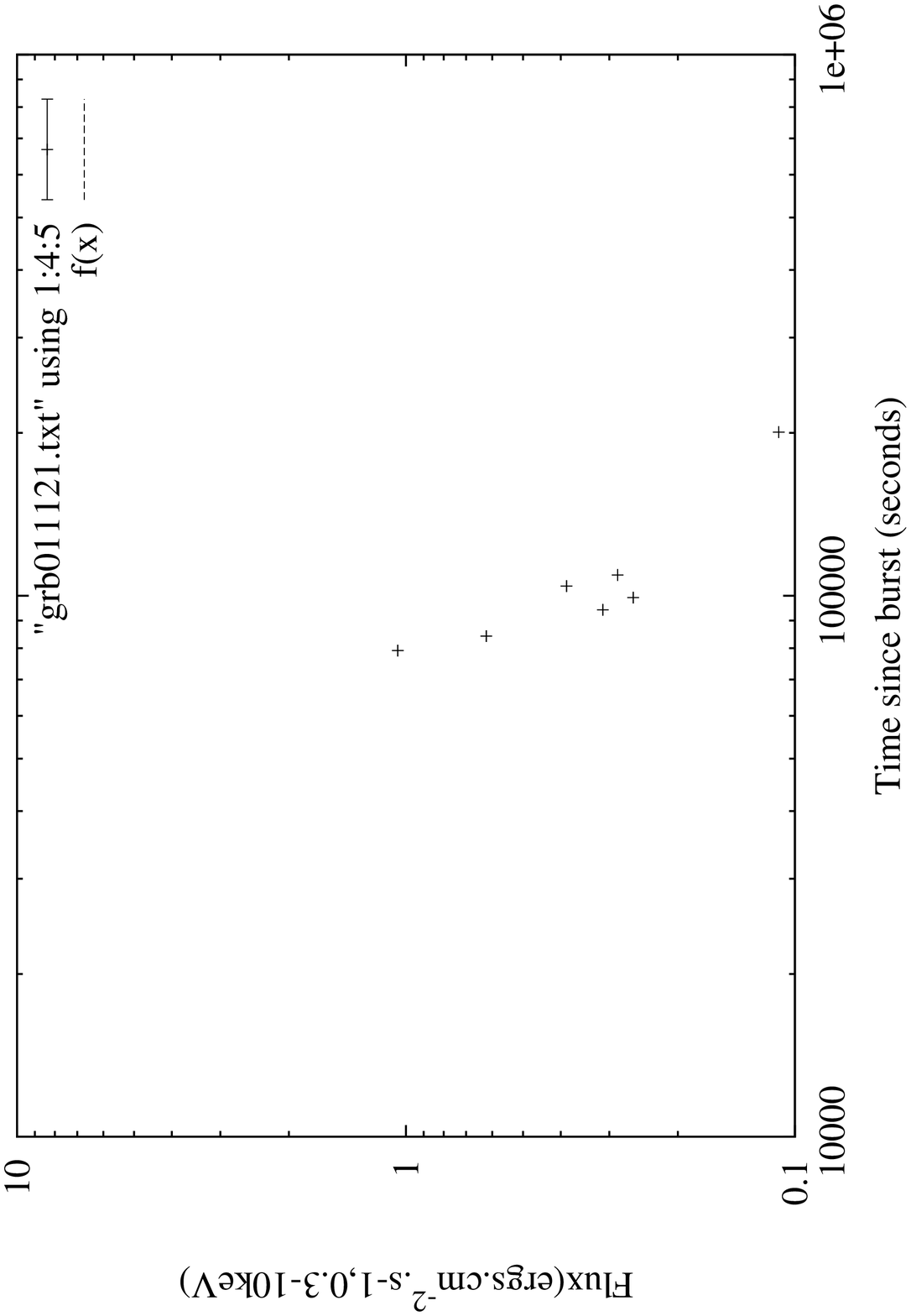}
\includegraphics[scale=0.2]{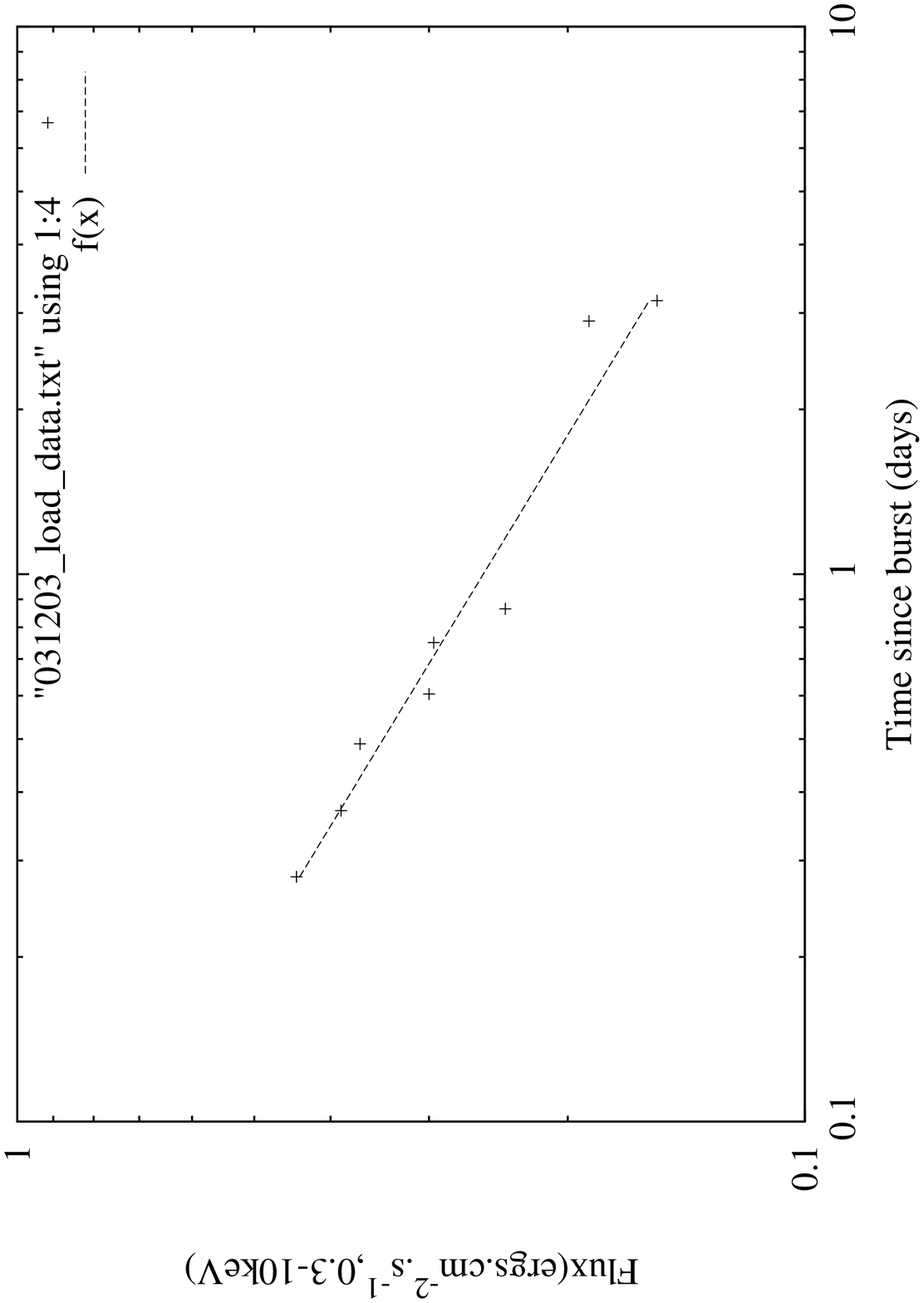}
\includegraphics[scale=0.2]{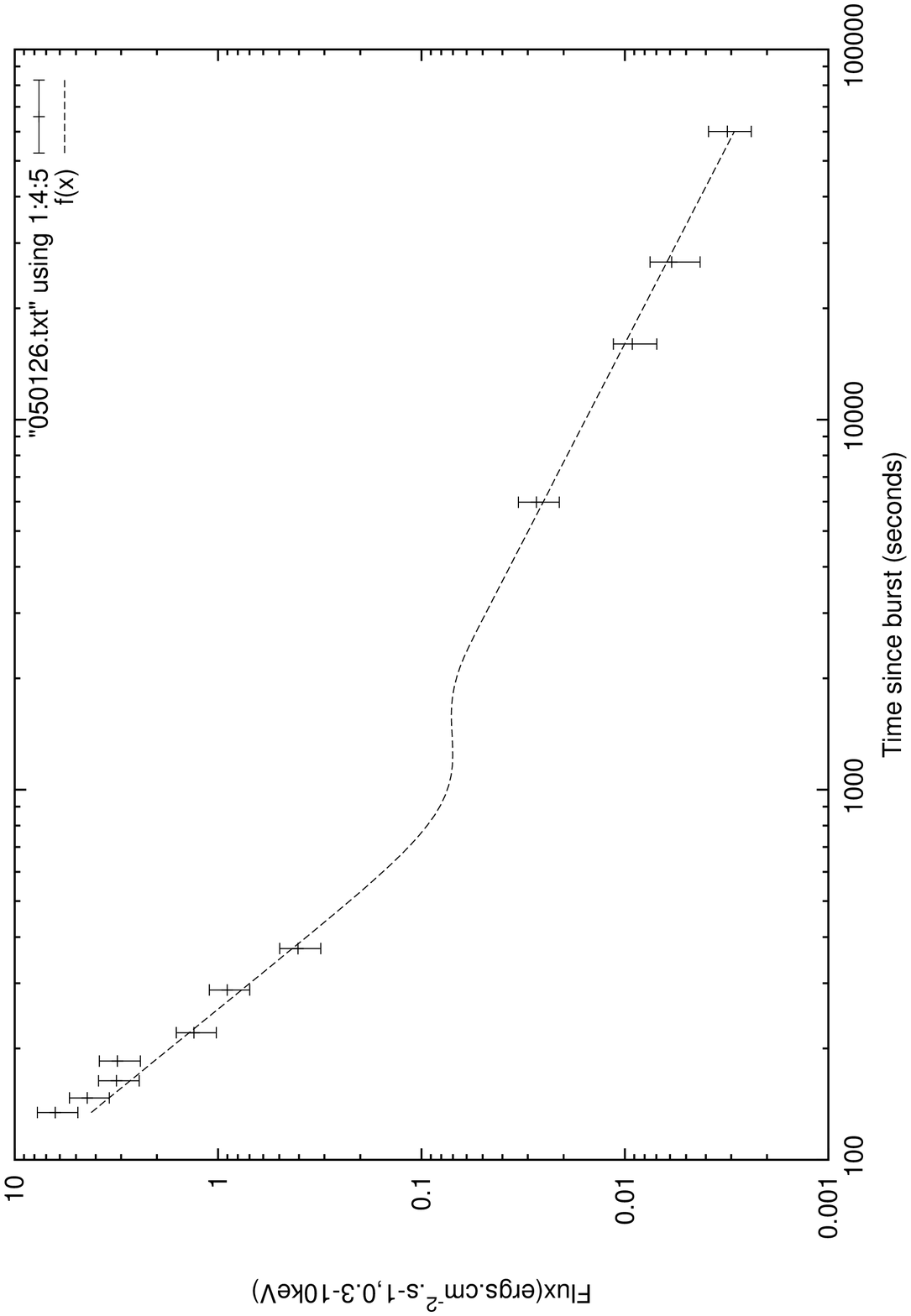}
\includegraphics[scale=0.2]{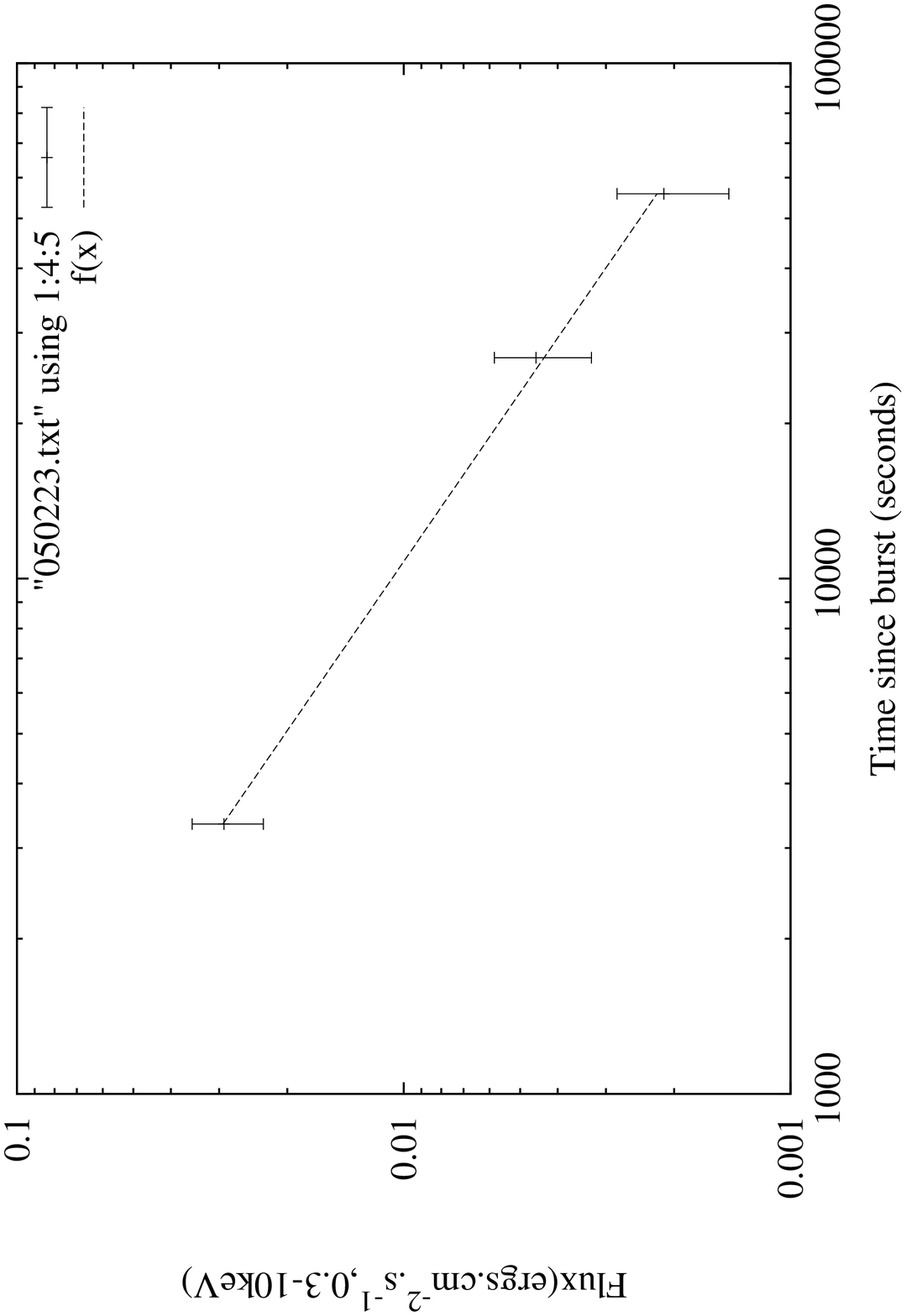}
\includegraphics[scale=0.2]{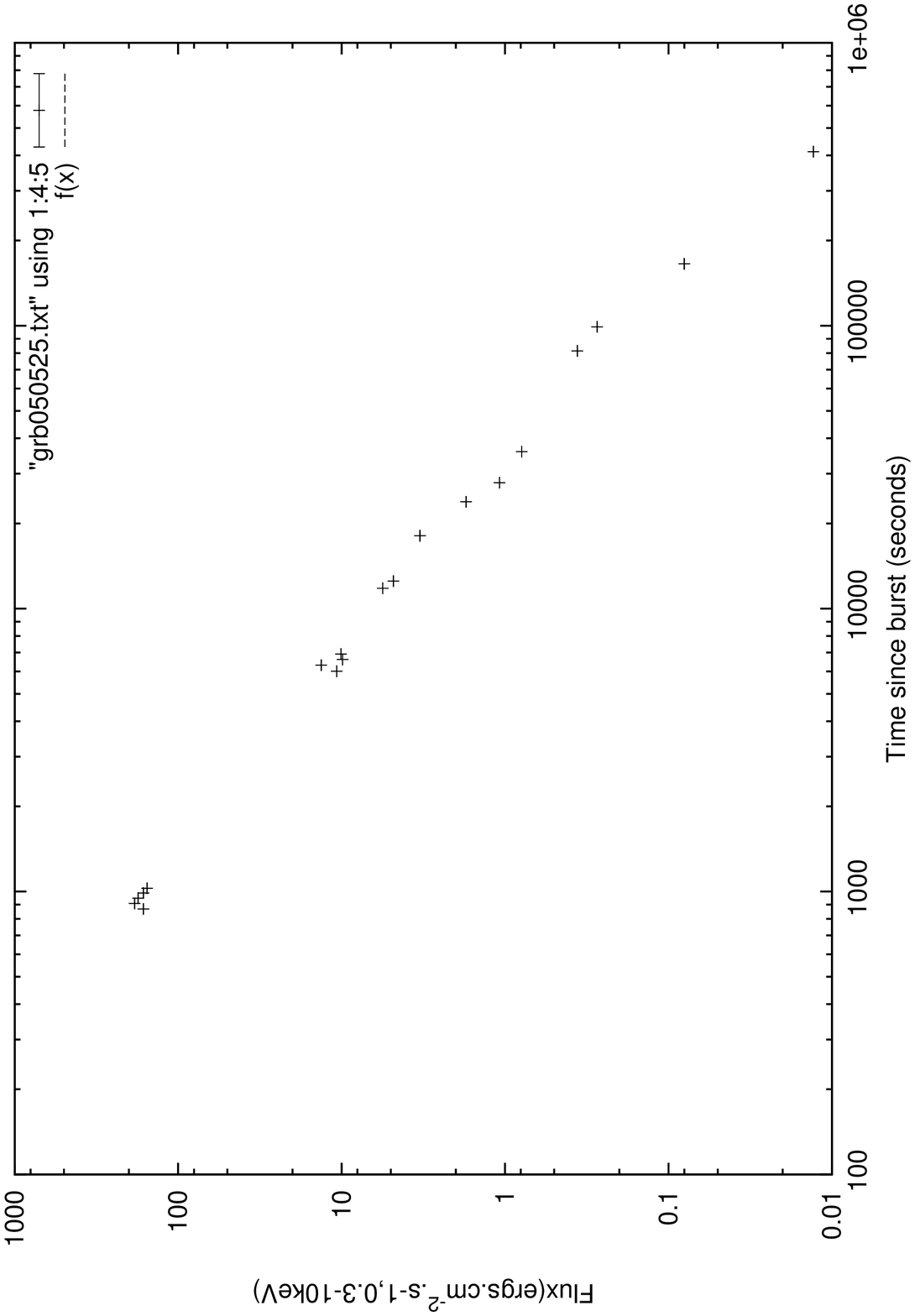}
\includegraphics[scale=0.2]{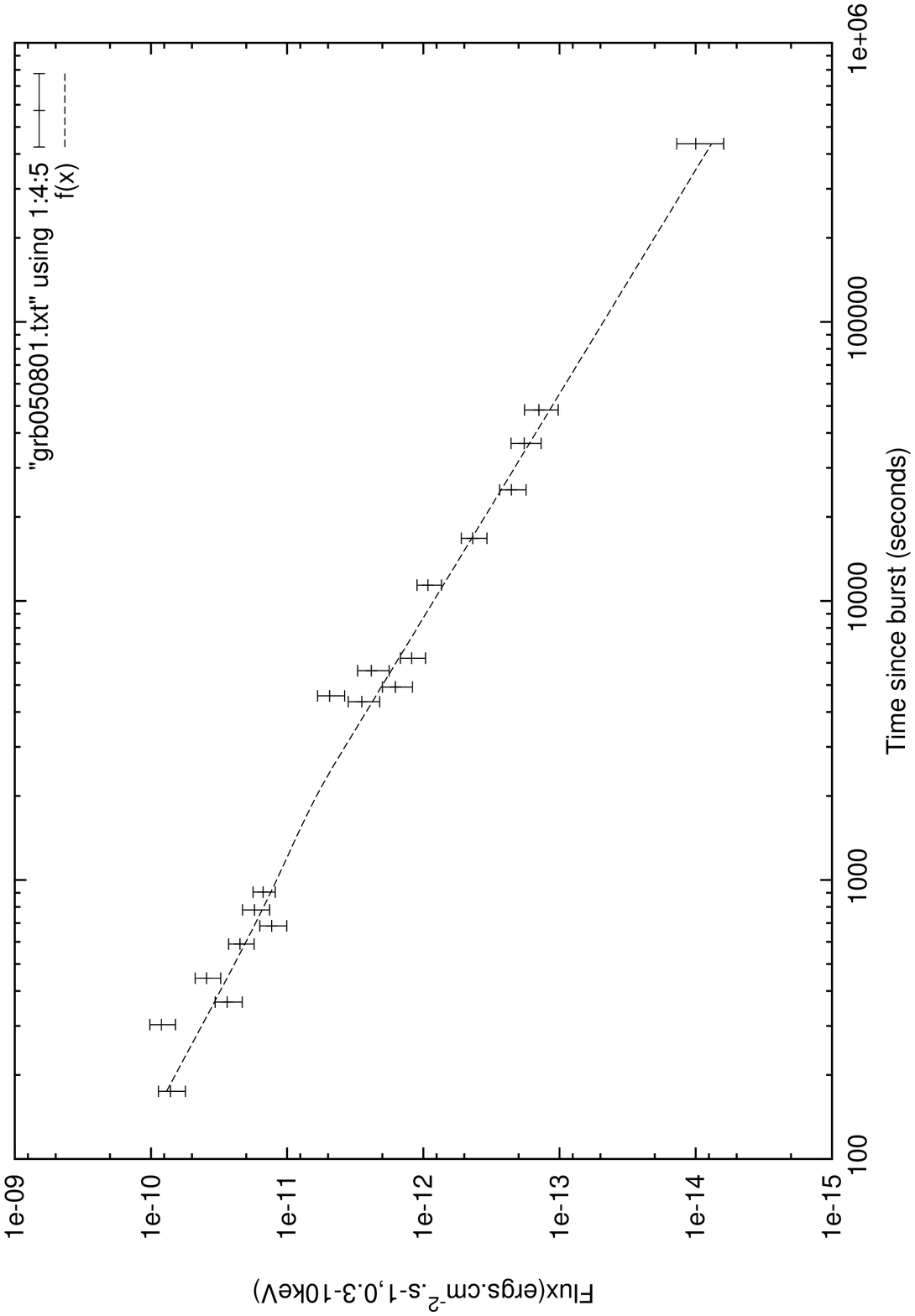}
\includegraphics[scale=0.2]{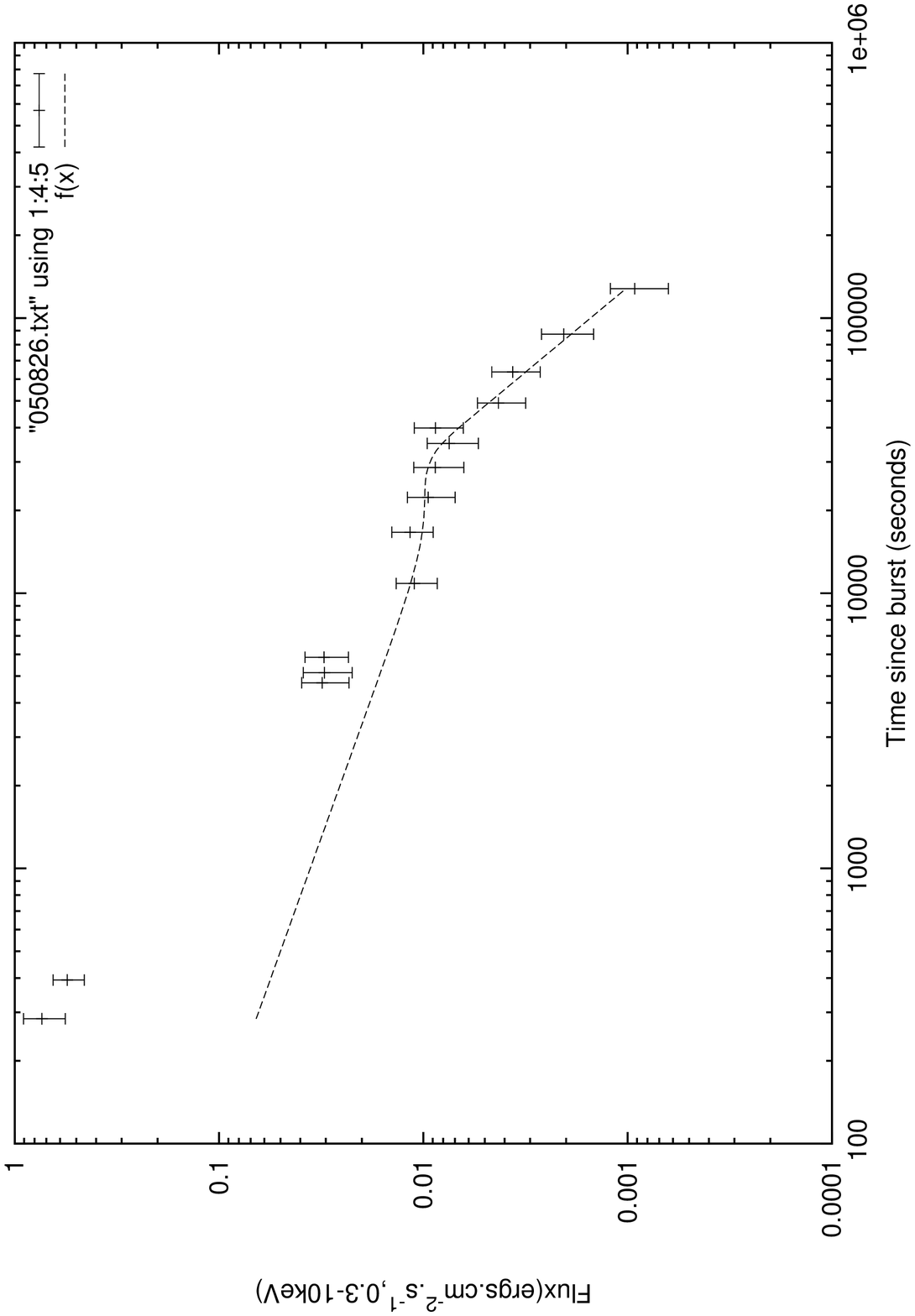}
\includegraphics[scale=0.2]{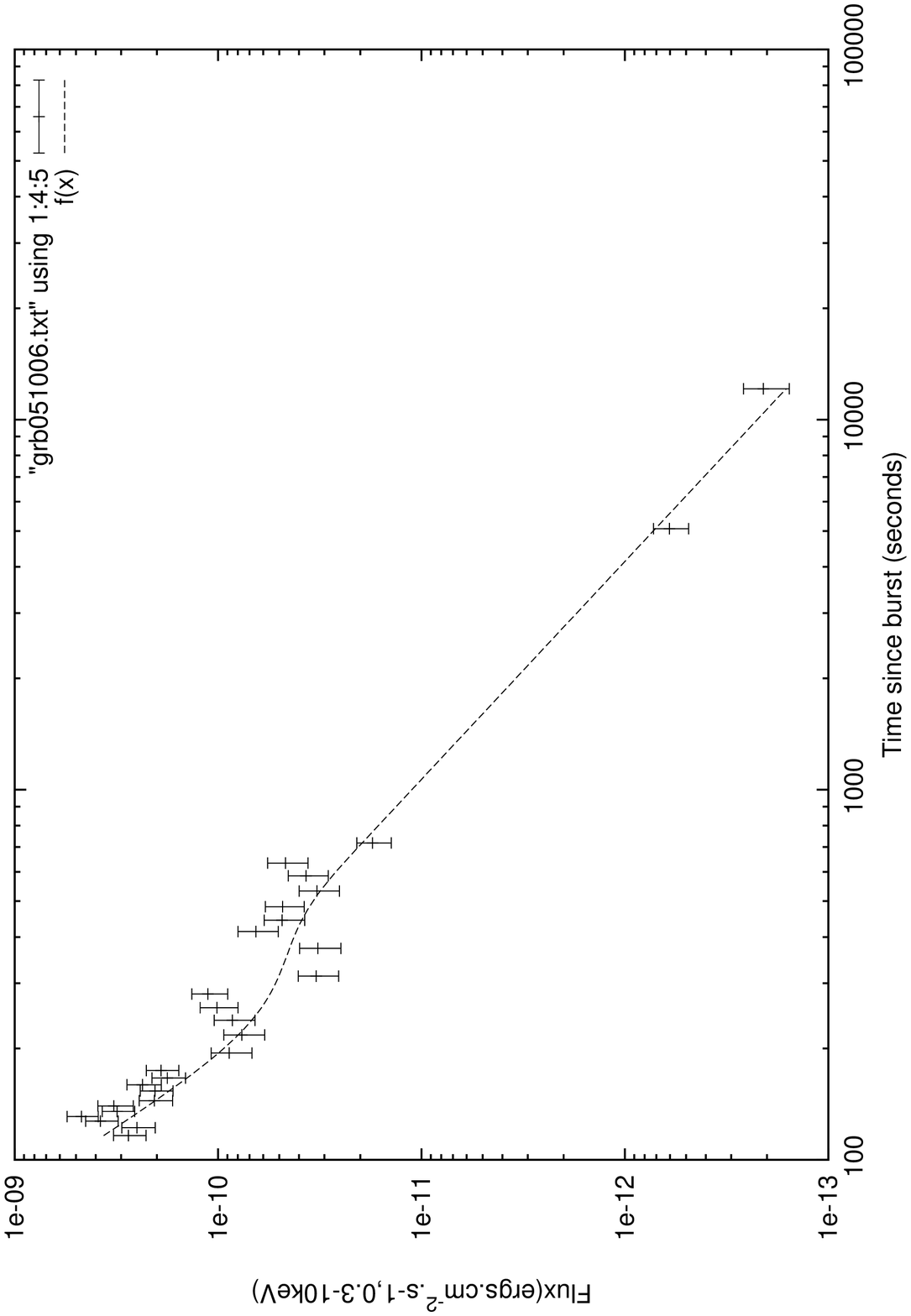}
 \includegraphics[scale=0.2]{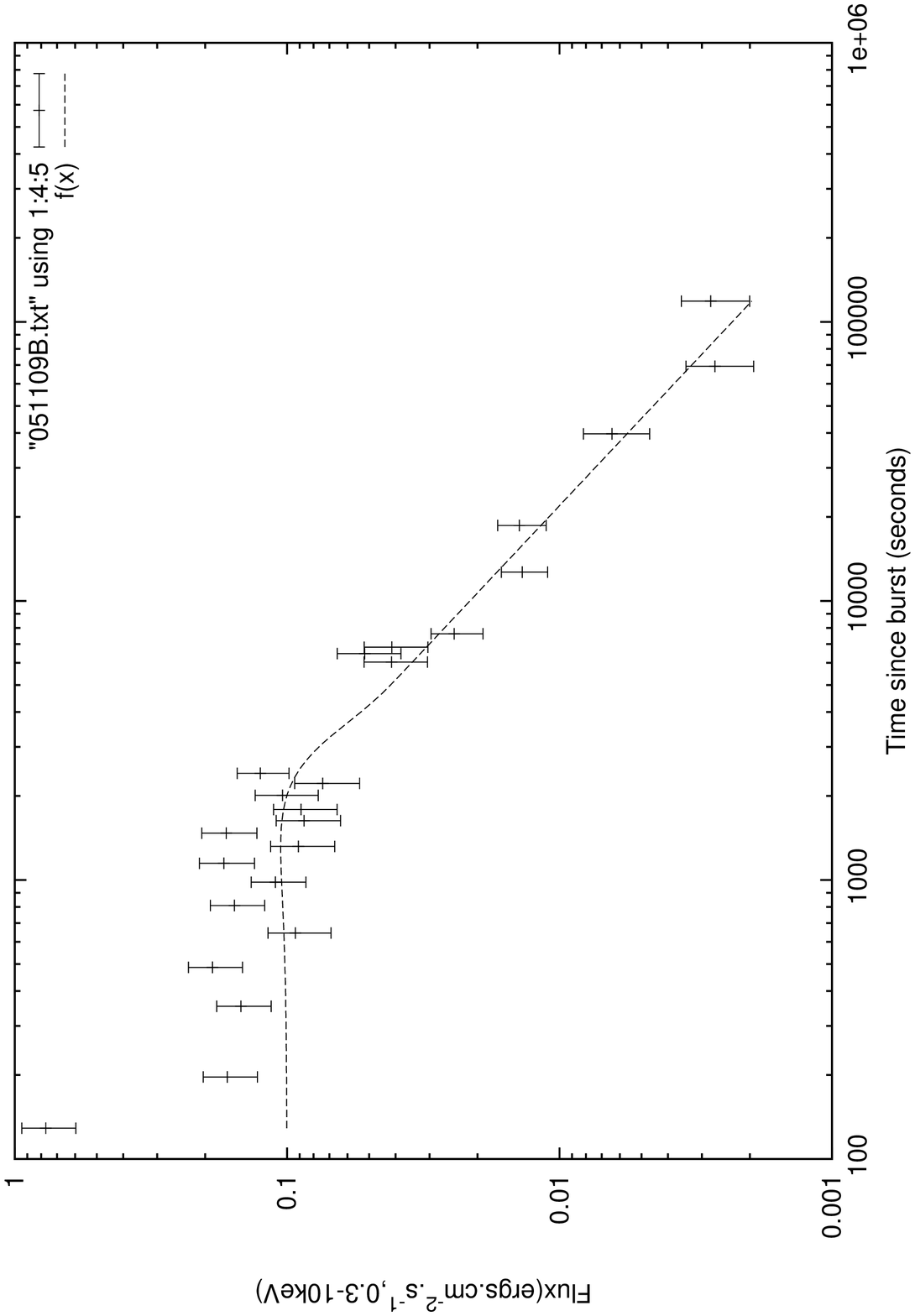}
 \includegraphics[scale=0.2]{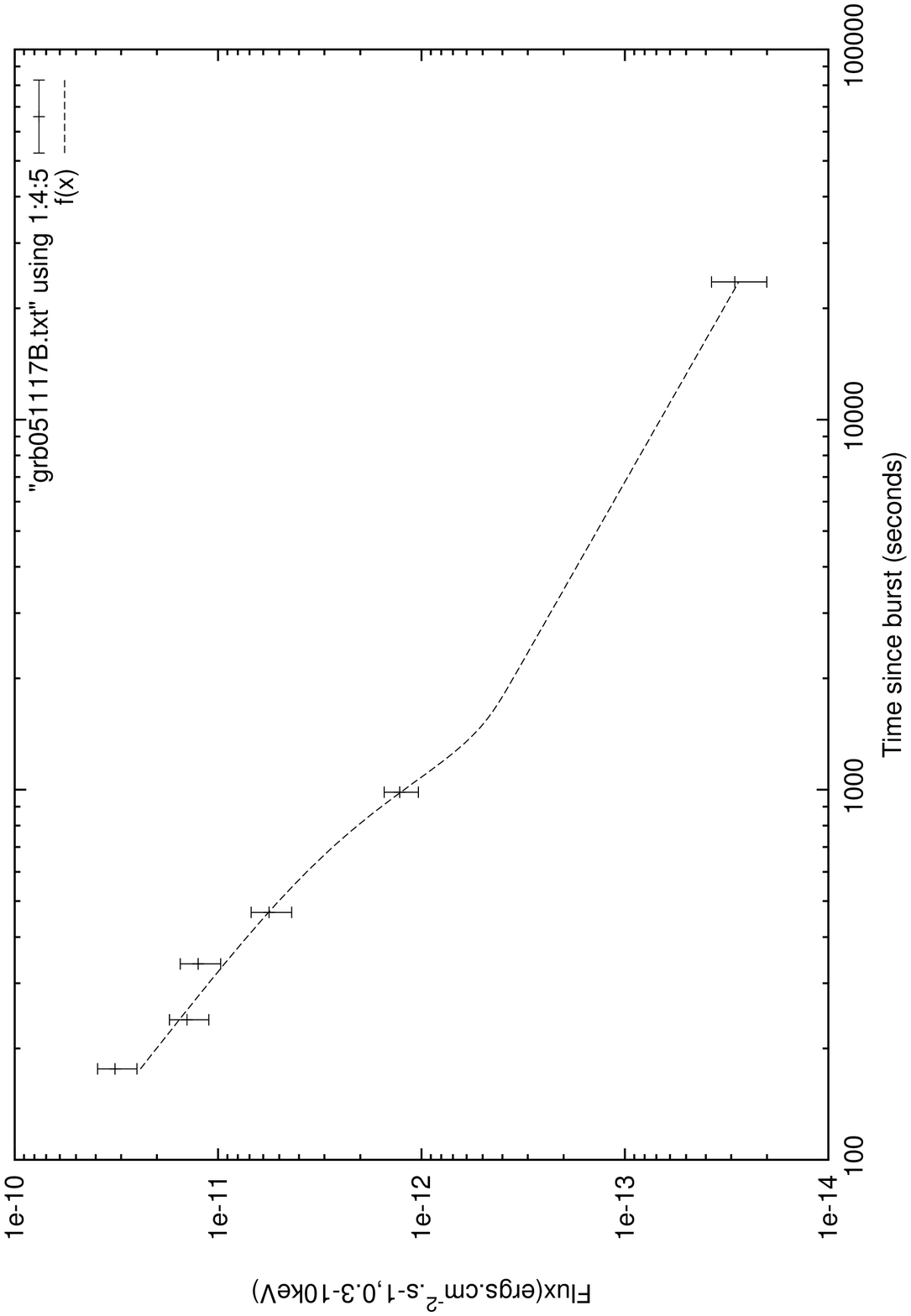}
 \includegraphics[scale=0.2]{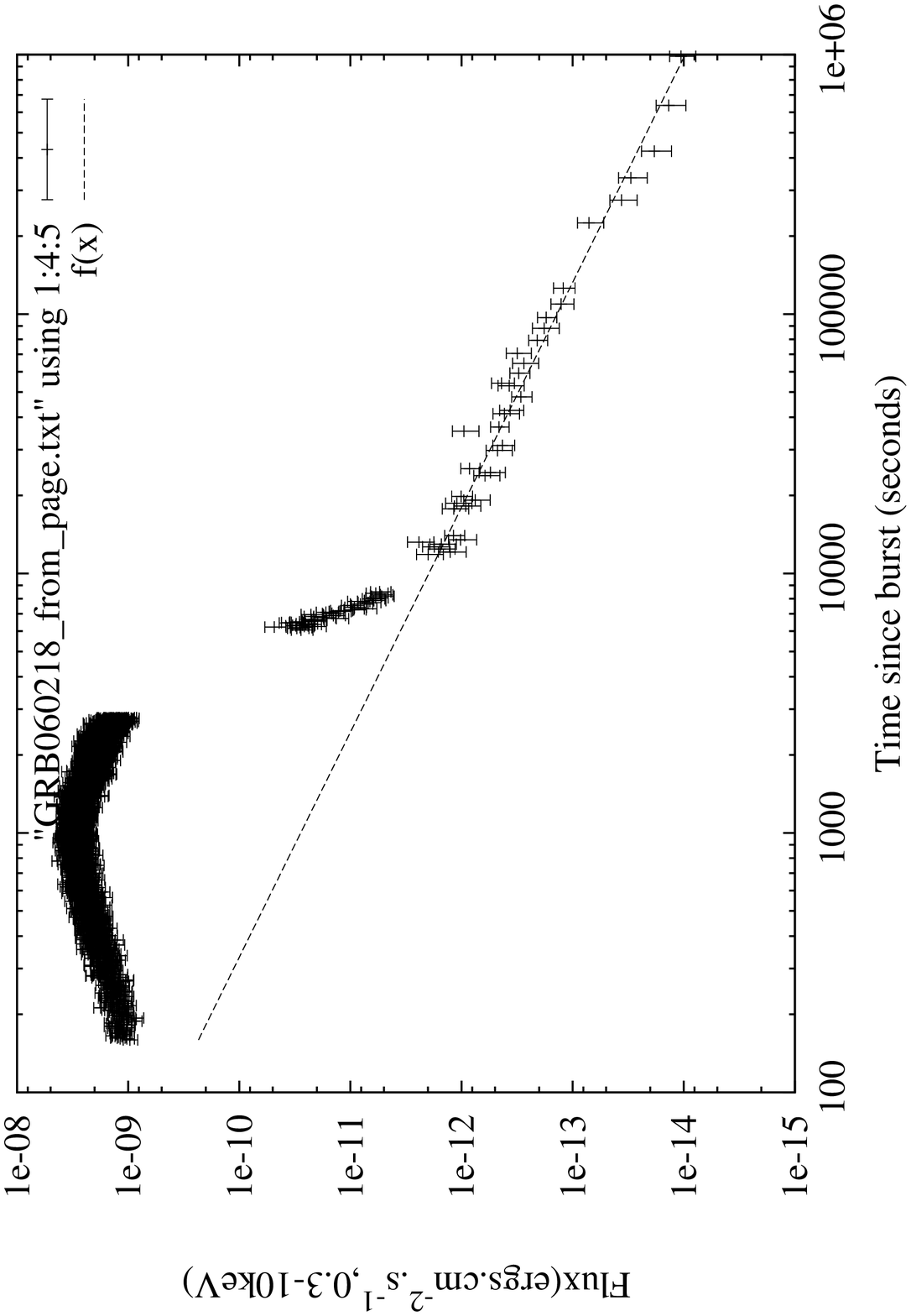}
\includegraphics[scale=0.2]{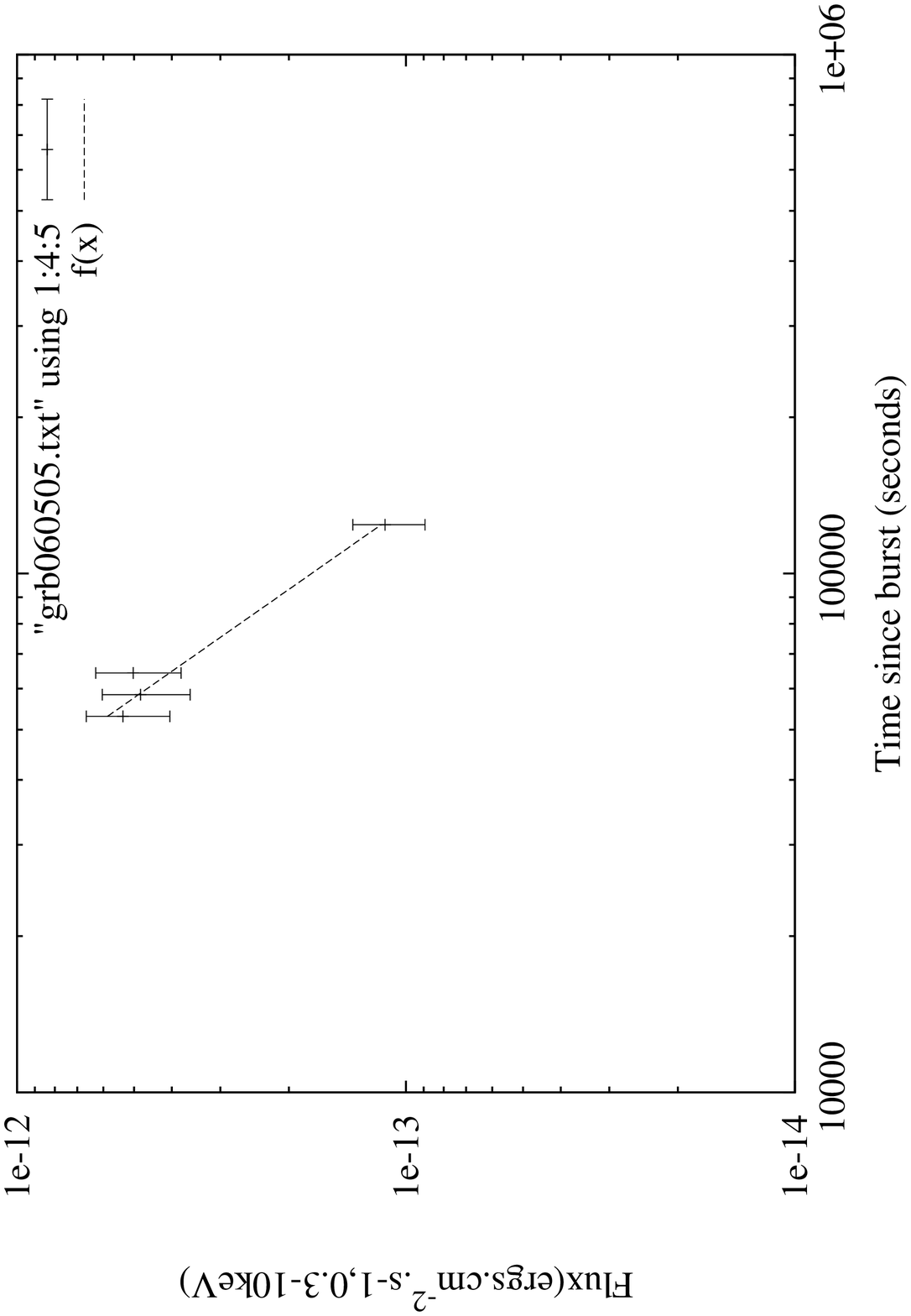}
 \includegraphics[scale=0.2]{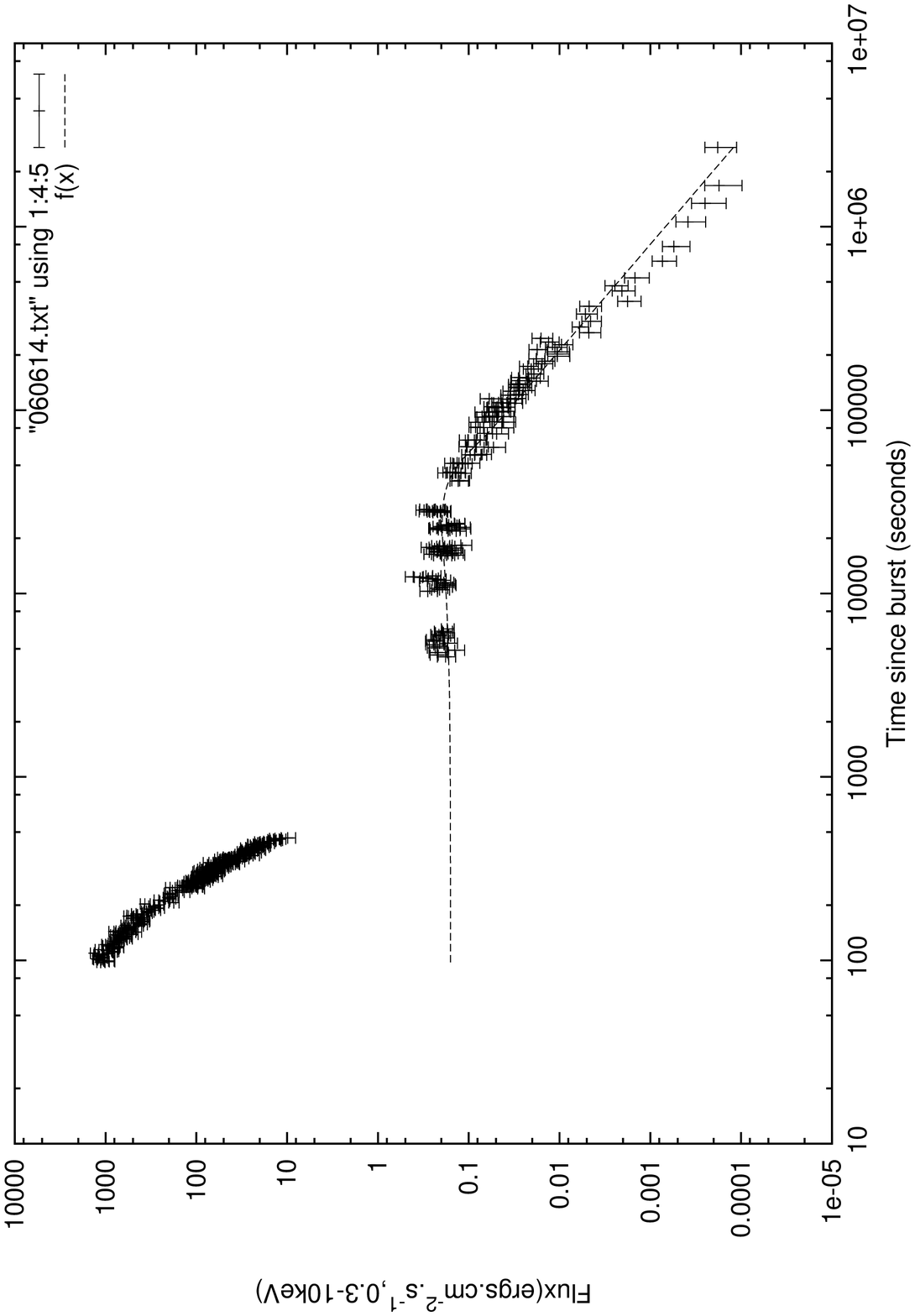}
 \includegraphics[scale=0.2]{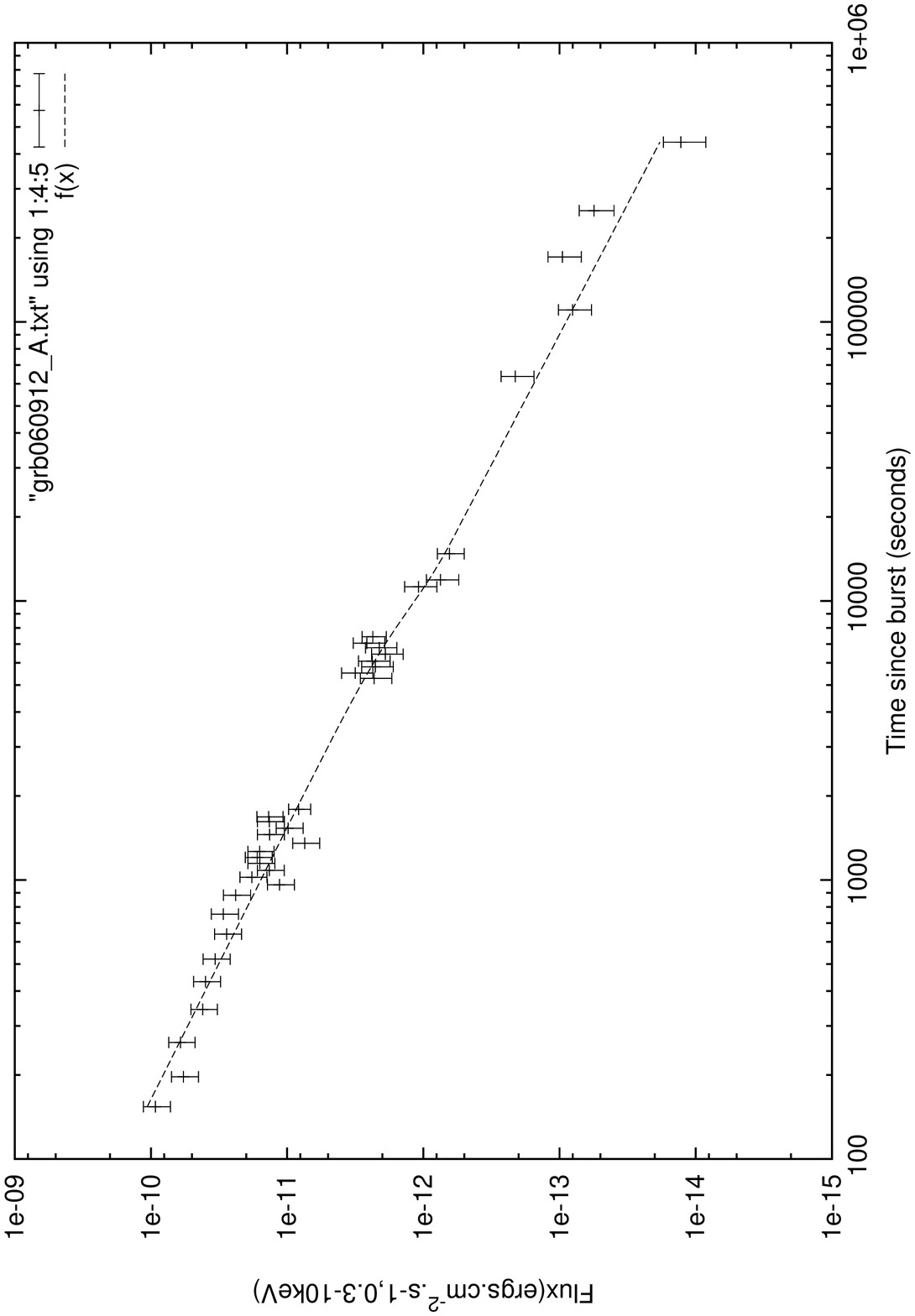}
 \includegraphics[scale=0.2]{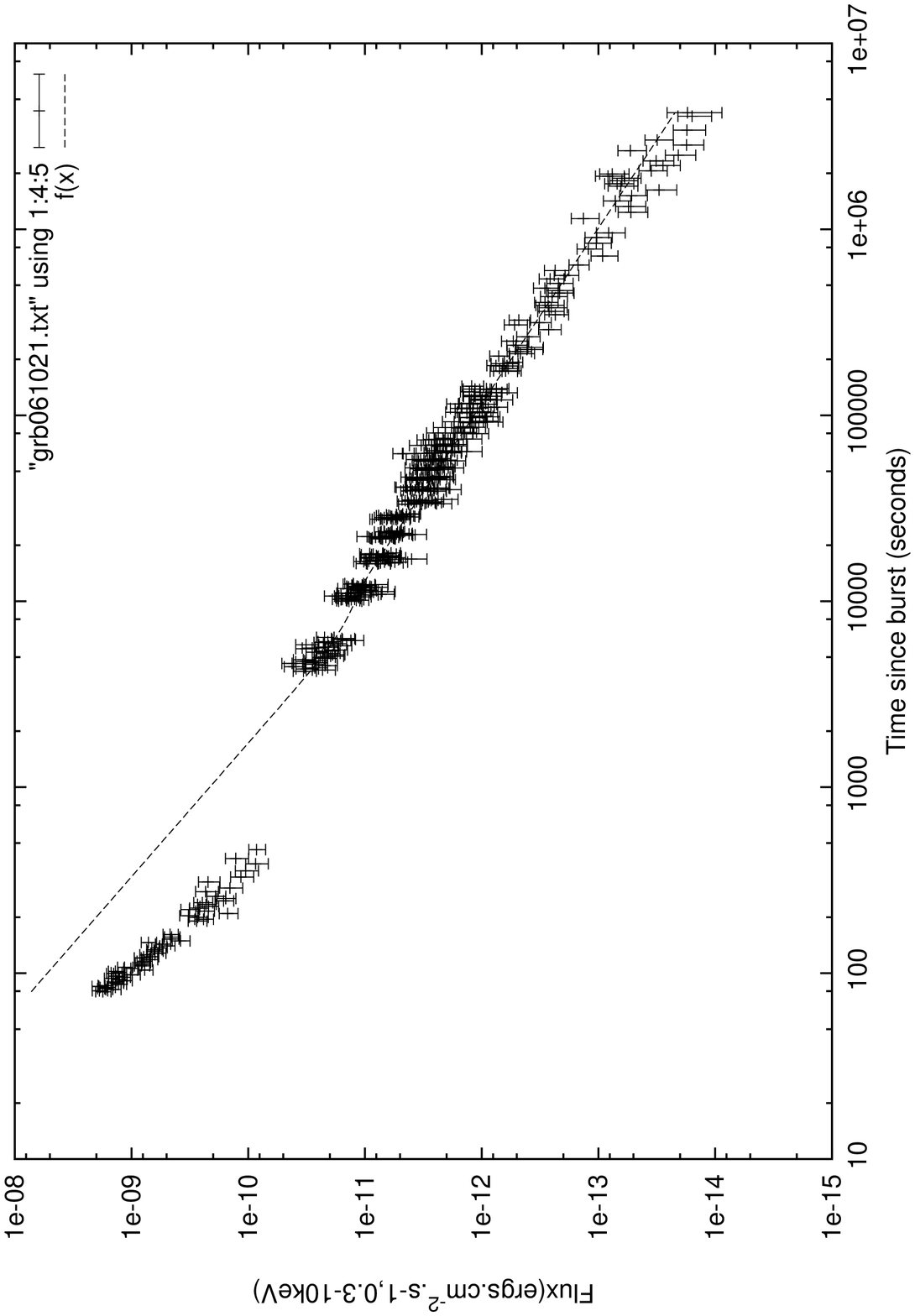}
\caption{The simple power law or smoothly broken power law fit results of 31 sources. \label{decay_index_fit}}
\end{center}
\end{figure*}

For the computation of the temporal decay index, the fit is performed either with a simple power law: 
\begin{equation}
f(x)=a \ x^b
\end{equation}
or a smoothly broken power law given by: 
\begin{equation}
\label{decay_SBPL}
f(x)=a\ x^b\ exp(-(x/x_b)^n)+c\ x^d\ (1-exp(-(x/x_b)^n)) \text{,}
\end{equation}
where $x_b$ is the position of the break. Since I am interested in a sharp transition and as $n$ is poorly constrained by the fit, it is set to the value $n = 3$.  The results are presented on column 7 of Table \ref{table_sample} and in Figure \ref{decay_index_fit}. Additionally, the $\chi^2$ and the degrees of freedom (d.o.f) are given in Table \ref{decay_index_fit_results}.

\begin{table}
\centering
  \caption{The temporal fit results of LLA GRBs.\label{decay_index_fit_results}}
  \begin{tabular}{llccccc}
  \hline
GRB & $z$ & First &Second &$\chi^2$ & d.o.f \\ 
        &     &index & index &(reduced) &  \\
    \hline
GRB~980425   & 0.0085   &$\cdots$  &0.26$\pm$0.062 &0.0025 &2 \\
GRB~011121 & 0.36 &  & & & \\
GRB~031203   & 0.105	& $\cdots$ &0.42$\pm$0.037 &0.0004 &6 \\
GRB~050126 & 1.29 &2.23$\pm$0.34 &0.94$\pm$0.09 & 3 & 0.32 \\
GRB~050223   & 0.5915	&$\cdots$  &0.91$\pm$0.025 &0.045 &1 \\
GRB~050525 & 0.606 & & & & \\
GRB~050801 &1.38 & 1.1$\pm$0.21 & 1.25$\pm$0.13 &14 & 1.73 \\
GRB~050826   & 0.297	&0.31$\pm$0.19  &1.64$\pm$0.19 &0.27 &6 \\
GRB~051006 & 1.059 & 2.8$\pm$3.69 & 1.69$\pm$0.13 & 11 & 2.88 \\
GRB~051109B  & 0.08	&0.0  &0.96$\pm$0.1&1.34 &13 \\
GRB~051117B & 0.481 & 1.41$\pm$0.62 & 1.03$\pm$0.50 & 1 &1.51 \\
GRB~060218   & 0.0331	& $\cdots$  &1.15$\pm$0.03 &0.75 &37 \\
GRB~060505 & 0.089 & $\cdots$ & 1.91$\pm$0.2 & 2 & 0.39  \\
GRB~060614   & 0.125	& 0.0 &1.73$\pm$0.05 &1.4 &147 \\ 
GRB~060912A & 0.937	&1.02$\pm$0.1  &1.07$\pm$0.08 &1.15 &23 \\ 
GRB~061021 & 0.3463	&1.38$\pm$0.33  &1.05$\pm$0.01 &0.94 &267 \\ 
GRB~061110A & 0.758	&1.79$\pm$0.55  &0.68$\pm$0.072 &1.12 &9 \\ 
GRB~061210 & 0.4095 & $\cdots$ & 1.67$\pm$0.85 & 1 & 0.81 \\
GRB~070419A & 0.97	        & $\cdots$  &0.56$\pm$0.0 &0 &0 \\ 
GRB~071112C & 0.823	& 2.86$\pm$0.49 &1.58$\pm$0.07 &0.57 &13 \\  
GRB~081007 & 0.5295 & 0.47$\pm$0.31 & 1.23$\pm$0.05 & 42 & 0.94 \\
GRB~090417B & 0.345	&1.62$\pm$5.7  &1.36$\pm$0.03 &1.03 &75 \\ 
GRB~090814A & 0.696	&2.3$\pm$0.27 &1.3$\pm$0.15& 0.72& 11 \\ 
GRB~100316D & 0.059	&$\cdots$ &1.42$\pm$0.06 &1.6e-28 &1\\ 
GRB~100418A & 0.6235	& 0.13$\pm$0.24&1.42$\pm$0.08 &1.74 &22  \\ 
GRB~101225A & 0.847 & 2.8$\pm$0.37  & 3.05$\pm$0.87 & 63 & 2.05 \\
 GRB~110106B & 0.618 & 0.63$\pm$0.06 & 1.35+/0.06 & 52 & 1.27 \\
GRB~120422A & 0.283	& 0.27$\pm$0.16&1.17$\pm$0.16 & 0.98& 6 \\ 
GRB~120714B & 0.3984	&$\cdots$ &1.89$\pm$0.02 &0.18 & 5 \\ 
GRB~120722A & 0.9586	& 0.27$\pm$0.6&1.99$\pm$0.66  &1.63 & 1 \\ 
GRB~120729A & 0.8	   &1.17$\pm$0.06 & 2.8$\pm$0.23 &1.02 & 62  \\ 
\hline
\end{tabular}
\end{table}

\begin{figure*}[!ht]
\begin{center}
 \includegraphics[scale=0.2]{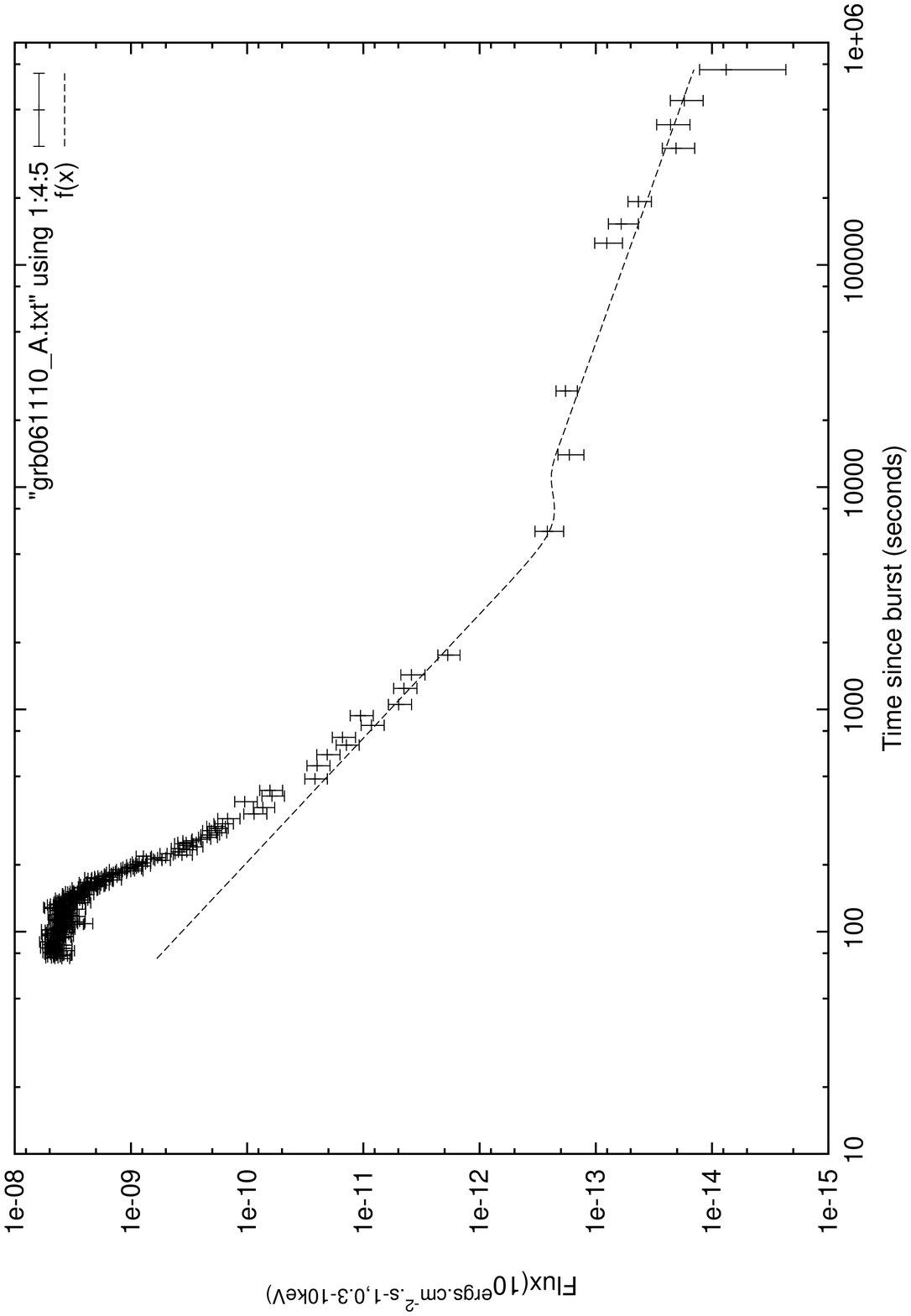}
\includegraphics[scale=0.2]{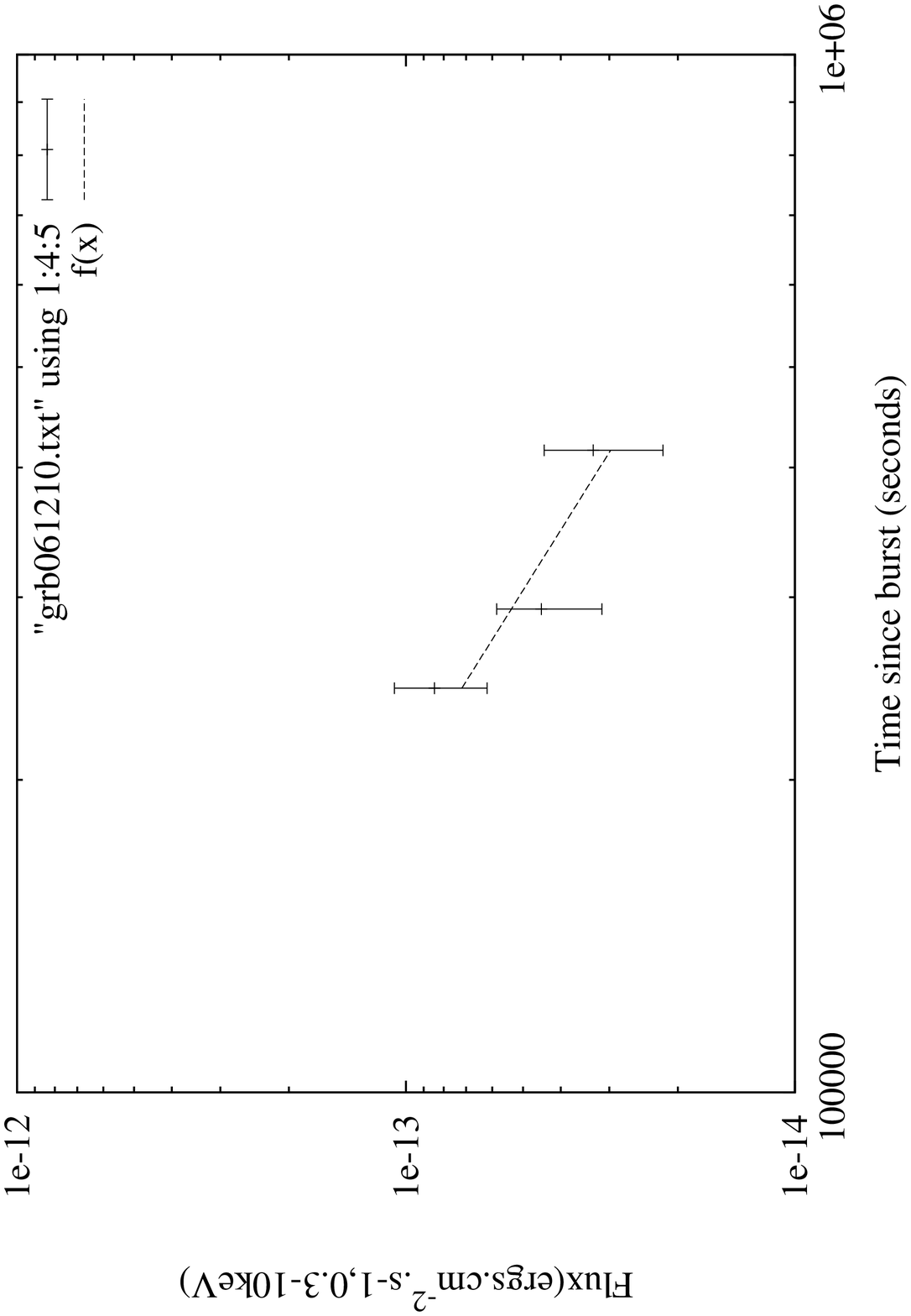}
 \includegraphics[scale=0.2]{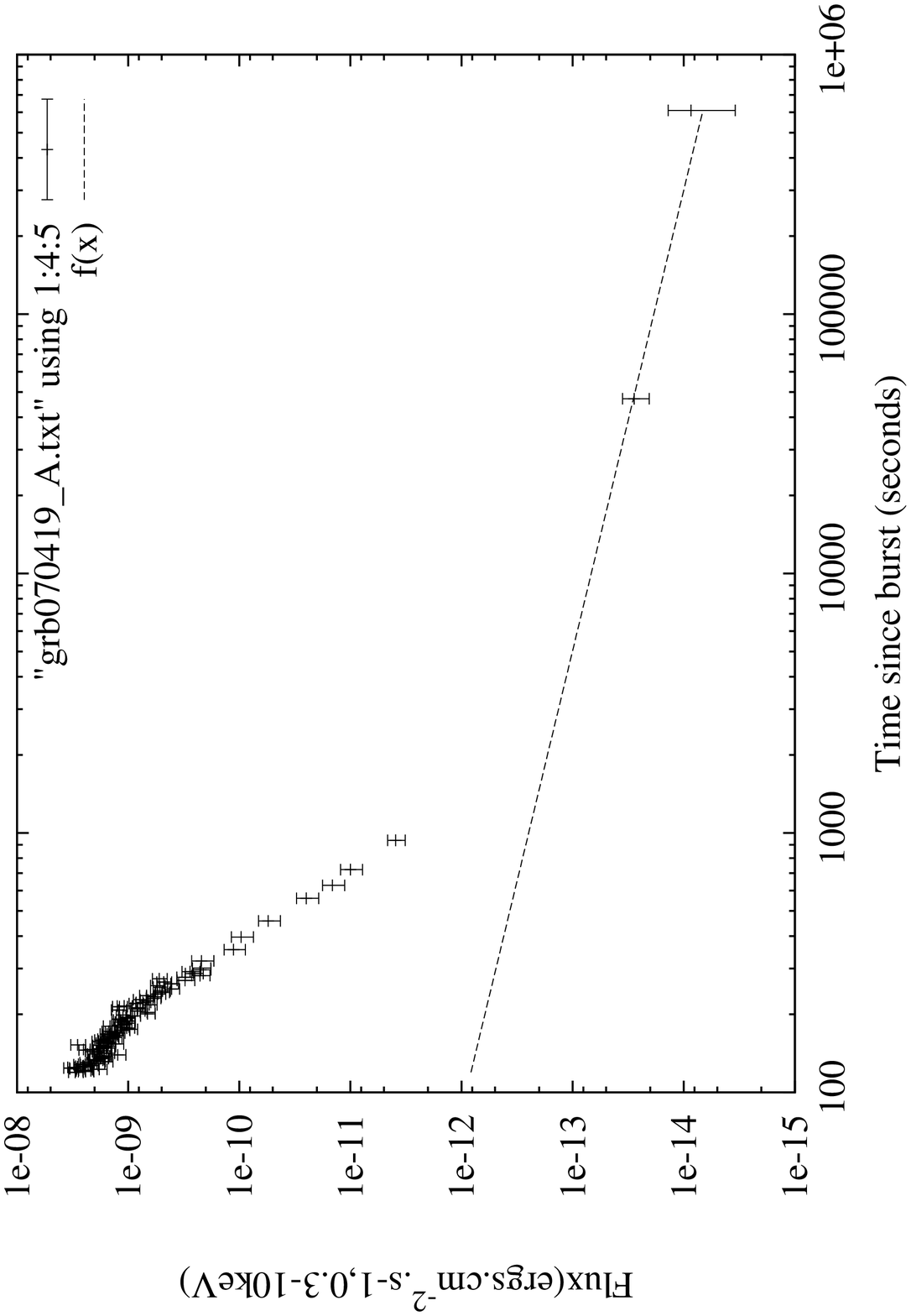}
 \includegraphics[scale=0.2]{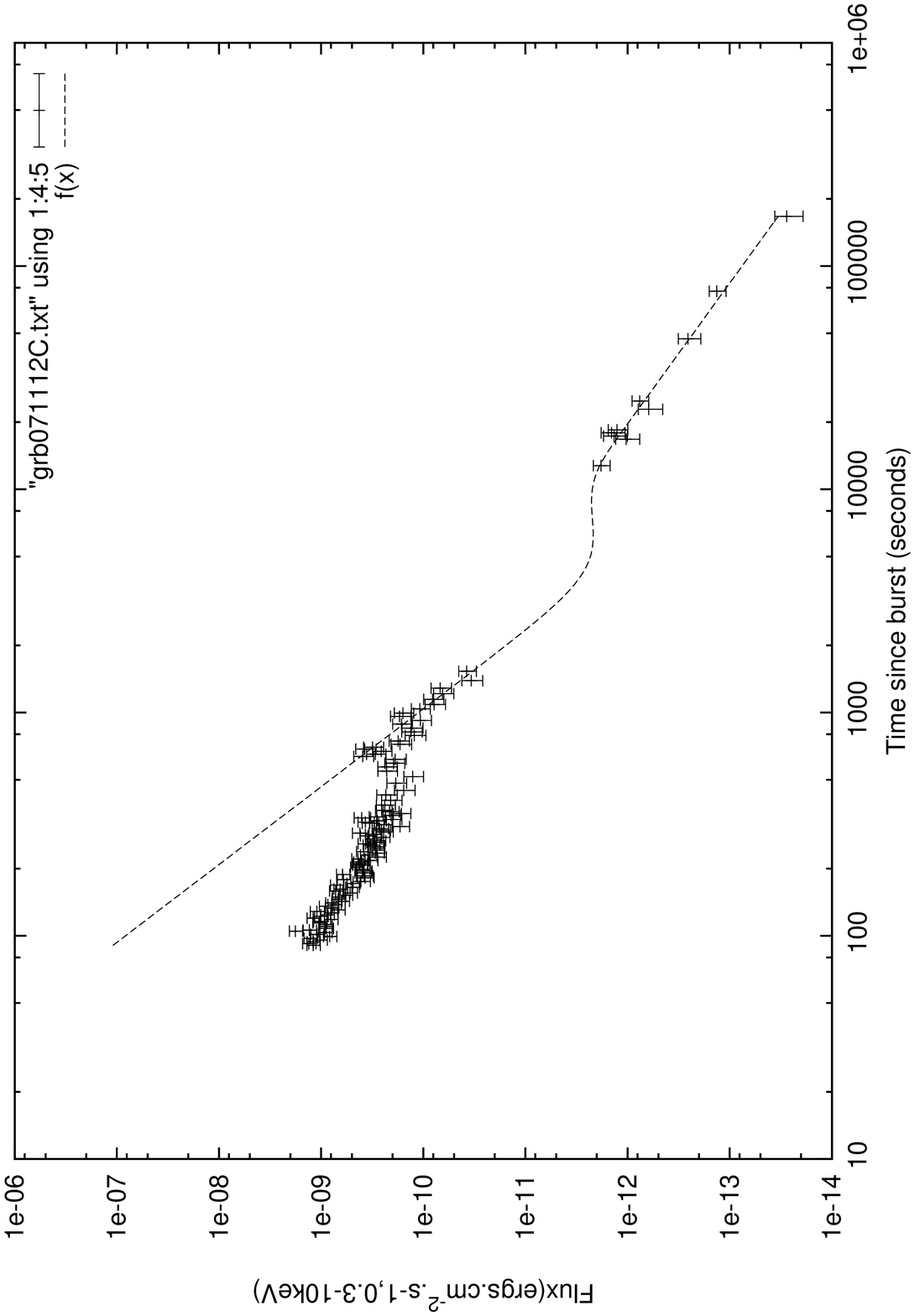}
 \includegraphics[scale=0.2]{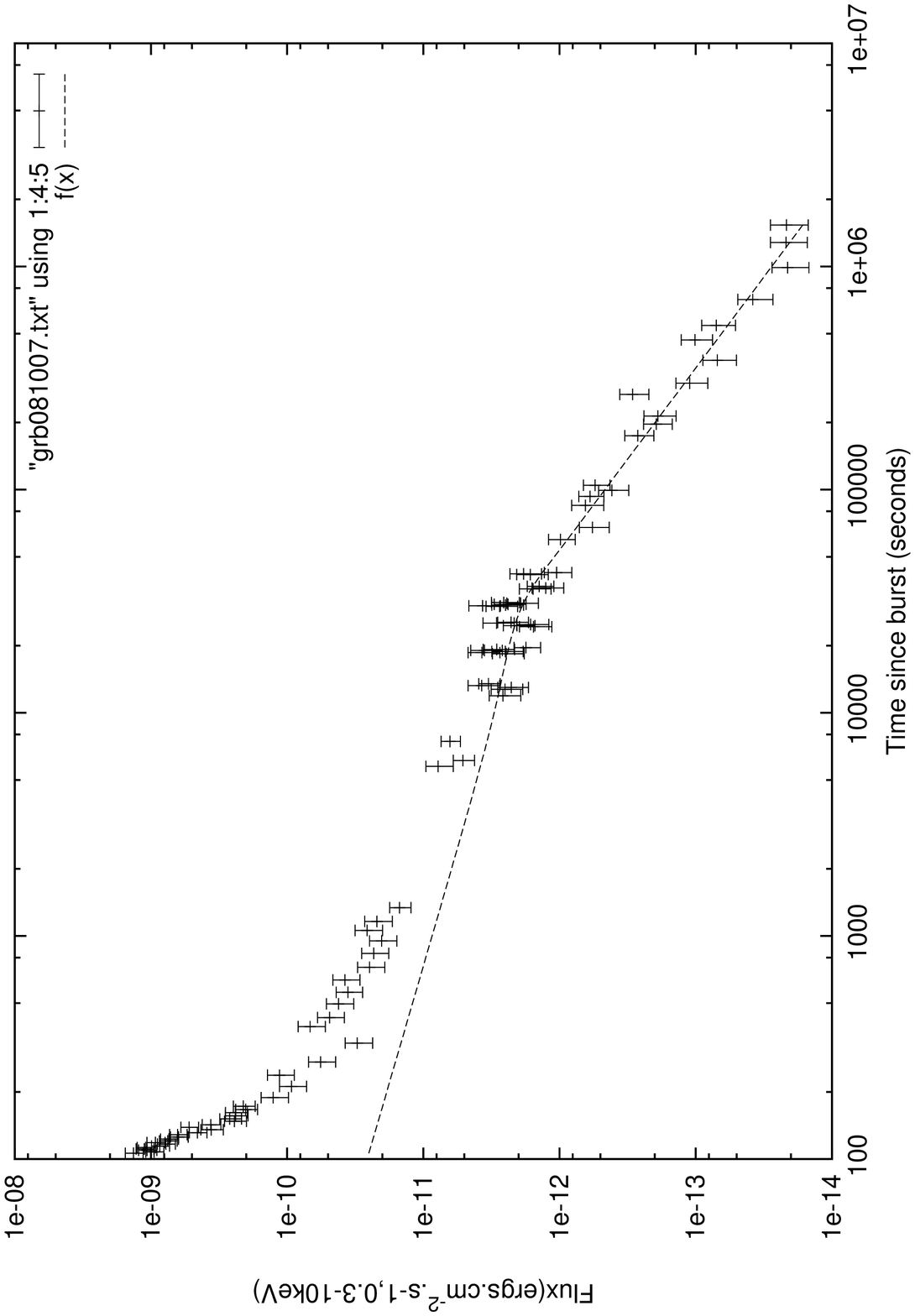}
 \includegraphics[scale=0.2]{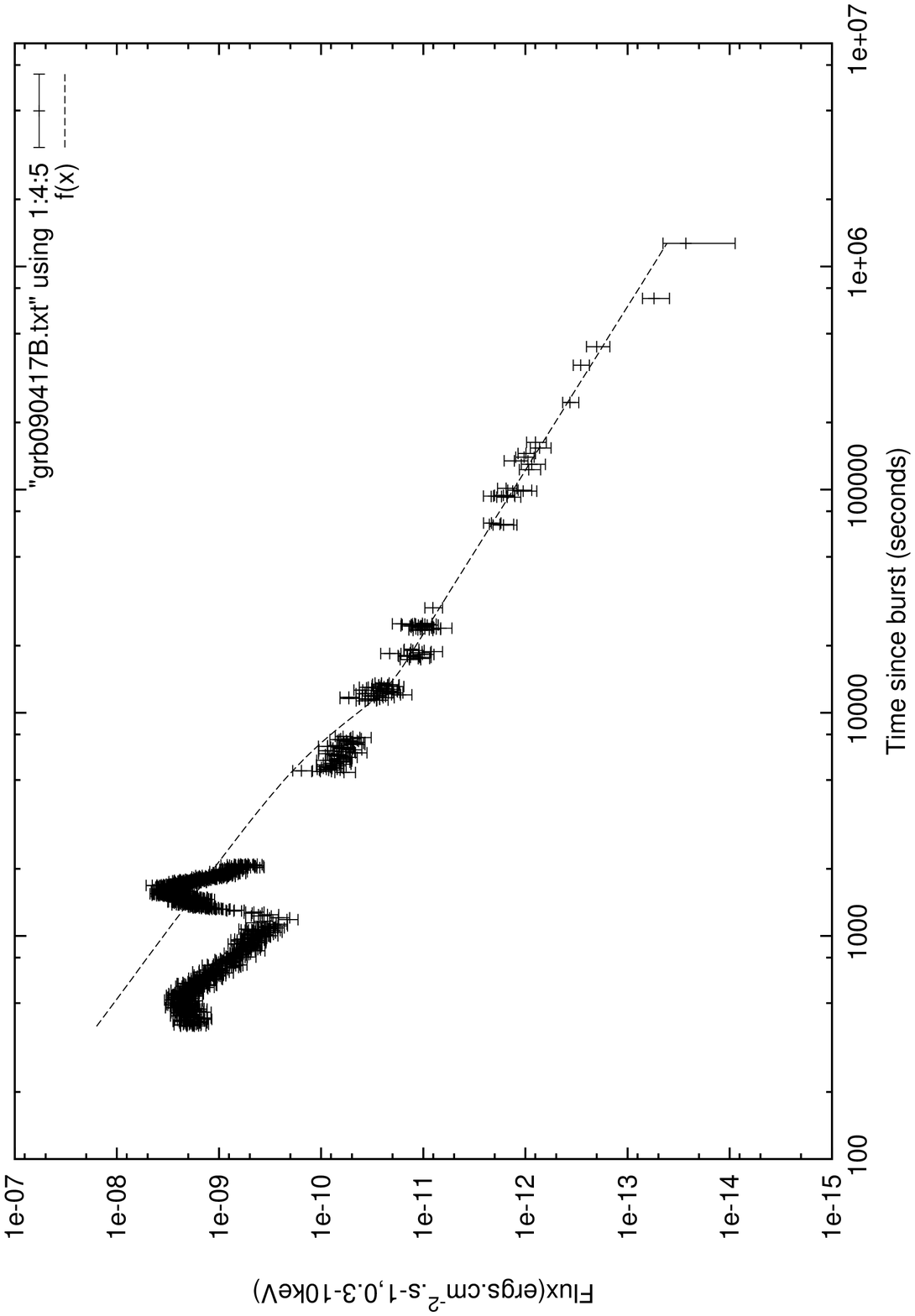}
 \includegraphics[scale=0.2]{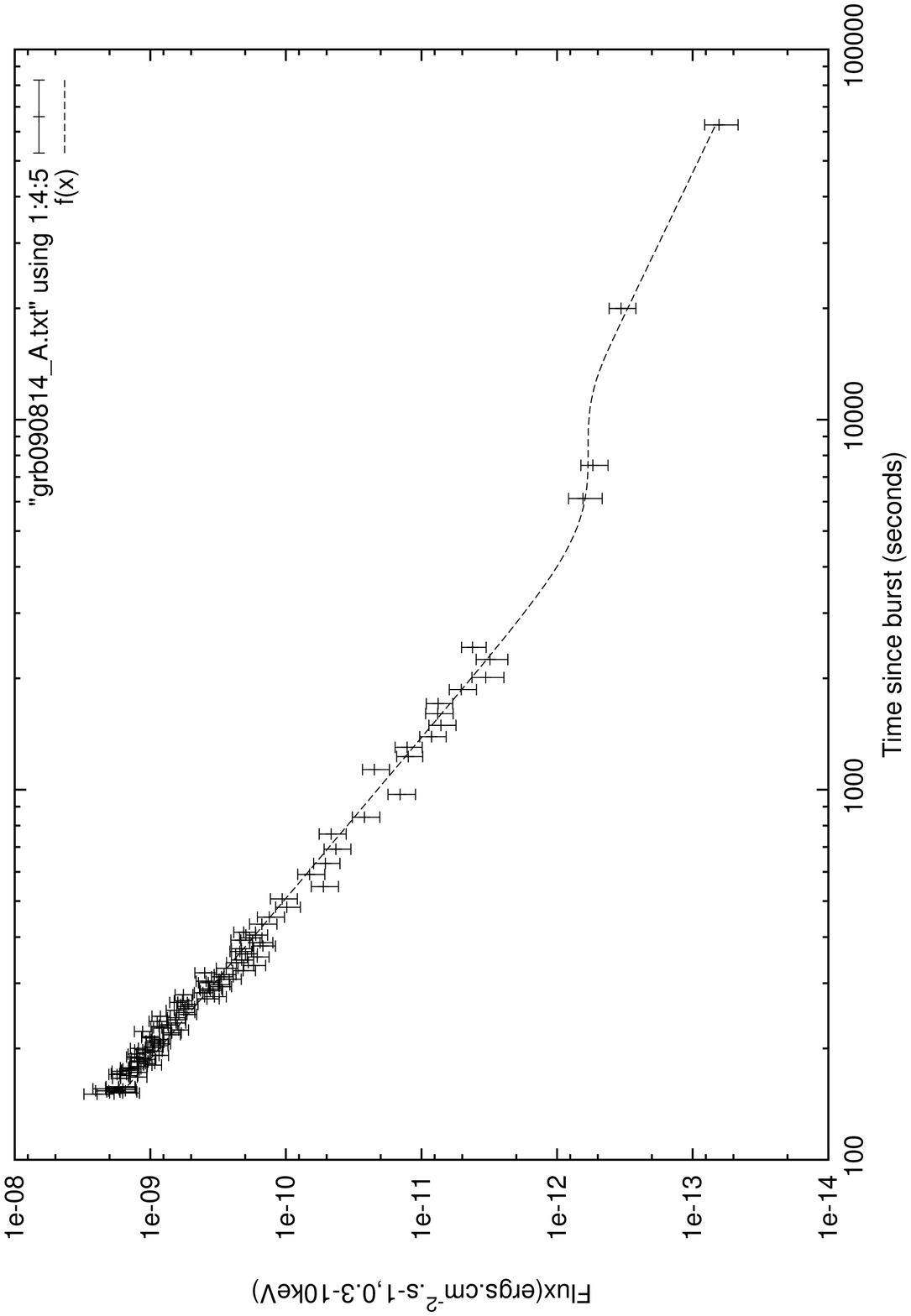}
 \includegraphics[scale=0.2]{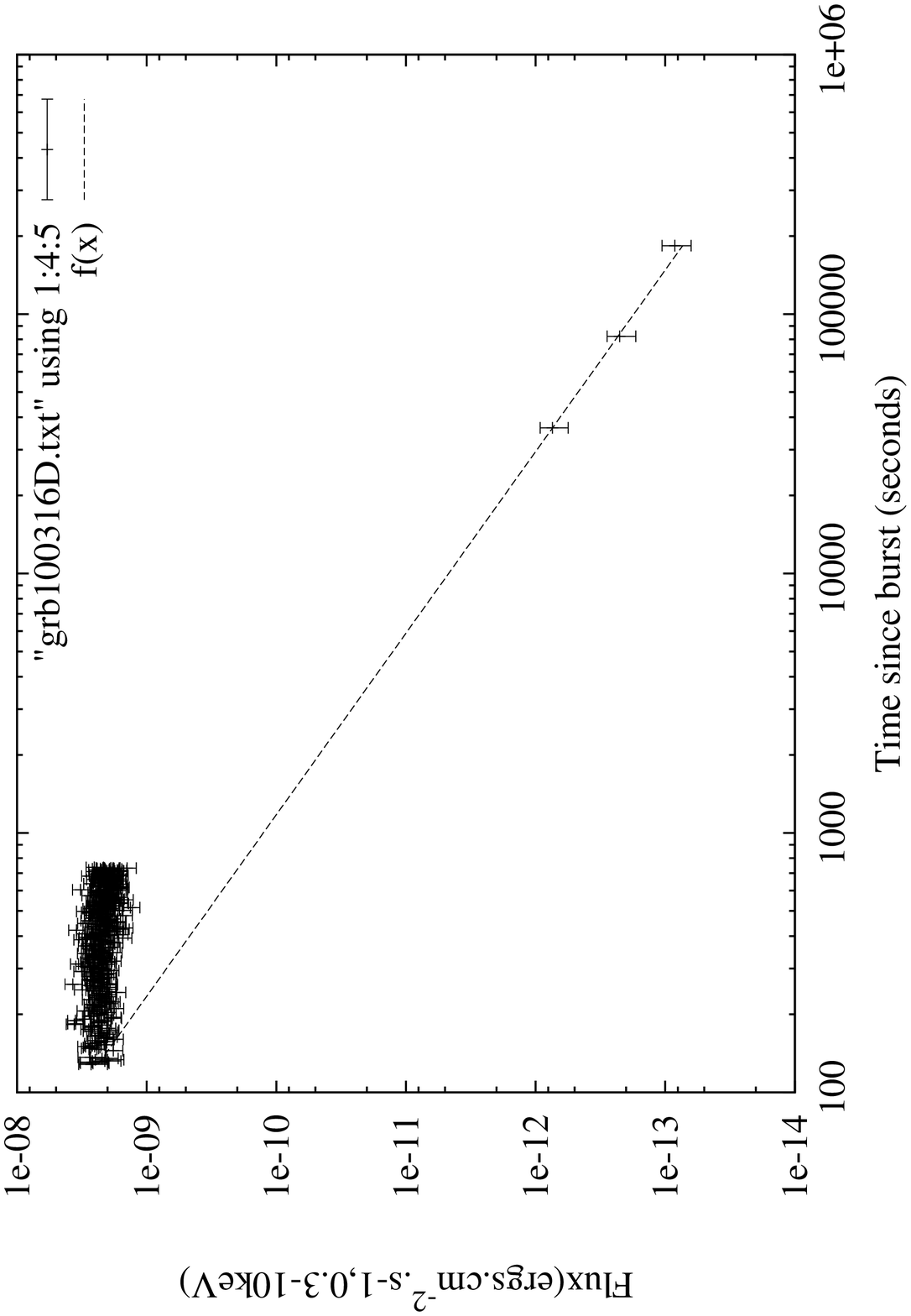}
 \includegraphics[scale=0.2]{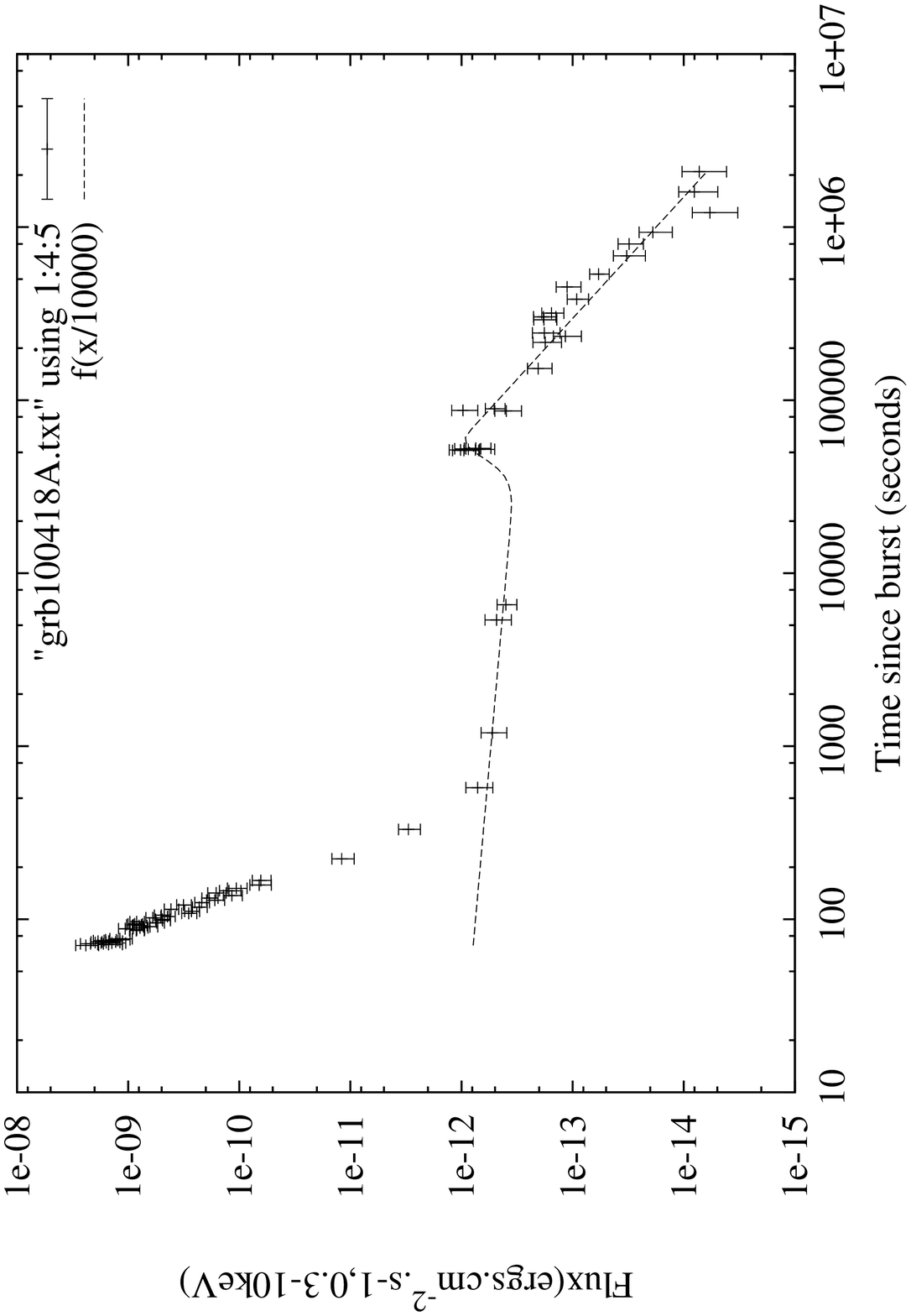}
 \includegraphics[scale=0.2]{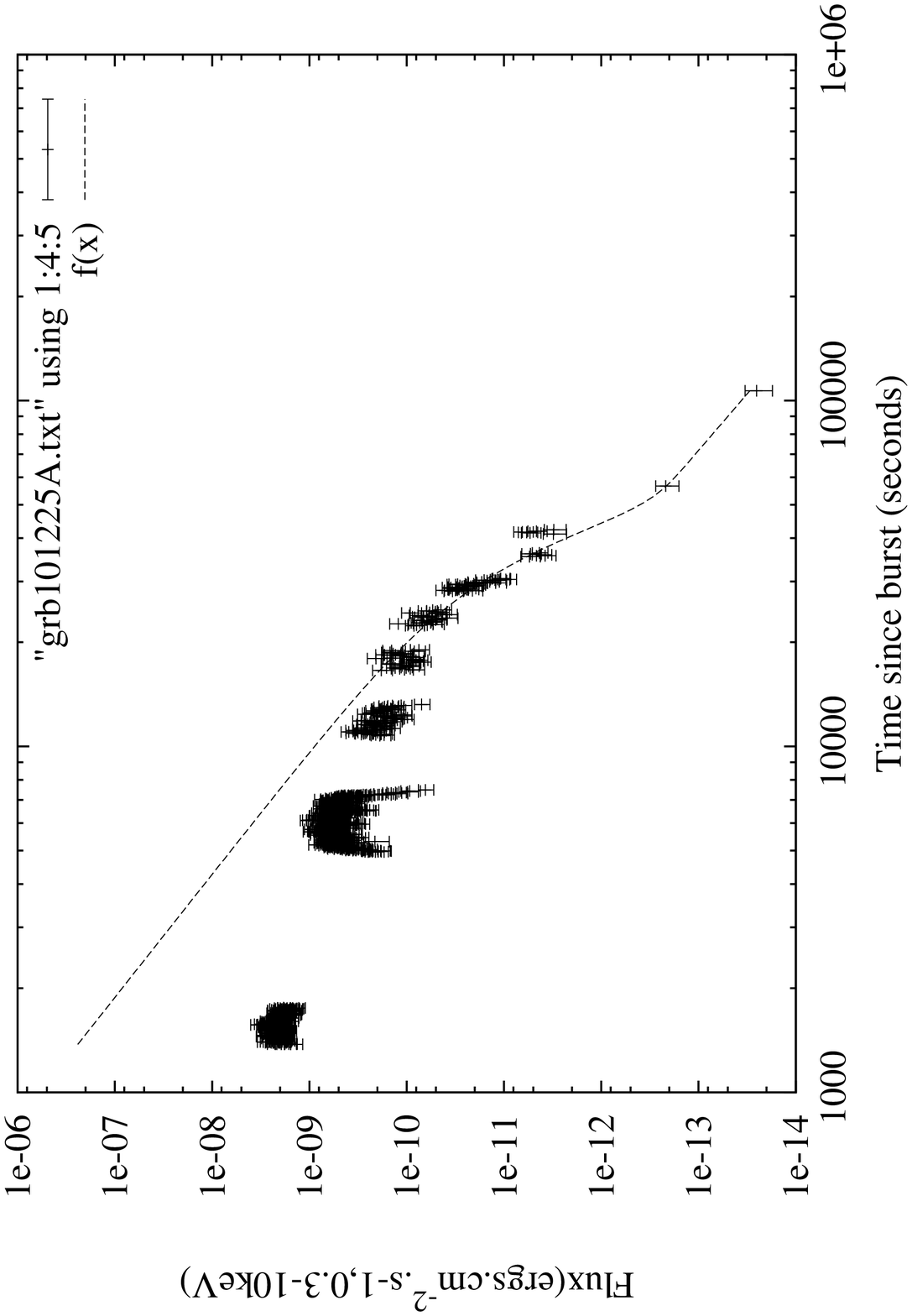}
 \includegraphics[scale=0.2]{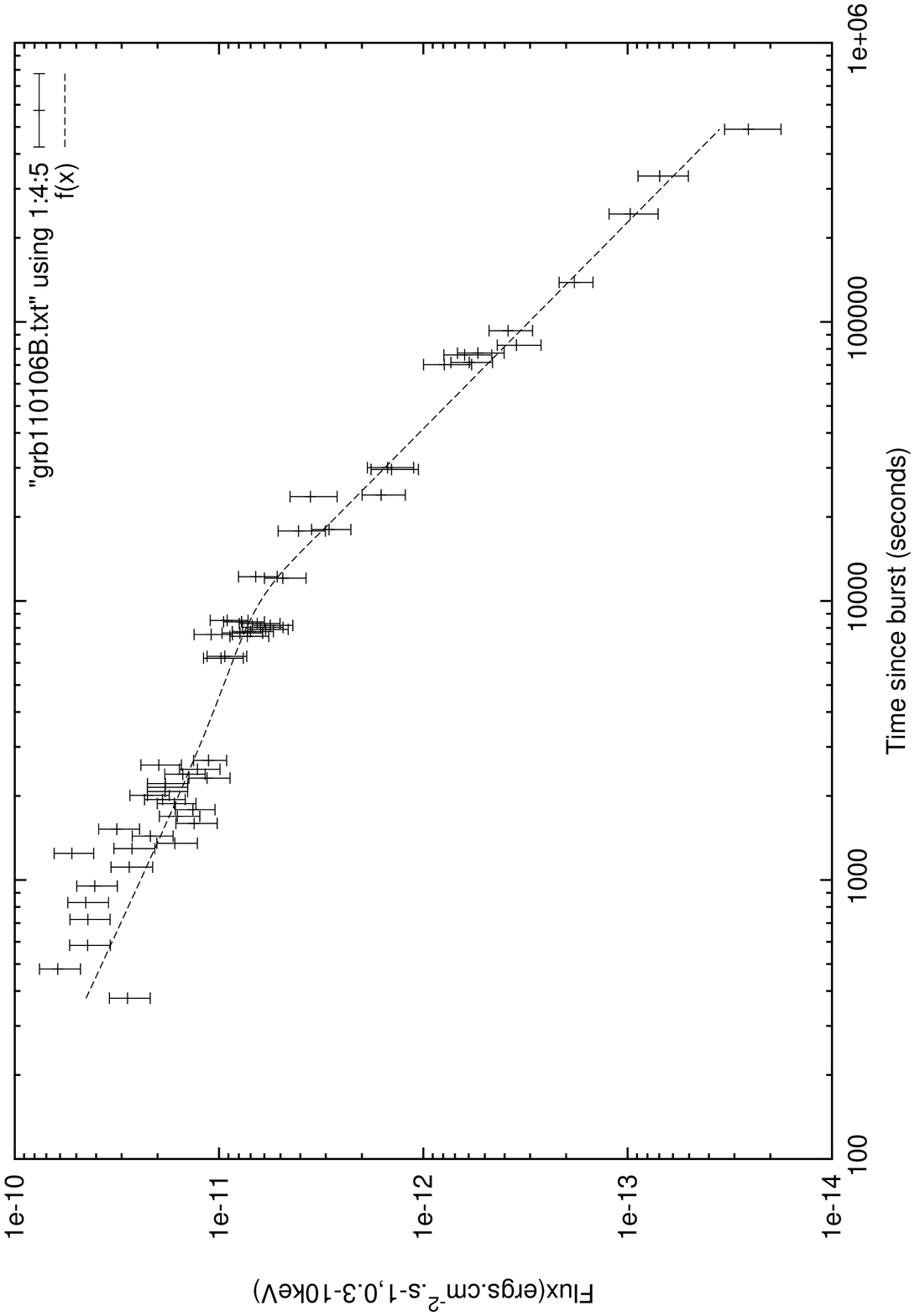}
\includegraphics[scale=0.2]{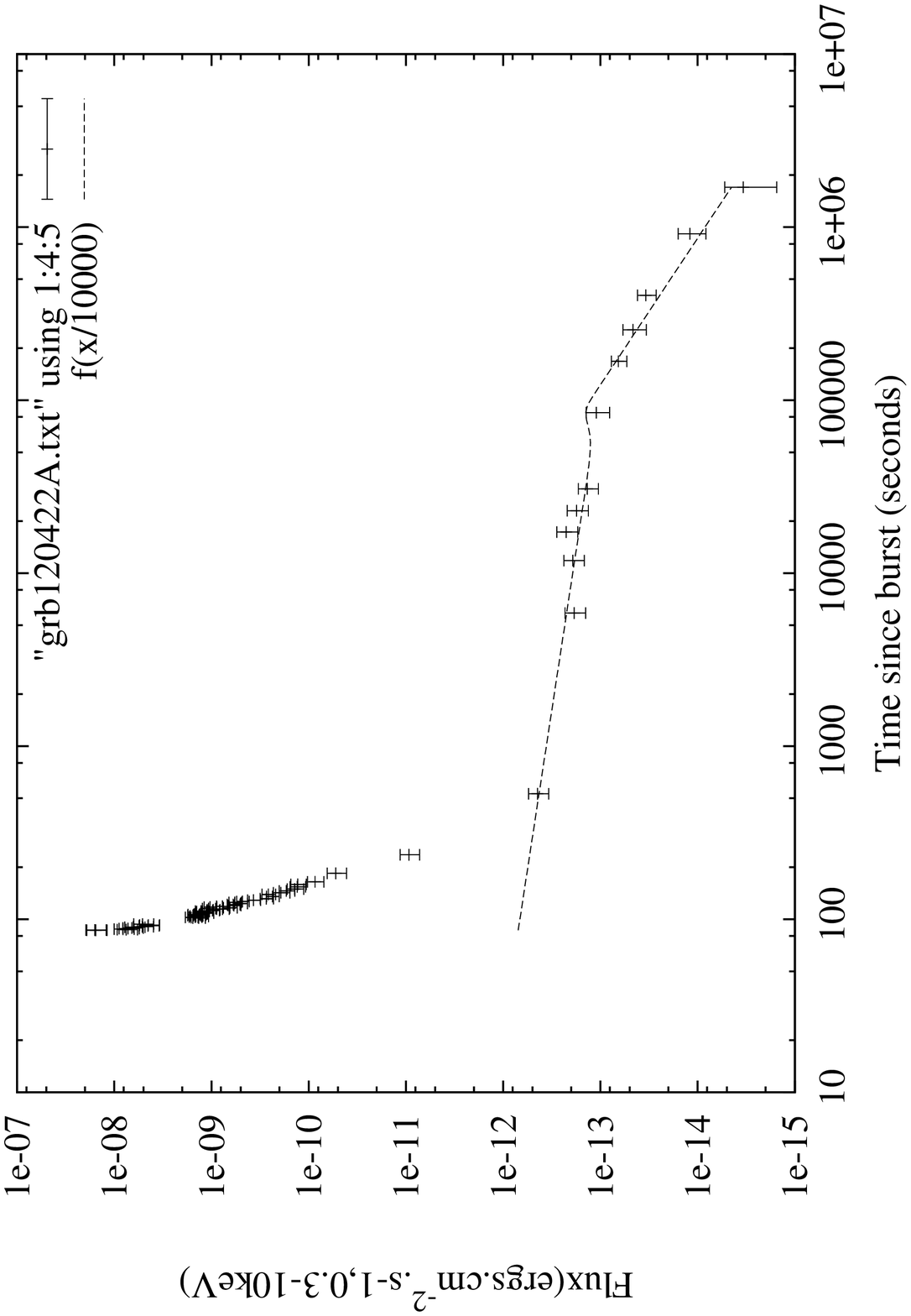}
\includegraphics[scale=0.2]{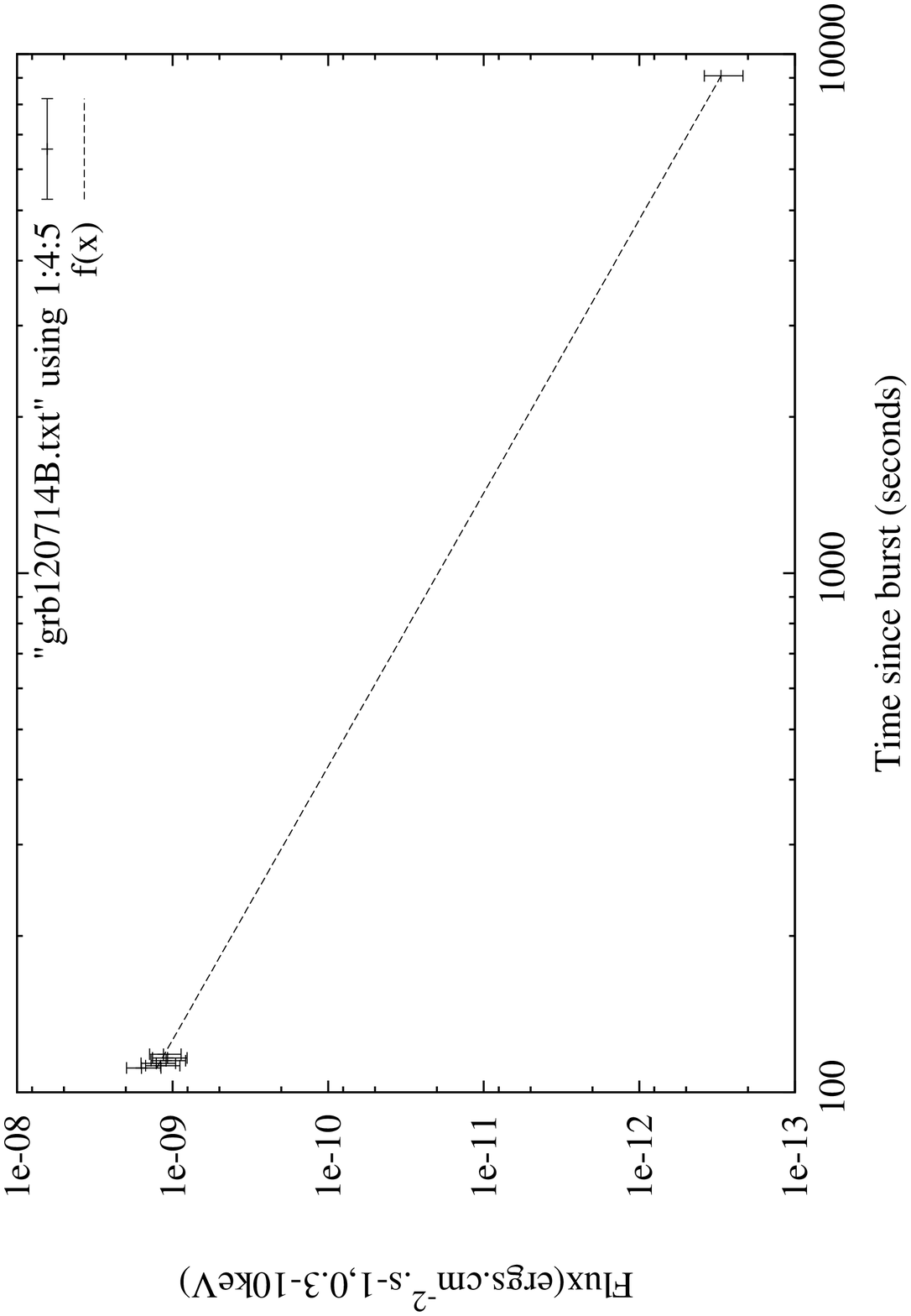}
\includegraphics[scale=0.2]{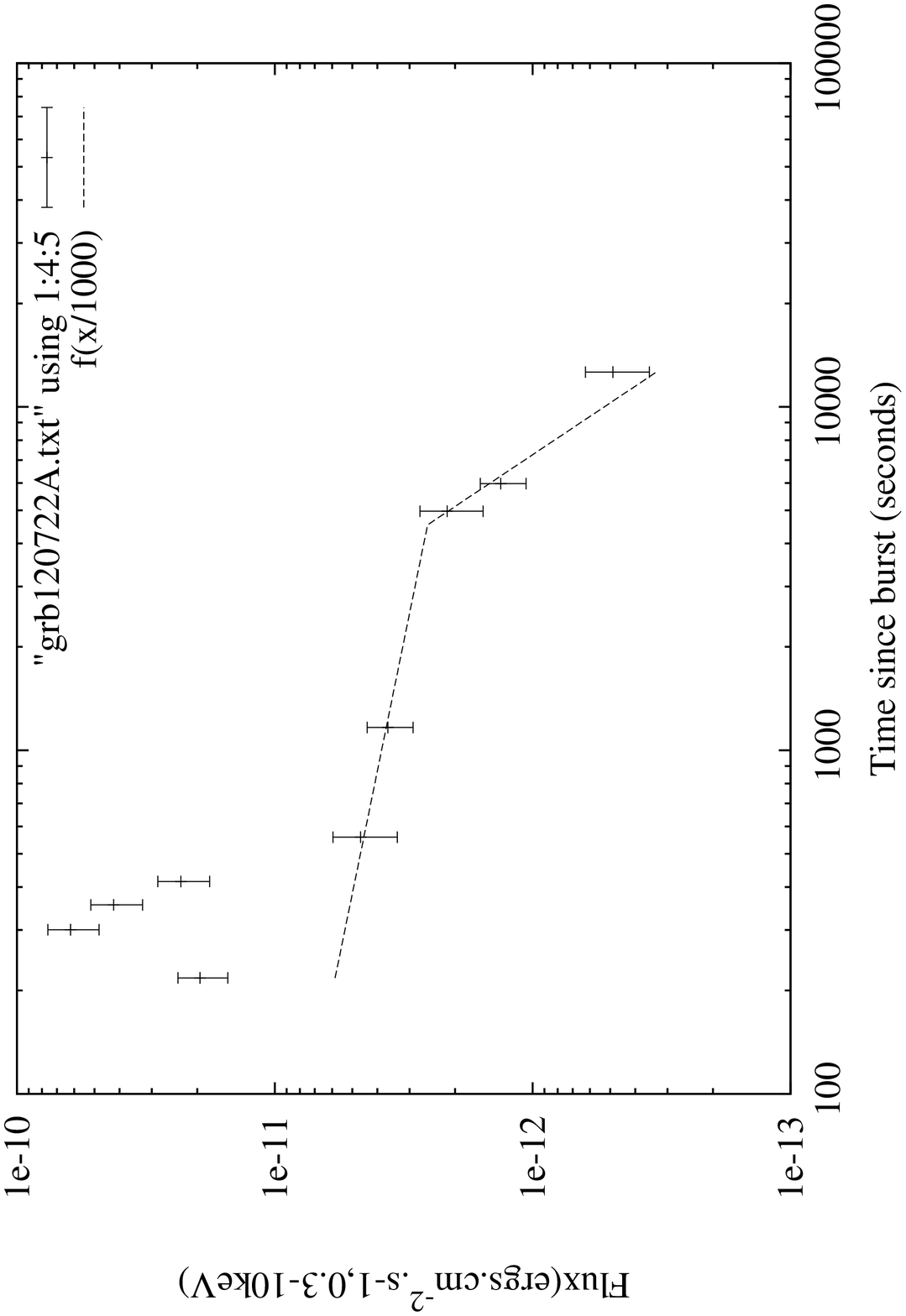}
\includegraphics[scale=0.2]{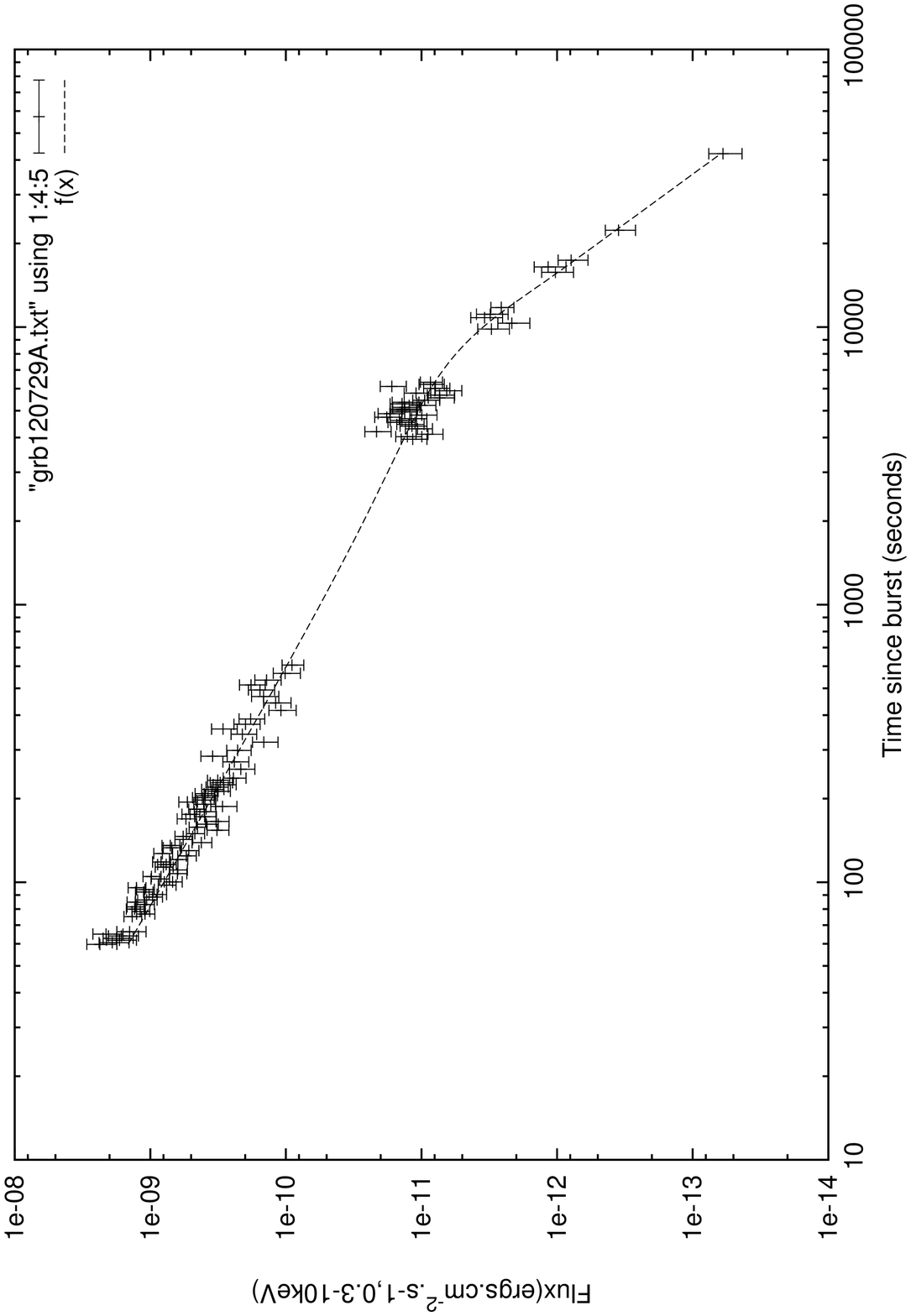}
\caption{Continue of Figure \ref{decay_index_fit} \label{decay_index_fit1}}
\end{center}
\end{figure*}

\section{Conclusion}
X-ray instruments are important to work on the early and late afterglow of GRBs. Using the X-ray data, the environment of the bursts can be identified and the different emission mechanisms can be constrained.
The X-ray afterglow light-curves show different properties: flares, injection of energy in the plateau phase and jet physics. Thanks to the increased number of observed GRBs in the past few years, statistical studies can now be performed. In the next chapter, I will present the statistical study which was performed on the properties of GRBs late afterglow. 

%% file: Chapter4.tex

\chapter[Low-Luminosity Afterglow GRBs]{\parbox[t]{\textwidth}{Low-Luminosity Afterglow GRBs}}
\chaptermark{LLA GRBs}
\label{Chapter4}

In this chapter, I present the result of my study of the intrinsic properties of LLA GRBs (spectral index, decay index, distance, luminosity, isotropic radiated energy and peak energy), to assess if LLA GRBs and the other brighter GRBs belong to different populations.

\section{Properties of Low-Luminosity Afterglow GRBs}
\label{sec_sample_prop}

\begin{figure*}[!ht]
\begin{center}
\includegraphics[width=0.9\textwidth]{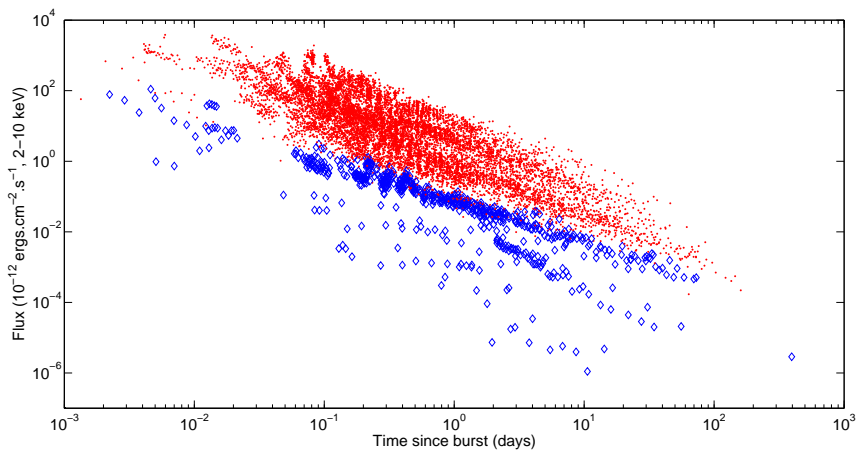}
\caption{The afterglow light-curves of all GRBs, rescaled at a common redshift $z = 1$. LLA events are shown by blue diamonds while the control sample is shown with red dots.\label{fig_sample}}
\end{center}
\end{figure*}

The 31 LLA GRBs are represented by blue diamonds in Figure \ref{fig_sample}. They represent about 12\% of all bursts with known redshift. The other bursts, 223~GRBs, represented by red points, are used as a control sample. 

The LLA GRBs are also listed in Table \ref{table_sample} which displays the GRB name, redshift, galactic and host absorption, $N_H$, galactic and host  extinction, $A_V$, the afterglow temporal and spectral indexes, the isotropic and peak energies, and the time during which 90\% of the energy of the prompt is emitted ($T_{90}$).

\begin{sidewaystable*} [!htbp] 
\centering
  \caption{LLA GRBs and their main characteristics (see text). The spectral and temporal indexes of GRBs which happened before August 2006 are taken from \cite{gendre2008}. \label{table_sample}}
  {\scriptsize
  \begin{tabular}{lllllllllllll}
  \hline
GRB        & z      & \multicolumn{2}{c}{$N_H$} & \multicolumn{2}{c}{$A_V$} & \multicolumn{2}{c}{Afterglow} & log$T_a$  & $E_{iso}$ &  $E_{p,i}$ 	&$T_{\rm 90}$ &Ref.\\
           &        & Gal & Host                &  Gal & Host            &  Temporal & Spectral             & (s)    &(10$^{52}$erg) & (keV)	  &	(s) & \\
           &        & \multicolumn{2}{c}{(10$^{21}$cm$^{-2}$)} & \multicolumn{2}{c}{(mag)} & index  &  index &       &               &          &     & \\
\hline
GRB~980425 & 0.0085 & 0.428 & $\cdots$ &0.071 &1.73 & 0.10$\pm$0.06 & (0.8) &$\cdots$ & (1.3$\pm$0.2)$\times10^{-4}$ &55$\pm$21 & 18 & (1), (13) \\ 
GRB~011121& 0.36 & 0.951 & $\cdots$ & 0.061 & 0.38 & 1.3$\pm$0.03 & (0.8) & $\cdots$ & 7.97$\pm$2.2 & 1060$\pm$275 & 28 & (1), (14), (15) \\
GRB~031203& 0.105	 & 6.21 & $\cdots$ & 0.117 & 0.03 & 0.5$\pm$0.1& 0.8$\pm$0.1 &$\cdots$ & (8.2$\pm$3.5)$\times10^{-3}$ & 158$\pm$51 &40 &(1), (14) \\
GRB~050126 & 1.29 & 0.551 & (0.0) & 0.182 & $\cdots$ &1.1$^{+0.6}_{-0.5}$ & 0.7$\pm$0.7 & $\cdots$ & [0.4 - 3.5] & $>$201 & 24.8 & (17) \\
GRB~050223 & 0.5915 & 0.729	 & (0.0) & 0.078 & $>$2 & 0.91$\pm$0.03 & 1.4$\pm$0.7 &$\cdots$ & (8.8$\pm$4.4)$\times10^{-3}$	 & 110$\pm$55 & 22.5 &(2), (18) \\ 
GRB~050525 & 0.606 & 0.907 & 0.38$^{+9.1}_{-0.38}$ & 0.221 & 0.36$\pm$0.05 & 1.4$\pm$0.1 & 1.1$\pm$0.4 & 3.8 & 2.3$\pm$0.5 & 129$\pm$12.9 & 8.8 & (5), (19) \\
GRB~050801 & 1.38 & 0.698 & (0.0) & 0.989 & 0.3$\pm$0.18 & 1.25$\pm$0.13 & 1.84$^{+0.56}_{-0.53}$ & 3.2 & [0.27 - 0.74] & $<$145 & 19.4 & (5), (17) \\
GRB~050826 & 0.297 & 2.17 & 8$^{+6}_{-4}$ & 2.398 & $\cdots$ & 1.13$\pm$0.04 & 1.1$\pm$0.4 & 4.04 & [0.023 - 0.249] & $>$37 & 35.5 & (17) \\
GRB~051006 & 1.059 & 0.925 & (0.0) & 2.345 & $\cdots$ & 1.69$\pm$0.13 & 1.5$^{+0.44}_{-0.46}$ & 2.77 & [0.9 - 4.3] & $>$193 & 34.8 & (17) \\
GRB~051109B & 0.08 & 1.3 & $<$2 & 0.3 & $\cdots$ & 1.1$\pm$0.3 & 0.7 $\pm$0.4 & 3.14 & $\cdots$ & $\cdots$ & 14.3 & \\
GRB~051117B & 0.481 & 0.46 & (0.0) & 0.321 & $<1.4$ & 1.03$\pm$0.5 & (0.8) & $\cdots$ & [0.034 - 0.044] & $<$136 & 9.0 & (11), (17) \\
GRB~060218 & 0.0331 & 1.14 & 6$\pm$2 & 0.437 & 0.5$\pm$0.3 & 1.3$^{+1.1}_{-0.6}$ & 0.51$\pm$0.05 & 5.0 & (5.4$\pm$0.54)$\times10^{-3}$ & 4.9$\pm$0.49 & $\sim$2100 & (1), (20) \\ 
GRB~060505 & 0.089 & 0.175 & (0.0) & 0.209 & 0.63$\pm$0.01 & 1.91$\pm$0.2 & (0.8) & $\cdots$ & (3.9$\pm$0.9)$\times10^{-3}$ &120$\pm$12 & $\sim$4 & (1), (21) \\
GRB~060614 & 0.125 & 0.313 & 0.5$\pm$0.4 & 0.068 & 0.11$\pm$0.03 & 2.0$^{+0.3}_{-0.2}$ & 0.8$\pm$0.2 & 4.64 & 0.22$\pm$0.09 & 55$\pm$45 &108.7 & (3), (21) \\ 
GRB~060912A & 0.937 & 0.420	& (0.0) & 1.436 & 0.5$\pm0.3$ & 1.01$\pm$0.06 & 0.6$\pm$0.2 & 3.3 & [0.80 - 1.42] & $>$211 & 5.0 & (4), (17) \\ 
GRB~061021 & 0.3463 & 0.452	& 0.6$\pm$0.2 & 0.185 & $<$ 0.10 & 0.97$\pm$0.05 & 1.02$\pm$0.06 & 3.63 & $\cdots$ & $\cdots$ & 46.2 & (3) \\ 
GRB~061110A & 0.758 & 0.494	& (0.0) & $<$0.10 & $<$ 0.10 & 1.1$\pm$0.2 & 0.4$\pm$0.7 & 3.68 & [0.35 - 0.97] & $>$145 & 40.7 & (3), (17) \\ 
GRB~061210 & 0.4095 & 0.339 & (0.0) & 0.489 & $\cdots$ & 1.67$\pm$0.85 & (0.8) & $\cdots$ & [0.10 - 0.33] & $>$105 & 85.3 & (17) \\
GRB~070419A & 0.97 & 0.24 & $<10$ &0.081 &$<$0.8 & 0.56$\pm$0.0 & (0.8) & $\cdots$ & [0.20 - 0.87] & $<$69 & 115.6 & (5), (17) \\
GRB~071112C & 0.823 & 0.852	& $<$5 & 0.203 & 0.20$^{+0.05}_{-0.04}$ & 1.43$\pm$0.05 & 0.8$^{+0.5}_{-0.4}$ & 3.0 & $\cdots$ & $\cdots$ & 15 & (4), (17) \\ 
GRB~081007 & 0.5295 & 0.143 & 0.97$^{+6.9}_{-0.97}$ & 0.196 & 0.36$^{+0.06}_{-0.04}$ & 1.23$\pm$0.05 & 0.99$^{+0.88}_{-0.43}$ & 4.5 & 0.18$\pm$0.02 & 61$\pm$15 & 10 & (7), (22) \\
GRB~090417B & 0.345 & 0.14 & 22$\pm$3 & 0.083 & 0.8$\pm0.1$ & 1.44$\pm$0.07 & 1.3$\pm$0.2 & 3.54 & [0.17 - 0.35] & $>$70 & $>$260  & (6), (17) \\ 
GRB~090814A & 0.696 & 0.461	& (0.0) & 0.15 & $<$0.2 & 1.0$\pm$0.2 & (0.8) & 3.5 & [0.21 - 0.58] & $<$114 & 80 &(7), (17) \\ 
GRB~100316D & 0.059 & 0.82 & (0.0) & 0.088 & 2.6 & 1.34$\pm$0.07 & 0.5$\pm$0.5 & $\cdots$ & (6.9$\pm$1.7)$\times10^{-3}$ & 20$\pm$10 & 292.8 & (8), (23) \\ 
GRB~100418A & 0.6235 & 0.584 & (0.0) & 0.623 & 0.0 & 1.42$\pm$0.09 & 0.9$\pm$0.3 & 4.82 & [0.06 - 0.15] & $<$50 & 7.0 & (9) (17) \\ 
GRB~101225A & 0.847 & 0.928 & (0.0) & 0.311 & 0.75 & $\cdots$ & (0.8) & 4.65 & [0.68 - 1.2] & $<$98 & 1088 & (12), (17) \\
GRB~110106B & 0.618 & 0.23 & (0.0) & 0.032 & $\cdots$ & 1.35$\pm$0.06 & 1.32$^{+0.67}_{-0.32}$ & 4.04 & 0.73$\pm$0.07 & 194$\pm$56 & 24.8 & (24) \\
GRB~120422A & 0.283 & 0.372 & (0.0) & 1.241 & 0.0 & 1.3$\pm$0.3 & 0.4$\pm$0.4 & 5.07 & [0.016 - 0.032] & $<$72 & 5.35 & (10), (25) \\ 
GRB~120714B & 0.3984 & 0.187 & (0.0) & 0.077 & $\cdots$ & 1.89$\pm$0.02 & (0.8) & $\cdots$ & 0.08$\pm$0.02 & 69$\pm$43 & 159 & (17) \\ 
GRB~120722A & 0.9586 & 0.298 &350$^{+230}_{-170}$ & 0.555 & $\cdots$ & 1.2$\pm$0.4 & 1.2$\pm$1.2 & $\cdots$ & [0.51 - 1.22] & $<$88 & 42.4 & (17) \\ 
GRB~120729A & 0.8 & 1.4 & (0.0) & 0.112 & $\cdots$ & 2.8$\pm$0.2 & 0.8 $\pm$0.3 & 3.9 & [0.80 - 2.0 ] & $>$160 & 71.5 & (17) \\ 
\hline
\end{tabular}

}
Note: for $A_V$ values: (1)~ \cite{savaglio2009}; (2) \cite{pellizza2006}; (3) \cite{zafar2011}; 
(4) \cite{schady2012}; (5) \cite{kann2010}; (6) \cite{perley2013}; (7) \cite{greiner2010}; (8) \cite{starling2011};
(9) \cite{marshall2011}; (10) \cite{cano2013} ; (11) \cite{michalowski2012}; (12) \cite{campana2011};
for $E_{iso}$ \& $E_{p,i}$ values: (13) \cite{pian1999}; (14) \cite{ulanov2005}; 
(15) \cite{amati2009}; (17) in this work; (18) \cite{cabrera2008}; (19) \cite{sakamoto2011}; 
(20) \cite{campana2006}; (21) \cite{amati2007}; (22) \cite{bissaldi2008}; (23) \cite{starling2011}; 
(24) \cite{bhat2011}; (25) \cite{melandri2012}
\end{sidewaystable*}

\section{Statistical Properties}
\label{sec_properties}

\subsection{The redshift distribution}
\label{redshift_dist}

\begin{figure*}[!ht]
\begin{center}
\includegraphics[width=0.45\textwidth]{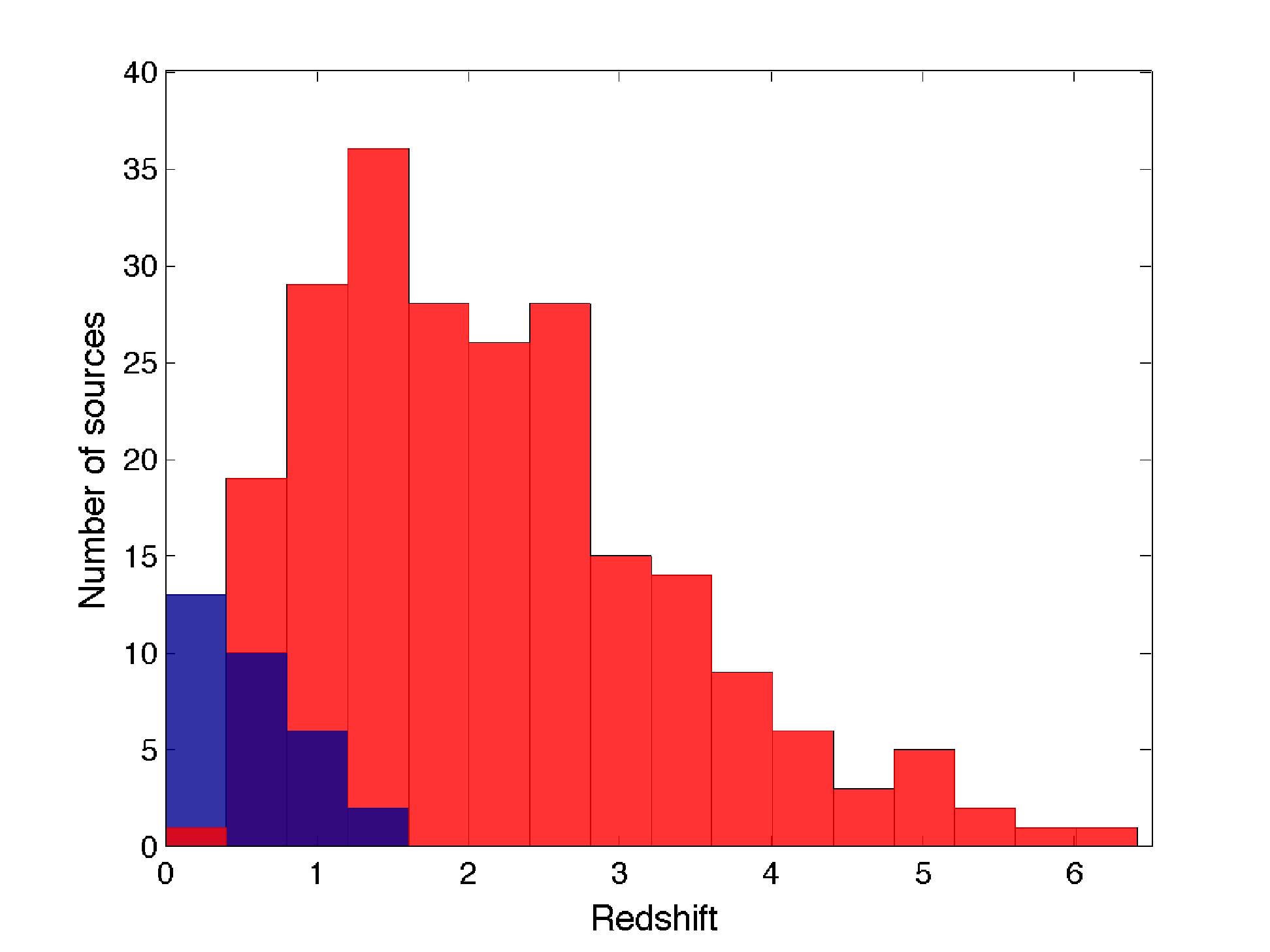}
\includegraphics[width=0.45\textwidth]{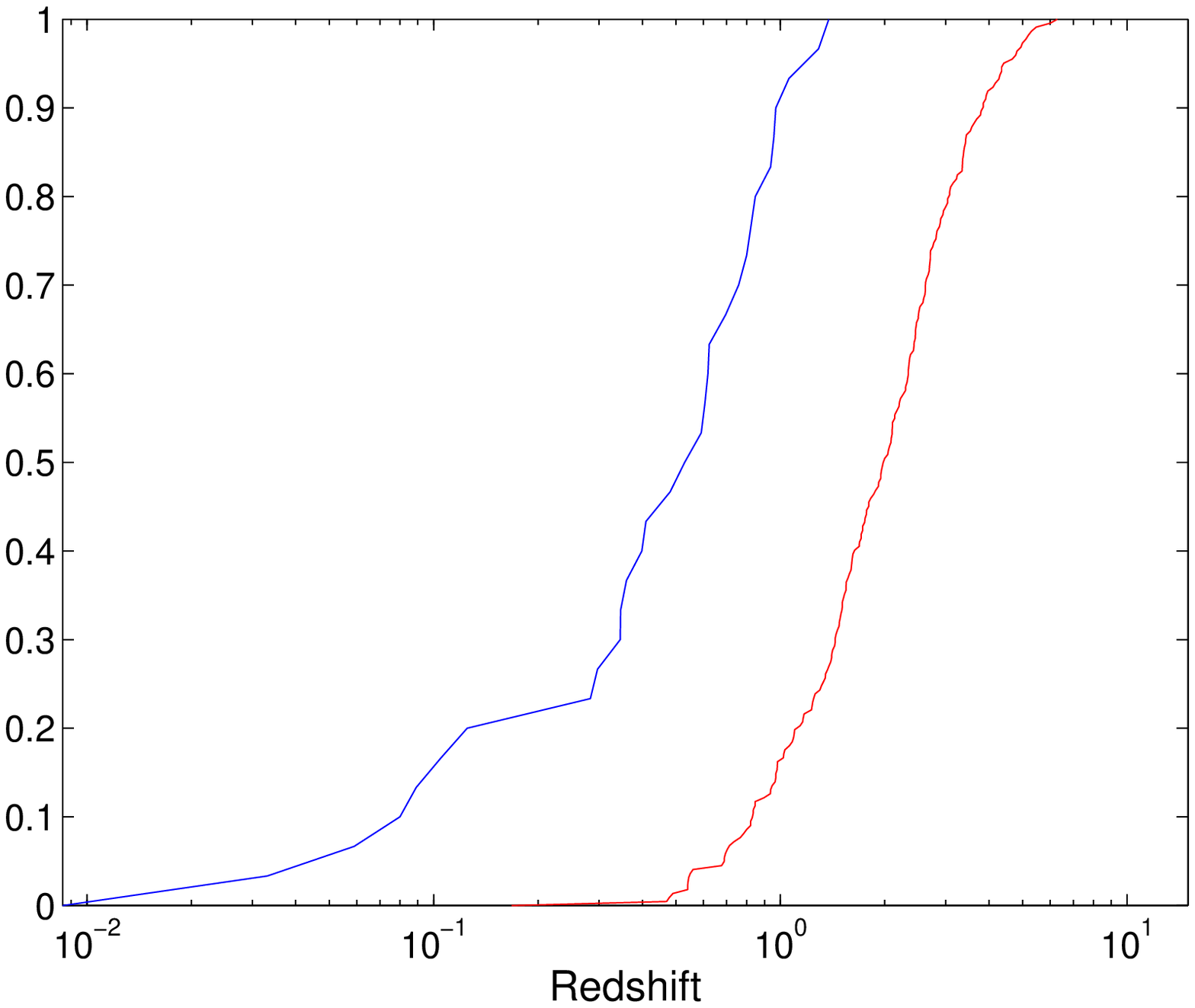}
\caption{Left: Redshift distribution of LLA GRBs (blue) compared to the normal lGRBs (red). Right: Cumulative distribution of the same samples.\label{fig_distri_z}}
\end{center}
\end{figure*}

The redshift distributions of LLA GRBs is shown in Figure \ref{fig_distri_z}, together with that of all lGRBs in the global sample.
From a rapid examination of  Figure \ref{fig_distri_z}, it is seen that LLA GRBs are closer than normal lGRBs, whose mean value is $z \sim$~2.2 (\textit{e.g.} \cite{jakobsson2006, coward2013}). Table \ref{table_fit_z}  displays the statistical parameters of the two distributions. A Kolmogorov-Smirnov test is performed on the two datasets which shows that the probability for the two distributions to be based on the same population is $1.1 \times 10^{-14}$.

\begin{table*}[!ht]
\centering
  \caption{Statistical parameters of the cumulative redshift distributions.\label{table_fit_z}}
   \begin{tabular}{lll}
  \hline
  Parameter& LLA GRBs & All lGRBs\\
    \hline
mean&0.5454&2.1785\\
median&0.5295&1.98\\
standard deviation&0.3746&1.187\\
\hline
\end{tabular}
\end{table*}

\begin{figure*}[!ht]
\begin{center}
\includegraphics[width=0.9\textwidth]{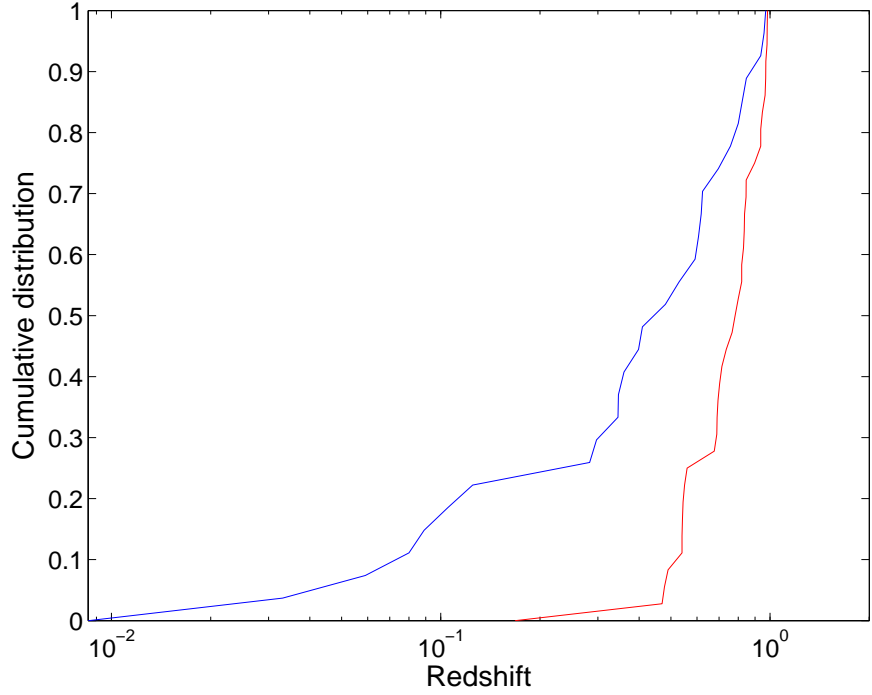}
\caption{Cumulative distribution of the redshift of LLA GRBs (blue) compared to that of all normal lGRBs (red) with redshifts $z < 1 $.\label{fig_distri_z<1}}
\end{center}
\end{figure*}

The difference between the redshift distribution of LLA GRBs and that of lGRBs can be due to a selection bias, or alternatively, it is an intrinsic property. Faint events are more difficult to detect than brighter ones. In section \ref{sec_prompt}, I will discuss the prompt properties of LLA GRBs. However, these events are barely more energetic than the detection threshold, see \textit{e.g.} \cite{nava2009}, and the measure of the redshift is based on spectroscopic observations at optical wavelengths. As the afterglow decays with time, the probability to obtain a successful measurement decreases as well. If the afterglow is faint from the very beginning of the event, then it is initially difficult to measure a redshift. 
Several of these bursts could be distant GRBs with a low luminosity afterglow and one hypothesis is that the large difference between the two distributions is due to distant events. In order to investigate this hypothesis, the distribution of LLA GRBs and of normal lGRBs are restricted to bursts with redshifts $z < 1$. At this distance, all bursts are detected independently of their intrinsic luminosity. I have recomputed the cumulative redshift distributions for this sub-sample (see Figure \ref{fig_distri_z<1}). The difference is still large, and from a Kolmogorov-Smirnov (K-S) test the probability that the two distributions are drawn from the same population is $9.4 \times 10^{-4}$. I can therefore conclude that the \textit{observed} LLA GRBs are nearby events because of their low intrinsic luminosity.

\subsection{Absorption and Extinction}
\label{absorp_extinc}

I constrained absorption (gas) and extinction (dust) on the line of sight of LLA GRBs. The X-ray absorption has little (if none) effect on my selection criteria since I used the flux in the 2.0~-~10.0 keV band, where absorption can be neglected \cite{morrison1983}. The intrinsic hydrogen column density $N_H$ can be linked to the host properties \cite{reichart2002}. In addition, the optical extinction can bias the distribution (for instance the well-known problem of dark bursts, \textit{e.g.} \cite{jakobsson2004}). It additionally decreases the probability of measuring the redshift. 

\subsubsection{Milky Way Galaxy}

\begin{figure*}[!ht]
\begin{center}
\includegraphics[width=0.9\textwidth]{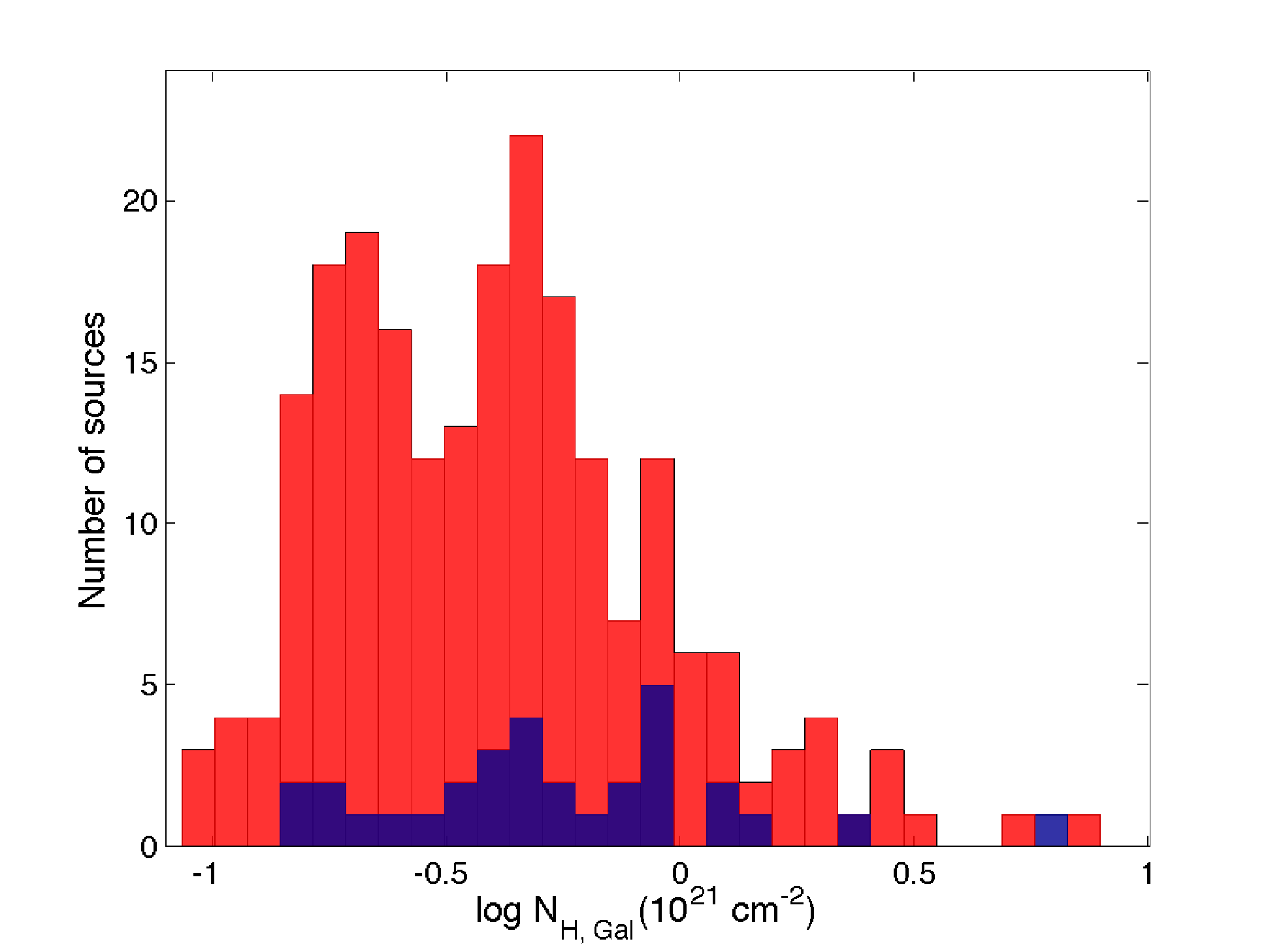}
\caption{Galactic HI column density distribution for LLA GRBs sample (blue) and normal long GRBs (red). \label{fig_Nh}}
\end{center}
\end{figure*}

\begin{figure*}[!ht]
\begin{center}
\includegraphics[width=0.9\textwidth]{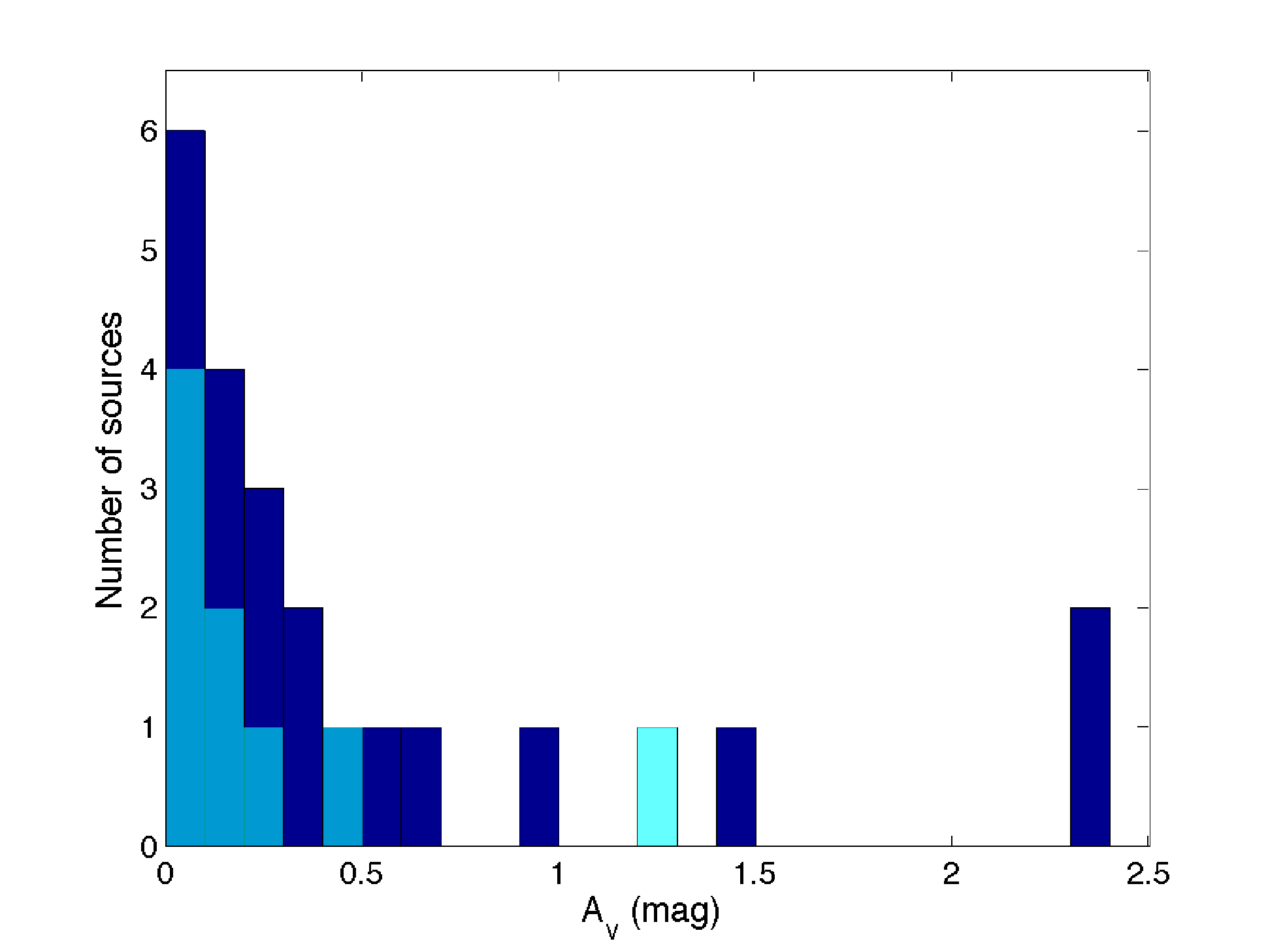}
\caption{Histogram of the optical extinction $A_V$ in the Milky Way Galaxy for LLA GRBs sample. GRB-SN associations are represented by cyan bars while other LLA GRBs are shown with blue bars.\label{fig_AV}}
\end{center}
\end{figure*}

For consistency, I checked that the distribution for the Milky Way Galaxy values of $A_{V,Gal}$ and $N_{H,Gal}$ (\textit{i.e.} the optical extinction and X-ray absorption parameters) is consistent with the whole sample of normal lGRBs. The $N_{H,Gal}$ was calculated as explained in section \ref{fit_spec}. The optical extinction was calculated using the NASA/IPAC extragalactic database\footnote{http://ned.ipac.caltech.edu/forms/calculator.html} for the  Landolt V-band measured by \cite{schlegel1998} for all bursts but except GRB~060904B and GRB~061110A. 
These two bursts are seen in projection on the galactic disk (\textit{i.e.} with a galactic latitude between $-5$ and 5 degrees), where the measures of \cite{schlegel1998} are highly variable with the position. For these two events, I will rely on the most accurate measurements of \cite{schady2012} and \cite{zafar2011} respectively. The results are reported in Figure \ref{fig_Nh} and Figure \ref{fig_AV}. As the two samples are consistent for $N_{h,gal}$ only, I will consider in the following that the absorption in the Milky Way Galaxy has not introduced a bias in the LLA GRBs sample.

\subsubsection{Host galaxy}

\begin{figure*}[!ht]
\begin{center}
\includegraphics[width=0.9\textwidth]{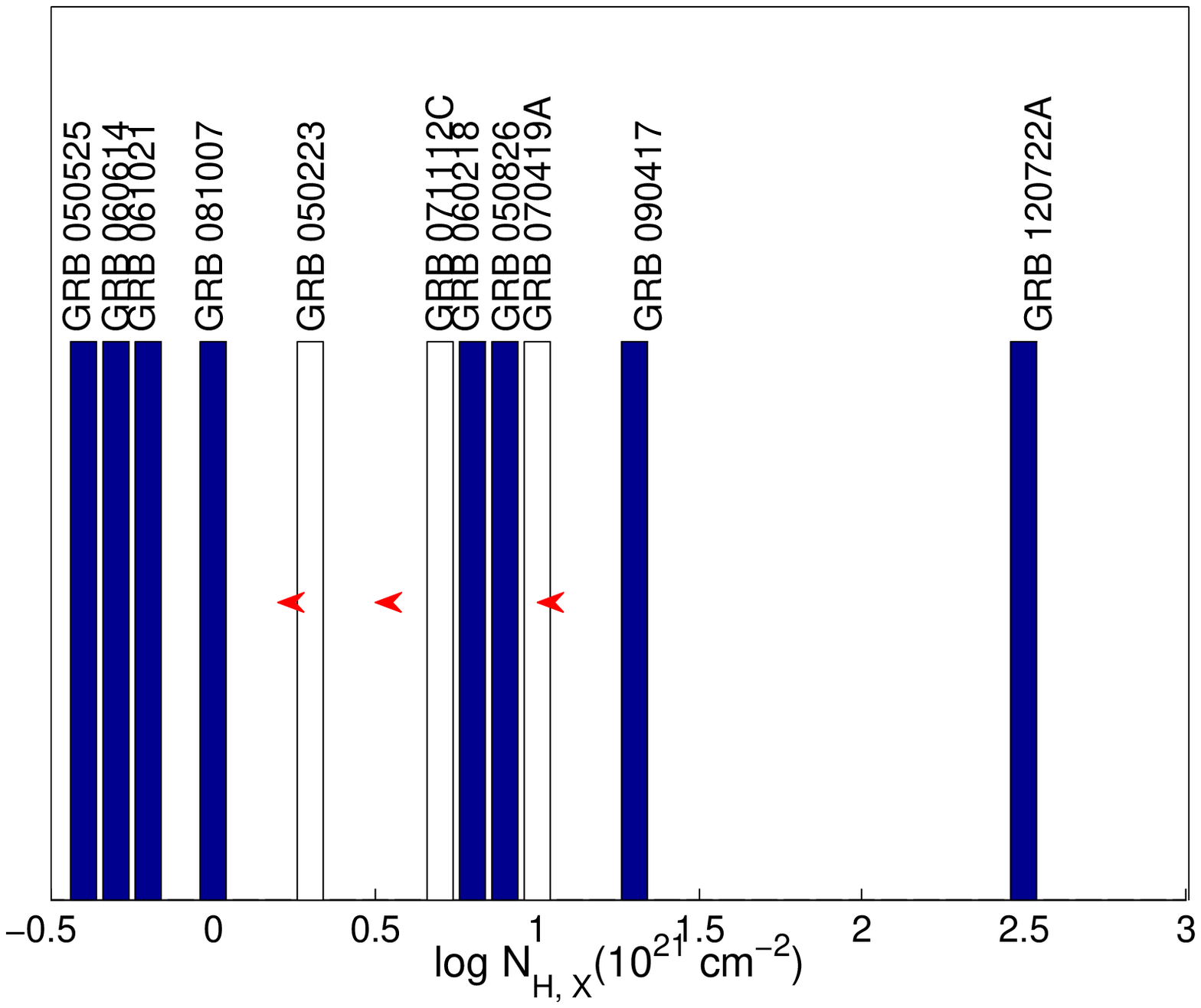}
\caption{Host galaxy HI column density, $N_{H,X}$, distribution for LLA GRBs sample. The blue bars are fitted values while the white ones with red arrows are upper limits. \label{fig_distri_NH}}
\end{center}
\end{figure*}

I obtained the $N_{H,X}$ values for the intrinsic host absorption from the X-ray data analysis of all LLA GRBs by using Xspec. They are displayed on Figure \ref{fig_distri_NH}. Most of them are compatible with little or no intrinsic absorption. Comparing Figure \ref{fig_Nh} and Figure \ref{fig_distri_NH}, I can see that for the bursts with a non-zero $N_{H,X}$, the absorption of the host galaxy is on average a factor 10 larger than in the Milky Way Galaxy, as already noted by \cite{starling2013}. At low redshift this effect was attributed to the gas in the host galaxy.

\begin{figure*}[!ht]
\begin{center}
\includegraphics[width=0.9\textwidth]{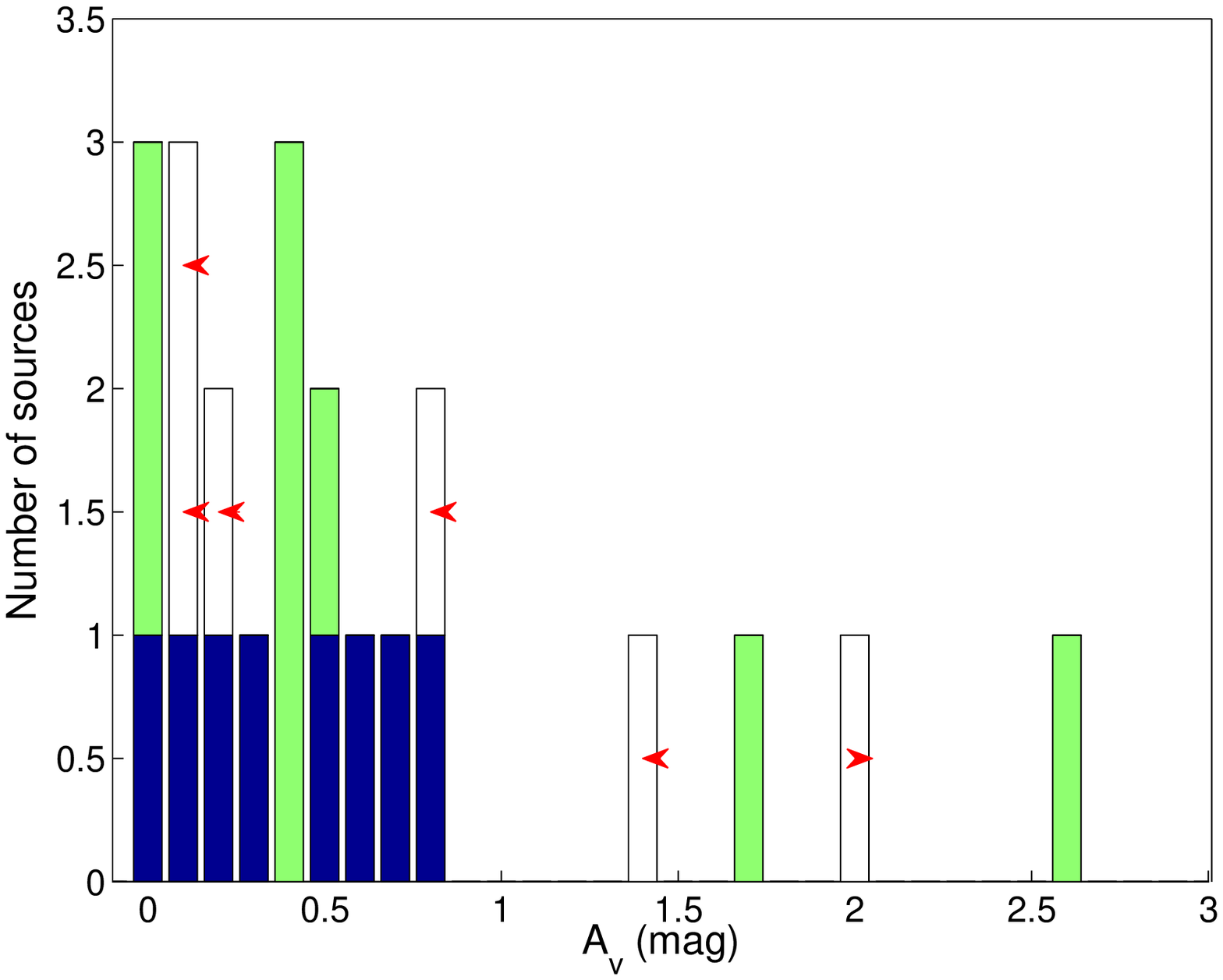}
\caption{Histogram of the host galaxy optical extinction $A_V$ for LLA GRBs. The green bars represent measured values of GRB-SN associations. On the other hand, the blue bars show measured values and the white bars with red arrows display upper and lower limits (depending on the direction of the arrow) burst with no SN detected.\label{fig_distri_AV}}
\end{center}
\end{figure*}

Optical extinction values are displayed on Figure \ref{fig_distri_AV}. The extinction for most sources can be fitted by the Small Magellanic Cloud (SMC) model, with the exception of GRB~060912A for which the Milky Way (MW) model was applied. I cannot see strong differences with the typical values reported for the Milky Way Galaxy, suggesting that bursts are not located in a dusty medium.

\section{Afterglow properties}
\label{sec_afterglow}

The late time afterglow emission corresponds to the decaying part after the plateau phase in the canonical X-ray afterglow light-curve \cite{nousek2006}, see also segments III and IV of Figure \ref{Swift_canonical_light_curve}. In this regime the flux $F_{\nu}$ is modelized by
\begin{align}
F_{\nu}\propto t^{-\alpha} \nu^{-\beta}\text{,} \label{eq_pow}
\end{align}
where ${\alpha}$ and ${\beta}$ are respectively the temporal and the spectral indexes.
\label{afterglow}

\subsection{Temporal decay index}

\begin{figure}
\begin{center}
\includegraphics[width=0.9\textwidth]{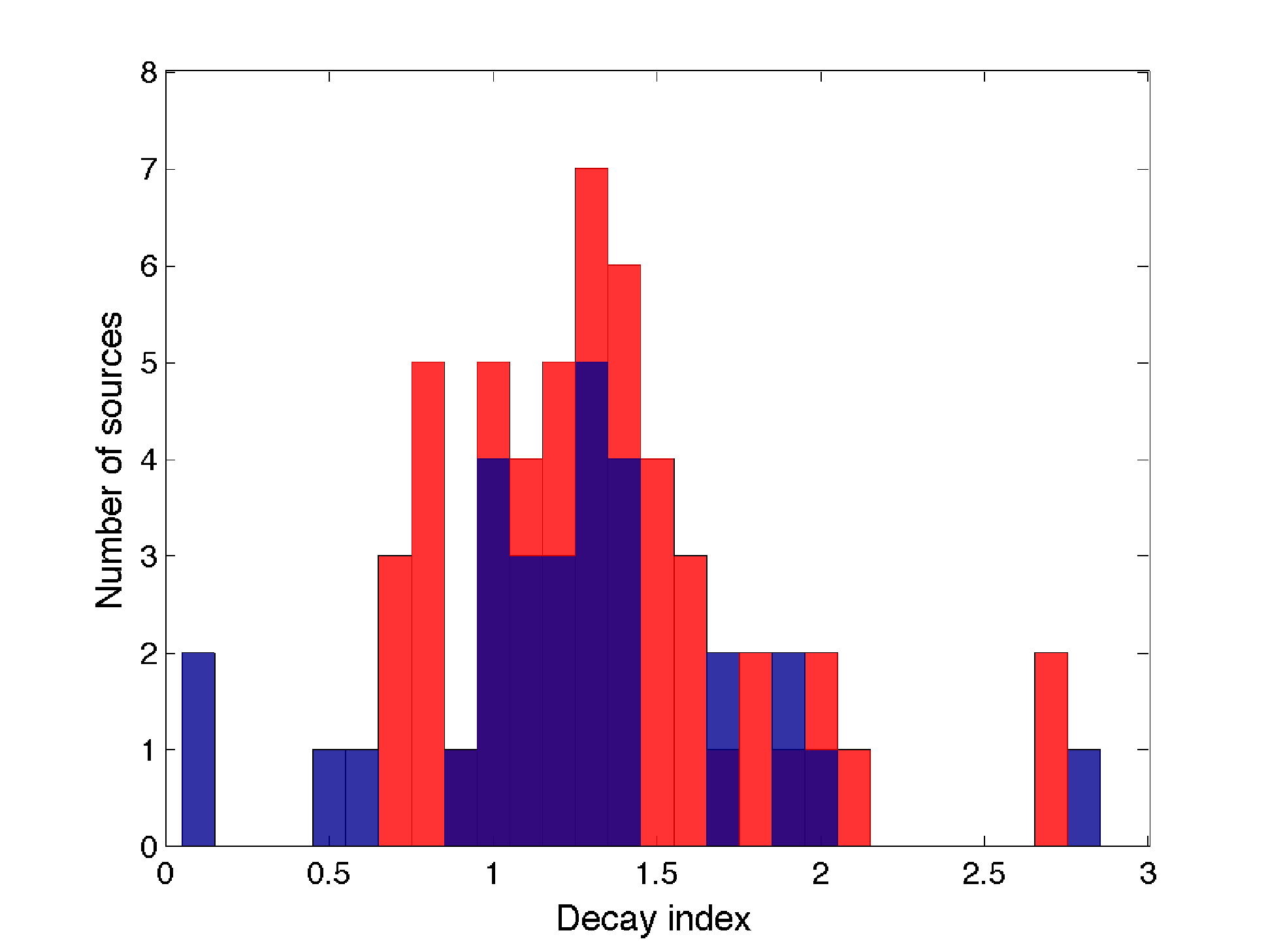}
\caption{Distribution of the decay indexes of LLA GRBs (blue) compared to the reference sample (red). The decay index reported in \cite{gendre2008} for GRB~060607A is incorrect, and not shown on this figure.\label{decay_index}}
\end{center}
\end{figure}

The distribution of the temporal decay indexes for LLA GRBs is presented in Figure \ref{decay_index}. In order to have a reference sample for comparison, I used a reference sample of bursts (higher luminous) listed in \cite{gendre2008} and not members of the LLA GRB subclass. Note that the decay index of GRB~060607A reported in this last article is incorrect and is not considered in the comparison. As it can clearly be seen, the two samples seem similar. This is confirmed by a K-S test (p = 0.88), that confirm the hypothesis that the two samples are drawn from same populations of bursts.

\subsection{Spectral index}

\begin{figure}
\begin{center}
\includegraphics[width=0.45\textwidth]{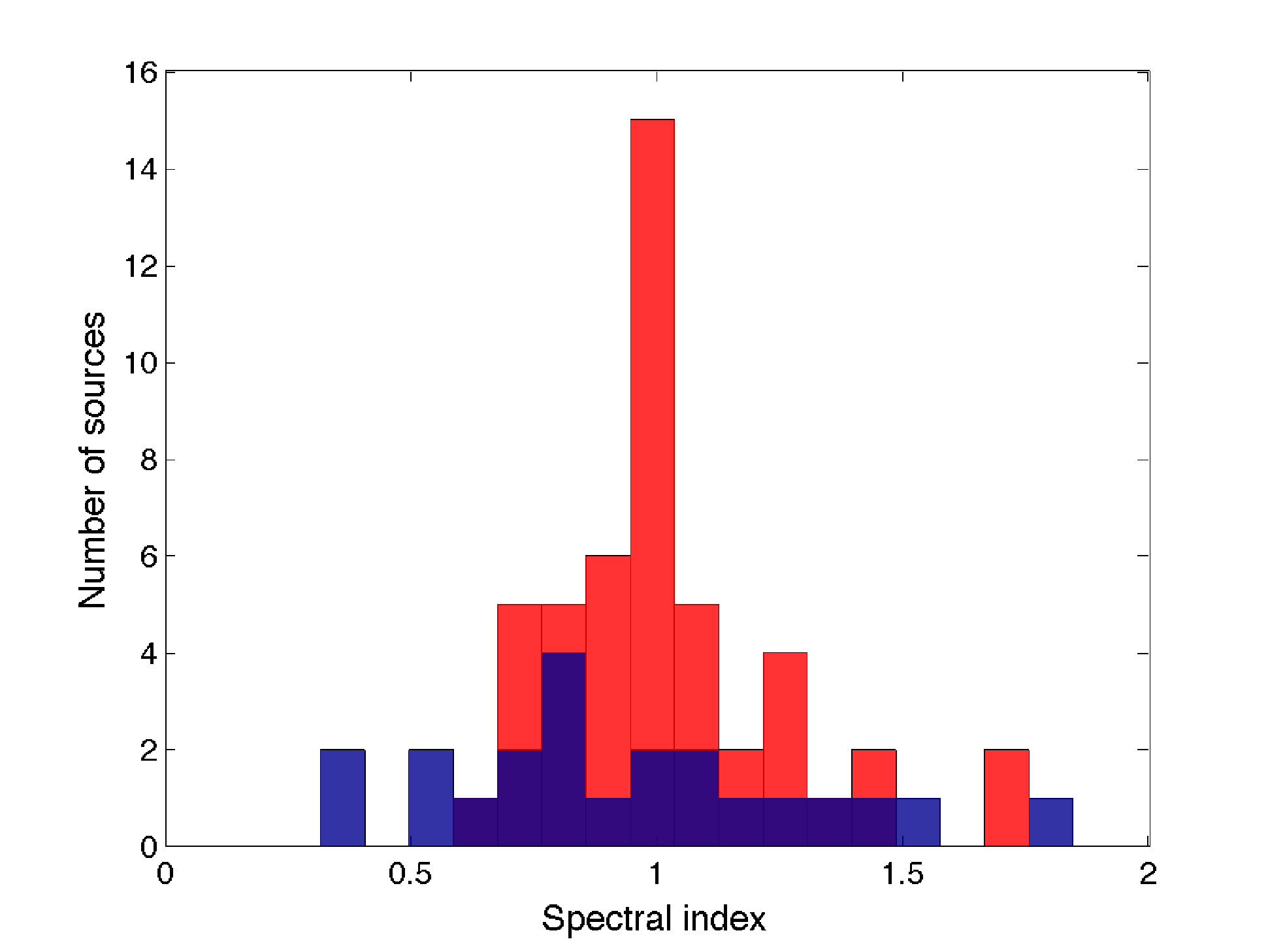}
\includegraphics[width=0.45\textwidth]{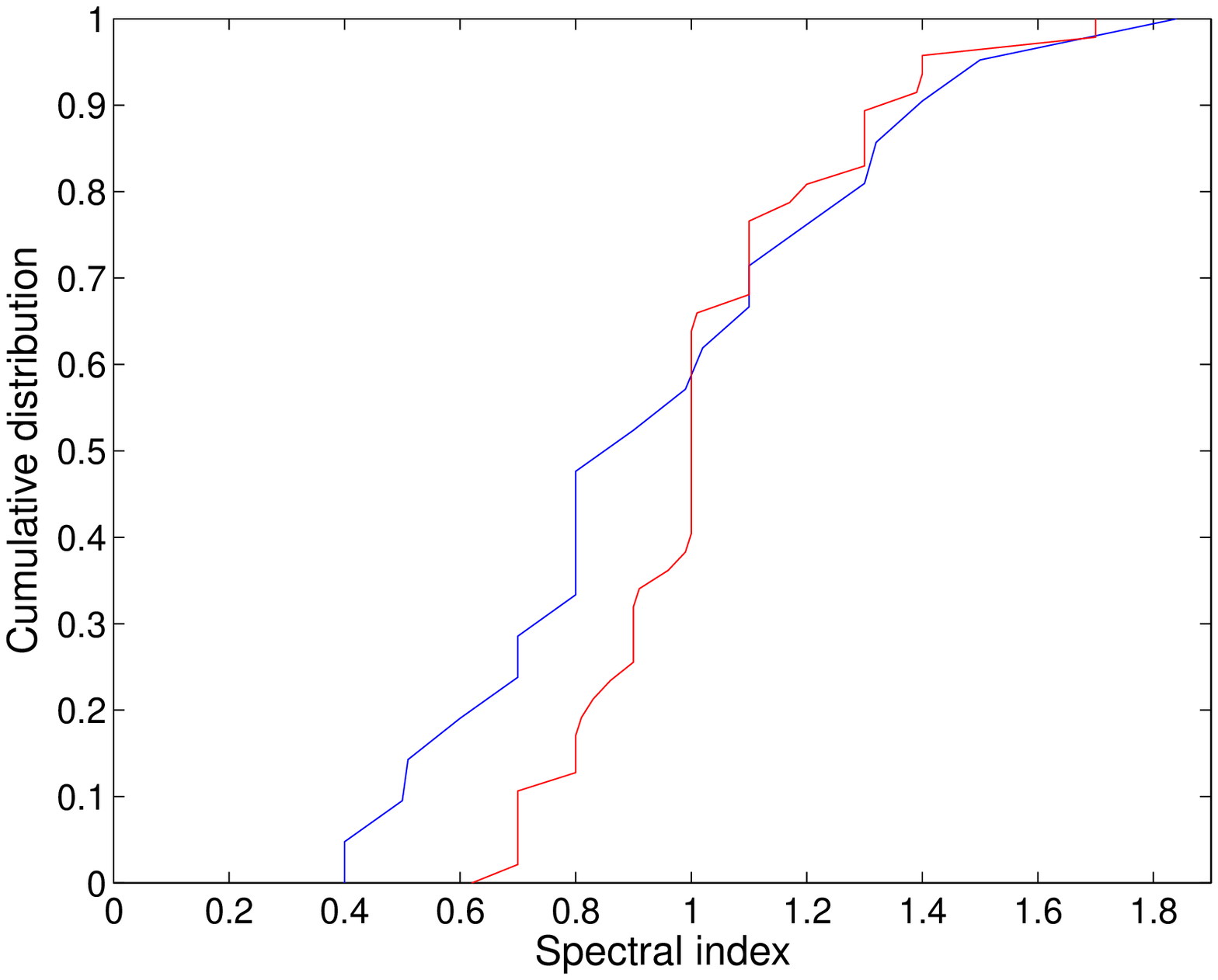}
\caption{Left: Distribution of the spectral indexes of LLA GRBs (blue) compared to the reference sample (red). Right: Cumulative distributions of the same samples.\label{spectral_index}}
\end{center}
\end{figure}

The distribution of the spectral indexes for LLA GRBs is presented in the left panel of Figure \ref{spectral_index}. I used again the reference sample (higher luminous ones \cite{gendre2008}) for comparison. Here, the two distributions seem to be different and it is confirmed by a K-S test: the hypothesis of a single population is confirmed with a probability of 0.084. 

\subsection{Closure relations in the X-ray band}

I used the closure relations \cite{meszaros1998, sari1998, sari1999a, chevalier2000, zhang2004}, see also section \ref{detailed_afterglow_theory}, to investigate the burst geometry, the fireball micro-physics, its cooling state and the surrounding medium. They are displayed on Figure \ref{closure} together with the spectral vs. decay indexes of each burst. Clearly, within the measurement errors,  most bursts can be explained by at least one of the closure relations, with two exceptions.

\subsubsection{GRB~120729A}

\begin{figure*}[!ht]
\begin{center}
\includegraphics[width=0.9\textwidth]{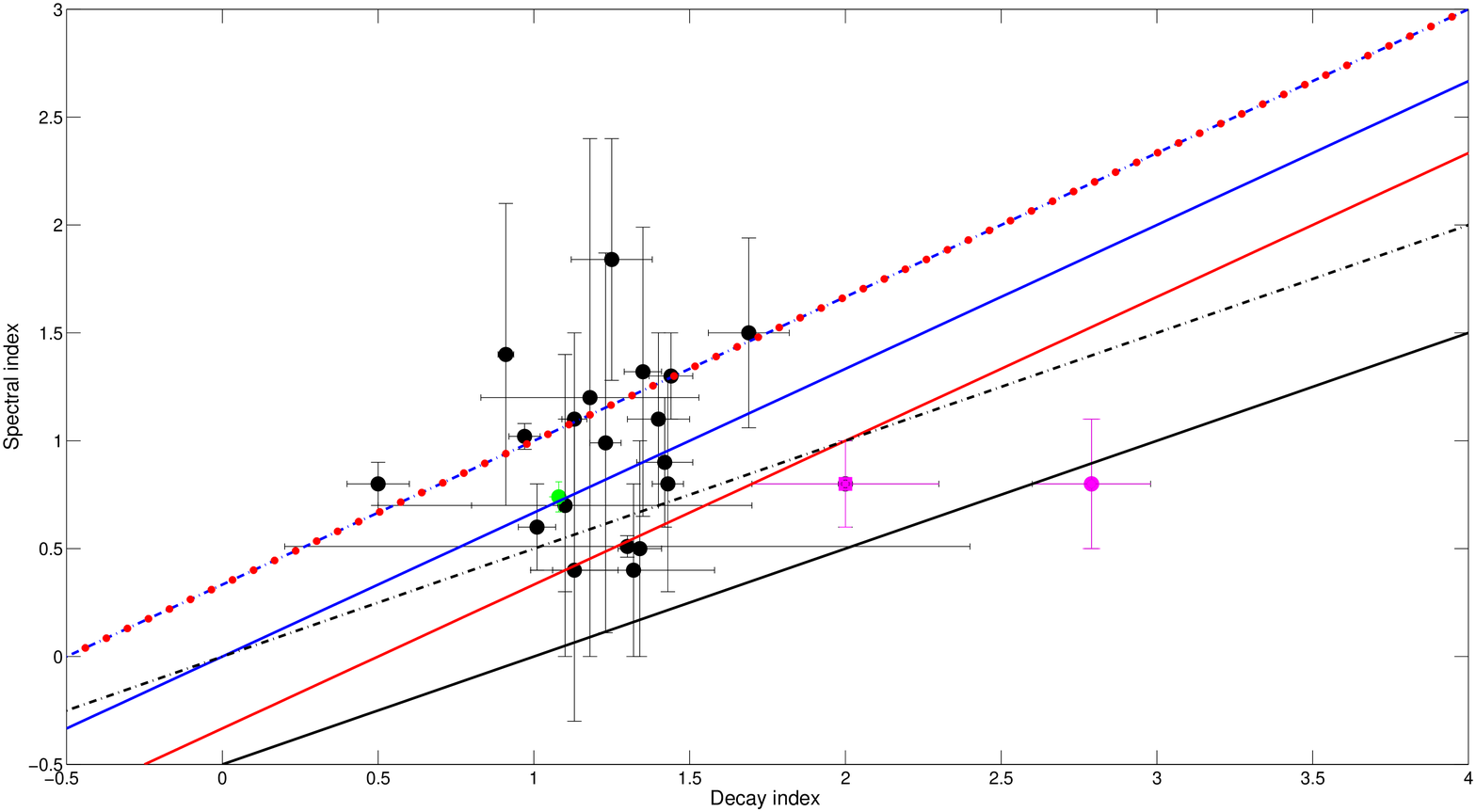}
\caption{X-ray decay index versus spectral index of LLA GRBs. The purple filled circle and square represent GRB~120729A and GRB~060614 respectively. The green dot represent GRB~120729A before the break at $t_b = 8.1$ ks. All closure relations, indicated by the lines, are computed for $p > 2$ in the slow-cooling phase. Solid and dash-dotted lines stand for $\nu_m < \nu < \nu_c$ and $\nu_c < \nu$ respectively. Blue, red and black lines stand for ISM, wind medium, and jet effect respectively. As it can be seen from Table \ref{table_closure}, the fast-cooling regime for ISM and wind $\nu > \nu_m > \nu_c$ has the same dependence with slow-cooling regimes of $\nu_m < \nu_c < \nu$. \label{closure}}
\end{center}
\end{figure*}

The non-jet closure relations are rejected for this event. In fact, this burst can be interpreted with jet in the slow-cooling phase, with the X-ray band located between the injection and cooling frequencies (\textit{i.e.} $\nu_m < \nu_{XRT} < \nu_c$), with an electron distribution index of $p = 2.8 \pm 0.2$. This hypothesis is strengthened by the presence of a break in the light-curve at $t_b = 8.1$ ks. Assuming that the afterglow part located before this break is the true standard afterglow, I recomputed the temporal and decay indexes: the new values agree with the non-jet closure relations (green point in Figure \ref{closure}).

\subsubsection{GRB~060614}

This event could be compatible with a jetted afterglow, with $p = 2.25 \pm 0.05$. However, the errors bars are large enough to accommodate some non-jetted closure relations. Thus, no firm conclusion can be reached on the jet hypothesis for this source based on the closure relations alone.

\subsection{Closure relations in the optical band}
\begin{figure*}
\begin{center}
\includegraphics[width=0.9\textwidth]{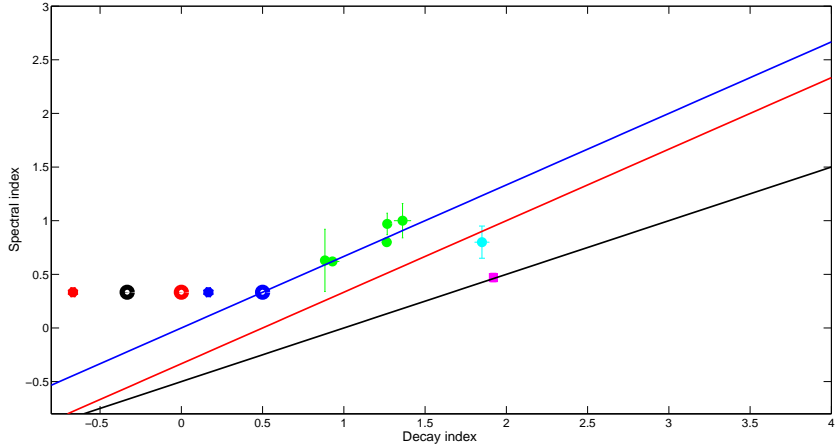}
\caption{Optical decay indexes versus spectral indexes of LLA GRBs. Blue, red and black colors stand for ISM, wind medium, and jet effect respectively. Solid lines stand for slow-cooling regime with $\nu_m < \nu < \nu_c$. Big circles stand for slow-cooling regime with $\nu_a < \nu < \nu_m$. Finally, small circles stand for fast-cooling regime with $\nu_a < \nu < \nu_m$. In green optical decay vs. spectral indexes of LLA GRBs. The purple filled square represents GRB~060614. The cyan filled circle represents GRB~011121. \label{closure_optical}}
\end{center}
\end{figure*}

The temporal decay and spectral indexes of 7 LLA GRBs at optical wavelength (see Table \ref{optical_properties}) are calculated and they are presented along with closure relations of $\nu_a < \nu < \nu_m$ and one of $\nu_m < \nu < \nu_c$, on Figure \ref{closure_optical}. The optical component of GRB~060614 can be explained by a jetted afterglow with $\nu_m < \nu_{optical} < \nu_c$, which strengthen the finding in the X-ray band. Finally, the five of LLA GRBs with optical measurements indicate a fireball in the slow-cooling regime with $\nu_m < \nu_{optical} < \nu_c$, which is expanding in a ISM. GRB~011121 is compatible with the slow cooled fireball expanding in a wind medium, compatible with constraints from the X-ray bands.

GRB~060614 also showed the same properties in the jetted closure relation like seen in X-ray data $\nu_m < \nu < \nu_c$. Five of the GRBs with optical measurements indicated that $\nu_m < \nu < \nu_c$ in the slow-cooling region of ISM. GRB~011121 indicated the same region but wind closure relation which is compatible with X-ray emission. 

\begin{table*}
\centering
  \caption{The temporal decay and spectral indexes of 7 LLA GRBs in the optical band.\label{optical_properties}}
  \begin{tabular}{llllll}
  \hline
  GRB & Redshift & Spectral &Decay \\
   name  &  & index & index \\
    \hline
GRB~011121 & 0.36 & 0.8$\pm$0.15 & 1.9$\pm$0.1 \\
GRB~050525A & 0.606 & 0.97$\pm$0.1 & 1.3 \\ 
GRB~050801 &1.38 & 1$\pm$0.16 & 1.4$\pm$0.05 \\ 
GRB~060614 & 0.125 & 0.47$\pm$0.04 & 1.9$\pm$0.01 \\	
GRB~060912A & 0.937 & 0.62 & 0.93$\pm$0.04 \\ 
GRB~070419A & 0.97 & 0.8 & 1.3$\pm$0.03 \\
GRB~071112C & 0.823 &0.63$\pm$0.29 & 0.9$\pm$0.02 \\
\hline
\end{tabular}
\end{table*}

\section{Prompt properties}
\label{sec_prompt}

The \textit{Swift}/BAT and the \textit{Fermi}-GBM datasets are re-analyzed for the LLA GRBs listed in Table \ref{table_sample}. They are combined with previously published results from Konus-Wind and BeppoSAX. About half of the events have a firm measurement of the prompt parameters ($E_{p,i}$ and $E_{iso}$), the other half present upper and lower limits. However, prompt parameters of some sources are missing in the table. Concerning GRB~051109B, it was only detection by \textit{Swift}/BAT and its spectrum is fitted by a simple power-law, which does not allow to constrain $E_{p,i}$. Concerning GRB~071112C, any values could not be derived because the dataset is incomplete. The prompt properties of GRB~061021 is given by \cite{golenetskii2006} obtained from the Konus-Wind data for the first pulse, $E_{p,i}$ is $1046\pm 485$ and $E_{iso}$ is $0.46\pm0.08 \times 10^{52} \text{ergs}$, while it has also a weak tail. So this source was not taken to account in the study of $E_{p,i}$-$E_{iso}$ correlation.

During the calculations for the lower and upper limits for $E_p$, the BAT spectrum was fitted by the power-law (PL) model. When the PL photon index is lower than $-1.7$, the low-energy index $\beta$ is set to $-2.3$ and the fit is performed with Band function in order to constrain the other parameters. This gives a lower limit on $E_{p,i}$. On the other hand, when the PL photon index is between $-1.7$ and $-1.9$, the situation is uncertain and any reliable limit on $E_{p}$ cannot be derived. However, in the opposite situation, when the PL photon index is larger than $-2$ the low-energy index $\alpha$ is set to -1, and the fit is performed with Band function in order to constrain the normalization, $\beta$ and $E_p$. This gives an upper limit of $E_{p,i}$. Both upper and lower limits are at 90\% confidence level and are computed using the error procedure in \textit{XSPEC}.

\begin{figure*}
\begin{center}
\includegraphics[width=0.9\textwidth]{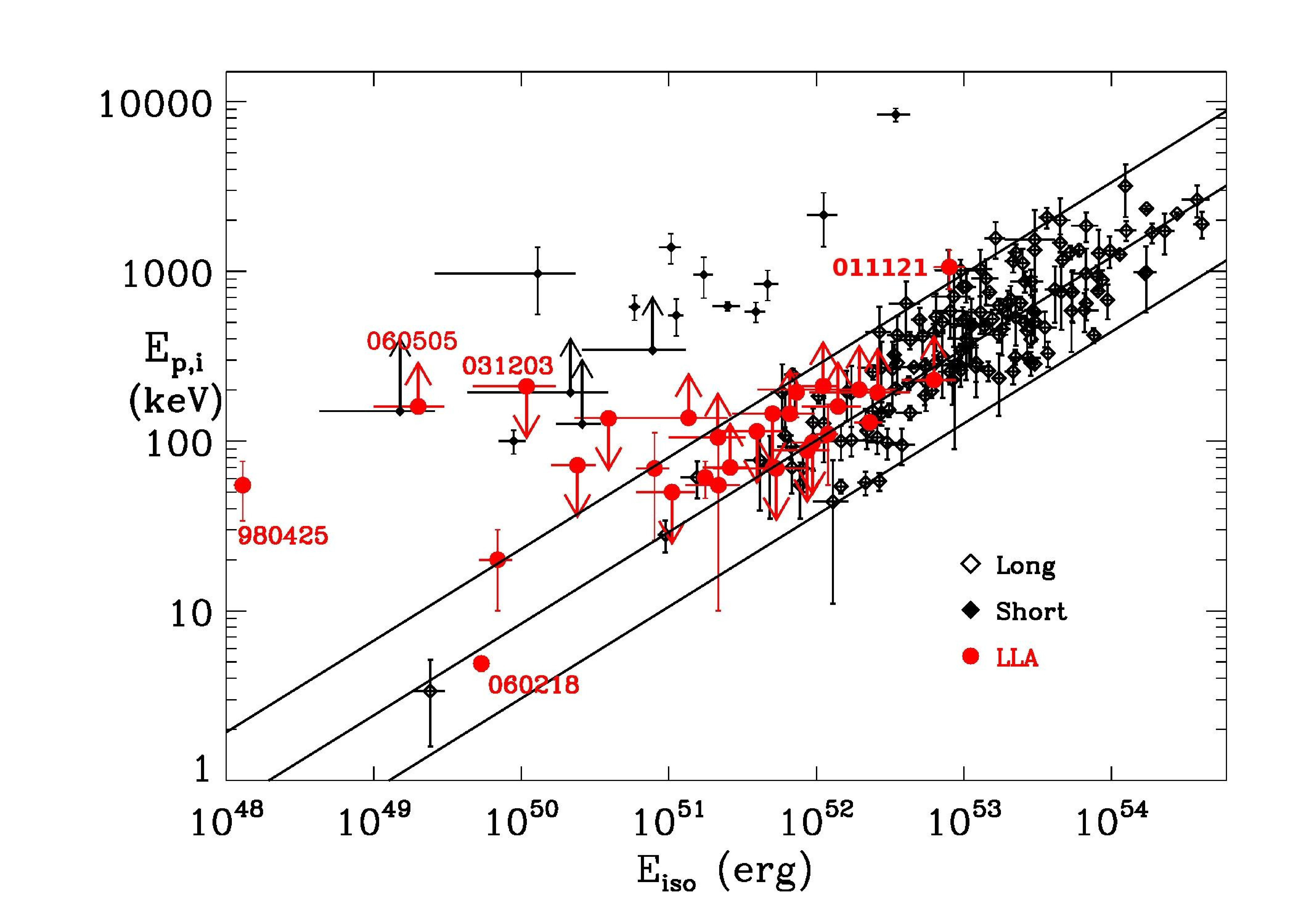}
\caption{Location in the $E_{p,i}-E_{iso}$ plane of LLA GRBs sample compared to both short and normal lGRBs. The best fitting power law, $E_{p,i} = a\times E_{iso}^{b}$, obtained by fitting the data accounting the sample variance. Derivation of $E_{p,i}$ from the best fitting power law of $\pm 2 \sigma \log E_{p,i}$ assuming that $\sigma \log E_{p,i}$ is the central value of the 90 percent confidence interval, they are computed by using the Reichart method as reported in \cite{amati2013}.}
\label{fig_amati_tot}
\end{center}
\end{figure*}

\begin{figure*}
\begin{center}
\includegraphics[width=0.45\textwidth]{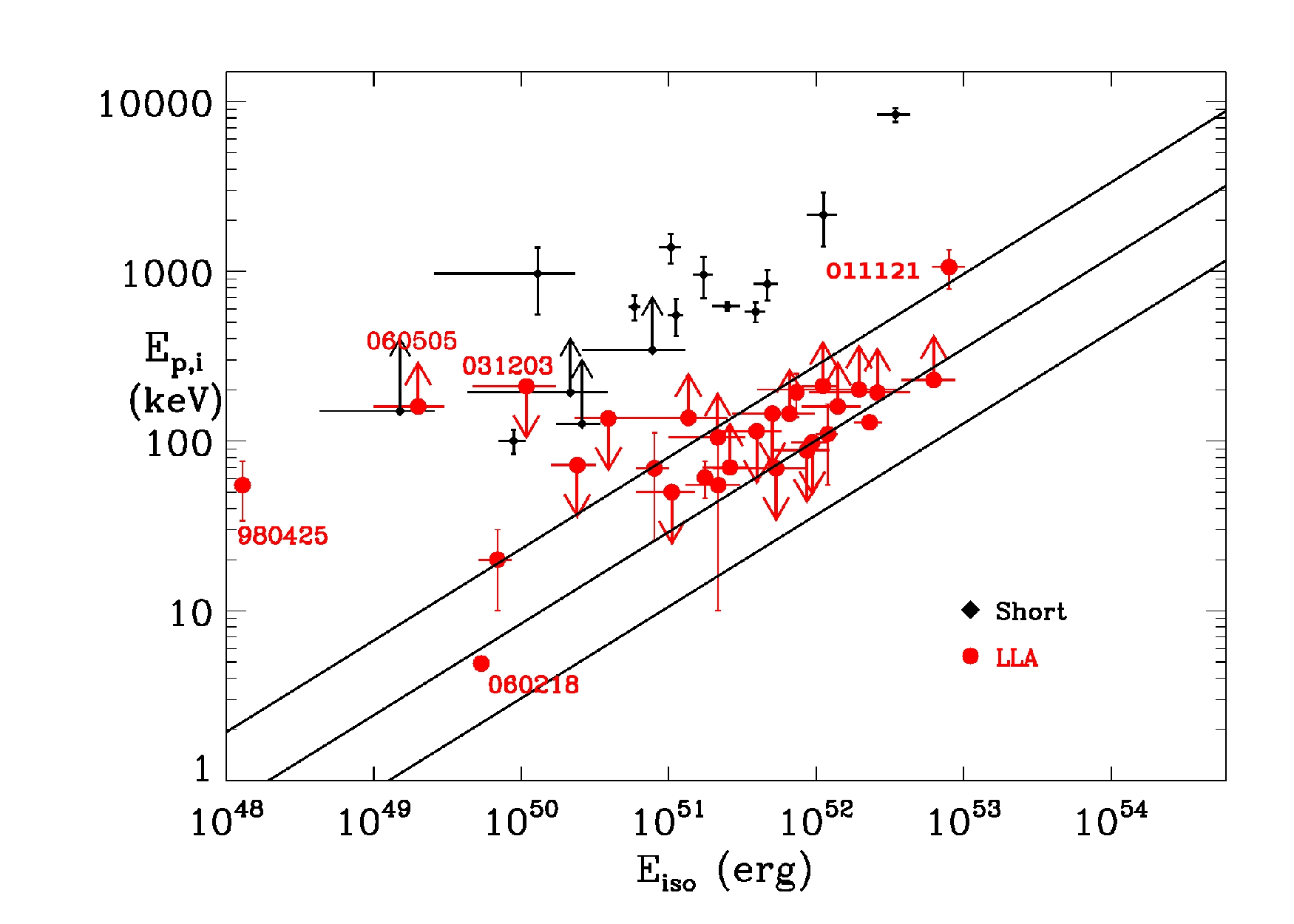}
\includegraphics[width=0.45\textwidth]{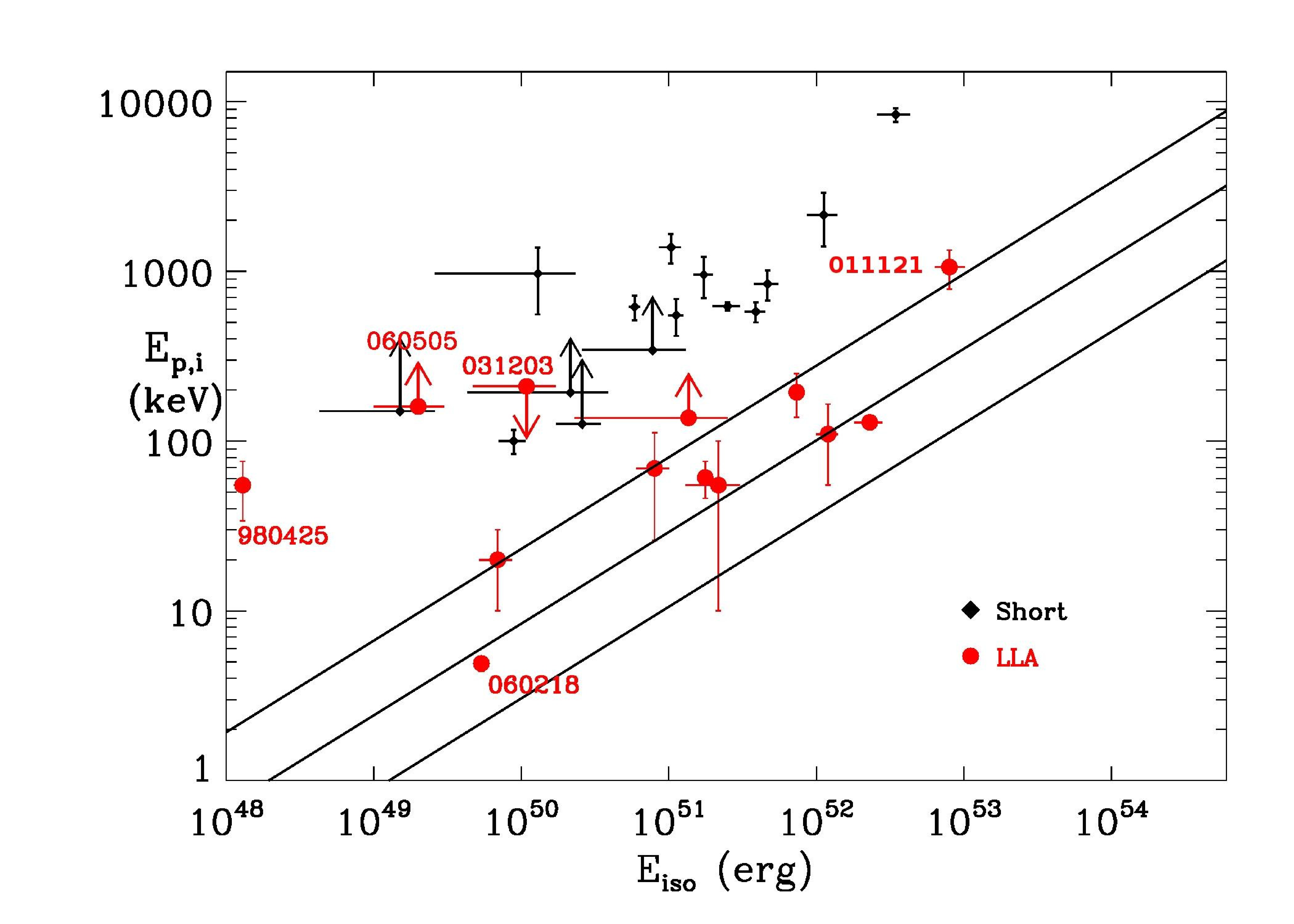}
\caption{Location in the $E_{p,i}-E_{iso}$ plane of LLA GRBs sample; on the left, compared to short events with the non-compatible lower limits on $E_{p,i}$ and $E_{iso}$; on the right, same comparison with firm measurements and non-compatible lower limits on $E_{p,i}$ only. \label{fig_amati_short}}
\end{center}
\end{figure*}

In order to estimate $E_{iso}$ and its uncertainty, first $E_{iso}$ is computed by assuming the best fit PL. Then, $E_{iso}$ is computed by performing the fit using the Band function \ref{Eq.band} with $\alpha$, $\beta$ and $E_p$ set at the values that they had when computing the lower or upper limit to $E_{p,i}$. In this way, a range of lower and higher value of $E_{iso}$ is obtained for each spectrum. $E_{iso}$ is taken as the average of lower and higher limits while the error is the higher value of $E_{iso}$ minus the average.

\begin{figure*}
\begin{center}
\includegraphics[width=0.45\textwidth]{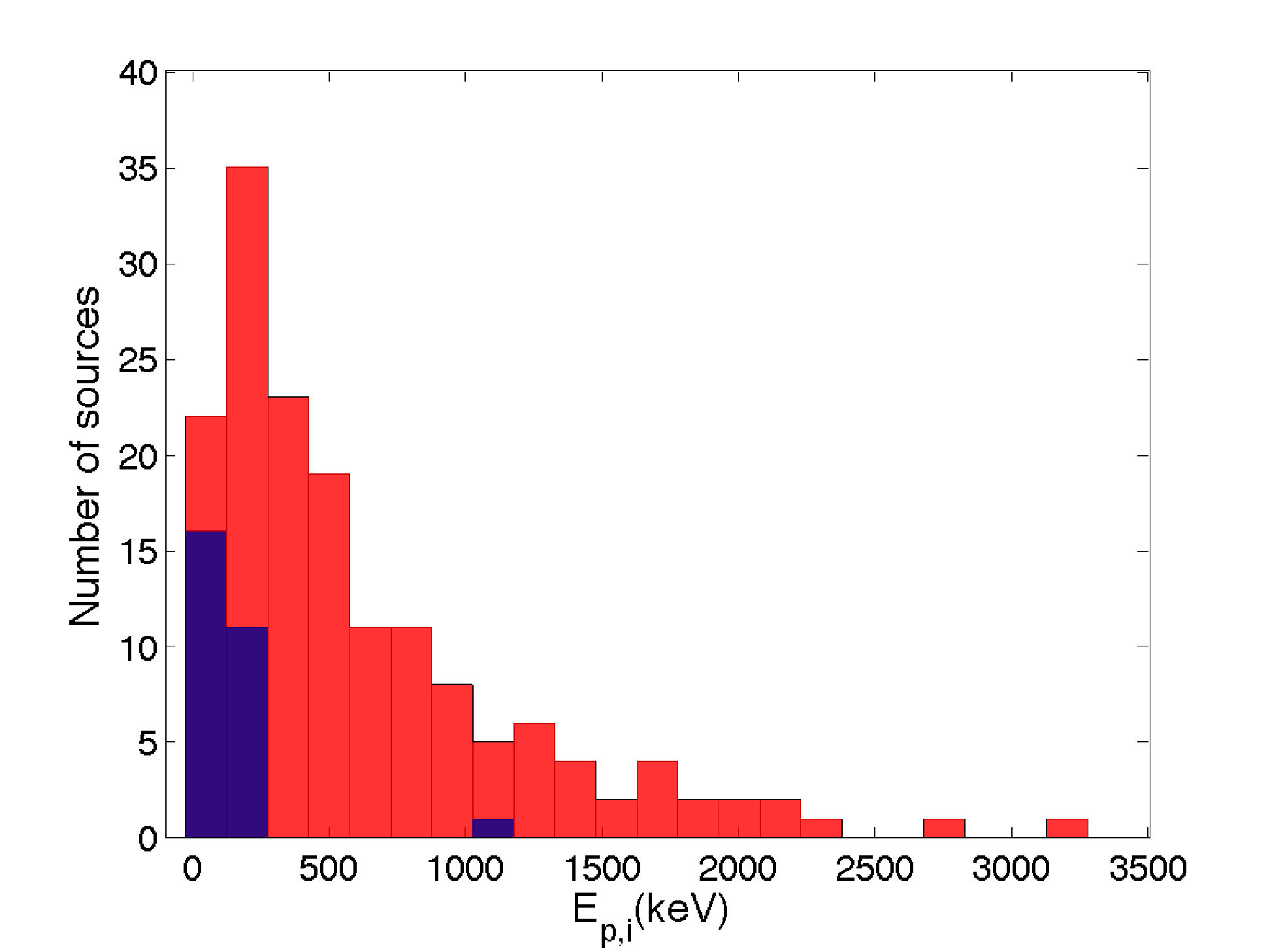}
\includegraphics[width=0.45\textwidth]{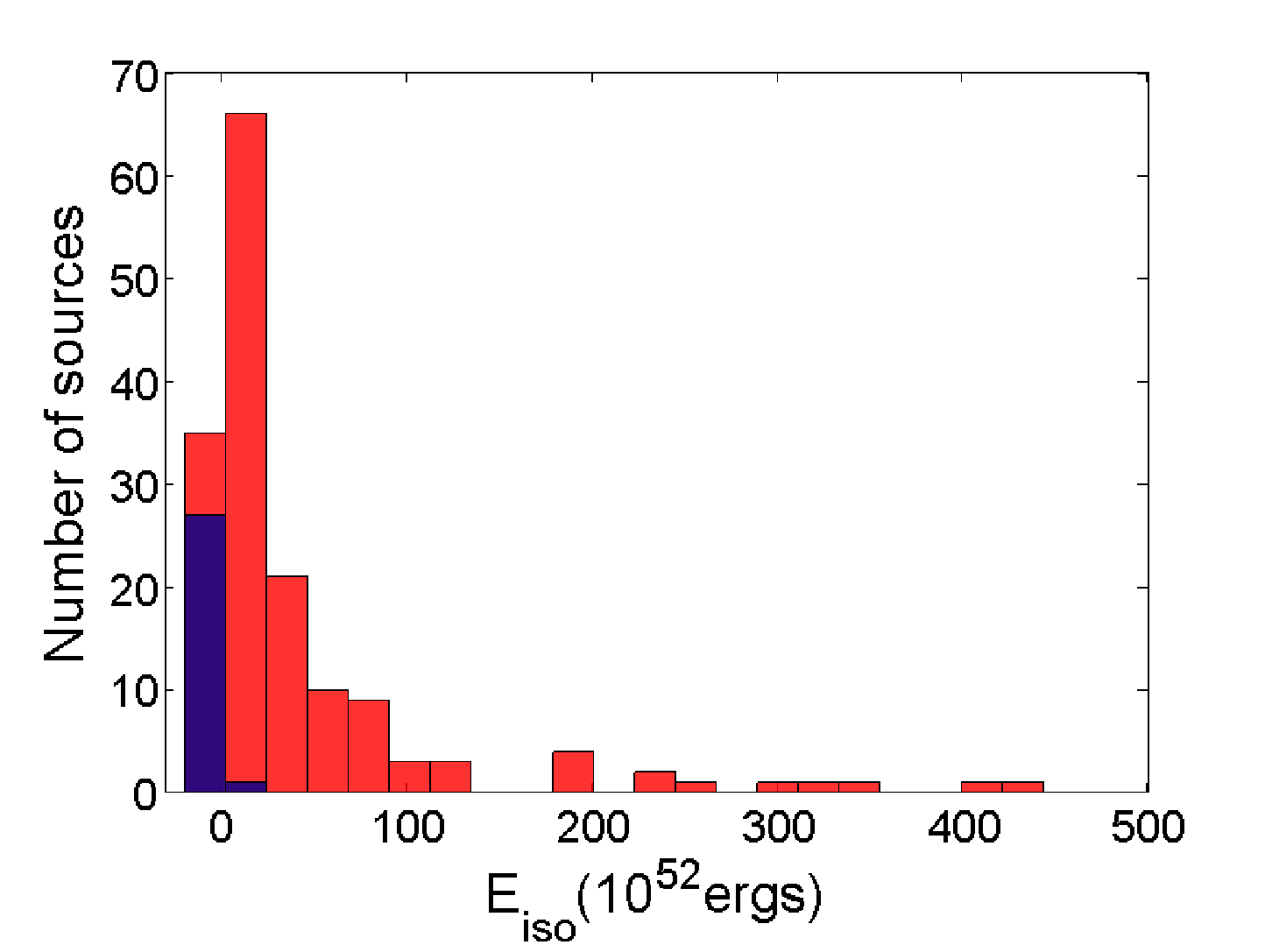}
\caption{Histogram of the $E_{p,i}$ (left) and $E_{iso}$ (right) of LLA GRBs (blue) and of normal IGRBs (red). When only a lower and upper limits were given, the avarage values are considered. \label{fig_hist_Ep_Eiso}}
\end{center}
\end{figure*}

It is found that the $E_{p,i}$ values cluster broadly within the 40-200 keV range, as expected. Figure \ref{fig_hist_Ep_Eiso} shows the histogram of the $E_{p,i}$ values for both LLA GRBs and normal lGRBs. 
However, the situation is different with $E_{iso}$. There is a clear shift in the $E_{iso}$ axis with respect to most normal lGRBs. LLA GRBs are less energetic during their prompt phase compared to normal GRBs. Figure \ref{fig_hist_Ep_Eiso} shows the histogram of $E_{iso}$ values for both LLA GRBs and normal lGRBs. 
By taking into account the median redshift of LLA GRBs, the high energy limit of BAT, the $E_{p,i}$ values, and the intrinsic scatter of the Amati relation, measurements up to $E_{iso} = 3 \times 10^{53}$ ergs would be expected, almost one order of magnitude larger than the BAT measurements listed in Table \ref{table_sample}. Thus, I conclude that this effect is not due to a bias, but is an evidence that LLA GRBs are intrinsically less energetic, both during the prompt and the afterglow phases, compared to normal lGRBs.

\subsubsection{Amati Relation}
The peak energy ($E_{p,i}$) in the rest frame of the burst of the $\nu F\nu$ spectrum is given:

\begin{equation}
E_{p,i} = E_{p,obs} \times (1+z)  \text{,}
\end{equation}
where $z$ is redshift, $E_{p,obs}$ is the observed peak energy:

\begin{equation}
E_{p,obs} = (2 + \alpha) \times E_0 \text{,}
\end{equation}
where $\alpha$ is the low-energy index and $E_0$ is the break energy. $E_{p,obs}$is the energy at which most of the power of the burst is emitted \cite{band1993}. The isotropic equivalent energy ($E_{iso}$) can be obtained by taking the integral of the spectrum and by using the redshift. It is given by:

\begin{equation}
E_{iso} = \frac{4 \pi d_L^2}{1+z} S_{bol} \text{,}
\end{equation}
where $d_L$ is the luminosity distance which is given by \cite{hogg1999}:

\begin{equation}
\label{luminosity_distance}
d_L = \frac{c}{H_0} (1+z) \int_0^{z} \frac{dz'}{\sqrt{\Omega_m (1+z')^3 + \Omega_k (1+z')^2 + \Omega_{\Lambda}}} \text{.}
\end{equation}

 Finally, $S_{bol}$ is the bolometric luminosity computed by: 

\begin{equation}
S_{bol} = S_{obs} \frac{\int_{E_1}^{E_{2}} E N(E) dE}{\int_{e_{1}(1+z)}^{{e_{2}(1+z)}} E N(E) dE} \text{,}
\end{equation}
where $[E_1 = 1~\text{keV}, E_2 = 10000~\text{keV}]$ is the bolometric gamma ray range, $[e_1, e_2]$ is the sensitivity band of the detector, $N(E)$ is the source photon spectrum which is fitted by the Band function, see Equation \ref{Eq.band} \cite{band1993}.

$E_{p,i}$ and $E_{iso}$ are correlated: $E_{p,i} \propto E_{iso}^{0.5}$. This was discovered in 2002 by using 11 BeppoSAX bursts \cite{amati2002}.  This correlation was later confirmed and extended to X-ray rich GRBs (XRRs) and X-ray flashes (XRFs) based on HETE-2 data \cite{amati2003}.

This correlation is used to constrain prompt emission mechanism, jet geometry and properties but also unify GRB and XRF etc. (see \cite{amati2006} for a review).

The main implications of the correlation include prompt emission mechanisms, jet geometry and properties, GRB/XRF unification models, identification and nature of sub-classes of GRBs (\textit{e.g.} sub-energetic, short) (see \cite{amati2006} for a review). I also used this relation to confirm that LLA GRBs are different from normal lGRBs when considering isotropy equivalent energy. 

\subsection{Jetted afterglows}

GRB~120729A can be accounted for by the closure relation of a jet. The opening angle is given by \cite{levinson2005} who extended the work of \cite{sari1999b} to account for the radiative efficiency of the prompt phase $\eta_\gamma$:
\begin{equation}
\label{eq_opening1}
\theta (t_{b}, E_{iso} ) = 0.161\left(\frac{t_{b,d}}{1+z}\right)^{3/8}n^{1/8}\left(\frac{\eta_\gamma}{E_{iso,52}} \right)^{1/8}\text{.}
\end{equation}  
In my computations, the standard values for the number density of the medium $n=1~\text{cm}^{-3}$ and the radiative efficiency $\eta_\gamma=0.2$ are used. I got $\theta = 2.7^\circ$. This is indeed a jet-break as it is seen at the same time in both the X-ray and optical bands \cite{maselli2012, davanzo2012}.

GRB~060614 is compatible with a jet, according to the closure relations. This burst also displays an achromatic break (around 36.6 ks) in X-ray, optical and UV \cite{mangano2007}. Before the jet-break, this burst features a plateau phase and not a standard afterglow. If I assume the presence of a jet, the corresponding jet opening angle is $6.3^\circ$.

A statistical study by \cite{ghisellini2012} shows that the mean opening angle of lGRBs is $\theta = 4.7^\circ$. The results for both GRBs are consistent with this value, and the jet, when detected, is not different from that of normal lGRBs.

\subsection{Environment of the bursts}

Most sources can be explained by either a wind environment or a constant ISM. As shown below, many of these sources are associated with SNe (see Table \ref{table_SNe}). This association would point towards a wind environment \cite{chevalier2004}. However, as shown by \cite{gendre2007}, the termination shock can lie very close to the star, and I cannot conclude firmly on the surrounding medium. The absence of strong dust extinction also does not  for a strong wind. 

One source deserves a more careful study: GRB~120729A. From the jet part of the light-curve, I derived $p = 2.8 \pm 0.2$. This value is not compatible with {\it both} the spectral and temporal decay indexes (1.08$\pm$0.03 and 0.74$\pm$0.072 respectively) of the pre-break part of the light-curve. Only the spectral index is marginally consistent with this value of $p$, assuming $ \nu_m < \nu_{XRT} < \nu_c$ and a constant ISM. The temporal decay is too flat (while expecting a value of at least 1.5). In order to reconcile all of these facts, I need to involve some late-time energy injection to flatten the light-curve \cite{hascoet2014}. This energy injection needs to be present during the pre-break part of the light-curve, but should stop during the post-break part. Let us  note that the sampling of the X-ray light-curve is not good during the jet break and allows for some non-simultaneity.  

\subsection{Microphysics of the fireball}

For LLA GRBs (see section \ref{afterglow}), when a closure relation indicates a given status of the fireball, the X-ray band is located {\it below} the cooling frequency. This fact is not common. Indeed, most late GRB afterglows are compatible with the X-ray band located above the cooling frequency \cite{gendre2006, depasquale2006}, see section \ref{sec_afterglow}. 

In the case of a constant ISM, the formula of the cooling frequency is given by Equation \ref{Eq:coolingfreqwind} \cite{panaitescu2000} and is reminded here:

\begin{equation}
\label{eq_nuc1}
\nu_c = 3.7 \times 10^{14} E_{53}^{-1/2} n^{-1} (Y+1)^{-2} \epsilon_{B,-2}^{-3/2} T_d^{-1/2} {\rm Hz}\text{,}
\end{equation}
where $E_{53}$ is the isotropic energy in units of $10^{53}$ ergs, $n$ is the number density of the medium in cm$^{-3}$ , $Y$ is the Compton parameter, $\epsilon_{B,-2}$ is the fraction of internal energy in the magnetic field, $T_d$ is the observed time expressed in days after the burst.

Instead in the case of a wind medium, the cooling frequency is given by \cite{panaitescu2000}: 
\begin{equation}
\label{eq_nuc2}
\nu_c = 3.5 \times 10^{14} E_{53}^{1/2} A_*^{-2} (Y+1)^{-2} \epsilon_{B,-2}^{-3/2} T_d^{1/2} {\rm Hz}\text{,}
\end{equation}
where $A_*$ is a constant normalizing number density of the wind.

I start by assuming that the fireball expands in the constant ISM. The X-ray band is from 0.3 to 10.0 keV, which correspond to $7.2 \times 10^{16}$ Hz and $2.4 \times 10^{18}$ Hz respectively. I assume that $\nu_c$ is above $3.7 \times 10^{18}$ Hz for simplicity. Equation \ref{eq_nuc1} simplifies to :
\begin{equation}
\label{eq_nuc3}
10^{-4} E_{53}^{-1/2} \epsilon_{B,-2}^{-3/2} < 1\text{,}
\end{equation}
when assuming the standard density $n = 1~\text{cm}^{-3}$, the Compton parameter $Y \ll 1$ and considering the observation time $T_d = 1$. From the prompt parameters, I find that $E_{53}$ can be as low as $10^{-5}$ for GRB~980424, I then finally obtain : $\epsilon_{B,-2} > 0.1$. A larger value of $E_{53}$ reduces the constraint on $\epsilon_B$. In any case, a constraint of $\epsilon_{B,-2} > 0.1$ is not really constraining, as typical values of $\epsilon_{B,-2}$ should be in the order of 1. I therefore conclude that, under the hypothesis of the fireball expanding in a constant ISM, the uncommon position of the cooling frequency is due to the small energy of the fireball.

The situation is similar when assuming a wind medium, for which Equation \ref{eq_nuc2} implies : 
\begin{equation}
\label{eq_nuc4}
10^{-4} E_{53}^{1/2} \epsilon_{B,-2}^{-3/2} < 1\text{,}
\end{equation}
when assuming standard density and Compton parameter $Y \ll 1$, and considering the observation time $T_d = 1$. From the prompt parameters, I find that $E_{53}$ can be as high as $0.797$ for GRB~011121, I thus finally obtain : $\epsilon_{B,-2} > 0.0012$, which is again not constraining. A smaller value of $E_{53}$ reduce the constraint on $\epsilon_B$. 

\subsection{Prompt properties of LLA GRBs}

In Figure \ref{fig_amati_tot}, it is clearly seen that all outliers to the Amati relation are LLA GRBs. Several explanations have been proposed to explain these events (see \cite{amati2006} and reference therein for details): GRB~060505 may be a short GRB (as its location in the $E_{p,i}$ - $E_{iso}$ plane in Figure \ref{fig_amati_short} may suggest); the $E_{p,i}$ value of GRB~061021 refers to the first hard pulse, while a soft tail is present in this burst (so the true $E_{p,i}$ may be lower); GRB~031203 may be much softer than measured by INTEGRAL/ISGRI as supported by dust echo measured by XMM. I notice, however, that the outliers are all located on the left part of the diagram, I will propose two alternative solutions.

\subsubsection{The Amati relation is valid for LLA events}

The left part of the $E_{p,i}$ - $E_{iso}$ plane in Figure \ref{fig_amati_short} relates to soft and faint events. In this part of the diagram, the usual gamma-ray instruments are not well-suited to measure the prompt properties. For instance, if only BAT measurements of GRB~060218 \cite{sakamoto2006} are taken into account, this events is more similar to GRB~980425, \textit{i.e.} a clear outlier. The reason can be lack of the softer part of emission due to limited energy band of the detector. However, the advantage of {\em Swift} compared to BeppoSAX is its ability to perform simultaneously X-ray and gamma-ray observations. Combining the XRT and BAT measurements allows to make this event (GRB~060218) fully consistent with the Amati relation. 

\subsubsection{The Amati relation is not valid for LLA events}

On the other hand, GRB~980425, GRB~060505 and GRB~050826 are not compatible at all with the Amati relation. Assuming that these measurements are correct, then the best-fit relation in the $E_{p,i}$ - $E_{iso}$ plane changes dramatically, being way more flatter. 
A flatter Amati relation has been foreseen as early as 2003 \cite{yamazaki2003}, using GRBs seen largely off-axis. This is not in contradiction with the results of the closure relations, as these relations apply only to events seen on-axis. For completeness, I also note that a similar explanation hold in case of the canonball model \cite{dado2005}. Being seen off-axis, these events are expected to be less luminous than normal lGRBs {\it even during the afterglow} (see \textit{e.g.} \cite{dalessio2006}). Thus, the discrepancies between the LLA GRBs and normal lGRBs are explained by the geometry of the events and not by the progenitor.

\section{Conclusions}

I present strong evidence that LLA GRBs belong to a population of nearby events. They represent about 12\% of the total population of long bursts, and their afterglow emission are faint. It is also found that the afterglow spectral and temporal properties are similar to those of the general population. I show that there is no more selection effects due to absorption and extinction introduced by the Milky-Way and host galaxies than for normal lGRBs.

However, these events are also faint during their prompt phase, and located in another area of the $E_{p,i}$-$E_{iso}$ plane in Figure \ref{fig_amati_tot}. They also include all outliers of the Amati relation. In addition, their redshift  distribution is different from that of normal lGRBs. Indeed there is a lack of luminous lGRBs at small distances, which should have been detected, since they are luminous.

The similar properties of LLA bursts and normal lGRBs point towards the same emission mechanism for the afterglow and might indicate that LLA GRBs are the tail of the luminosity distribution of normal long GRBs. However, it does not explain the lack of luminous GRBs in the local universe. This tension might be solved by further GRBs monitoring and the observation of luminous nearby sources. Instead this tension might increase with the observation of several additional low-luminous nearby sources, which might be the indication for a different kind of bursts.

%% file: Chapter5.tex

\chapter[Additional Properties of LLA GRBs]{\parbox[t]{\textwidth}{Additional Properties of LLA GRBs}}
\chaptermark{Additional Properties}

\section{Supernovae associated to LLA GRBs}

Nine LLA GRBs are firmly associated to SNe by spectral and photometric optical observations. There are also two other sources (GRB~070419A, GRB~100418A) that may be associated to SNe. The first one shows a faint bump in its light-curve similar to the one of GRB~980425 \cite{hill2007} while for the second one, a bright host galaxy may prevent to observe the signature of a faint SN, which cannot be brighter than $-17.2$, comparable to the magnitude of the faintest Ic SN \cite{niino2012}. However, two sources (GRB~060505, GRB~060614) are firmly not associated to SNe (while they are nearby). All referenced GRB-SN pairs are listed in Table \ref{table_SNe}. It also contains the pairs for which the burst is not an LLA GRB. As it can be seen for several well-known associations, the burst is an LLA GRBs. For example, GRB~980425/SN~1998bw, GRB~031203/SN~2003lw, and GRB~060218/SN~2006aj. If the former was thought to be due to a hypernovae \cite{iwamoto1998}, the latter has been proposed to be a neutron star experiencing an SN \cite{mazzali2006}.
Being nearby objects, it is worth wondering why not all LLA GRBs are associated to SNe. The most trivial solution is a large optical extinction preventing the detection of the SN. However, when looking at Figure \ref{fig_distri_AV}, which displays the distribution of host galaxy extinction $A_V$, the non-associated events are not more affected by a larger extinction than the associated ones. The same is true for the Galactic extinction, see Figure \ref{fig_AV}. Other observational problems, such as a bright host, can mask out a faint underlying SN signal. 
However, a large fraction of LLA GRBs are associated with SNe (64\%). This can lead to a potential problem: if LLA GRBs are intrinsically different (as seen in the $E_{p,i}$ - $E_{iso}$ plane in Figure \ref{fig_amati_tot}) from normal bursts, the SN-GRB connection might not be representative of all lGRBs.

\begin{table*}[!ht]
 \centering\scriptsize
  \caption{GRB-SN events with some physical parameters. The last part of the table shows the firm associations of bursts which are not LLA GRBs. The isotropic luminosity $L_{ iso}$ is calculated from $E_{iso}$ by dividing it by $T_{90}$, see Table \ref{table_sample} \label{table_SNe}}.
  \begin{tabular}{llllllllL{0.08\textwidth}}
  \hline
  GRB & & SN & SN & SN &Host & LLA &$L_{iso}$ & \\
   name & redshift & identification & name & type& type & GRB& (GRB) & Ref. \\ 
   &&&&&&&10$^{49}$erg.s$^{-1}$& \\
 \hline
GRB~980425 & 0.0085 &spectral&SN~1998bw & BL-lc &dwarf spiral & yes & 0.033 & (1) \\
&&& & & (SbcD) & &  \\
GRB~011121 & 0.36 &spectral &SN~2001ke & IIn & no & yes &387  & (2) \\
GRB~031203 & 0.105 &spectral&SN~2003lw & BL-Ic &Irr & yes & 0.56 & (3), (4) \\
&&& & &Wolf-Rayet & &  \\
GRB~050525 & 0.606 &spectral &SN~2005nc & $\sim$Ic& no & yes  & 417 & (5), (6) \\
GRB~060218 & 0.0331 &spectral&SN~2006aj & BL-Ib/c & dwarf Irr & yes & 0.02  & (7), (8), (9) \\
GRB~081007 & 0.5295 &bump &SN~2008hw  & Ic & no & yes & 30 & (10), (11) \\
GRB~100316D & 0.059 &spectral &SN~2010bh & BL-Ic & Spiral blue & yes & 0.056 & (12), (13) \\
GRB~120422A & 0.283 &spectral &SN~2012bz & Ib/c & no & yes & 0.44 & (14), (15) \\
GRB~120714B & 0.3984 &spectral& SN~2012eb & I & no & yes &0.7 & (16), (17) \\ 
 \hline
GRB~070419A  & 0.97 &bump& no & & no & yes & 6.6 & (18) \\
GRB~100418A & 0.6235 & no detection & $\cdots$ & & dwarf blue & yes & 24.4 & (19) \\
 \hline
 GRB~060505 & 0.089 &no association & $\cdots$ & & Spiral(SbcD)  & yes & 0.82 & (20) \\
GRB~060614 & 0.125 & no association & $\cdots$ &  & no & yes & 2.2 & (21), (22), (23) \\
\hline
GRB~021211& 1.01 &spectral &SN~2002lt & $\sim$Ic & no & no & 969 & (24), (25), (26) \\
GRB~030329& 0.168 &spectral &SN~2003dh & BL-Ic & no & no & 75 & (27), (28), (29) \\
GRB~091127& 0.49 &bump&SN~2009nz & BL-Ic & no & no & 345 & (30), (31) \\
GRB~101219B & 0.55 &spectral &SN~2010ma & Ic & no & no & 29 & (32), (33) \\
GRB~130215 & 0.597 &spectral &SN~2013ez & Ic & no & no & 75 & (34), (35) \\
\hline
\end{tabular}

Note: for GRB-SN associations: (1)~ \cite{galama1998}; (2) \cite{bloom2002}; (3) \cite{soderberg2003}; 
(4) \cite{tagliaferri2004}; (5) \cite{dellaValle2006a}; (6) \cite{blustin2006}; (7) \cite{cobb2006}; (8) \cite{campana2006};
(9) \cite{soderberg2006};(10) \cite{soderberg2008}; (11) \cite{markwardt2008}; (12) \cite{chornock2010}; (13) \cite{sakamoto2010};
(14) \cite{melandri2012}; (15) \cite{barthelmy2012}; (16) \cite{klose2012}; (17) \cite{cummings2012}; (18) \cite{hill2007};
(19) \cite{holland2010}; (20) \cite{haislip2006}; (21) \cite{fynbo2006}; (22) \cite{dellaValle2007}; (23) \cite{gal-yam2006};
(24) \cite{crew2002}; (25) \cite{dellaValle2003}; (26) \cite{vreeswijk2003}; (27) \cite{golenetskii2003}; (28) \cite{kawabata2003};
(29) \cite{stanek2003}; (30) \cite{cobb2009}; (31) \cite{wilson-hodge2009}; (32) \cite{vanderHorst2010}; (33) \cite{sparre2011}; 
(34) \cite{deUgartePostigo2013}; (35) \cite{younes2013}; 
\end{table*}

In general, types of SNe which are associated to GRBs are Ic, broad-line~Ic and Ib/c, as seen in column 5 of Table \ref{table_SNe}, except SN~2001ke associated to GRB~011121 which is of type IIn: hydrogen can be find in its spectrum. This GRB and the other LLA GRBs have different properties. It is the brightest one ($7.97\pm2.2 \times 10^{52}$~ergs) with the highest peak energy (1060$\pm$275~keV). It is accommodated by the closure relation of a wind in both X-ray band (see Figure \ref{closure}) and optical band (see Figure \ref{closure_optical}).

\section{The signature of an SN in the GRB light-curve}

For the GRB-SN associations, most SNe are identified by spectroscopic analysis as seen in column 3 of Table \ref{table_SNe}. On the other hand, there is another way to detect an SN in GRB light-curves, which is to identify a bump in the optical afterglow light-curve. This bump is expected between 10 to 20 days after the GRB explosion. For such an identification, the light-curve of GRB~980425 is considered as a template.

\begin{figure*}[!ht]
\begin{center}
\includegraphics[width=0.3\textwidth, angle=270]{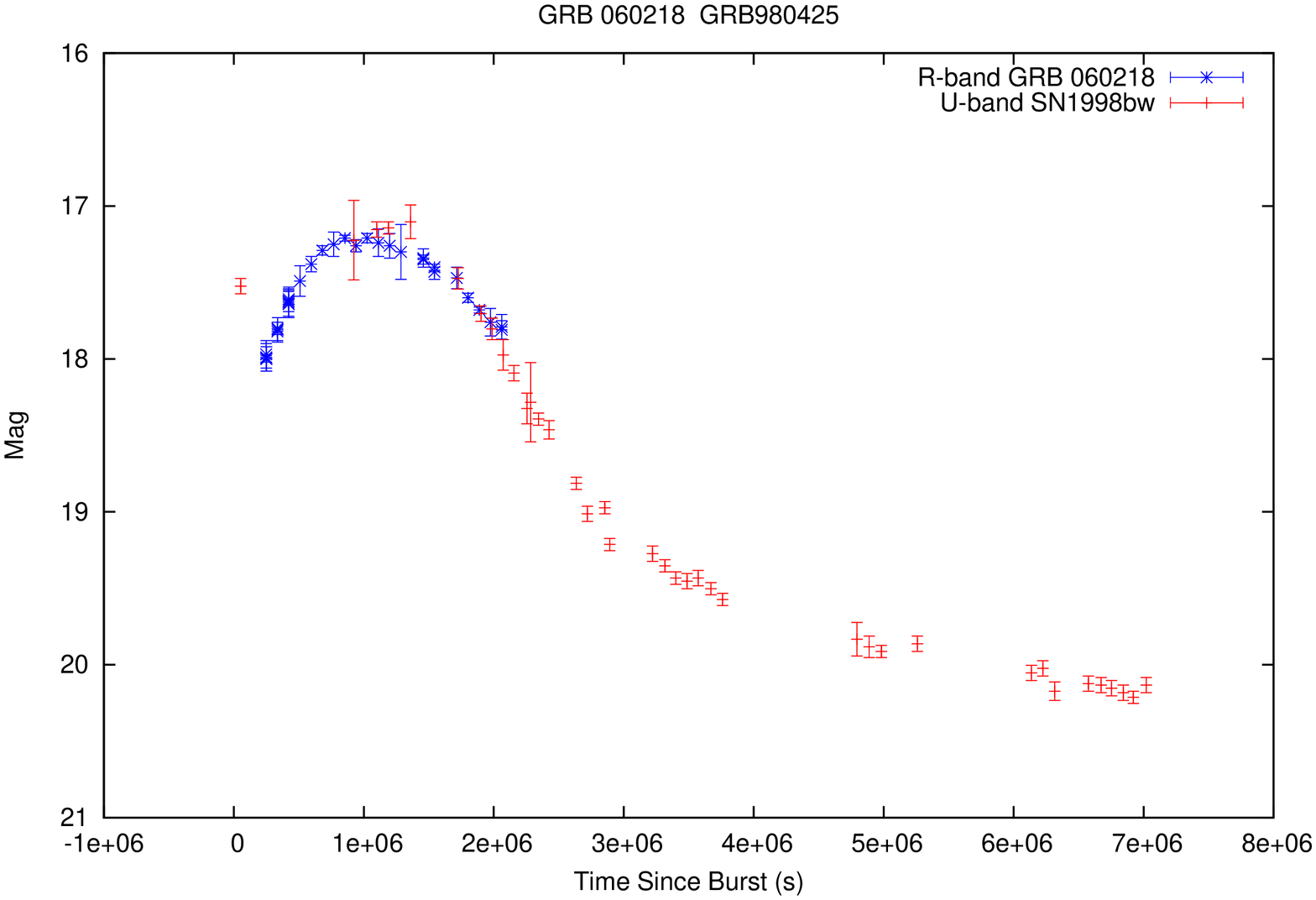}
\includegraphics[width=0.3\textwidth, angle=270]{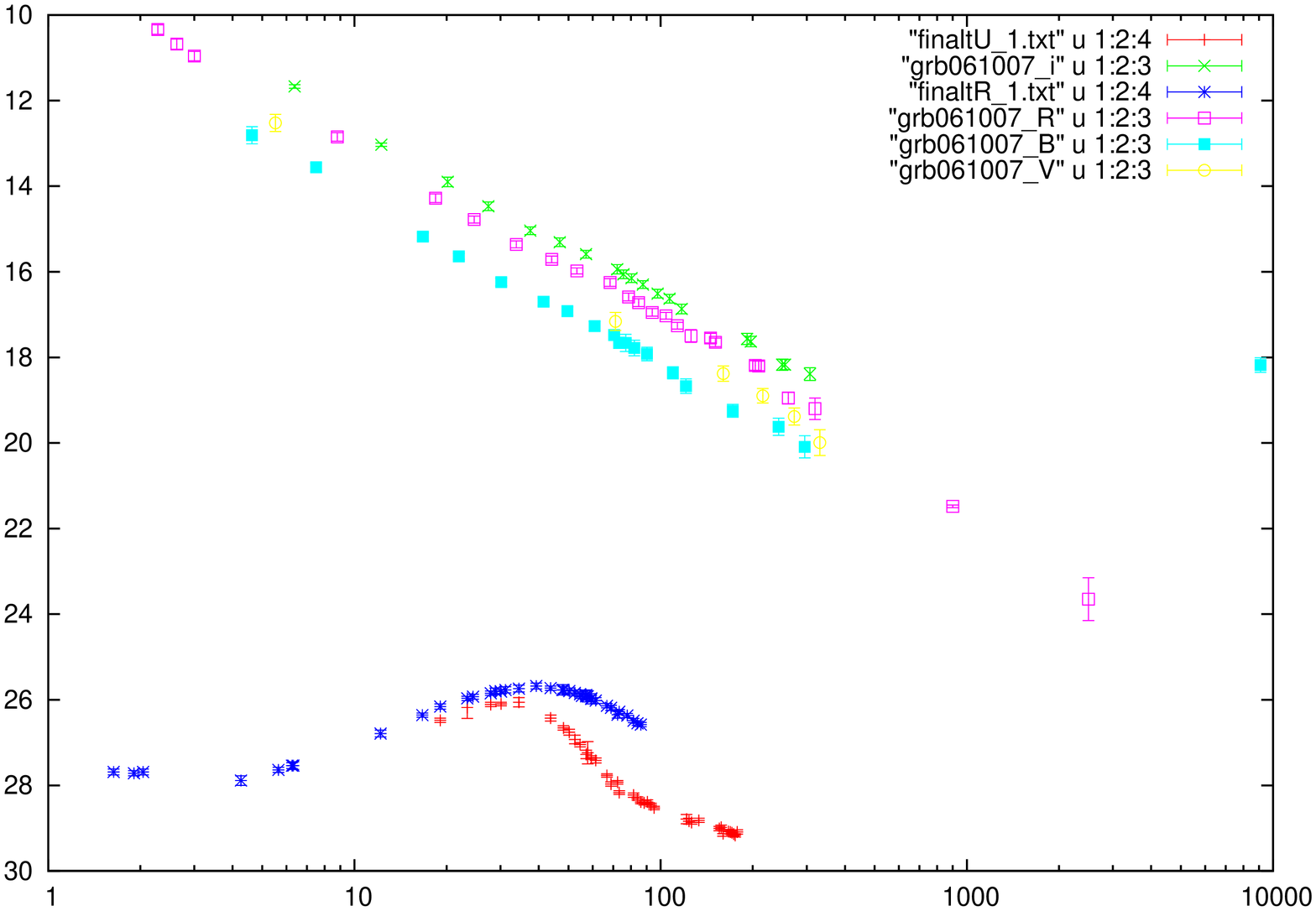}
\caption{Left: U-band observations of GRB~980425 transposed to the redshift of GRB~060218. Right: The R and U-band observations of GRB~980425 transposed to the redshift of GRB~061007. They are compared with the multi-band observations of GRB~061007.
\label{bump_060218_061007}}
\end{center}
\end{figure*}

I used this template by transforming the U-band light-curve of GRB~980425 at redshift 0.0085 to the redshift of GRB~060218 ($z = 0.0331$), to show the SN bump in its light-curve. This light-curve is compared in Figure \ref{bump_060218_061007} (left) with the one of GRB~060218. This association is also confirmed by spectral observations.

To perform the transformation, first of all, I used the classical conversion of apparent magnitude $m_1$ at redshift $z_1$ to the apparent magnitude $m_2$ at redshift $z_2$:
\begin{equation}
m_2 = m_1 - 5 \ \log_{10} \left(\frac{d_L(z_1)}{d_L(z_2)}\right) \textbf{,}
\end{equation}
where $d_L$ is the luminosity distance and it is given by Equation \ref{luminosity_distance}. Finally, $z_1$ and $z_2$ are the redshifts of the GRB which used as a template and of the tested GRB respectively. 

Then, the time correction is applied:

\begin{equation}
t_2 = t_1 \frac{1+ z_1}{1+z_2} \textbf{,}
\end{equation}
where $t_1$ and $t_2$ are times measured at redshift $z_1$ and $z_2$ respectively.

However, the bump of the SN (assuming it exist) cannot be seen in most GRB light-curves which might be because of the bright afterglow. This is shown by GRB~061007 for example, see Figure \ref{bump_060218_061007} (right). The template for the SN bump is transformed at the redshift GRB~061007 and compared to each optical band (BVRI).

\begin{figure*}[!ht]
\begin{center}
\includegraphics[width=0.4\textwidth, angle=270]{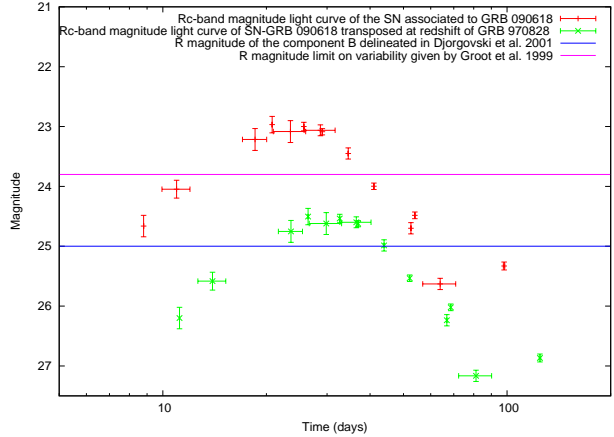}
\caption{The transposed Rc-band light-curve of GRB~090618 at the redshift of GRB~970828 (green) compared with Rc-band (red) light-curve of GRB~090618. The purple and blue lines represent the limit given in the deep images for GRB~970828 by \cite{groot1998} and \cite{djorgovski2001} respectively.}
\label{bump_090618}
\end{center}
\end{figure*}

There is also another way to figure out a possible bump and determine the time of the SN in the extrapolated light-curve of GRB without optical observation. To do it, I used the light-curve of GRB~090618, which shows an SN bump ten days after the explosion without a spectroscopic confirmation \cite{cano2011}. It is applied to GRB~970828 \cite{ruffini2013}. I transformed the Rc-band light-curve of GRB~090618 at the redshift of GRB~970828 following the same method. This light-curve is compared with the limits which were given by \cite{groot1998} (purple) and \cite{djorgovski2001} (blue) on Figure \ref{bump_090618}: the possible SN bump would have been visible from $\sim$ 20 to $\sim$ 40 days after the burst triggered by neglecting local absorption while optical observations were present for up to 7 days from the GRB triggers, reaching a R band magnitude limit $\sim$ 23.8 \cite{groot1998} and subsequent deeper images after $\sim$ 60 days \cite{djorgovski2001}.

\section{LLA GRBs without SNe in the sample}

LLA GRBs are weak enough during the afterglow phase (by definition) to allow the detection of any coincident SN. However, only 9 GRBs out of 31 LLA GRBs are firmly associated to an SN. Several reasons can be given to explain this discrepancy. First, optical follow up observations may be in sufficient during the time at which SN appear depend on the weather conditions, position of the bursts and the capacity of the instruments. Second, the contribution of the host galaxy can be important, and allow the signal of a weak supernova to be hidden. For instance, the host galaxy of GRB~090417B is quite bright. Additionally, GRB~061210 is in a galaxy cluster which even prevents the exact identification of its host. On the other hand, I showed that even for bright bursts, the afterglow is often decaying fast enough to allow for the SN identification.

The third argument is the distance. Among all GRB-SN pairs, the most distant one (GRB~021211) is at redshift $z=1.1$ \cite{dellaValle2003}, while the most distant pair (GRB~050525) for LLA GRB is at comparable redshift $z=0.606$. Evidence of GRB-SN associations should also be found at larger redshift. However, the spectroscopic identification of SNe becomes difficult at large redshift because they are faint. Additionally, optical contaminations from the host galaxy and the afterglow of GRBs reduce the signal-to-noise ratio. Thus, detection of SNe associated to GRBs are not expected to be successful above $z \sim 0.7$. Discarding all sources at larger redshift (10 LLA GRBs) and those that are associated to SNe leaves 12 LLA GRBs. Additionally, GRB~090417B (bright galaxy) and GRB~061210 which is in a galaxy cluster are removed to leave only 10 GRBs.

Interestingly, three very nearby bursts, GRB~060505 \cite{fynbo2006a}, GRB~060614 \cite{dellaValle2006} and GRB~100418A \cite{holland2010}, are not associated to SNe. They are in high metallicity host galaxies (\text{i.e.} evolved ones) and were already described as long ($T_{90}$) short GRBs, because no supernovae were found.

To conclude, many LLA GRBs are firmly associated to SN (64\% of all firm associations), the remaining being too far to give a firm conclusion. However, three nearby sources are not associated to SNe, despite their proximity, which questions their classification in the long GRB category, or questions the assumption that all long GRBs are associated to SNe. More observations of nearby LLA GRBs are required to give a conclusion on the GRB-SN association, and might even lead to the creation of a new class of events: nearby, low-luminous and not associated to a supernova.

\section{Host galaxies of LLA GRBs}

The host of long GRBs are generally low metallicity, low mass and high star formation rate galaxies \cite{levesque2010, graham2013, wang2014}. In this section, the differences between the host galaxies of LLA GRBs and those of normal lGRBs are studied.

\subsection{Mass and metallicity}

\begin{figure*}[!ht]
\begin{center}
\includegraphics[width=0.45\textwidth]{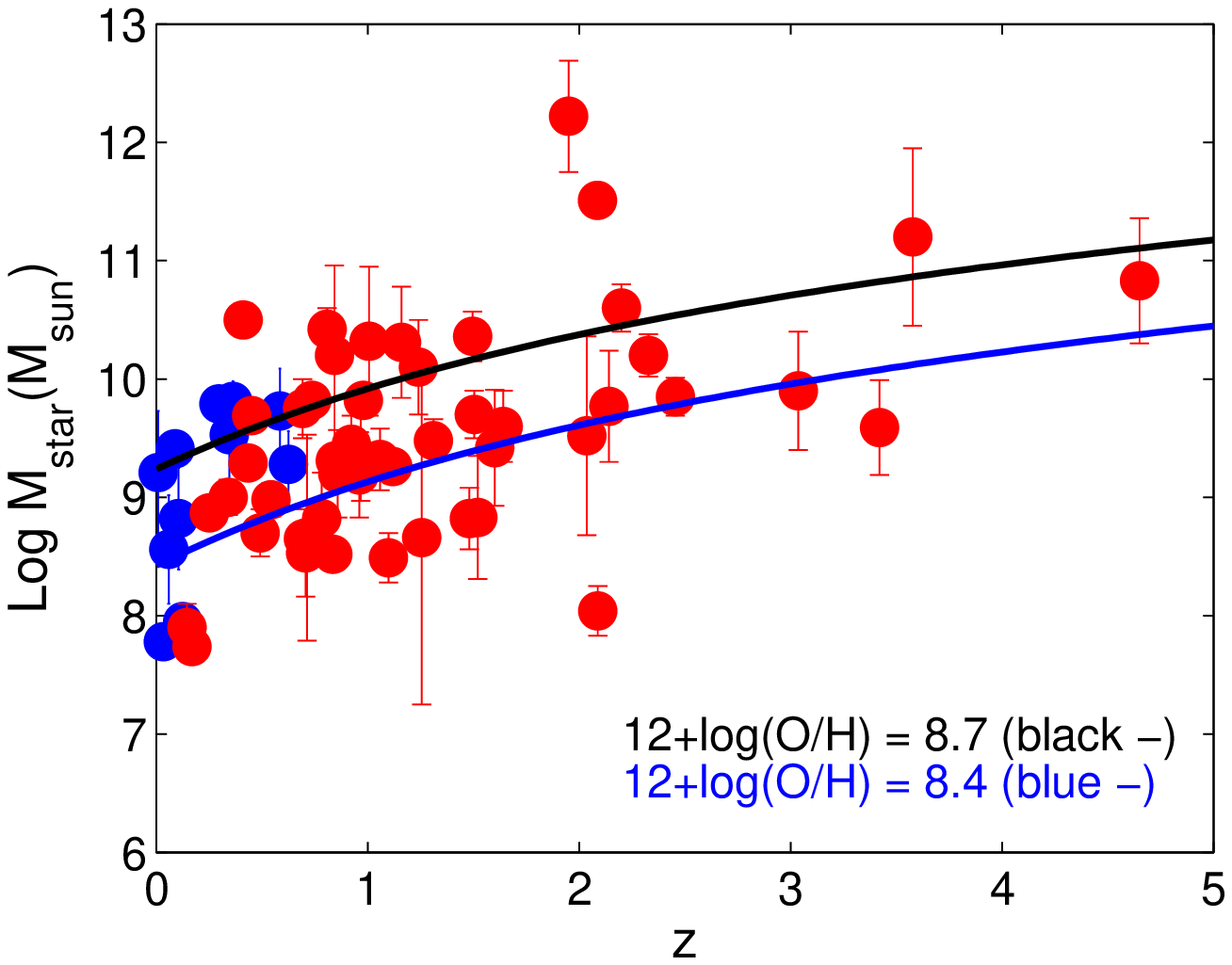}
\includegraphics[width=0.45\textwidth]{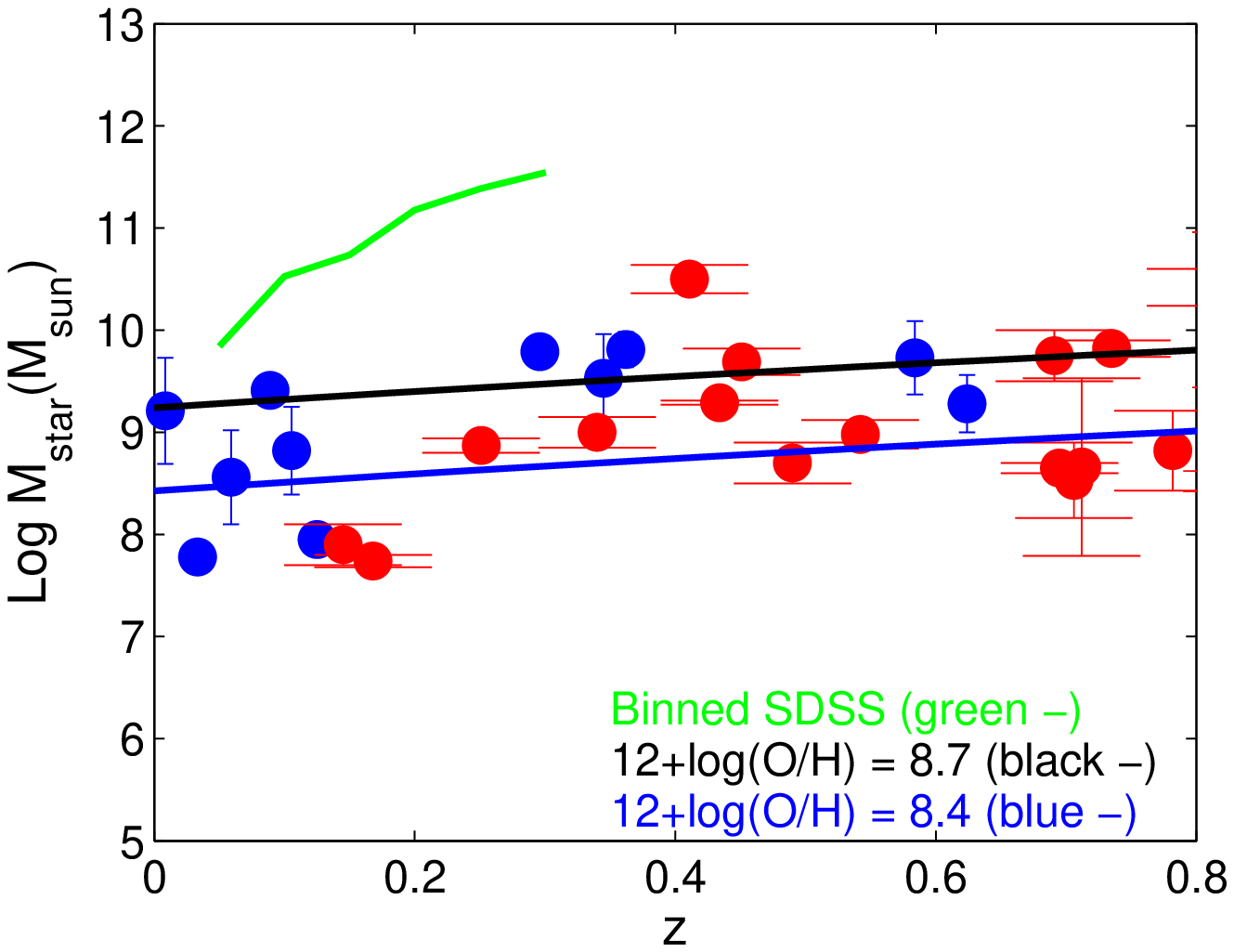}
\caption{Left: The mass-redshift distributions of LLA GRBs (blue filled circles) and lGRBs
(red filled circles) host galaxies. The solid lines represent the mean value of the
metallicity $12 + \log \text(O/H)_{\text{KK04}} = 8.7$ (black) of SDSS galaxies \cite{savaglio2005} and the mean value of the metallicity $12 + \log \text(O/H)_{\text{KK04}}  = 8.4$ (blue) of LLA GRB host. Right: Same mass-redshift distribution rescaled to redhifts from 0 to 0.8. The green solid line
represent the average binned mass of SDSS galaxies. 
 \label{host_mass_z}}
\end{center}
\end{figure*}

First, the mass and metallicity of host galaxies are studied. Figure \ref{host_mass_z} displays the position of the bursts (both LLA GRBs and normal lGRBs) in the mass-redshift plane. No large difference can be found between the two samples. Indeed, the mean value of the host masses is $10^{9.5\pm 0.32} M_{\odot}$ for normal lGRBs, while it is slightly smaller for LLA GRBs, the mean value being $10^{9.1\pm 0.27} M_{\odot}$. However, the masses of the host galaxies of GRBs are in average smaller than those of the galaxies in the Sloan Digital Sky Survey (SDSS) sample, whose average mass is represented by the green line on Figure \ref{host_mass_z} (right) \cite{wang2014}.

In order to discuss the metallicity, a relation between the redshift, the mass of the galaxy and the metallicity is given:

\begin{equation}
\label{mass_metallicity}
\begin{aligned}
 12 + \log \text(O/H)_{\text{KK04}}  =& - 7.5903 + 2.5315 \log M_{\star} - 0.09649 \log^2 M_{\star} & \\
 &+ 5.1733 \log \ t_H - 0.3944 \log^2 t_H& \\
&-0.4030 \log \ t_H \log \ M_{\star} \text{,}& \\
\end{aligned}
 \end{equation}
where KK04 represents the metallicity scale of Kobulnicky \& Kewley (2004) \cite{kobulnicky2004}, $t_H$ is the Hubble time at redshift $z$ in Gyr and $M_{\star}$ is the galactic stellar mass in unit of solar mass. The Hubble time at redshift $z$ is given by: 

\begin{equation}
t_H(z)=\frac{1}{H_0} \int_{z}^{\infty} \frac{dz\prime}{(1+z\prime)\sqrt{\Omega_m(1+z\prime)^3+\Omega_{\Lambda}}}.
\end{equation}
Cosmological parameters are considered, $\Omega_m$ = 0.27, $\Omega_{\Lambda}$ = 0.73 and $H_0$ =71 km s$^{-1}$ Mpc$^{-1}$ in order to be consistent with \cite{wang2014}.

The mean value of the metallicity is close to $12+ \log \text(O/H)_{\text{KK04}} = 8.7$ \cite{savaglio2005}, while the mean value for the host of LLA GRBs is $12+ \log \text(O/H)_{\text{KK04}} = 8.4$, to be compared to  $12+ \log \text(O/H)_{\text{KK04}} = 8.35$ for the host galaxies of normal lGRBs. Thus, no significant difference between LLA GRBs and normal lGRBs is found for the metallicity. 

However, only a small fraction of LLA GRBs (17\%) and of lGRB (30\%), are below the line representing a metallicity $12+ \log \text(O/H)_{\text{KK04}} = 8.4$ through Equation \ref{mass_metallicity}, obtained by fitting all SDSS galaxies. Interestingly, it implies that the host galaxies of GRBs have larger masses than those in the SDSS survey at a constant metallicity level, and that the normal lGRB host masses are slightly larger than the LLA GRB host masses at a constant metallicity level.

It was proposed that the progenitor of long GRBs be massive metal-poor stars, which are rare in massive galaxies. The creation of such stars requires particular conditions, and especially low metal abundance in the galaxy (or at least in the star formation region).

\subsection{Metallicity and brightness}

\begin{figure*}[!ht]
\begin{center}
\includegraphics[width=0.7\textwidth]{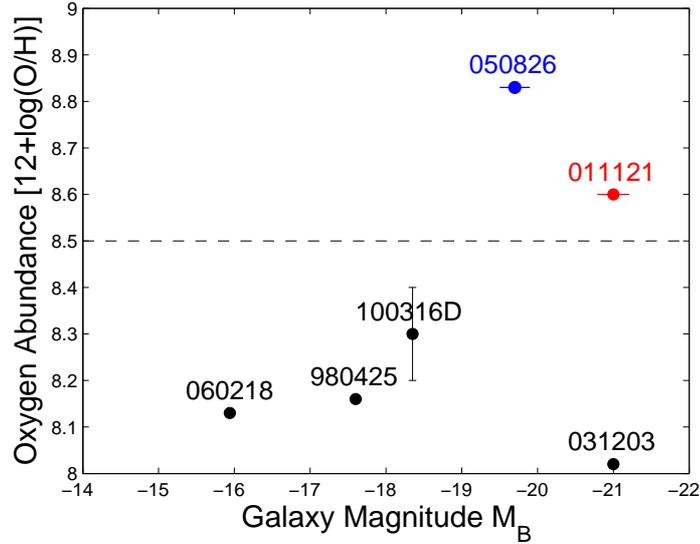}
\caption{The distribution of metallicity-B magnitude LLA GRBs. The barred line present the cut-off limit of metallicity at $12 + \log \text{(O/H)}_{\text{KD02}} = 8.5$ \cite{modjaz2008a}. GRBs associated to broad-line Ic SNe are presented by black points while GRB~011121 which is associated to type IIn SN~2001ke and GRB~050826 (no association to SN) are presented by red and blue points respectively. \label{host_mag_metallicity}}
\end{center}
\end{figure*}

Obtaining the magnitude of the host galaxies of GRBs in the optical band is extremely challenging. Thus, the information is available only for a few bursts in my sample. Figure \ref{host_mag_metallicity} represents the metallicity of the host galaxy as a function of its magnitude for LLA GRBs.

Additionally, all bursts represented by the black dots are associated to broad-line Ic SN. A cut-off in metallicity was obtained to differentiate the SNe Ic associated to GRBs from those that are not $12+ \log \text{(O/H)}_{\text{KK04}} = 8.5$ \cite{modjaz2008a}. Below this metallicity, all SNe are associated to a GRB, while above none of them is seen with a GRB. Interestingly in my sample, one burst (GRB~011121) is associated to SN and its host metallicity is larger than $12+ \log \text{(O/H)}_{\text{KK04}} = 8.5$. However, this SN is of type IIn (hydrogen is found in its spectrum), which indicates a smaller initial main sequence star, around 20~$M_{\odot}$. Moreover, GRB~011121 is the most luminous LLA GRB. Thus, the energy budget required to produce both the SN and the GRB might not be fulfilled from a small star, indicating that the progenitor might be in a binary system, with the GRB and the SN produced by two different objects.

In addition, GRB~050826 is also above $12+ \log \text{(O/H)}_{\text{KK04}} = 8.5$, however it is not associated to an SN, due to the lack of observations. 

Finally, I did not compare the hosts galaxy properties LLA GRBs with those of normal lGRBs, as the information required is difficult to find. This is one of the open question that I wish to continue for a future project.

\subsection{Star formation rate and metallicity}

\begin{figure*}[!ht]
\begin{center}
\includegraphics[width=0.7\textwidth]{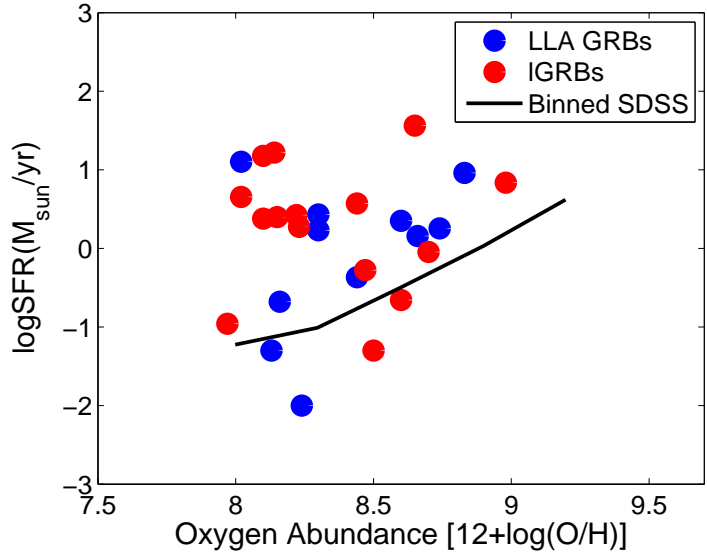}
\caption{The SFR-metallicity distribution of LLA GRBs (blue circles) and lGRBs (red circles) host galaxies. The black solid line represents the binned metallicity of SDSS galaxies in metallicity range \cite{modjaz2008a}.\label{host_metallicity_SFR}}
\end{center}
\end{figure*}

Figure \ref{host_metallicity_SFR} displays the star formation rate (SFR) against the metallicity for the hosts of normal and LLA GRBs. The mean values for the star formation rate are  $2.94 M_{\odot}.\text{yr}^{-1}$ for LLA GRBs and $6.29 M_{\odot}.\text{yr}^{-1}$ for normal lGRBs \cite{wang2014}. The reason for this difference is that since less stars are formed as a whole, less very massive stars are formed as well. Thus, the possible progenitor of a GRB would be less massive, and as a result the GRB less energetic.

It is also clearly seen that the SFR of GRBs hosts is larger compared to that of SDSS galaxies, as found in a previous study \cite{wang2014}.

\section{Rate density of LLA GRBs}

\subsection{Observed rate density}
\label{subsec:const_rate}
Within 31 LLA GRBs, 28 GRBs were observed by the \textit{Swift} satellite. The three remaining bursts, GRB~980425, GRB~011121 and GRB~031203, respectively observed by BATSE, INTEGRAL and Ulysses, before the \textit{Swift} launch, are excluded from this study since each instrument has its own properties: flux threshold, redshift measurement efficiency, operation time of satellites, etc.

Except one source GRB~101225A, called the ``Christmas burst", all other sources have a firm measured peak flux, see column 3 of Table \ref{table_rate_density}. I excluded GRB~101225A of this computation even if it has a peak flux measurement after 1372 s \cite{thone2011}.

The observed rate density of one event is given by \cite{guetta2007}:

\begin{equation}
\label{Eq:rate_density_obs}
R_{GRB, obs} = \frac{1}{V_{max}} \frac{1}{S_{cov}} \frac{1}{T} \text{,}
\end{equation}
where $S_{cov}$ is the fractional sky coverage (0.17 for the \textit{Swift} satellite), $T$ is the time during which the satellite was observing. Here, it is the difference between the time of the first observation by the \textit{Swift} and the time of the latest burst in the sample, \textit{i.e.} 8~years.
However, when I take into account the times when the Swift is not observing because of the South Atlantic Anomaly ($\sim$ \% 10 of all observations), satellite slewing and technical problems ($\sim$ \%10 of all observations) this time is 6.4 years. So the telescope is actually performing the \% 80 of all observations \cite{lien2014}.

 Finally, $z_{max}$ is the maximum redshift such that the fluence of a burst would be above the threshold detectability. Correspondingly, $V_{max}$ is the maximum volume, is given by \cite{coward2012}: 

\begin{equation}
\label{eq.v_max}
V_{max} = \int_0^{z_{max}} \frac{dV}{dz} dz \text{,}
\end{equation}
where $dV/dz$ is a volume element factor, which is given by \cite{howell2013}:

\begin{equation}
\label{eq.dv/dz}
\frac{dV}{dz} = \frac{4\pi c}{H_0} \frac{dL^2(z)}{(1+z)^2 \ h(z)} \text{,}
\end{equation}
where $d_L(z)$ is the luminosity distance, calculated by assuming a flat-$\Lambda$ universe ($\Omega_k = 1- \Omega_M - \Omega_{\Lambda}$) with $H_0$ = 72 km.s$^{-1}$.Mpc$^{-1}$, $\Omega_M = 0.3$ and $\Omega_{\Lambda} = 0.7$. For this precise computation, the fit given by Equation \ref{fit_luminosity_distance} is considered as It prevents the computing of a double integral.


The $h(z)$ factor appearing in Equation \ref{eq.dv/dz} is the normalized Hubble parameter:

\begin{equation}
h(z) = H(z)/H_0 = [\Omega_M(1+z)^3 + \Omega_{\Lambda}]^{1/2} \text{.}
\end{equation}

The maximum redshift $z_{max}$ appearing in Equation \ref{eq.v_max}, corresponds to the maximum luminosity distance $d_{max}$ such that the burst would have been detected by satellites. $z_{max}$ is calculated using the luminosity distance given by Equation \ref{luminosity_distance} and $d_{max}$ is given by \cite{coward2012}: 
\begin{equation}
d_{max} = \sqrt{\frac{F_p \ k(z)}{F_{Lim} \ k(z_{Lim})}} \ d_L(z) \text{,}
\end{equation}
where $F_p$ is the observed peak flux (given in the third column of Table \ref{table_rate_density} for each burst), $k(z)$ and $k(z_{Lim})$ the k corrections for a burst at the measured and maximum limiting distances respectively. $F_{Lim}$ = 0.4 ph.s$^{-1}$.cm$^{-2}$ is the theoretical \textit{Swift}/BAT sensitivity.

I used the largest redshift value in my sample ($z = 1.38$ for GRB~050801 and detection peak flux is 1.46 $\pm$0.14 ph.s$^{-1}$.cm$^{-2}$) for  $z_{Lim}$ because of the large uncertainty on the luminosity function of the sample. Actually, $z_{max}$ should be consistently computed by finding the solution to: 
 
 \begin{equation}
d_{max}(z_{max}) = \sqrt{\frac{F_p \ k(z)}{F_{Lim} \ k(z_{max})}} \ d_L(z) \text{.}
\end{equation}

However, the computation is done using $z_{Lim}$, since the error is expected to be small in any case. The first reason is that if $z_{max}$ was larger than $z_{Lim}$, additional bursts should be observed. The second reason is that since LLA GRBs are intrinsically sub-luminous and they are close to the instrumental threshold, it would be difficult to observe them at a distance much larger than $z_{Lim}$. 

The k correction, which accounts for the downshift of $\gamma$-ray energy from the burst to the observer's reference frame, is computed using Equation \ref{k_corr}, considering $[e_1, e_2]$ is 15 - 150 keV, which is the sensitivity band of \textit{Swift}/BAT. Additionally, $N(E)$, the source photon spectrum is assumed to be given by the Band function, see Equation \ref{Eq.band}, using the average GRB spectral indexes $\alpha = -1$, $\beta = -2.3$ and a rest frame power-law break energy at 511 keV.

The rate density of each burst is given in column 5 of Table \ref{table_rate_density}. The total observed rate density, obtained by summation of each single rate, is 0.37~Gpc$^{-3}$.yr$^{-1}$. It does not take into account the two sources GRB~060218 (63.6~Gpc$^{-3}$.yr$^{-1}$) and GRB~100316D (43.7~Gpc$^{-3}$.yr$^{-1}$), since their peak flux is below the instrumental threshold as seen in the lower part of Table \ref{table_rate_density}. On the other hand, they have compatible values with the result of  \cite{howell2013}, when assuming $F_{Lim}$ to be 0.15 ph.s$^{-1}$.cm$^{-2}$. Moreover, from the summation of each GRB, two more GRBs are excluded: GRB~051109B (1.69~Gpc$^{-3}$.yr$^{-1}$) and GRB~060505 (0.17~Gpc$^{-3}$.yr$^{-1}$) because of their huge rate density which are higher than all others. The reason can be their small redshifts.

\begin{table*}[!ht]
\centering
  \caption{The rate density results of LLA GRBs sample. GRBs with their corresponding redshift (1, 2 respectively), 
photon peak Flux between $15 - 150$~keV: BAT 1 - sec Peak (http://swift.gsfc.nasa.gov/archive/grb\_table/) (3), isotropic luminosity (4), observed GRB rate (5), and intrinsic GRB rate (beaming angle taken into account) (6). \label{table_rate_density}}
  {\scriptsize
  \begin{tabular}{lllllllllllll}
  \hline
GRB & $z$ & Peak flux & $L_{iso}$ & $R_{GRB,obs}$ & $R_{GRB,int}$ & \\
name & & ph.s$^{-1}$.cm$^{-2}$ & ergs.s$^{-1}$ & Gpc$^{-3}$.yr$^{-1}$ & Gpc$^{-3}$.yr$^{-1}$&\\ 
\hline
GRB~050126 & 1.29   & 0.71$\pm$0.17& 1.7$\times10^{51}$ & 0.002379  & 2.143  & \\
GRB~050223 & 0.5915 & 0.69$\pm$0.16 & 8.8$\times10^{50}$  & 0.00975325  & 8.78575  & \\
GRB~050525 & 0.606  & 41.70$\pm$0.94& 4.2$\times10^{51}$  & 0.0005637 & 0.5077875 & \\
GRB~050801 & 1.38   & 1.46$\pm$0.14 & 6.2$\times10^{50}$  &0.00132513  & 1.1936875 & \\
GRB~050826 & 0.297  & 0.38$\pm$0.13 & 4.9$\times10^{49}$  &0.0858775   & 77.35875  & \\
GRB~051006 & 1.059  & 1.62$\pm$0.30 & 1.5$\times10^{51}$  & 0.00178275  & 1.605875  & \\
GRB~051109B& 0.08   & 0.55$\pm$0.13 & $\cdots$            &1.693375     & 1525.375  & \\
GRB~051117B& 0.481  & 0.49$\pm$0.14 &6.4$\times10^{49}$  & 0.02108125   & 18.99  & \\
GRB~060505 & 0.089  & 2.65$\pm$0.63 & 8.2$\times10^{48}$  & 0.1696875    & 152.8625  & \\
GRB~060614 & 0.125  & 11.50$\pm$0.74& 2.2$\times10^{49}$  & 0.01536125   & 13.8375   & \\
GRB~060912 & 0.937  & 8.58$\pm$0.44 & 4.3$\times10^{51}$  &0.00077015 & 0.69375   & \\
GRB~061021 & 0.3463 & 6.11$\pm$027  & $\cdots$            & 0.00393413  & 3.543875  & \\
GRB~061110 & 0.758  & 0.63$\pm$0.12 & 2.8$\times10^{50}$  & 0.00652538  & 5.878  & \\
GRB~061210 & 0.4095 & 5.31$\pm$0.47 &  3.6$\times10^{49}$  & 0.00330775  & 2.979625  & \\
GRB~070419 & 0.97   & 0.20$\pm$0.10 & 6.6$\times10^{49}$  &0.0115675   & 10.42   & \\
GRB~071112C& 0.823  & 8$\pm$1       & $\cdots$            & 0.00092668 & 0.83475  & \\
GRB~081007 & 0.5295 & 2.6$\pm$0.4   &3.0$\times10^{50}$   &0.00377013  & 3.396125  & \\
GRB~090417B& 0.345  & 0.3$\pm$0.1   & 1.2$\times10^{49}$  & 0.0783825   & 70.6075  & \\
GRB~090814 & 0.696  & 0.6$\pm$0.2   & 8.4$\times10^{49}$  &0.00803413  & 7.237125  & \\
GRB~100418 & 0.6235 & 1.0$\pm$0.2   & 2.4$\times10^{50}$  & 0.00628113  & 5.658  & \\
GRB~110106B& 0.618  & 2.1$\pm$0.3 & 4.8$\times10^{50}$   & 0.0034325   & 3.092  & \\
GRB~120422A & 0.283  & 0.6$\pm$0.2   & 4.4$\times10^{48}$  & 0.0574575   & 51.75875  & \\ 
GRB~120714B& 0.3984 & 0.4$\pm$0.1  & 7.0$\times10^{48}$  & 0.04009   & 36.1125  & \\
GRB~120722A& 0.9586 & 1.0$\pm$0.3   & 4.0$\times10^{50}$  & 0.00294663  & 2.65425  & \\
GRB~120729A& 0.8    & 2.9$\pm$0.2   & 1.3$\times10^{50}$  & 0.0018005  & 1.621875  & \\
\hline
\hline
GRB~060218 & 0.0331 & 0.25$\pm$0.11 & 2.6$\times10^{46}$  & 63.64375     & 45864   & \\
GRB~100316D& 0.059  & 0.1$\pm$0.0 & 5.6$\times10^{46}$  & 43.73875     & 39400   & \\
\hline
\end{tabular}
}
\end{table*}

\subsection{Estimation of upper limit density}

The intrinsic rate density is obtained by multiplying Equation \ref{Eq:rate_density_obs} with a term which takes into account that if GRBs are collimated, the number of observed bursts is reduced as some of them will not have their jet aligned with the detection. Then, Equation \ref{Eq:rate_density_obs} becomes:

\begin{equation}
\label{Eq:rate_density_int}
R_{GRB,int} = \frac{1}{V_{max}} \frac{1}{S_{cov}} \frac{1}{T} \ B(\theta) \text{,}
\end{equation}
where $B(\theta) = [1- \text{cos}(\theta)]^{-1}$ and $\theta$ is the beaming angle. In the LLA GRB sample, two sources have been shown to be collimated in jet, with a corresponding beaming angle 2.7$^{\circ}$ for GRB~120729A and 6.3$^{\circ}$ for GRB~060614. The smallest value (2.7$^{\circ}$) is used in the computation of $B(\theta)$ to give an upper limit to $\theta$ for all bursts with an unknown beaming angle.

The results are presented in the last column of the Table \ref{table_rate_density}. The total intrinsic rate density obtained is 330.9~Gpc$^{-3}$.yr$^{-1}$ following the same way as in subsection \ref{subsec:const_rate}.

\subsection{Redshift effect on the rate density}
In order to consider all bursts without redshift, I should take into account the efficiency ($\eta_z$) of obtained redshift, as applied by \cite{howell2013}. So the Equation \ref{Eq:rate_density_obs} becomes:

\begin{equation}
\label{Eq:rate_density_RGRB}
R_{GRB} = \frac{1}{V_{max}} \frac{1}{S_{cov}} \frac{1}{T} \frac{1}{\eta_z} \text{,}
\end{equation}
where $\eta_z$ is 0.3 for lGRBs observed by the \textit{Swift} satellite. The total observed rate density is 1.23~Gpc$^{-3}$.yr$^{-1}$. When the beaming of GRBs outflows is taken into account, the total intrinsic rate density is 1103.04~Gpc$^{-3}$.yr$^{-1}$. However, the value of $\eta_z$ = 0.3 is biased towards low-luminosity bursts. Indeed, when the optical instruments are fast enough and the sources are luminous enough, the redshift measurements are highly efficient \cite{coward2009}. Conversely, if the sources are less luminous, such efficiency decreases, as it is difficult to get a spectrum of quality. As the bursts considered here are intrinsically less luminous (both during the prompt and the afterglow phase), it is expected that the redshift measurements be on average less efficient than for all other lGRBs. 

\subsection{Evolution of the rate density with the redshift}
In the previous sections, I assumed that the rate density is constant with redshift. However, it is unlikely to be the case at high redshift. One way to deal with the problem would be to consider the observed redshift distribution of GRBs but it is known to be biased since GRBs are thought to be linked to the death of massive star. Another solution is to the consider star formation rate and to re-normalize it. In this case $R_{GRB}(z)$ is given by: 

\begin{align}
\begin{split}
R_{GRB}(z) = \rho_0 \frac{R_{SF}(z)}{R_{SF}(z=0)} \text{,} \\
R_{SF}(z) = \frac{0.02 + 0.12z}{1 + (z/3.23)^{4.66}} \text{,} \\
\end{split}
\end{align}
where $R_{SF}(z=0) = 0.02$ is the local star formation rate, $\rho_0$ is the re-normalization constant, interpreted as the local rate density of GRBs. 

For normal lGRBs, it was found by \cite{howell2013} that $\rho_0$ = 0.09 $\pm$ 0.01~Gpc$^{-3}$.yr$^{-1}$ obtained by the best fit of the differential peak flux distribution of \textit{Swift}. Indeed, taking into account the star formation rate, the number of bursts per unit of time, redshift and luminosity, is given by \cite{howell2013}: 

\begin{equation}
\frac{dN}{dt \ dz \ d_L} = \frac{dV(z)}{dz} \ \frac{R_{GRB}(z)}{1+z} \ dz \ \phi(L)
\end{equation}
where $\phi(L)$ is the luminosity distribution function. Integrating over redshift and luminosity gives the rate at which bursts are observed normalized by $\rho_0$, which is found by a direct comparison to the observed value. As the number of sources considered in  \cite{howell2013} is a few hundreds, the authors could self-consistently determine $\rho_0$ and the luminosity distribution function of lGRBs.

However, as the LLA GRBs sample only contains 31 bursts, the computation cannot be reproduced since the luminosity distribution function of LLA GRBs cannot be constrained. 

\subsection{Summary}
The total intrinsic and observational rate density of the LLA GRBs are presented in Table \ref{table_total_rate_density}. The effect of opening angle (made more than 100 times) and redshift efficiency (made more than 3 times) are not negligible.

\begin{table}[!ht]
\centering
  \caption{The total rate densities in each conditions. The observed and intrinsic total rate density are computed by using the parameters of 23 LLA GRBs. They are presented on the first two columns; the last two show the results of computation made by taking $\eta_z$ into account.
   \label{table_total_rate_density}}
  \begin{tabular}{lllllllllllll}
  \hline
 $R_{GRB, obs}$ & $R_{GRB, int}$ & $R_{GRB, obs}$  & $R_{GRB, int}$ &  & \\
 Gpc$^{-3}$.yr$^{-1}$& Gpc$^{-3}$.yr$^{-1}$ & Gpc$^{-3}$.yr$^{-1}$  & Gpc$^{-3}$.yr$^{-1}$ & &\\ 
\hline
  &   &  &   &  & \\
  0.37 & 330.9  & 1.23 & 1103.04  &  & \\
\hline
\end{tabular}
\end{table}

\section[Discussion: comparison of rate density]{Discussion: comparison between the rate of low-luminosity GRBs and LLA GRBs}
\chaptermark{Comparison of rate density}

The rate density of low-luminosity GRBs was already considered in many studies \cite{coward2005, guetta2007, virgili2009, howell2013}. These bursts are selected based on their faint prompt emission, while in my sample they are selected based on the luminosity of their afterglow. Most of these studies rely on GRB~980425, GBR~031203 and GRB~060218. However, the first two were not considered in my study as they were not observed by the \textit{Swift} satellite.

First, a study showed that the three GRBs were taken separately, form a different sub-class of low-energy close bursts \cite{coward2005}. There, the rate was found to be 220~Gpc$^{-3}$.yr$^{-1}$. As the bursts were connected to SNe~Ib/c, a comparison with the rate of these events showed that only 0.6\% of the type Ib/c~SNe might be associated to GRBs.

Second, an independent study \cite{guetta2007} based on the luminosity function of GRBs (they took into account bursts of luminosity $L<10^{49} \text{erg.s}^{-1})$ and on a bandwidth of [1~keV - 10$^3$~keV] (in order to compare bursts detected by different satellites) showed that the local rate of these three  GRBs is as high as $100-1800$~Gpc$^{-3}$.yr$^{-1}$, compatible with the result of \cite{coward2005}. Note that in this computation, the possible jet structure is not taken into account, and thus not corrected for. In my computation, when beaming is not taken into account, the rate density for GRB~060218 is equal to 50.91~Gpc$^{-3}$.yr$^{-1}$, to be compared to 380~Gpc$^{-3}$.yr$^{-1}$ in \cite{guetta2007}.

Such large differences between the results can be explained by the different energy range considered in their computation [1~keV - 10$^3$~keV]. The flux was transformed in this band by extrapolating the low-energy spectrum to high energies, which might not be adequate. Indeed the low-energy slope is extended up to 1~MeV, while the average peak energy of the spectrum is on average 214~keV \cite{ghirlanda2009}. Note that the rate of normal lGRBs was also estimated to be as high as 100~Gpc$^{-3}$.yr$^{-1}$ \cite{guetta2007}, which has to be compared to the rate found in a more recent study: 1~Gpc$^{-3}$.yr$^{-1}$ \cite{virgili2009}.

Finally, in a global study of low-luminosity GRBs and of normal lGRBs \cite{howell2013}, in which the luminosity function of GRBs is explicitly taken into account (see Equation \ref{eq.dv/dz}), the rate of low-luminosity GRBs is found to be in the order of 200, close to my results. However, their analysis consider only the extension of the distribution function of all GRBs up to low-luminosity, with an imposed cut-off at $10^{49} \text{erg.s}^{-1}$, in order to take into account only low-luminous bursts. In other words, the distribution function of low-luminosity GRBs is not fitted for.

To conclude, I found that the under-luminous and the LLA GRBs have comparable local rate densities, while being selected on different criteria (prompt versus afterglow luminosity). However, this is not surprising as the rate density in my computation is dominated by GRB~060218 and GRB~100316D. This is because these two bursts are extremely close ($z<0.1$) and their maximum flux is below the theoretical flux detection threshold. Interestingly, if I remove these bursts from the computation, I found that the local rate density of LLA GRBs is a factor two bigger than that of normal lGRB. 

\subsection{Comparison with Short GRBs rate}

Another comparison is performed between LLA GRBs and short bursts. This is important as some LLA GRBs are compatible with the short GRBs in the $E_p-E_{iso}$ plane (see section \ref{sec_prompt} of chapter \ref{Chapter4}). 

The rate of short GRBs was estimated to be  8$^{+5}_{-3}$~Gpc$^{-3}$.yr$^{-1}$ if the emission is not collimated in a jet, and 1100$^{+700}_{-470}$~Gpc$^{-3}$.yr$^{-1}$ if it is \cite{coward2012}. The estimation was performed using 8 short GRBs observed by the \textit{Swift} satellite. Interestingly, 5 GRBs in my sample are at the exact position of short bursts: GRB~050223, GRB~051117B, GRB~060505, GRB~120422A and GRB~031203. However, this last burst does not take into account the computation of the rate density LLA GRBs as it was not observed by the \text{Swift} satellite. Summing the rate density of these GRBs (without taking into account the collimation and GRB~031203) gives the rate density of these bursts: 0.25~Gpc$^{-3}$.yr$^{-1}$, which is much smaller (by more than one order of magnitude) than that of short GRBs.

Additionally, in the computation of the rate of short bursts, $F_{lim}$ was set to 1.5~counts.s$^{-1}$, while in mine, it was set to  $F_{lim}=0.4$~counts.s$^{-1}$. A larger flux limit reduces the maximum volume $V_{max}$ and thus increases the rate. However, looking at the peak flux of the four bursts under consideration shows that the 
peak flux is much below the experimental detection threshold of short bursts, which means that they would have not been detected if their duration was not long enough. Only GRB~060505 with a peak flux of 2.65~counts.s$^{-1}$ would have been detected. Performing the computation for this burst only gives a rate of 0.91~Gpc$^{-3}$.yr$^{-1}$, comparable to the average rate (1~Gpc$^{-3}$.yr$^{-1}$) of individual short bursts. Actually, this burst was already thought to be a short burst of duration slightly larger than 2~s ($T_{90}=4~s$) \cite{ofek2007}.

Thus, I can conclude that the LLA GRBs (except GRB~060505) which are at the position of short GRBs in the $E_p-E_{iso}$ plane cannot be considered as short bursts with a duration larger than 2~s, as their rate is too small compared to that of short GRBs. Concerning GRB~060505, its rate density is in agreement with a short burst of duration longer than 2~s.

\section{Discussion for possible progenitor}

LLA GRBs are found to be intrinsically different compared to normal lGRBs. They are closer as indicated by their redshift distribution, and their prompt emission is much less energetic. 

However, the rate density does not allow to give a firm conclusion. Indeed, it is clearly seen that LLA GRBs are separated in two groups: on the one hand GRB~060218 and GRB~100316D (to which GRB~980425 and GRB~031203 can be added \cite{coward2005}), whose rate density is much larger than the one of normal lGRBs, and on the other hands the other LLA GRBs whose rate density is similar to that of lGRBs.  If it is assumed that this distinction is not relevant, then it is possible to conclude that the rate density of LLA GRBs is much larger than that of normal lGRBs. Conversely, if such distinction is relevant and is an intrinsic property of these bursts, then no conclusion can be obtained. 

Finally, it has to be noted that no clear difference between LLA GRBs and normal lGRBs was found when studying their host galaxy. However, such study is delicate, as it is difficult to obtain optical measurements of the host galaxies of GRBs and reliable constraints on their mass, metallicity and star formation rate.

Since clear differences can be found between LLA GRBs and normal lGRBs for some of their properties, two situations can be discussed. First, the progenitor of LLA GRB is different from that of normal lGRB. In this case, the differences have to be explained either by considering different emission mechanisms or properties of the outflow. Conversely, the progenitor of LLA GRBs can be different than that of normal lGRBs.

\subsection{Same progenitor}

\subsubsection{Instrumental effect}

First, if I consider that the differences between LLA GRBs and normal lGRBs are only due to instrumental effects, I can say that the progenitor is assumed to be the same, as well as the emission mechanisms. It implies that the LLA GRBs are the low-luminosity part of the distribution of GRB luminosity.

First, it is found that LLA GRBs are observed at closer distances than normal lGRBs. This is not really surprising as their intrinsic faint luminosity does not allow LLA GRBs to be above the threshold detection of the X-ray detectors. Additionally, it is argued that when comparing all bursts at redshift smaller than one, the cumulative distribution functions are different. However, such difference can be explained by the fact that only bursts with known redshift are taken into consideration in this study. As the redshift is taken in the late afterglow phase of GRBs (after several hours), the probability of obtaining the redshift is biased towards low-luminosity sources. Indeed, the LLA GRBs are by definition low luminous during the afterglow phase. It means that the probability of successfully obtaining the redshift for a burst at intermediate redshift (a few tens) is biased towards intermediate redshift.

Additionally, the fact that some bursts in the sample do not fulfil the Amati correlation can be explained by the fact that the X-ray instruments are not designed to determine the prompt properties of very faint sources.

However, it is difficult to explain the rate density of LLA GRBs (at least for GRB~980425, GRB~031203, GRB~060218 and GRB~100316D) only with instrumental effects. Indeed, if the LLA GRBs are only the low-energy extremity of the distribution function of all GRBs, it is reasonable to expect that the rate density as a function of luminosity be continuous. This means that the rate density should not increase towards low luminosity. However, this is not the case when considering bursts of intermediate luminosity.

To conclude, it is difficult to explain all differences between LLA GRBs and normal lGRBs by instrumental effects only.

\subsubsection{Environment of bursts}

The burst environment can strongly impact the properties of a GRB. LLA GRBs, being weak events, might even be strongly affected. Firstly, I showed that the distributions of gas and dust in the host galaxy and in the Milky Way for LLA GRBs are not different from their counterparts for normal lGRBs. Thus, absorption and extinction do not favor the same progenitor assumption. Only one of the LLA GRBs (GRB~090417B) is a dark burst, which means that the soft X-rays are strongly absorbed: its hydrogen column density is the second largest of the whole sample. However, one burst cannot explain the statistical differences.

Secondly, by using the closure relation, constraints on the surrounding burst environment can be given. I showed that some LLA GRBs are compatible with a wind environment (for instance GRB~011121). However, no clear differences were found with normal lGRBs: most bursts are explained by a fireball expanding in the constant interstellar medium, while a few expand in a wind-type environment. Interestingly, it was found that two LLA GRBs are jetted, with an opening angle compatible to that of normal lGRBs.

Lastly, the host galaxies properties (type, metallicity, mass and SFR) of LLA GRBs are not different compared to that of normal lGRBs. To conclude, it is unlikely that all differences between LLA GRBs and normal lGRBs could be explained only by the properties of the environment.

\subsubsection{Geometry of bursts}

\begin{figure*}[!ht]
\begin{center}
\includegraphics[width=0.5\textwidth]{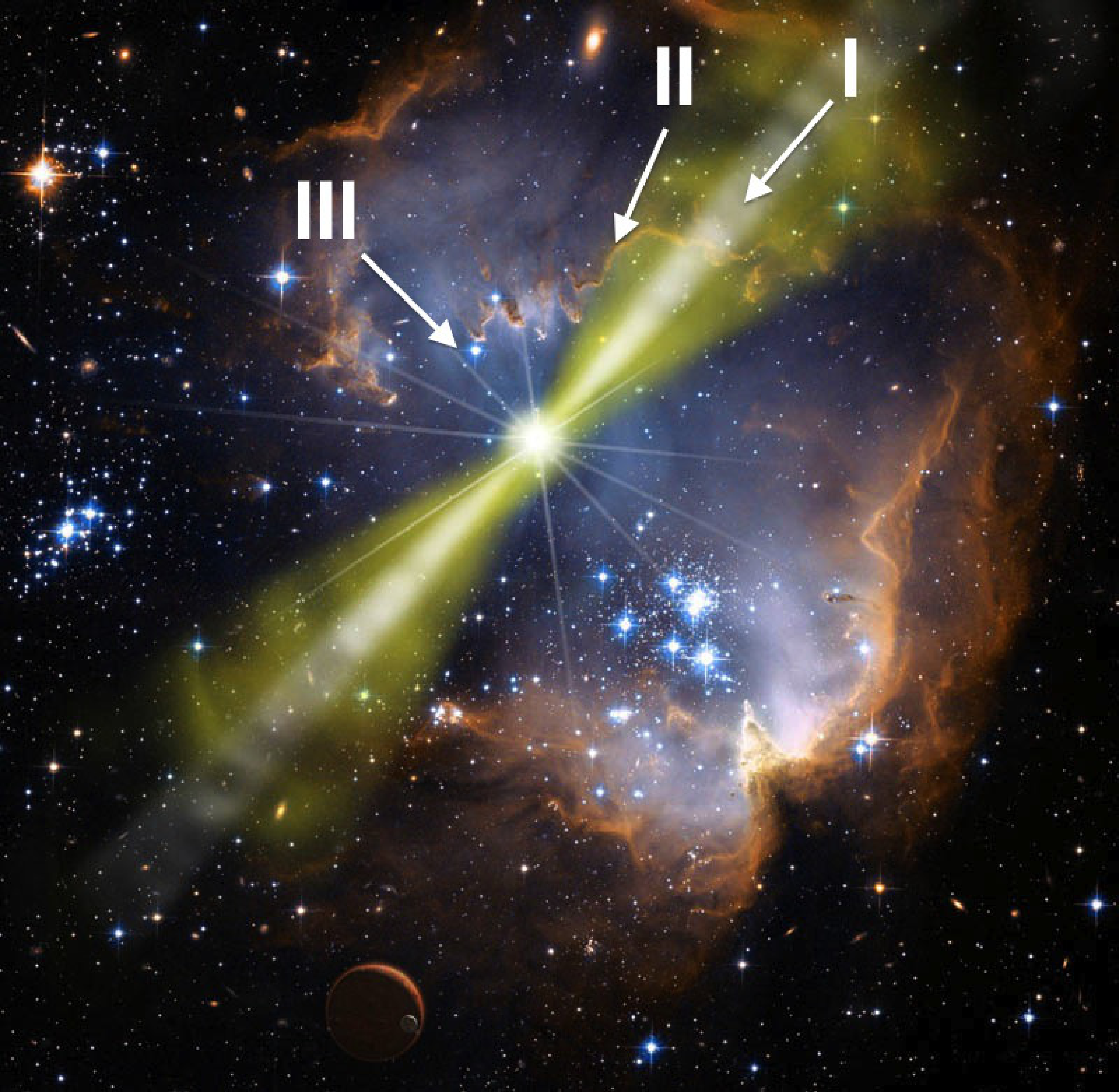}
\caption{Artistic view of a bright GRB occurring in a star-forming region. The possible positions of the observer: on the axis on which the jet beamed (I), on jet edge (II) and off-axis (III). Image Credit: NASA/GSFC. \label{on_off_axis}}
\end{center}
\end{figure*}

As attested by the identification of two jet breaks in the LLA GRBs, it is expected that the emission in GRB be collimated in a jet, mainly in order to reduce the energy budget. Such non-spherical geometry is displayed in Figure \ref{on_off_axis}. If the observer is in the emission cone (region labeled I in the figure), the effects of the jet are noticeable only during the late afterglow, and they manifest themselves by the jet break. However, if the observer is outside of the emission cone (region labeled III), the GRB is not observable. But the SN is visible as its explosion is almost spherically symmetric.

In the unlikely situation in which the observer is on the edge of the jet (region labeled II), the resulting observed GRB is expected to be weak, as the Lorentz factor is smaller at the edge than in the emission cone. However in this case, no clear jet break is expected, which is in contradiction with the two jet breaks observed. Additionally, the transition between the jet and the interstellar medium is expected to be sharp. Thus, the rate density of events seen on the edge of the jet is expected to be smaller than that of events seen on-axis. Interestingly, I found that the rate of LLA GRBs is larger than that of normal lGRBs, which shows that LLA GRBs are not even seen on the edge of a jet.

However, it is possible to consider stratified jets \cite{zhang2003}: in this model, the Lorentz factor of the burst is assumed to be a smooth function of the angle between the direction of the observer the central axis of the jet. In this model the transition from the central part of the jet to the ISM is smooth and the central part of the jet is often assumed to be narrow ($\theta_j \sim 1/ \Gamma$).

Interestingly in this model, a jet seen at angle larger than the jet opening angle is less luminous, as the Lorentz factor is smaller. Its peak energy is also expected to be lower, as the Lorentz boost is smaller. And finally, the rate density of GRBs seen at angles larger than $\theta_j$ is expected to be larger than that of GRBs seen on axis, since the jets are often assumed to be narrow. However, it is expected that the rate density should be a smoothly varying function of the luminosity, which is not the case, thus challenging the assumption that LLA GRBs are seen off-axis from a stratified jet.

\subsection{Content of Ejecta}

In the classical fireball model, the dynamics is mediated by the dimensionless entropy $\eta = L/ (\dot{M}c^2)$. In particular, the Lorentz factor of the outflow after the acceleration phase is equal to $\eta$. Interestingly, there is a correlation which links the luminosity to the Lorentz factor of the outflow $\Gamma_0 \simeq 249 \ L_{\text{iso},52}^{0.3}$ \cite{lu2012}. As an example, the coasting Lorentz factor of GRB~980425 deduced from the correlation is as low as $\eta = \Gamma_0 = 7.14$.

Additionally, there are many correlations between the afterglow luminosity, its kinetic energy, prompt luminosity and Lorentz factor. Assuming that these correlations are also valid for LLA GRBs (as their progenitor is assumed to be the same as the one of normal lGRBs), it is possible to show that the luminosity of the afterglow correlates to the Lorentz factor, so to the baryon load.

Indeed, the Lorentz factor correlates to the isotropic luminosity \cite{lu2012}:
\begin{align}
\Gamma_0 \simeq 249 L_{\text{iso},52}^{0.3},
\label{eq:corrgammaLiso}
\end{align}
which correlates to the X-ray luminosity of the afterglow at 1 day \cite{davanzo2012a}:
\begin{align}
L_{X,1\text{day}} \propto L_{\text{iso}}^{0.74}. 
\label{LX_Liso}
\end{align}

Combining these two equations leads to:
\begin{align}
\Gamma_0 \propto L_{X,1\text{day}}^{0.3}.
\end{align}

I have shown that LLA GRBs also have low-luminous prompt phase, which indicates that they fulfill (at least partially) the correlation given by Equation \ref{LX_Liso}. However, the high-baryon load argument does not allow to explain the sharp transition in the rate density between the LLA GRBs and the normal lGRBs. Actually, a smooth transition would be expected.

\subsection{Different progenitors}
Here, it is assumed that LLA GRBs and normal lGRBs originate from different progenitors by considering the result seen in the redshift distribution and $E_{p,i}$-$E_{iso}$ plane in Figure \ref{fig_amati_tot}. This could be the case as the rate of LLA GRBs is larger than that of normal lGRBs. It implies that the progenitor of LLA GRBs be more common than that of normal lGRBs or that the explosion mechanism has more chance of success.

It was shown by \cite{woosley1993} that the W-R stars able to produce a GRBs should fulfill very specific conditions: they should be fast rotating (needed to create the accretion torus around the black-hole, which powers the GRB), have a massive core (energy reservoir) and have lost their hydrogen envelop (to allow the jet to expands easily once it has drill its way through the core of the progenitor). If initially the star is single, only a few percent of O and B stars might have such properties once they are about to collapse. However, these conditions are more easily fulfilled if the progenitor is in a binary system. For instance, the outer hydrogen layers can be removed in the binary interaction after the common envelop stage.

In the case of LLA GRBs, the accompanying SN (when present) are of type Ib/c, which means that the progenitor has lost its hydrogen envelop. If it is further assumed that the LLA GRBs are \textit{intrinsically} less energetic (the energy put in the relativistic outflow is smaller) than normal lGRBs, the difference might be explain by a smaller core before collapse, which means less matter to be accreted by the black-hole during fall-back. This implies a progenitor with a smaller initial mass.

However, a lower mass progenitor would not necessarily lose all its hydrogen envelop. Such problem can be solved by considering that the progenitor is in a binary system. Actually, it is a reasonable assumption as 70 percent of massive stars are found in binary systems \cite{abt1990, mason1998}. The lower-mass progenitor is also favored by the rate density of LLA GRBs that is larger. Indeed, by using the Salpeter mass relation, the initial mass function of stars can be fitted as:
\begin{align}
\frac{dN}{dm} \propto  m^{-2.25}. 
\label{salpeter}
\end{align}
Applied to a $40M_{\odot}$ progenitor for a normal lGRB and a $15M_{\odot}$ progenitor for a LLA GRB gives a ratio in the number of progenitors equal to 9. The ratio of progenitors can even be increased when considering the other requirements to have an efficient collapsar.

However, if the progenitor of LLA GRBs is not in a binary system, it should be massive enough to lose all its hydrogen and a part of its helium envelops. Thus, the energy extracted during the accretion onto the black-hole and used to power the bursts is expected to be comparable in the case of normal lGRBs and LLA GRBs. Thus, LLA GRBs would not be \textit{intrinsically} low-luminous, the difference being obtained by the efficiency of the radiative process. The correlation given by Equation \ref{eq:corrgammaLiso}, indicates that the lower the luminosity of the burst emitted during the prompt phase, the lower the Lorentz factor, \textit{i.e.} the larger the baryon load. Assuming a constant energy conversion by the accretion, it means that the ratio of the core mass by that of the outer layers is larger in the case of LLA GRBs. Thus, it implies that the progenitor of LLA GRBs are initially more massive, than the progenitor of normal lGRBs. It could be a WO or WC W-R star.

However, a more massive progenitor is more difficult to obtain (see the Salpeter mass function Equation \ref{salpeter}), and thus the rate of LLA GRBs should be smaller than that of normal lGRBs. This disfavors the model of a single star for the progenitor of LLA GRBs. So, the purposed progenitor for LLA GRBs can be a low mass CO core star in a binary system.

%% file: Chapter6.tex

\chapter{Conclusion and Perspectives}

\section{Conclusion}
My thesis deals with the classification of GRBs and their tight connection to SNe. I presented evidence that the LLA GRBs are a subclass of long GRBs. In the global context, the present association of some long GRBs to SNe indicate that (some) long GRBs originate from the core collapse of a massive star. This idea motivated the study of SNe in my thesis.

By using the photometric and spectroscopic data of the type IIb supernova~2004ex during the transition $(30~\text{days} \le t \le60~\text{days})$ and nebular ($t \ge 60~\text{days}$ past explosion) phases obtained by the 1.82~m Ekar, 3.5~m TNG and 2.2~m WFI telescopes, the light-curve of SN 2004ex was completed. By fitting the light-curve with a third order polynomial function, I found that the peak magnitude is $-17.27$, reached at 21.5 days after the explosion.

The results were compared with the prototypes of type~IIb SNe:  SN~1993J and SN~2008ax respectively.  I found that the light-curve of SN~2004ex is very similar to the two prototypical light-curves. However, it is more similar to the one of SN~2008ax. On the other hand, its helium lines are still well-detected 4 weeks after the explosion. And the hydrogen (H$_{\alpha}$) line at 6286~$\angstrom$ is weaker than the one found for SN~1993J at a similar period. I confirmed that SN~2004ex is of type IIb with the rapidly declining light-curve.

By using the similarities to those prototypes, I assumed that the mass of the $^{56}\text{Ni}$ can be in the order of $ 0.07 - 0.15 M_{\odot}$, while the mass of the ejecta is $2 - 5 M_{\odot}$ and its kinetic energy can be $1-6 \times 10^{52} \text{erg}$. The velocity calculation of the H$_\alpha$ line (14000 km/s) has shown that the value is in the expected range for type IIb SNe. 

An additional comparison was performed between type IIb SN~2004ex and types Ib and Ic SNe to find similar properties. This idea came from the fact that SNe might form a continuous sequence being explained mainly by the envelope mass. A similar behavior is seen in the light-curves of type~IIb and type~Ib SNe.

The detailed classification of SNe and the increased number of observed GRBs motivated the research of a more precise classification of GRBs. Indeed, GRBs are poorly classified into two groups (short and long bursts) based on their duration. Thus, starting from the separated groups of long GRBs based on the X-ray afterglow luminosity \cite{gendre2008}, I increased the number of faint sources by presenting and discussing the selection method of LLA GRBs. To prepare a list of bursts with known redshift observed before February 15$\textsuperscript{th}$, 2013, I removed the short bursts from the sample, considering $T_{90} < 2$~s in general. This new list consists of 254 GRBs. Then the different corrections (energy and cosmological) were presented in details. In addition, I have precisely defined how only the late afterglow was selected from the light-curves of GRBs. And finally, I showed how the LLA GRBs were selected from the remaining sample of bursts, by considering the faintness of their afterglow. The final sample consists of the 12\% faintest X-ray afterglows from the total population of long GRBs with known redshift. In addition, all LLA GRB afterglows were re-analysed manually. Their spectral and temporal characteristics were determined precisely and these results are presented in Table \ref{table_sample}.

By making the statistical analysis of LLA GRBs both for their prompt and afterglow properties, I found that they are on average closer and less energetic than normal lGRBs, possibly some evidences for a different progenitor. I also showed that this sample is not more affected from the dust and gas compare to normal lGRBs.

I studied the environment of the progenitor using the closure relations both in X-ray and optical bands. Most of them can be described by a fireball in the slow-cooling regime expanding in the constant ISM. Some sources (\textit{e.g.} GRB~011121) can be described by a fireball in the slow-cooling regime expanding in a stellar wind environment which is expected in the case of W-R star progenitor. The few outliers (GRB~060614, GRB~120729A) can be accounted for by an early jet break. These two sources show achromatic jet breaks (seen in X-ray, optical, radio bands). This achromatic break is interpreted as the result of a jet. The time of the break can be used to measure the opening angle of the jet, which give information about the efficiency of a matter collimation.

I showed evidence with the Amati correlation that LLA GRBs are also intrinsically fainter during their prompt phase. Actually, some events do not follow at all the $E_{\rm peak}$ - $E_{\rm iso}$ correlation. This can be explained either by the viewing angle or the energy budget of the event. I thus concluded that LLA GRBs are intrinsically different from normal lGRBs.

Additional properties of LLA GRBs were also considered. I firstly noted that LLA GRBs include many of the supernovae associated with GRBs (64\%), including the best known associations, GRB~980425, GRB~031203, GRB~060218. This means that the conclusion drawn on the general GRB-SN association is based on a sub-sample of the LLA GRBs which might not be representative of such association. I stressed the need to confirm this point and the previous work on GRB-SN associations using different spectral and light-curve templates. I further gave reasons to explain why some LLA GRBs are not associated to SNe pointing the brightness of their host or the large redshift $z>0.6$ making the observation difficult.

Secondly, the host galaxy properties of LLA GRBs are discussed. In the general contents of GRBs, the host properties are an additional argument to differentiate between long bursts (whose hosts are associated to young, low metallicity and actively star forming galaxies) and short bursts (whose hosts are associated to elliptical, low or no star forming rate, metal poor galaxies). Here, in the case of LLA GRBs, I found that the difference with the host properties of normal lGRBs are small. However, since optical constraints on the host galaxy of GRBs are difficult to obtain, this reduces the sample to only very few LLA GRBs. With this result, GRBs associated to the broad line Ic SNe show lower metallicity than the other type of SN association (\textit{i.e.} GRB~011121) and without SNe associations (\textit{i.e.} GRB~050826). Thus, more observations of the host would give tighter constraints and allow to conclude on the differences (if any) between the host galaxies of  LLA GRBs and normal lGRBs.

Finally, I computed the rate density of LLA GRBs and it is found to be much larger than that of normal lGRBs. However, in the sample some of events have a very high rate density. Even removing them I found that the rate of LLA GRBs is still two times larger than that of long GRBs. This can be used as an argument to differentiate these two populations.

As a conclusion, I briefly discussed the possible progenitors of LLA GRBs, trying to explain the observational differences either by environmental or instrumental effects (which can not fully explain them) or by an intrinsically different progenitor. By using the rate of LLA GRBs, I indicated that the initial mass function of stars points towards lower mass progenitor for LLA GRBs. However, as their accompanying SN is of type I, they should be in a binary system in order to have lost their hydrogen envelope during the common envelope phase.

In general conclusion, this LLA GRBs sample is a first step towards a more precise classification of GRBs, taking into account the luminosity of the afterglow in addition to the duration.

~

\section{Perspectives}

In my thesis, I investigated if there is a distinct subclass of low-energy GRBs. However, there are many perspectives for my work: multi-wavelengths comparison between LLA GRBs and normal lGRBs, additional constraints from the host galaxies with more observations, more precise determination of the rate density, etc.

Firstly, since the spectra of the afterglow in the X-ray band are peculiar, additional differences may also be found at other wavelengths. During next years, I will proceed on following ways by selecting the sources with optical data and/or radio data, then by performing a statistical study of the LLA GRB properties at such wavelengths and at the end I might assess the consistency of the results with the constraints imposed by the X-ray emission in the context of the fireball model: slow-cooled fireball expanding in a constant interstellar medium or wind. This can help to constrain the properties of progenitor if it is a very massive star which can blow off its outer layers with a wind or if it is a binary system. This comparisons will be more important when early and late follow-up multi-wavelength observations of the afterglow emission will increase and improve with new missions.

I additionally showed that the prompt emission properties are different. Actually, the Amati correlation is based on the band function which is not physically motivated. For instance, in recent years, a model with three components (Band, black-body and high-energy power-law) was used to fit the spectra of GRBs, and one may wonder if the properties of LLA GRBs remain incompatible with that of normal lGRBs. Using the unique capacity of \text{Swift} which enables observations between the prompt and the afterglow phases, I might find some clues in this particular time which could allow to further constrain the LLA GRBs. Thus, not only the study of the afterglow but also the study of the prompt emission proved to be of high interest in constraining LLA GRBs.

Secondly, I completed the computation of the rate density of LLA GRBs, finding that it is consistent with that of low-luminosity GRBs. However, the computation assumed that the rate is constant over redshifts, which is unlikely. A more precise computation can be done by considering that the rate density of LLA GRBs follow the star formation rate history. However, additional difficulties appear as the luminosity distribution function of LLA GRBs should also be constrained to perform this computation. It is possible to constrain the luminosity distribution, together with the rate density of LLA GRBs by Monte Carlo simulation, as done in previous works for all long GRBs \cite{virgili2009}. 
The rate density calculation is important because it can first precise how many events per unite time unit volume are expected, second show if there are different kinds of GRBs, and third be compared to the rate of other extreme  events ($\textit{e.g.}$ supernovae). Additionally, it can be used to constrain the population type of the progenitor \cite{bromm2006}. However, since LLA GRBs are at low redshift, their progenitors are expected to be population I stars.

Thirdly, many LLA GRBs are associated to SNe Ib/c, including the three prototypes of the GRB-SNe association (GRB 980425, GRB 031203, and GRB 060218), and one may wonder if it is a global property of LLA GRBs. From the previous comparison, it is known that GRBs are more concentrated in the brightest regions of their host galaxies than are core collapse SNe, suggesting that GRBs are formed from the most massive stars which formerly arose from metal-poor environments. Moreover, GRBs host galaxies are fainter and smaller than SNe hosts. A comparison between the host galaxies of the GRBs associated to SNe Ib/c and those of the SNe Ib/c that are not associated to GRBs could constrain the progenitors and/or the environments of both GRBs and SNe. Especially, metallicity measurements can give more information about the stellar evolution.  Moreover, it is thought that most of the long GRBs make a SN, which is out-shined by the afterglow emission (resulting from the deceleration of the relativistic afterglow by the ISM). This might be an explanation for the absence of optical bright SN for some LLA GRBs. However, this has to be contrasted with the results of GRB~060505 and GRB~060614 which do not show any clue for SNe associations even with deep studies.

The common question in GRBs area is what is the origin(s)/progenitor(s) of these events. Is there only one kind? or more? There are several proposed candidates for the progenitor of GRBs. The Wolf-Rayet (W-R) star is one of the example which is strongly supported by the collapser model. There are three possible subclasses of W-R star: with no hydrogen envelope, with a small hydrogen envelope (blue supergiant star) which is one candidate for ultra long GRBs, and with a large hydrogen envelope (red supergiant star) which is one candidate for the bursts characterized by a super-soft late time X-ray radiation, a large X-ray absorption, and an exceptionally long prompt $\gamma$-ray emission duration \cite{margutti2014}. In this last model, the afterglow radiation is reprocessed by the surrounding material which was ripped out from the progenitor during its evolution. This kind of study, like the one performed in my thesis, provides tight constraints on both the afterglow model and on the progenitor of GRBs.

In the case of ultra-long GRBs, the blue supergiant progenitor can explain the super-soft late time X-ray radiation and long-prompt $\gamma$-ray emission duration but, it can not explain the large X-ray absorption since it is found to be compatible with the standard afterglow model. Moreover, this progenitor can explain why some stars produce GRBs but not SNe. For example, the thermal component on the afterglow of the ultra-long GRB~111209A is too weak compared to the expectations of a shock breakout (for instance GRB~060218 or GRB~100316D). Adding to this argument the lack of SN observation challenges the W-R progenitor for this event. The long duration might be explained by the accretion of the remaining hydrogen layers onto the central compact object powering the GRB for $10^4$ seconds. The presence of the hydrogen layer can be explained by a single low metallicity star \cite{gendre2013, gendre2013a}. 

In the case of LLA GRBs, I propose that low mass CO core star (WC or WO stars) in a binary system be their progenitor. They can lose their outer layer by interaction with the companion star without losing their angular momentum. The idea is supported by the fact that type Ibc SNe form from relatively close binaries rather than single star \cite{vanDyk1996} and considering that 30\% of LLA GRBs are associated to SNe. Indeed, if it is a single star, the CO core can be created by burning its hydrogen and helium layers or by ripping out the outer layers with strong rotating and magnetic field. In this case, there is a risk of high metallicity (\textit{e.i.} $ Z > 0.3 Z_{\odot}$) which can remove too much angular momentum or high baryon loading in the outflow which can prevent to have relativistic jet \cite{yoon2005}. Moreover, there are example that a single massive star might produce GRB with high metallicity (\textit{e.g.} GRB~980425, $Z > 0.7 Z_{\odot}$). However, this issue is still doubtful and open issue to be work on.

Moreover, another three possible progenitor scenarios proposed by \cite{fryer2005} can be considered and applied to have different kind of GRBs: single star which has insufficient angular momentum, the merger of a compact remnant with its binary companion called He merger, and the merger of two helium cores called He star/ He star merger which can have too much angular momentum. The important differences between these three possible progenitors is the distribution of redshift. Because the last two models can occur at large redshift while single star progenitors can be found at lower redshift. Indeed, the number of single star progenitor decrease with their lower metallicity (so increased with redshift) because they can not blow off their hydrogen envelope without metal. All these possible scenarios can be considered to explain the diversity of LLA GRBs.

And finally, I could try to increase the number of LLA GRBs by considering the bursts without redshift. First of all, I am planning to constrain the range of possible redshift for a given burst and to establish if it can be a LLA GRB. The method depends on the computation of the flux at one day in the $2 - 10$~keV band, and on the mean value of the flux at one day for each group.  Furthermore, to determine if a burst is a LLA GRB, additional points of comparison which do not depend upon the redshift are needed. The first additional indication is the value of the spectral index, which should indicate a fireball in the slow-cooling state. The second one may be a possible clustering between the luminosity in optical and in X-ray wavelengths, $L_{optical}-L_{X-ray}$, as already studied in the context of short bursts \cite{berger2014} . 
 
The study of GRBs without redshift is important to contrast the population of long GRBs. Indeed, as the afterglow luminosity of LLA GRBs is low by definition, it is expected that the efficiency of measuring their redshifts be lower than for other GRBs. As a result, many bursts without redshift might be LLA GRBs. Thus, determining the group of GRBs without redshift would increase the number of sources in the LLA sample, allowing for stronger constraints of their properties.

%% file: appendix1.tex
\chapter[The table of short GRBs]{\parbox[t]{\textwidth}{The table of short GRBs}}
\chaptermark{The table of short GRBs}
\label{A}

\begin{table*}[!ht]
\centering
  \caption{The list of short GRBs removed from the global sample. Most sources are presented in \cite{siellez2014}. \label{table_short_GRBs}}
  \begin{tabular}{lllllll}
  \hline
  Names  & Redshift &  $T_{90}$ (s) &$T_{90, rest}$ (s) & ref. \\
    \hline
GRB 050406 & 2.44 & 5.40 & 1.57 & \\
GRB 050416 & 0.653  & 2.5 & 1.51 & \\
GRB 050922C & 2,198 & 4.5 & 1.41 & \\
GRB 051221A & 0.547 & 1.4 & 0.9  & \\
GRB 060206 & 4.045 & 7.6 &1.51 & \\
GRB 060801 & 1.131 & 0.49 & 0.23  &\\
GRB 060926 & 3.208 & 8 & 1.9 & \\
GRB 061006 & 0.4377 & 1.04 & 0.72 & T. Sokomoto 2011 \\
GRB 070506 & 2.31 & 4.3 & 1.3 & \\
GRB 070714B & 0.92 & 2 & 1.04 & GCN 6637 \\
GRB 070724A & 0.457 & 0.4 & 0.27 & \\
GRB 071020 & 2.142 & 4.2 & 1.34  & \\
GRB 071227 & 0.383 & 1.728 & 1.25 & \\
GRB 080520 &  1.545 & 2.8 & 1.1 & \\
GRB 080905A & 0.1218 & 1.0 & 0.47  & GCN 8187\\
GRB 080913 & 6.7 & 8.0 & 1.04 & \\
GRB 090205 & 4.65 & 8.8 & 1.56 & \\
GRB 090423 & 8.0 & 10.3 & 1.14 & \\
GRB 090426 & 2.609 & 1.2 & 0.33 & \\
GRB 090429B & 9.2 & 5.5 & 0.5 & \\ 
GRB 090510 & 0903 & 0.3 & 0.16 & GCN 9337 \\
GRB 090809 & 2.737 & 5.4 & 1.45 & \\
GRB 090927 & 1.37 & 2.20 & 0.93 & \\
GRB 100117A & 0.92 & 0.3 & 0.16  & GCN 10338\\ 
GRB 100206A & 0.41 & 0.12 & 0.09 & \\ 
GRB 100316B & 1.18 & 3.8 &1.74 & \\
GRB 100724A & 1.288 & 1.4 & 0.61 & \\
GRB 100816A & 0.8035 & 2.9 & 1.61 & GCN 11111\\
GRB 101219A & 0.718 & 0.6 & 0.35 & GCN 14164 \\  
GRB 130131B & 2.539 & 4.3 & 1.21 & \\                                                       
\hline
\end{tabular}
\end{table*}

%% file: appendix2.tex
\chapter[The table of the global sample]{\parbox[t]{\textwidth}{The table of the global sample}}
\chaptermark{The table of global sample}
\label{B}

\begin{center}
 \begin{longtable}[c]{{lc|c|c|clcl}}
 \caption{The global sample list. \label{table_global_sample_parameter}} \\
 \hline
 \textbf{GRB Name} & \textbf{Redshift} & \textbf{ECF} & \textbf{log ($T_a$)}  & \textbf{Spectral index} \\
 \hline
 \endfirsthead
 \multicolumn{5}{c} {} \\
 {\tablename\ \thetable\ -- \textit{Continued }} \\
\hline
\textbf{GRB Name} & \textbf{Redshift} & \textbf{ECF} & \textbf{Log (Ta)}  & \textbf{Spectral index} \\
\hline
 \endhead
 \hline
 \endfoot
 
 \hline
 \multicolumn{5}{c}{End of Table}\\
 \hline
 \endlastfoot
GRB970228  & 0.695  & 1         & 0             & 1.0    \\
GRB~970508  & 0.835  & 1         & 0             & 1.4    \\
GRB~971214  & 3.42   & 1         & 0             & 1.1    \\
GRB~980425  & 0.0085 & 1         & 0             & 1.0    \\
GRB~980613  & 1.096  & 1         & 0             & 1      \\
GRB~980703  & 0.966  & 1         & 0             & 1.77   \\
GRB~990123  & 1.60   & 1         & 0             & 1.0    \\
GRB~990510  & 1.619  & 1         & 0             & 1.2    \\
GRB~991216  & 1.02   & 1         & 0             & 0.7    \\
GRB~000210  & 0.846  & 1         & 0             & 0.9    \\
GRB~000214  & 0.47   & 1         & 0             & 1.0    \\
GRB~000926  & 2.066  & 1         & 0             & 0.7    \\
GRB~010222  & 1.477  & 1         & 0             & 0.97   \\
GRB~011121  & 0.36   & 1         & 0             & 1.45   \\
GRB~011211  & 2.14   & 1         & 0             & 1.3    \\
GRB~020405  & 0.69   & 1         & 0             & 1.0    \\
GRB~020813  & 1.25   & 1         & 0             & 0.83   \\
GRB~021004  & 2.33   & 1         & 0             & 1.01   \\
GRB~030226  & 1.98   & 1         & 0             & 0.9    \\
GRB~030328  & 1.52   & 1         & 0             & 1.1    \\
GRB~030329  & 0.168  & 1         & 0             & 1.0    \\
GRB~031203  & 0.105  & 1         & 0             & 0.8    \\
GRB~050126  & 1.29   & 28.67     & 2.34          & 0.7    \\
GRB~050223  & 0.5915 & 18.85     & 0             & 1.4    \\
GRB~050315  & 1.949  & 1         & 0             & 0.91   \\
GRB~050319  & 3.240  & 1         & 0             & 0.96   \\
GRB~050401  & 2.9    & 1         & 0             & 1.0    \\
GRB~050505  & 4.27   & 21.51     & 4.39          & 1.0    \\
GRB~050525  & 0.606  & 1         & 0             & 1.1    \\
GRB~050603  & 2.821  & 26.61     & 4.83          & 0.7    \\
GRB~050730  & 3.967  & 30.12     & 4.13          & 0.62   \\
GRB~050802  & 1.71   & 1         & 3.96          & 0.81   \\
GRB~050814  & 5.3    & 26.66     & 3.93          & 0.7    \\
GRB~050824  & 0.83   & 1         & 0             & 0.82   \\
GRB~050826  & 0.297  & 39.18     & 0             & 1.1    \\
GRB~050908  & 3.344  & 1         & 0             & 0.65   \\
GRB~051016B & 0.9364 & 1         & 0             & 0.91   \\
GRB~051109A & 2.346  & 26.79     & 3.93          & 1.0    \\
GRB~051109B & 0.08   & 41.03     & 3.67          & 0.7    \\
GRB~051111  & 1.55   & 22.78     & 0             & 1.1    \\
GRB~060108  & 2.7    & 1         & 0             & 0.93   \\
GRB~060115  & 3.53   & 1         & 3.86          & 1.3    \\
GRB~060123  & 0.56   & 14.88     & 0             & 1      \\
GRB~060210  & 3.91   & 20.67     & 4.46          & 1.0    \\
GRB~060218  & 0.0331 &  1        & 0             & 0.51   \\
GRB~060223A & 4.41   & 24.02     & 2.73          & 1.4    \\
GRB~060418  & 1.489  & 28.43     & 3.44          & 0.8    \\
GRB~060510B & 4.94   & 1         & 4.55          & 1.7    \\
GRB~060512  & 2.1    & 1         & 0             & 0.91   \\
GRB~060522  & 5.11   & 1         & 2.86          & 1.1    \\
GRB~060526  & 3.221  & 35.72     & 3.84          & 0.7    \\
GRB~060604  & 2.1357 & 1         & 4.55          & 1.3    \\
GRB~060605  & 3.78   & 1         & 4.16          & 1.0    \\
GRB~060607A & 3.082  & 29.25     & 4.75          & 0.86   \\
GRB~060614  & 0.125  & 23.11     & 5.00          & 0.8    \\
GRB~060707  & 3.425  & 12.98     & 3.58          & 1.2    \\
GRB~060714  & 2.711  & 1         & 3.20          & 1.4    \\
GRB~060729  & 0.54   & 11.48     & 5.11          & 1.39   \\
GRB~060814 &  1.9229&  0.543e12 &  4.0532469599 &  1.133  \\
GRB~060904B & 0.703  & 0.6095e12 & 3.7715299824  & 0.99   \\
GRB~060906  & 3.686  & 0.6112e12 & 4.1314970448  & 1.01   \\
GRB~060908  & 1.8836 & 0.4535e12 & 3.8975334343  & 1.44   \\
GRB~060912A & 0.937  & 0.6532e12 & 3.0099962952  & 0.63   \\
GRB~060927  & 5.47   & 0.5562e12 & 3.7187572914  & 0.97   \\
GRB~061007  & 1.261  & 0.5823e12 & 3.3505464558  & 1.011  \\
GRB~061021  & 0.3463 & 0.5648e12 & 3.6227865016  & 1.02   \\
GRB~061110A & 0.758  & 0.5641e12 & 4.4328432186  & 0.42   \\
GRB~061110B & 3.44   & 0.6004e12 & 3.5675232493  & 1.19   \\
GRB~061121  & 1.314  & 0.6363e12 & 4.0298207737  & 0.914  \\
GRB~061126  & 1.1588 & 0.6118e12 & 4.1824847413  & 0.971  \\
GRB~061210  & 0.4095 & 0.4696e12 & 0             & 1.86   \\
GRB~061222A & 2.088  & 0.7206e12 & 4.0310314327  & 0.946  \\
GRB~061222B & 3.355  & 0.5378e12 & 3.4243226011  & 1.1    \\
GRB~070103  & 2.6208 & 0.405e12  & 4.0593104896  & 1.34   \\
GRB~070110  & 2.352  & 0.4636e12 & 4.0859957897  & 1.122  \\
GRB~070125  & 1.547  & 0.5669e12 & 5.0697655652  & 1.03   \\
GRB~070129  & 2.3384 & 0.4375e12 & 4.2584421857  & 1.32   \\
GRB~070208  & 1.165  & 0.5754e12 & 4.2082998908  & 1.16   \\
GRB~070306  & 1.4959 & 0.7208e12 & 4.2584129662  & 0.951  \\
GRB~070318  & 0.836  & 0.6172e12 & 3.8568756445  & 1.13   \\
GRB~070411  & 2.954  & 0.4989e12 & 4.032195801   & 1.22   \\
GRB~070419A & 0.97   & 0.5485e12 & 2.9713147154  & 0.8    \\
GRB~070419B & 1.9591 & 0.7011e12 & 4.0618597885  & 0.599  \\
GRB~070521  & 1.35   & 0.7965e12 & 3.8303392967  & 1.03   \\
GRB~070529  & 2.4996 & 0.7026e12 & 3.0613187686  & 1.06   \\
GRB~070611  & 2.04   & 0.5533e12 & 4.338033476   & 0.83   \\
GRB~070714XA& 1.58   & 0.7443e12 & 2.9565630412  & 0.6    \\
GRB~070721B & 3.626  & 0.6799e12 & 2.9853458163  & 0.62   \\
GRB~070802  & 2.45   & 0.5441e12 & 4.0060330287  & 1.17   \\
GRB~070810A & 2.17   & 0.6204e12 & 2.6508042691  & 1.15   \\
GRB~071003  & 1.60435& 0.6344e12 & 4.534876254   & 0.91   \\
GRB~071010A & 0.98   & 0.6197e12 & 4.7217273524  & 1.33   \\
GRB~071010B & 0.947  & 0.5252e12 & 3.8440873614  & 1.04   \\
GRB~071021  & 2.452  & 0.4835e12 & 4.0443553103  & 1.13   \\
GRB~071025  & 5.2    & 0.4511e12 & 3.5554040155  & 1.2    \\
GRB~071031  & 2.692  & 0.2161e12 & 3.0605638372  & 1.69   \\
GRB~071112C & 0.823  & 0.6617e12 & 3.1851790301  & 0.83   \\
GRB~071117  & 1.331  & 0.6174e12 & 4.0192183178  & 1.09   \\
GRB~071122  & 1.14   & 0.6675e12 & 2.6026019902  & 0.8    \\
GRB~080129  & 4.349  & 0.7908e12 & 4.0039076255  & 1      \\
GRB~080207  & 2.0858 & 0.3859e12 & 4.0382671125  & 1.46   \\
GRB~080210  & 2.641  & 0.3777e12 & 4.0733785039  & 1.64   \\
GRB~080310  & 2.42   & 0.5529e12 & 3.8609640866  & 0.95   \\
GRB~080319B & 0.937  & 0.5794e12 & 3.6972373233  & 0.82   \\
GRB~080319C & 1.95   & 0.7391e12 & 3.3446106373  & 0.61   \\
GRB~080330  & 1.51   & 0.4636e12 & 1.9732429172  & 1.25   \\
GRB~080411  & 1.03   & 0.618e12  & 4.2267829447  & 0.976  \\
GRB~080413A & 2.433  & 0.5878e12 & 2.4878096887  & 1.02   \\
GRB~080413B & 1.1    & 0.5646e12 & 3.4313692834  & 0.966  \\
GRB~080430  & 0.767  & 0.5758e12 & 4.0568712478  & 1.058  \\
GRB~080514B & 1.8    & 0.5657e12 & 4.5848295892  & 0.88   \\
GRB~080515X & 2.47   & 0.6162e12 & 0             & 0.77   \\
GRB~080603A & 1.688  & 0.4593e12 & 4.2058511056  & 1.38   \\
GRB~080603B & 2.69   & 0.5479e12 & 3.7969714195  & 0.85   \\
GRB~080604  & 1.416  & 0.4719e12 & 3.9040295248  & 1.18   \\
GRB~080605  & 1.6398 & 0.6634e12 & 4.0155375741  & 0.84   \\
GRB~080607  & 3.036  & 0.5385e12 & 3.0003204137  & 1.13   \\
GRB~080707  & 1.23   & 0.5928e12 & 3.7949566769  & 1.07   \\
GRB~080710  & 0.845  & 0.5394e12 & 4.3152632845  & 1.02   \\
GRB~080721  & 2.591  & 0.6025e12 & 4.1104208036  & 0.943  \\
GRB~080804  & 2.2045 & 0.5079e12 & 3.8919994881  & 0.96   \\
GRB~080805  & 1.505  & 0.6098e12 & 3.7601262322  & 1.09   \\
GRB~080810  & 3.35   & 0.4696e12 & 3.7178179322  & 1.13   \\
GRB~080905B & 2.374  & 0.6215e12 & 3.8131821575  & 1.03   \\
GRB~080906  & 2.1    & 0.5513e12 & 4.0923362429  & 1.09   \\
GRB~080916A & 0.689  & 0.6877e12 & 4.1299475192  & 1.07   \\
GRB~080916C & 4.35   & 0.6342e12 & 0             & 0.96   \\
GRB~080928  & 1.692  & 0.5205e12 & 3.8278494596  & 1.139  \\
GRB~081007  & 0.5295 & 0.6393e12 & 4.1299852479  & 1.04   \\
GRB~081008  & 1.9685 & 0.5619e12 & 4.0729994937  & 1.04   \\
GRB~081028  & 3.038  & 0.5057e12 & 4.0889469511  & 1.031  \\
GRB~081029  & 3.8497 & 0.5177e12 & 4.2187489895  & 0.97   \\
GRB~081118  & 2.58   & 0.3785e12 & 4.7580541615  & 1.46   \\
GRB~081121  & 2.512  & 0.555e12  & 4.0471333686  & 0.944  \\
GRB~081203A & 2.05   & 0.5148e12 & 3.7887993238  & 1.04   \\
GRB~081221  & 2.26   & 0.6105e12 & 3.814703017   & 1.29   \\
GRB~081222  & 2.77   & 0.5057e12 & 3.8580866173  & 1.03   \\
GRB~081228  & 3.4    & 0.5125e12 & 2.4061523322  & 1.15   \\
GRB~081230  & 2.0    & 0.4672e12 & 4.0145379775  & 1.05   \\
GRB~090102  & 1.547  & 0.658e12  & 3.0665015789  & 0.787  \\
GRB~090113  & 1.7493 & 0.7088e12 & 2.6966321612  & 1.23   \\
GRB~090313  & 3.375  & 0.5577e12 & 4.9087300153  & 1.08   \\
GRB~090323  & 3.57   & 0.4941e12 & 0             & 0.98   \\
GRB~090328A & 0.736  & 0.7362e12 & 0             & 0.65   \\
GRB~090407  & 1.4485 & 0.4106e12 & 4.8448295739  & 1.39   \\
GRB~090417B & 0.345  & 0.7331e12 & 4.0802294785  & 1.34   \\
GRB~090418A & 1.608  & 0.6169e12 & 3.0009580801  & 1.016  \\
GRB~090424  & 0.544  & 0.6525e12 & 3.7459348683  & 0.96	 \\
GRB~090516  & 4.109  & 0.5184e12 & 4.0767204034  & 1.114  \\
GRB~090519  & 3.85   & 0.733e12  & 3.3217838938  & 0.61   \\
GRB~090529  & 2.625  & 0.6211e12 & 4.0894524006  & 0.67   \\
GRB~090530  & 1.3    & 0.5320e12 & 4.0827279454  & 1.23   \\
GRB~090618  & 0.54   & 0.6417e12 & 3.9244392323  & 0.92   \\
GRB~090715B & 3      & 0.5081e12 & 2.6819454853  & 1.05   \\
GRB~090726  & 2.71   & 0.403e12  & 0             & 1.45   \\
GRB~090812  & 2.452  & 0.5321e12 & 2.7811412941  & 1.11   \\
GRB~090814A & 0.696  & 0.4624e12 & 3.3846541723  & 0.8    \\
GRB~090902B & 1.822  & 0.581e12  & 0             & 0.85   \\
GRB~090926A & 2.1062 & 0.482e12  & 0             & 1.09   \\
GRB~090926B & 1.24   & 0.6527e12 & 3.3139980222  & 1.05   \\
GRB~091003  & 0.8969 & 0.6413e12 & 0             & 0.77   \\
GRB~091018  & 0.971  & 0.5217e12 & 4.2666547508  & 1.09   \\
GRB~091020  & 1.71   & 0.5118e12 & 3.8616918296  & 1.106  \\
GRB~091024  & 1.092  & 0.8777e12 & 3.730564648   & 0.72   \\
GRB~091029  & 2.752  & 0.2289e12 & 4.0740501254  & 1.752  \\
GRB~091109A & 3.076  & 0.5619e12 & 2.8168354041  & 0.98   \\
GRB~091127  & 0.49   & 0.6189e12 & 4.3205476385  & 0.79   \\
GRB~091208B & 1.063  & 0.6663e12 & 2.7696012151  & 0.94   \\
GRB~100219A & 4.6667 & 0.7035e12 & 4.0605956206  & 0.69   \\
GRB~100302A & 4.813  & 0.5558e12 & 4.0658649322  & 0.87   \\
GRB~100316D & 0.059  & 0.3586e12 & 2.866936236   & 0.54   \\
GRB~100414A & 1.368  & 0.6586e12 & 0             & 0.63   \\
GRB~100418A & 0.6235 & 0.7839e12 & 5.1833269981  & 0.87   \\
GRB~100425A & 1.755  & 0.5639e12 & 4.0192053741  & 1.03   \\
GRB~100513A & 4.772  & 0.9701e12 & 3.8596097877  & 1.27   \\
GRB~100621A & 0.542  & 0.7175e12 & 4.0602318088  & 1.41   \\
GRB~100728B & 2.106  & 0.5473e12 & 3.6972773917  & 1.09   \\
GRB~100814A & 1.44   & 0.7192e12 & 5.0099537069  & 0.44   \\
GRB~100901A & 1.408  & 0.5579e12 & 4.0820697543  & 1.094  \\
GRB~100906A & 1.727  & 0.6905e12 & 3.85720183    & 1.026  \\
GRB~101219B & 0.55   & 0.6678e12 & 4.8061895761  & 1.28   \\
GRB~110106B & 0.618  & 0.578e12  & 4.0867643399  & 0.95   \\
GRB~110128A & 2.339  & 0.5421e12 & 2.8943272861  & 0.84   \\
GRB~110205A & 2.22   & 0.4697e12 & 3.2664328792  & 1.117  \\
GRB~110213A & 1.46   & 0.6721e12 & 3.169913591   & 0.991  \\
GRB~110213B & 1.083  & 0.3251e12 & 0             & 1.74   \\
GRB~110422A & 1.77   & 0.6329e12 & 3.8718316614  & 0.932  \\
GRB~110503A & 1.613  & 0.5438e12 & 4.274612385   & 0.943  \\
GRB~110715A & 0.82   & 0.8073e12 & 4.8743891583  & 0.84   \\
GRB~110731A & 2.83   & 0.6599e12 & 3.8544184927  & 0.86   \\
GRB~110801A & 1.858  & 0.5464e12 & 4.2204990554  & 1.08   \\
GRB~110808A & 1.348  & 0.3578e12 & 4.3816692333  & 1.62   \\
GRB~110818A & 3.36   & 0.533e12  & 3.9012167041  & 1.02   \\
GRB~110918A & 0.982  & 0.4209e12 & 0             & 1.34   \\
GRB~111008A & 4.9898 & 0.5459e12 & 3.8555201095  & 0.91   \\
GRB~111107A & 2.893  & 0.5213e12 & 2.9246505814  & 1.25   \\
GRB~111209A & 0.677  & 0.6168e12 & 4.86          & 0.81   \\
GRB~111211A & 0.478  & 0.5431e12 & 0             & 1.1    \\
GRB~111228A & 0.714  & 0.5868e12 & 4.0364244956  & 1.024  \\
GRB~111229A & 1.3805 & 0.6459e12 & 3.8157725195  & 0.86   \\
GRB~120119A & 1.728  & 0.7543e12 & 3.2088038442  & 0.59   \\
GRB~120326A & 1.798  & 0.625e12  & 4.6995329121  & 0.864  \\
GRB~120327A & 2.813  & 0.7006e12 & 3.8480091823  & 0.85   \\
GRB~120404A & 2.876  & 0.51e12   & 3.1307370462  & 1.09   \\
GRB~120422A & 0.283  & 0.5342e12 & 4.9278600406  & 0.42   \\
GRB~120711A & 1.405  & 0.7138e12 & 0             & 0.849  \\
GRB~120712A & 4.1745 & 0.4122e12 & 3.1173024842  & 1.37   \\
GRB~120714B & 0.3984 & 0.5039e12 & 2.0733622952  & 0.8    \\
GRB~120722A & 0.9586 & 0.8533e12 & 3.6963001094  & 0.17   \\
GRB~120724A & 1.48   & 0.6491e12 & 4.3783121399  & 0.7    \\
GRB~120729A & 0.8    & 0.652e12  & 3.8010512275  & 0.79   \\
GRB~120802A & 3.796  & 0.3508e12 & 4.0072350438  & 1.11   \\
GRB~120811C & 2.671  & 0.4633e12 & 3.1677811811  & 1.24   \\
GRB~120815A & 2.358  & 0.6605e12 & 4.3429908712  & 0.79   \\
GRB~120907A & 0.97   & 0.6342e12 & 4.0950336752  & 0.81   \\
GRB~120909A & 3.93   & 0.5606e12 & 3.6971794351  & 1.099  \\
GRB~121024A & 2.298  & 0.5962e12 & 4.0013035671  & 1.01   \\
GRB~121027A & 1.773  & 0.4523e12 & 5.141031371   & 1.45   \\
GRB~121128A & 2.2    & 0.6575e12 & 3.1466561085  & 1.12   \\
GRB~121201A & 3.385  & 0.4466e12 & 3.755895154   & 1.17   \\
GRB~121211A & 1.023  & 0.5648e12 & 3.6240103554  & 1.06   \\
GRB~121229A & 2.707  & 0.5948e12 & 4.6602363951  & 1.08	 \\
GRB~050215B & 2.62   & 0.645e12  & 4             & 0.62   \\
GRB~050318  & 1.44   & 0.542e12  & 4             & 0.91   \\
GRB~050801  & 1.38   & 0.643e12  & 3.77815125    & 0.79   \\
GRB~050819  & 2.5043 & 0.548e12  & 4.477121255   & 1      \\
GRB~050820A & 2.612  & 0.595e12  & 5.812913357   & 0.89   \\
GRB~050822  & 1.434  & 0.508e12  & 4.230448921   & 1      \\
GRB~050904  & 6.29   & 0.615e12  & 5             & 0.85   \\
GRB~050915A & 2.5273 & 0.494e12  & 3.966000858   & 1.01   \\
GRB~051001  & 2.4296 & 0.408e12  & 4             & 1.2    \\
GRB~051006  & 1.059  & 0.249e12  & 4             & 3.4    \\
GRB~051117B & 0.481  & 0.789e12  & 4             & 0.32   \\
GRB~060111A & 2.32   & 0.531e12  & 3.62797999    & 0.97   \\
GRB~060124  & 2.296  & 0.658e12  & 4.785329835   & 0.91   \\
GRB~060202  & 0.783  & 0.358e12  & 4.477121255   & 1.98   \\
GRB~060319  & 1.172  & 0.546e12  & 4.184691431   & 1.12   \\
GRB~060502A & 1.51   & 0.627e12  & 4.421603927   & 0.91   \\
GRB~060505  & 0.089  & 0.763e12  & 0             & 0.4    \\
GRB~060708  & 1.92   & 0.487e12  & 3.906011625   & 1.02   \\
GRB~060719  & 1.532  & 0.643e12  & 3.909983695   & 1.23   \\
GRB~080825B & 4.3    & 0.662e12  & 0             & 1.17   \\
GRB~081109  & 0.9787 & 0.699e12  & 4.477121255   & 1.00   \\
GRB~100424A & 2.465  & 0.782e12  & 4             & 0.6    \\
GRB~100615A & 1.398  & 0.893e12  & 4.278753601   & 1.23   \\
GRB~100728A & 1.567  & 0.795e12  & 4.477121255   & 0.73   \\
GRB~101225A & 0.847  & 0.711e12  & 5             & 0.7    \\
GRB~111123A & 3.1516 & 0.412e12  & 4.608526034   & 1.38   \\
GRB~120118B & 2.943  & 0.723e12  & 3.431041945   & 1.04   \\
GRB~120521C & 6      & 0.448e12  & 4.301029996   & 1.3    \\
GRB~120922A & 3.1    & 0.627e12  & 3.385963571   & 0.91   \\
GRB~050408  & 1.2357 & 0.655e12  & 4             & 1.06   \\
GRB~051022  & 0.8    & 0.869e12  & 4             & 1.06   \\
GRB~070508  & 0.82   & 0.741e12  & 4             & 0.83   \\      
\end{longtable}
\end{center}

%% file: appendix3.tex
\chapter[Data analysis process]{\parbox[t]{\textwidth}{Data analysis process}}
\chaptermark{One example of data analysis process}
\label{C}

\section{XRT modes}
The XRT instrument can operate in auto and manual states. The auto state is the normal operating mode in which XRT can automatically select the science mode according to the count rate of the source, while the manual state is used for calibration and for a given observation in which the science mode can be selected. 

There are four science modes:
\begin{itemize}
\item Image long and short (IM): images are taken and the source position is calculated. 
\item Low rate (LR) and Piled-up photodiode (PU), which is designed for very bright sources: fluxes up to 60 Crab, and high time resolution: 0.14~ms.
\item Window timing (WT) mode is to obtain by binning 10 rows in serial register and readout only the control 200 columns of the CCD. Its time resolution is 1.7~ms and its image capability is one dimensional. This mode is used when the source flux ranges 1 to 600~mCrab. 
\item Photon Counting (PC) works with full image and spectroscopic resolution. In this mode the time resolution is limited to 2.5 seconds. It is used for the source flux below 1~mCrab. It can be piled-up if there are more than 2 counts per bin. This is the mode that I used to extract the spectral parameters of LLA GRBs.    
\end{itemize}

\section{Directory of data}
Here, I only introduce the information about the PC mode. However, more information about the PC mode and other modes can be found in the {\em Swift} XRT data reduction Guide \footnote{http://swift.gsfc.nasa.gov/analysis/xrt\_swguide\_v1\_2.pdf} and at the UK Swift science data centre \footnote{http://www.swift.ac.uk/analysis/xrt/}.  The data are mainly taken from the Swift archive download aortal of Leicester \footnote{http://www.swift.ac.uk/swift\_portal/} \cite{evans2007, evans2009} and sometimes from the NASA webpage for the faintest GRBs \footnote{http://heasarc.gsfc.nasa.gov/cgi-bin/W3Browse/swift.pl}.

There can be one or several observations for one GRB depending on its brightness. Each observation has its own  ID number and it contains several directories: 
\begin{itemize}
\item  ~~~ /auxil ~~~ /xrt ~~~ /bat ~~~ /uvot ~~~ /log ~~~ /tdrss
\item  ~~~ /event ~~~ /image ~~~ /hk ~~~ /productions
\end{itemize}
Inside each observation, there is an event fits file which contains 

\begin{itemize}
\item the primary header,
\item the event extension: the binary table which contains the observation time, position of sources, pulse-height information, 
\item the good time interval extension 
\item the bad pixel table 
\end{itemize}
The guide gives information about all parameters, however, the event extension is in the form of a binary table. It is called events, which contains the observation time, position of sources, pulse-height information. 

\section{Structure directory in xrtpipeline}
\begin{figure*}[!ht]
\begin{center}
\includegraphics[width=0.9\textwidth]{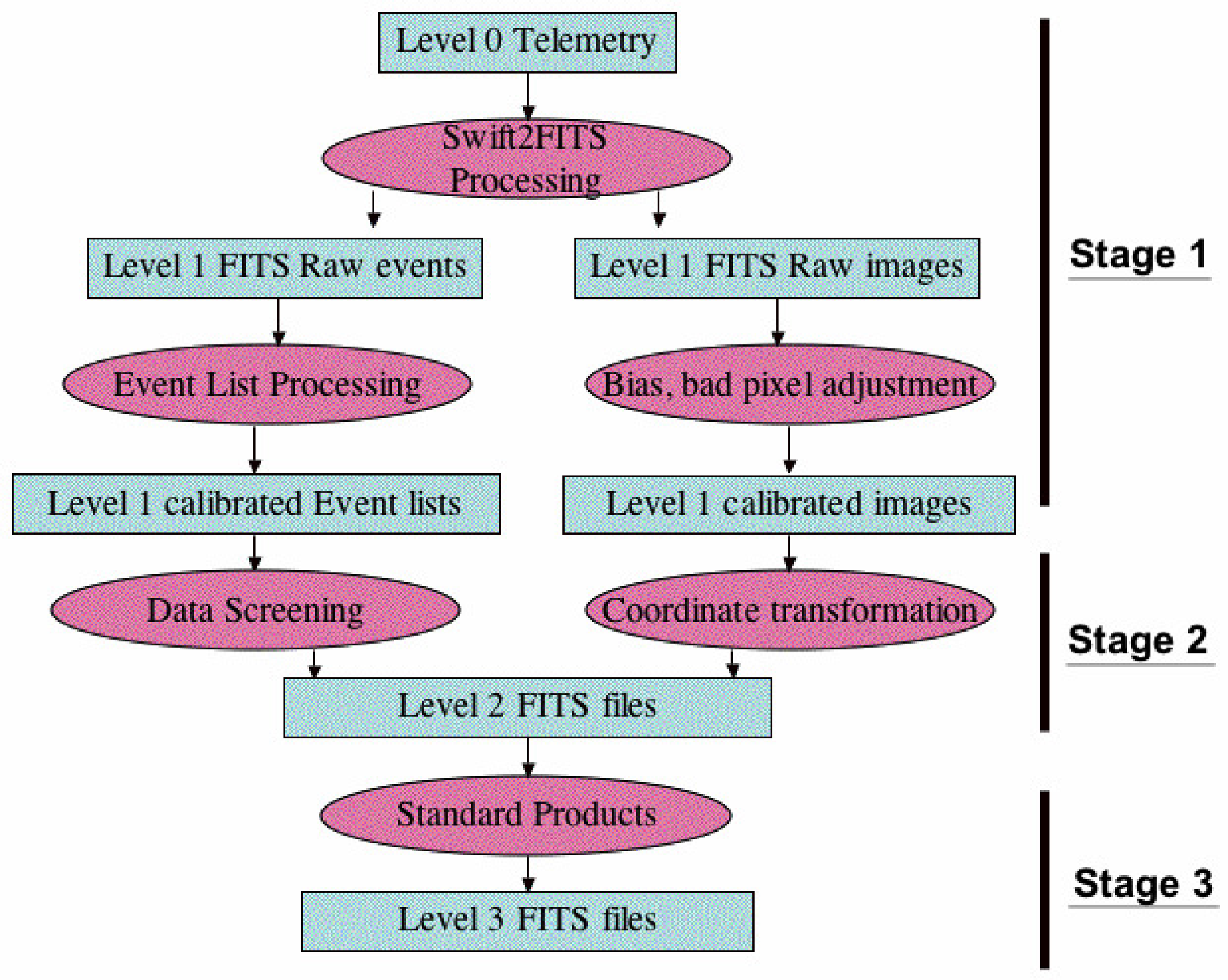}
\caption{Diagram of the {\em Swift} data processing in pipeline. \label{fig_XRT_data_process}}
\end{center}
\end{figure*}

The processing produces two different kind of files, Level 1 and Level 2, by following the structure of the diagram seen in Figure \ref{fig_XRT_data_process}. The Level~1 file is the first file which is produced without information loss. Additional information is calculated and added to this file during the Stage 1, as it can be seen in  Figure \ref{fig_XRT_data_process}. In this file, data is taken in the low-rate photodiode mode during the slews and the settling is stored. The intermediate level of files, called Level~1a, is used by the photodiode mode and the window timing mode by taking into account the timing mode. In this stage, bad pixels or bad columns, coordinates information, time tagging of events, computation of the PHA and PI values and elimination of the pile-up frames and partially exposed pixels were identified. However, in stage 2, Level 2 file  is created by calibration and screening. This file is processed following standard procedures in the stage 3. After stage 3, an event file is obtained and used in \textit{XSELECT}. It is read and processed to extract images, light-curves, and spectra of events. All stages can be processed in the \textit{xrtpipeline}.

\section{XRT file name}
XRT file names for the event and image files are built as follows:
\begin{itemize}
\item sw[obs\_id]x[mm][pp]\_[lew].ext
\item sw[obs\_id]x[mm]\_[lew].ext
\end{itemize}
where sw indicates the mission name (\textit{Swift}), obs\_id is a 11 digit number which identifies the observation, x represents the XRT instrument, mm is the instrument operating mode (pc, for more information see guide), ww identifies the window settings of the CCD (between w1 and w4 for pc mode, for more information see guide), pp identifies if the event data is taken in pointing mode \textit{po}, during a slew \textit{sl} or during a settling  phase \textit{sd} (I used pointing mode \textit{po}), the lew is the file level which is cl for Level 2 event file. Finally, ext represent file extension: .evt or .img. 

Moreover, the housekeeping file name is :
\begin{itemize}
\item sw[obs\_id]x[hh].hk
\end{itemize}
where hh can be \textit{hd} which means that information from the science-packed data is stored in the header.



\section{Example of XRT data analysis process}
The process to obtain high-level products from level 2 screened event file is explained step by step below for GRB~050801. The data analysis process is follows: 

\begin{itemize}
\item \$ heainit
\item \$ cd $<$path of data$>$
\item \$ cd GRB050801
\item \$ ls
00148522000 ~~~~~~~~~~~~~~~~ 00148522000.tar ~~~~~~~~~~~~~~~~ swxpc0to12s0\_20070101v012.rmf

where .rmf is the XRT calibration file.  

\textbf{Note}: Before running \textit{xrtpipeline}, the position of the sources has to be prepared. Since the sources considered in the thesis have known redshifts, the positions obtained from optical observations have been chosen with known redshifts. it is better to use the positions obtained from optical observations as they are more precise, even if the differences are usually small compared to the positions obtained by \textit{Swift}. Specially, the last observation from \textit{Swift} shows that the error radius is around 4 arcsec for XRT, while it is around 2 arcsec UVOT.

	\begin{itemize}
	 \item Swift page\footnote{http://www.swift.ac.uk/grb\_region/}: 
 
 		\begin{enumerate}
		 \item RA (J2000): 13h 36m 35.34s (204.14724)
 		\item Dec (J2000): -21\degre 55' 42.8" (-21.92854)  
 		\item 90\% Error radius: 1.5"
		\end{enumerate}

	 \item Optical from GCN 3736:
 
	RA, Dec = 13:36:33.657, -21:55:29.55

	\end{itemize}
	
\subsection{XRTPIPELINE}
All files are recreated, starting from Level 1 files to Level 2 files as well as light-curves, spectra, etc., for each different mode if there is an observation.  
\item \$ xrtpipeline

===================================

                Running SWIFT XRT pipeline
 Task: xrtpipeline Version: 0.12.6 Release Date: 2010-07-01
 
===================================

		\begin{enumerate}
		\item Source RA position (degrees or hh mm ss.s) or POINT or OBJECT[110.80875] 204.14724
		\item Source DEC position (degrees or dd mm ss.s) or POINT or OBJECT[9.50541] -21.92854
		\item Target Archive Directory Path[./00158593000] ./00148522000
		\item Stem for FITS input files [i.e. sw00000000000][sw00158593000] sw00148522000
		\item Directory for outputs[./00158593000\_xrt] ./00148522000\_xrt
		\end{enumerate}

===================================

xrtpipeline\_0.12.6: Exit with no errors - Sun May 25 23:52:43 CEST 2014

===================================

\subsection{Photon counting mode in XSELECT}
I start from Level 2 files in pc mode to tighten the range of values for parameters which are used in the standard screening criteria. 

\item \$ xselect

		\begin{enumerate}
		\item $>$ Enter session name $>$[xsel4628] pc
		\item pc:SUZAKU $>$ read event
		\item $>$ Enter the Event file dir $>$[./00148522000\_xrt//] ./00148522000\_xrt
		\item $>$ Enter Event file list $>$[sw00148522000xwtw2po\_cl.evt] sw00148522000xpcw4po\_cl.evt
		\item $>$ Reset the mission $?$ $>$[yes] 
		 \end{enumerate}
 
\item \$ $>$ extract image

                 Total    \  Good   \ Bad:Time  \   Phase  \   Grade    \   Cut
          
           \  5617  \  \ \  \   5617    \   \  \  \  \ \   \     0  \ \  \  \  \   \  \  \  \  \  0   \  \  \ \  \  \  \  \  0  \  \  \  \  \  \   0

===================================
    
 Image            has     5617 counts for 0.2430     counts/sec

\subsection{Filtering}
The filter command allows to set the different types of filters: regions, time, intensity filter, grade, energy, phase. It can be used with the \textit{extract} command while it effects are reset by the command \textit{clear}. 
\subsubsection{Region filtering}
\textbf{Note}: To create a region filter in \textit{ds9}, I need to choose the source and background regions by inserting right essential (RA) and declination (Dec) of the source obtained from optical data. The shape of the region should be a circle for a point-like source. For the source region, a 30 pixel radius is considered while a 60 pixel radius is chosen for the background. Eventually, the central part of the source region can be removed if the count rate is above 0.6 counts.s$^{-1}$.

\item \$ $>$ plot image 
		\begin{enumerate}
		\item RA   13:36:35.34 
		\item DEC  -21:55:42.8
		\item WCS
		\item sexagesimal
		\item for sources: circle(497.31858,542.88596,30)
		\item for background: circle(445.31796,400.88717,60) 
		\item save ``source\_pc.reg''
		\item save ``back\_pc.reg''
		 \end{enumerate}
 
 \subsubsection{Energy filtering}
 \textbf{Note}: This process is applied to select the lower and upper channel boundaries. The energy to channel conversion is $PI = 100 \times E$ where $E$ is the energy in keV. It is used to select the energy from 0.5~keV to 10~keV and to remove lowest energies.
 
\item \$ $>$ filter pha\_cut
		\begin{enumerate}
		\item$>$ Lower cutoff for PHA $>$[30] 50
		\item$>$ Upper cutoff for PHA $>$[1000] 
 		\end{enumerate}

\subsubsection{Grade filtering}		
The grade is the shape of spread charge, which is created by an event in a the CCD. It is 0-12 for pc mode. It is suggested to use full grade for the spectral analysis. 
\item \$ $>$  filter grade 0 - 4
\item \$ $>$ extract image

 \subsubsection{Binning of data}
\textbf{Note}: The number of bins is important to determine the pile-up region. When the source is faint, I used 100 counts/bin, otherwise 500 counts/bin. 

\item \$ $>$  filter region source\_pc.reg
\item \$ $>$  clear grade
\item \$ $>$ set bin 500
\item \$ $>$ extract curve

          Total    \  Good   \ Bad:Time  \   Phase  \   Grade    \   Cut
          
           \  790  \  \ \  \   703   \   \  \  \  \ \   \     0  \ \  \  \  \   \  \  \  \  \  0   \  \  \ \  \  \  \  \  0  \  \  \  \  \  \   87
            
===================================

 Fits light curve has      703 counts for 3.0410E-02 counts/sec
 
\item \$ $>$ cpd /xw
 Plotting device chosen: /xw
\item \$ $>$ plot curve
\item \$ PLT$>$ q

 \subsubsection{Time filtering}
\textbf{Note}: There are different methods to create time filters. Since I am interested with the late afterglow, I need to chose a time interval of interest by considering $T_a$. So I used the command \textit{filter time scc}, where scc is the spacecraft clock time. It needs a start and stop time. First, TSTART and TRIGTIME are taken from the header of the event file of the observation which is the good time interval file. Second, they are subtracted from $T_a$ with half of the time step.

	\begin{itemize}
	\item TSTART  = 1.446137710726234E+08 / time start =144613771.0726234
	\item TRIGTIME=  144613681.728 / [s] MET TRIGger Time for Automatic Target
	\item Tstart - Trigtime= 89.3446234
	\item Ta - Tstart - Trigtime - half of the bin =7000 - 89.3446234 - 250 =6660.6553766  
	\end{itemize}


\item \$ $>$ filter time scc
		\begin{enumerate}
		\item PLT$>$ q
		\item Enter start and stop times $>$ i 6660.6553766,60000
		\item Enter start and stop times $>$ x

		Writing timing selections to file cp\_keybd\_gti001.xsl

		\end{enumerate}
 \subsection{Extraction of light-curve}
\item \$ $>$ extract curve

  Total    \  Good   \ Bad:Time  \   Phase  \   Grade    \   Cut
     
           \  790  \  \ \  \   183   \   \  \  \  \ \   \     591 \ \  \  \  \   \  \  \  \  \  0   \  \  \ \  \  \  \  \  0  \  \  \  \  \  \   16
            
===================================

 Fits light curve has      183 counts for 9.1812E-03 counts/sec

\item \$  $>$ save curve source\_pc.lc

Wrote FITS light curve to file source\_pc.lc

 \subsection{Extraction of spectrum}
\item \$  $>$ clear pha\_cut

\item \$  $>$ extract spectrum

              Total    \  Good   \ Bad:Time  \   Phase  \   Grade    \   Cut

           \  790  \  \ \  \   199   \   \  \  \  \ \   \     591 \ \  \  \  \   \  \  \  \  \  0   \  \  \ \  \  \  \  \  0  \  \  \  \  \  \   0
            
===================================

 Spectrum         has      199 counts for 9.9839E-03 counts/sec
 
\item \$ $>$ save spectrum source\_pc.pi

 \subsection{Extraction of the background}
\item \$ $>$ clear reg
\item \$ $>$ filter reg back\_pc.reg
\item \$ $>$ extract curve

         Total    \  Good   \ Bad:Time  \   Phase  \   Grade    \   Cut

           \  242  \  \ \  \   190   \   \  \  \  \ \   \     52 \ \  \  \  \   \  \  \  \  \  0   \  \  \ \  \  \  \  \  0  \  \  \  \  \  \   0
            
===================================

 Fits light curve has      190 counts for 9.5324E-03 counts/sec
 
 \item \$ $>$ save curve back\_p.lc

\item \$ $>$ extract spectrum

          Total    \  Good   \ Bad:Time  \   Phase  \   Grade    \   Cut

           \  242  \  \ \  \   190   \   \  \  \  \ \   \     52 \ \  \  \  \   \  \  \  \  \  0   \  \  \ \  \  \  \  \  0  \  \  \  \  \  \   0
            
===================================

 Spectrum         has      190 counts for 9.5324E-03 counts/sec
 
\item \$  $>$ save spectrum back\_pc.pi

\item \$  $>$ exit
\item \$  $>$ Save this session? $>$[yes] 

Session saved, goodbye

 \subsection{Creation of the ancillary response file (ARF)}
\textbf{Note}: The ancillary response file is called \textit{ARF} and it is used to make the conversion between photon numbers and energy. To prepare this file, the exposure map which is obtained after running \textit{xrtpipeline}, and the spectrum which is obtained from the manual extraction, are used. The exposure map is used to correct the \textit{ARF} file for hot columns, bad pixels, any lost of counts caused by using an annular extraction region. 

\item \$ xrtmkarf expofile=00148522000\_xrt/sw00148522000xpcw4po\_ex.img

\begin{enumerate}
		\item Name of the input PHA FITS file[source\_pc.pi] 
		\item Apply PSF correction (used if extended=no)?(yes/no)[yes] 
		\item Name of the output ARF FITS file[source\_pc\_exp.arf] 
		\item Source X coordinate (SKY for PC and WT modes, DET for PD mode)(used if extended=no):[-1] 
		\item Source Y coordinate (SKY for PC and WT modes, DET for PD mode)(used if extended=no):[-1] 
\end{enumerate}

-----------------------------------------------------------------------------------

		Running 'xrtmkarf\_0.6.0'
		
-----------------------------------------------------------------------------------

		 Input Parameters List: 
	 \begin{itemize}
	\item Name of the input RMF file                    :'CALDB'
	\item Name of the input mirror effective area file  :'CALDB'
	\item Name of the input filter transmission file    :'CALDB'
	\item Name of the input arf file                    :'CALDB'
	\item Name of the input exposure map file           :'00148522000\_xrt/sw00148522000xpcw4po\_ex.img'
	\item Name of the input vignetting file             :'CALDB'
	\item Name of the input spectrum file               :'source\_pc.pi'
	\item Name of the input PSF file                    :'CALDB'
	\item Name of the output ARF file                   :'source\_pc\_exp.arf'
	\item Source SKYX                                   :'-1.000000'
	\item Source SKYY                                   :'-1.000000'
	\item Source off-axis angle (arcmin)                :'-99.000000'
	\item Extended source?                              : no
	\end{itemize}
-----------------------------------------------------------------------------------

xrtmkarf\_0.6.0: Exit with success.

-----------------------------------------------------------------------------------

 \subsection{Binning the spectrum}
\textbf{Note}: To bin the spectrum and to write the header I used \textit{grppha}. When I wanted to use the $\chi^2$, I had to bin the data such that there were at least 20 counts per bin in the 25 group bin. 
If the flux too low, I used the cash statistic which requires at least one count per bin. 

Furthermore, I need to use at least 15 group bins to make $\chi^2$ statistic. Otherwise, I need to use the cash statistic which required 1 group bin so the spectrum would have one count per bin. 

With the \textit{grppha} command, 30 channels are removed since XRT data should not be fitted below 0.3 keV. ARF, RMF, and background files have been set to make the spectrum suitable to be used in \textit{XSPEC}. 

\item \$ grppha

		\begin{enumerate}
		\item [] source\_pc.pi
		\item [] source\_pc\_15bin.pi
		
-----------------------------------------------------------------------------------
  
		\item GRPPHA[bad 0-29] 
		\item GRPPHA[exit] group min 15
		\item GRPPHA[group min 15] chkey backfile back\_pc.pi
		\item GRPPHA[chkey backfile back\_pc.pi] chkey ancrfile source\_pc\_exp.arf
		\item GRPPHA[chkey ancrfile source\_pc\_exp.arf] chkey respfile swxpc0to12s0\_20070101v012.rmf
		\item GRPPHA[chkey respfile swxpc0to12s0\_20070101v012.rmf] exit
		\end{enumerate}

 ** grppha 3.0.1 completed successfully
\subsection{XSPEC}
\textit{XSPEC} is used to fit the spectrum. It is part of \textit{FTOOLS}\footnote{http://heasarc.gsfc.nasa.gov/docs/software/ftools/ftools\_menu.html} which is a collection of utility programs to create, examine or modify data files in the FITS (Flexible Image Transport System) format. It is distributed with the HEASOFT package\footnote{http://heasarc.gsfc.nasa.gov/heasoft/}. 
\item \$ xspec

\item XSPEC12 $>$ data source\_pc\_15bin.pi

1 spectrum  in use

	\begin{itemize}
	\item Spectral Data File: source\_pc\_15bin.pi  Spectrum 1
	\item Net count rate (cts/s) for Spectrum:1  7.635e-03 +/- 7.314e-04 (76.2 
 	\item Assigned to Data Group 1 and Plot Group 1
 	\item Noticed Channels:  1-382
	\item Telescope: SWIFT Instrument: XRT  Channel Type: PI
  	\item Exposure Time: 1.985e+04 sec
 	\item Using fit statistic: chi
 	\item Using Background File                back\_pc.pi
  	\item Background Exposure Time: 1.985e+04 sec
 	\item Using Response (RMF) File            swxpc0to12s0\_20070101v012.rmf for Source 1
 	\item Using Auxiliary Response (ARF) File  source\_pc\_exp.arf
	\end{itemize}

\item XSPEC12$>$cpd /xw
\item XSPEC12$>$setplot energy
\item XSPEC12$>$ignore bad

ignore:   370 channels ignored from  source number 1

\item XSPEC12$>$ignore **-0.3 10.0-**

    30 channels (1-30) ignored in spectrum \#     1
    
    25 channels (358-382) ignored in spectrum \#     1

\subsubsection{Galactic $N_H$ calculation}
\textbf{Note}: I wanted to fix the Galactic $N_H$ value so I used XSPEC to find the value from the following catalog:

\item XSPEC12$>$nh

Equinox (d/f 2000)[2000] 

		\begin{enumerate}
		\item RA in hh mm ss.s or degrees[110.80875] 204.14724
		\item DEC in dd mm ss.s or degrees[9.50541] -21.92854
		\end{enumerate}
		
	\begin{itemize}		
 	 \item $>$ $>$ Leiden/Argentine/Bonn (LAB) Survey of Galactic HI
  	\item LAB $>$ $>$ Average nH (cm**-2)  6.42E+20 ---$>$ 0.642 ---$>$0.64
  	\item LAB $>$ $>$ Weighted average nH (cm**-2)  6.34E+20
 	\end{itemize}

\textbf{Note}: So $N_{H,Gal}$ is 0.64 and the redshift of the source is $z = 0.606$

\subsubsection{Source and background models for the spectrum}
\item XSPEC12$>$mo wabs*zwabs*pow

Input parameter value, delta, min, bot, top, and max values for ...
	
              	1      0.001(      0.01)          0          0     100000      1e+06
	
	\begin{itemize}
	\item 1 :wabs:nH$>$0.64,-1
	
              	1      0.001(      0.01)          0          0     100000      1e+06
	
	\item 2 :zwabs:nH$>$
	
              	0      -0.01(      0.01)     -0.999     -0.999         10         10
	
	\item 3 :zwabs:Redshift$>$0.606
	
             	 1       0.01(      0.01)         -3         -2          9         10
	 
	\item 4 :powerlaw:PhoIndex$>$
	
              	1       0.01(      0.01)          0          0      1e+24      1e+24
	
	\item 5 :powerlaw:norm$>$
	\end{itemize}
	
\subsubsection{Fitting process}
\item XSPEC12$>$fit 

\item XSPEC12$>$fit 1000 0.001

===================================

 Model Model Component  Parameter  Unit     Value

 	\begin{itemize}
  	\item  1    1   wabs       nH         10$^{22}$    0.640000     frozen
   	\item  2    2   zwabs      nH         10$^{22}$    2.29906E-09  +/-  -1.00000     
   	\item  3    2   zwabs      Redshift            0.606000     frozen
   	\item  4    3   powerlaw   PhoIndex            2.54628      +/-  0.268456     
   	\item  5    3   powerlaw   norm                1.69195E-04  +/-  3.24698E-05  
  	\end{itemize} 
-----------------------------------------------------------------------------------

 Chi-Squared =         26.630 using 12 PHA bins.
 
 Reduced chi-squared =         2.9589 for      9 degrees of freedom 
 
 Null hypothesis probability =   1.610149e-03

\item XSPEC12$>$error 2 4

Cannot do error calc: Reduced Chi$^2$ (= 2.95886) $>$ maximum (2)

\subsubsection{$\chi^2$ statistic vs. cash statistic}

\textbf{Note}:  Since the reduced chi-squared  is too large, it should be 0.8 $\leq$ $\chi^2$ $\leq$ 1.1 as seen in Table \ref{chi_sqaure_distribution}, 
I had to change the number of bins which should contain at least 15 to 20 counts each. Whenever not possible, I applied the cash statistic which requires only one count per bin. The disadvantage of this statistic is that it does not assess the quality of the fit. When the first observation was not enough to get statistical significance, I combined the other observations until a good statistic firstly with $\chi^2$ or even with cash statistic was obtained.

\begin{figure*}[!ht]
\begin{center}
\includegraphics[width=0.6\textwidth]{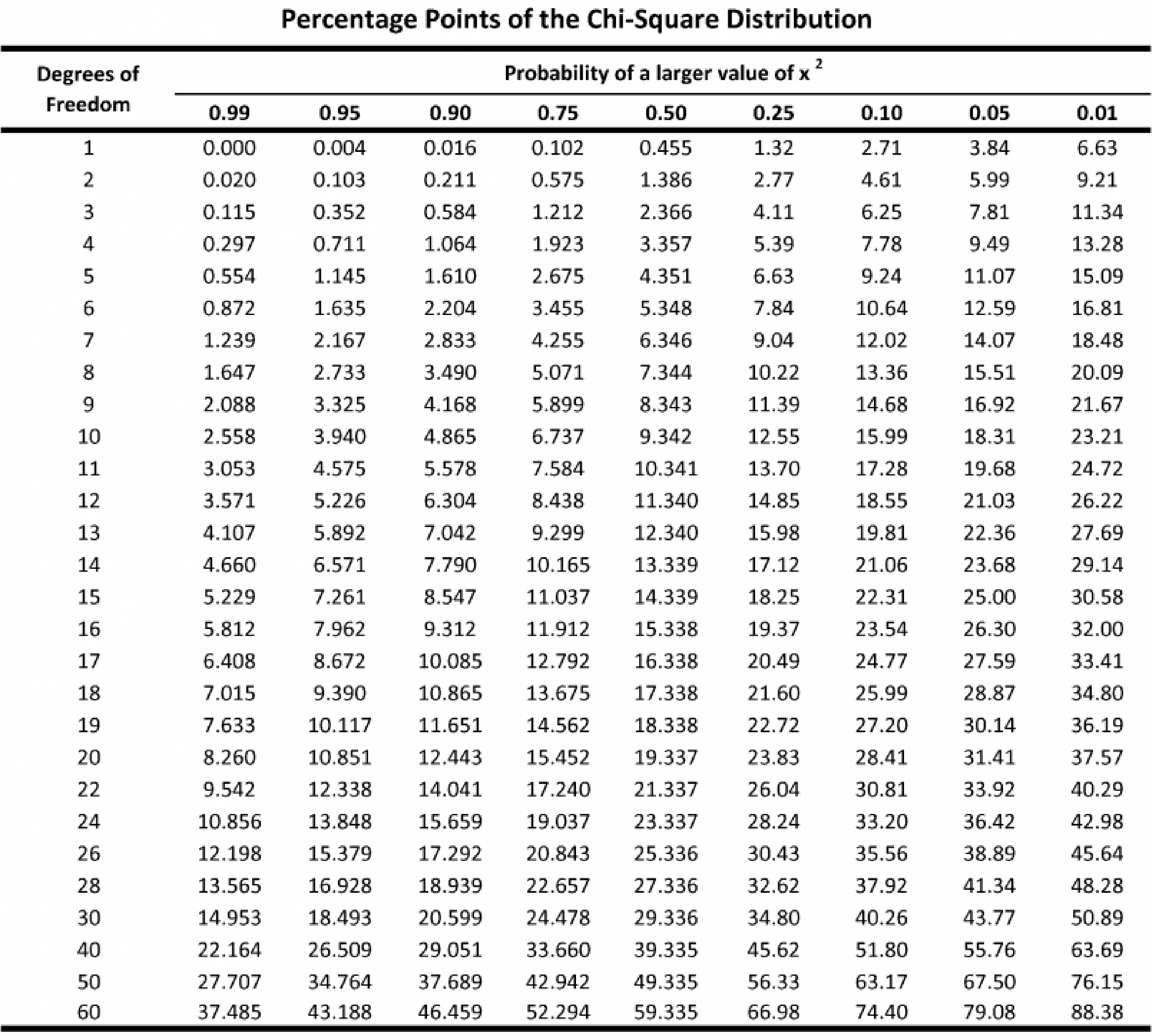}
\caption{The $\chi^2$ distribution. Finding the probability value for a chi-square of 2.9589 with 9 degrees of freedom. First read down column 1 to find the row for 9 degrees of freedom row and then go to the right close to the value 2.9589. This corresponds to a probability of less than 0.99 but greater than 0.95. \label{chi_sqaure_distribution}}
\end{center}
\end{figure*}


\item \$ grppha

		\begin{enumerate}
		\item [] source\_pc.pi
		\item [] source\_pc\_cash.pi
		
-----------------------------------------------------------------------------------
  
		\item GRPPHA[bad 0-29] 
		\item GRPPHA[exit] group min 1
		\item GRPPHA[group min 15] chkey backfile back\_pc.pi
		\item GRPPHA[chkey backfile back\_pc.pi] chkey ancrfile source\_pc\_exp.arf
		\item GRPPHA[chkey ancrfile source\_pc\_exp.arf] chkey respfile swxpc0to12s0\_20070101v012.rmf
		\item GRPPHA[chkey respfile swxpc0to12s0\_20070101v012.rmf] exit
		\end{enumerate}

 ** grppha 3.0.1 completed successfully
 
\item \$ xspec

\item XSPEC12 $>$ data source\_pc\_cash.pi

\textbf{Note}: The processes are the same as above. The only difference is the goal of the cash statistic.

\subsubsection{Cash statistic}
\item XSPEC12$>$stati cstat

Default fit statistic is set to: C-Statistic

\item  XSPEC12$>$fit 1000 0.001

===================================

 Model Model Component  Parameter  Unit     Value

 	\begin{itemize}
  	\item  1    1   wabs       nH         10$^{22}$    0.640000     frozen
   	\item  2    2   zwabs      nH         10$^{22}$    1.71028E-07  +/-  6.11886E-02      
   	\item  3    2   zwabs      Redshift            0.606000     frozen
   	\item  4    3   powerlaw   PhoIndex            2.61753      +/-  0.184373   
   	\item  5    3   powerlaw   norm                1.96417E-04  +/-  3.12063E-05
  	\end{itemize} 
	
-----------------------------------------------------------------------------------

 C-Statistic =        154.967 using 157 PHA bins and 154 degrees of freedom.

\textbf{Note}: Since I cannot constrain $N_{H,int}$ I fixed its value to ``0.0'' then I fitted the spectrum again to find the good spectral index. This $N_{H,int}$ value is also compared with the value from the \textit{Swift} web-page \footnote{http://www.swift.ac.uk/xrt\_spectra/00148522/} :
	 \begin {itemize}
	\item PC mode. Mean photon arrival: T0+15995 s
	\item NH (intrinsic): 0 (+4 $\times$ 10$^{20}$, -0) cm-2
	 \end{itemize}
	 

\item XSPEC12$>$mo wabs*zwabs*pow

Input parameter value, delta, min, bot, top, and max values for ...

              	1      0.001(      0.01)          0          0     100000      1e+06
	\begin{itemize}
	\item 1:wabs:nH$>$0.64,-1
	
              	1      0.001(      0.01)          0          0     100000      1e+06
	
	\item 2:zwabs:nH$>$0
	
              	0      -0.01(      0.01)     -0.999     -0.999         10         10
	
	\item 3:zwabs:Redshift$>$0.606
	
             	 1       0.01(      0.01)         -3         -2          9         10
	 
	\item 4:powerlaw:PhoIndex$>$
	
              	1       0.01(      0.01)          0          0      1e+24      1e+24
	
	\item 5:powerlaw:norm$>$
	\end{itemize}
	
\subsubsection{Fit results}
\item XSPEC12$>$fit 1000 0.001

===================================

 Model Model Component  Parameter  Unit     Value

 	\begin{itemize}
  	\item  1    1   wabs       nH         10$^{22}$    0.640000     frozen
   	\item  2    2   zwabs      nH         10$^{22}$    0.0          +/-  0.836612      
   	\item  3    2   zwabs      Redshift            0.606000     frozen
   	\item  4    3   powerlaw   PhoIndex            2.84846      +/-  0.337235    
   	\item  5    3   powerlaw   norm                2.31398E-04  +/-  7.81863E-05 
  	\end{itemize} 
	
-----------------------------------------------------------------------------------

 C-Statistic =        154.411 using 157 PHA bins and 154 degrees of freedom.

\textbf{Note}: C-Statistic/d.o.f = 154.967/154=0.8600462962963 = 1

\subsubsection{Error calculation}
\item XSPEC12$>$error 4

 Parameter   Confidence Range (2.706)
 
     4      2.31647      3.41198    (-0.531971,0.563545)

\subsubsection{Flux calculation}
\item XSPEC12$>$flux 0.3 10

 Model Flux 8.5685e-05 photons (2.8001e-13 ergs/cm$^2$/s) range (0.30000 - 10.000 keV)
 
\item XSPEC12$>$flux 2.0 10.0

 Model Flux 2.9385e-05 photons (1.6579e-13 ergs/cm$^2$/s) range (2.0000 - 10.000 keV)
 
\textbf{Note}: I calculated the energy correction factor from the hardness ratio since I use a flux light-curve.

\textbf{Note}: If I had used the count light-curve, I would have had to use "Net count rate " to find the ECF.

\item XSPEC12$>$show all

Net count rate (cts/s) for Spectrum:1  7.685e-03 +/- 7.266e-04 (77.4 
 
Spectral data counts: 197
 
 Model predicted rate: 6.41970E-03

 \end{itemize}
